\documentclass{icrc2009}
\pdfoutput=1
\usepackage{graphicx}   
\usepackage[font=footnotesize]{subfig}
\usepackage{fixltx2e}
\usepackage{url}
\usepackage{icrc2011_CML}
\usepackage[scaled=.90]{helvet}
\usepackage{amssymb}
\usepackage{amsmath}
\usepackage{float}
\usepackage{upgreek}
\usepackage{colortbl}

\newcommand{\etal}{\MakeLowercase{\textit{et al. }}} 
\newcommand {\fig}[1]{Fig.~\ref{#1}}

\newcommand {\sect}[1]{Sect.~\ref{#1}}

\newcommand {\ns}    {\mbox{${\rm ns}$}}
\newcommand {\tev}   {\mbox{${\rm TeV}$}}

\newcommand {\mev}   {\mbox{${\rm MeV}$}}

\newcommand {\m}     {\mbox{${\rm m}$}}

\newcommand {\dg}    {\mbox{$^\circ$}}
\newcommand {\hz}    {\mbox{${\rm Hz}$}}
\newcommand {\khz}   {\mbox{${\rm kHz}$}}
\def\cheren       {\v{C}erenkov}
\newcommand{\comment}[1]{}
\newcommand{\ave}[1]{\langle#1\rangle}

\begin{document}
\shorttitle{The ANTARES collaboration}
\begin{onecolumn}
\begin{center}
\begin{figure}

\includegraphics[width=0.20\textwidth]{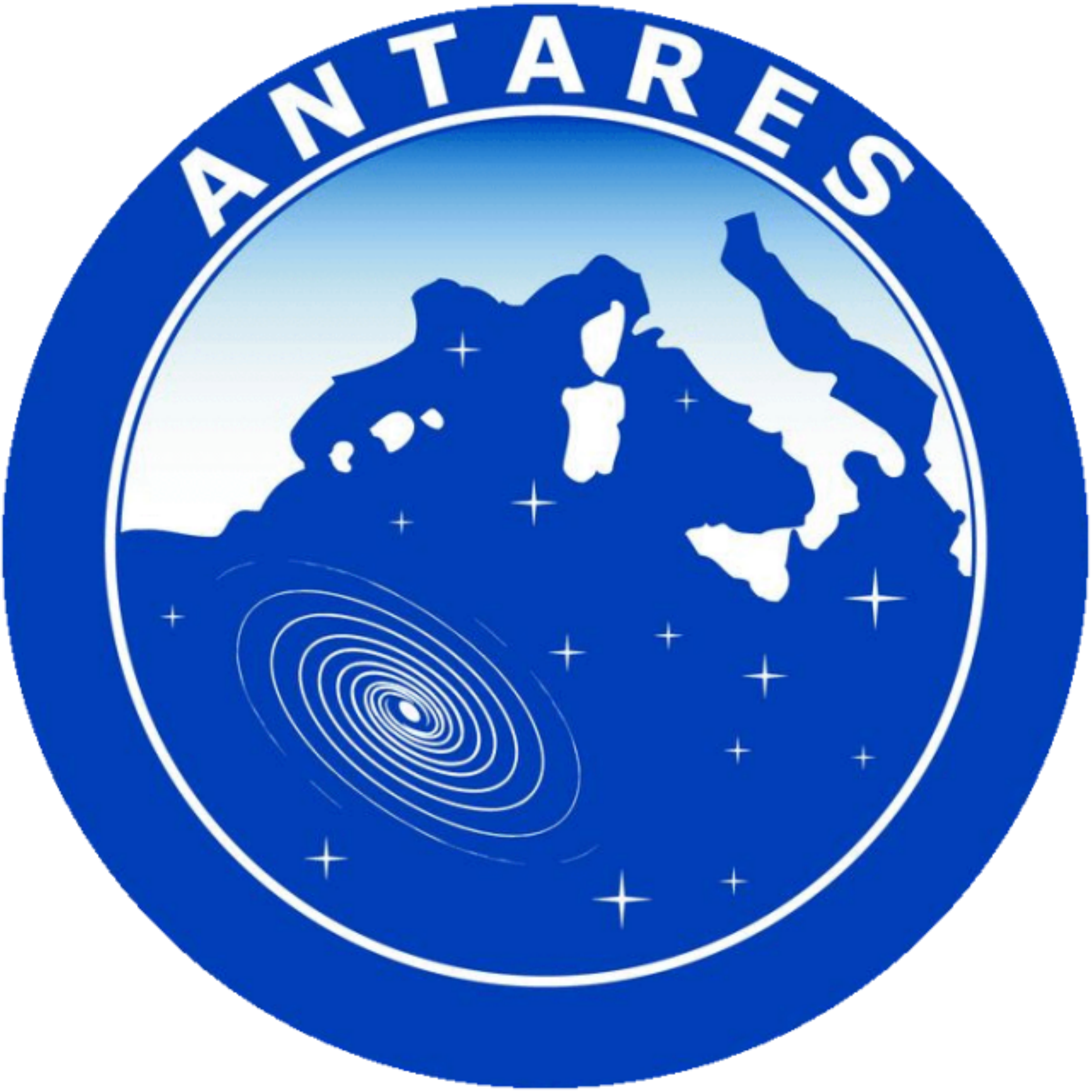} \hfill  \includegraphics[width=0.25\textwidth]{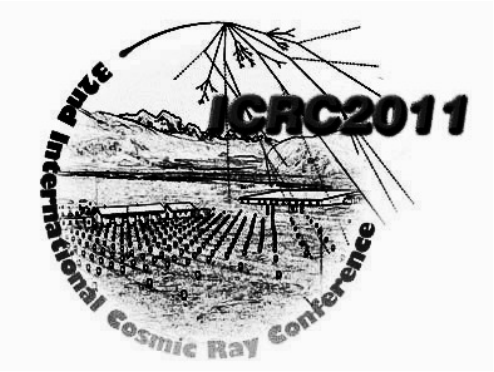}
\end{figure}

{\Huge \bf The ANTARES Collaboration:  \\
\vskip 0.5cm
 \Large contributions to the\\
 32$^{st}$  International Cosmic Ray Conference (ICRC 2011), \\
 Beijing, China,\\
 August 2011}
 \end {center}
 \vspace{2cm}
 \begin{center}
 {\large \bf Abstract}
\\
The ANTARES detector, completed in 2008, is the largest neutrino telescope in the Northern hemisphere. It is located at a depth of 2.5 km in the Mediterranean Sea, 40 km off the Toulon shore. The scientific scope of the experiment is very broad, being the search for astrophysical neutrinos the main goal. In this paper we collect the 22 contributions of the ANTARES collaboration to the 32nd International Cosmic Ray Conference (ICRC 2011). At this stage of the experiment the scientific output is very rich and the contributions included in these proceedings cover the main physics results (steady point sources, correlations with GRBs, diffuse fluxes, target of opportunity programs, dark matter, exotic physics, oscillatinos etc.)  and some relevant detector studies (water optical properties, energy reconstruction, moon shadow, accoustic detection, etc.)

 \end{center}
\newpage

\begin{center}
{\large \bf ANTARES Collaboration}
\end{center}
{\large

S.~Adri\'an-Mart\'inez$^1$,
J.A. Aguilar$^2$,
I. Al Samarai$^3$,
A. Albert$^4$,
M.~Andr\'e$^5$,
M. Anghinolfi$^6$,
G. Anton$^7$,
S. Anvar$^8$,
M. Ardid$^1$,
A.C. Assis Jesus$^9$
% ...
}

\scriptsize {

$^1$Institut d'Investigaci\'o per a la Gesti\'o Integrada de les Zones Costaneres (IGIC) - Universitat Polit\`ecnica de Val\`encia. C/  Paranimf 1 , 46730 Gandia, Spain\\ 
$^2$IFIC - Instituto de F\'isica Corpuscular, Edificios Investigaci\'on de Paterna, CSIC - Universitat de Val\`encia, Apdo. de Correos 22085, 46071 Valencia, Spain\\ 
$^3$CPPM, Aix-Marseille Universit\'e, CNRS/IN2P3, Marseille, France\\ 
$^4$GRPHE - Institut universitaire de technologie de Colmar, 34 rue du Grillenbreit BP 50568 - 68008 Colmar, France\\ 
$^5$Technical University of Catalonia, Laboratory of Applied Bioacoustics, Rambla Exposici\'o,08800 Vilanova i la Geltr\'u,Barcelona, Spain\\ 
$^6$INFN - Sezione di Genova, Via Dodecaneso 33, 16146 Genova, Italy\\ 
$^7$Friedrich-Alexander-Universit\"at Erlangen-N\"urnberg, Erlangen Centre for Astroparticle Physics, Erwin-Rommel-Str. 1, 91058 Erlangen, Germany\\ 
$^8$Dir. des Sciences de la Mati\`ere - Inst. de Rech. sur les lois Fondamentales de l'Univers - SEDI, CEA Saclay, 91191 Gif-sur-Yvette Cedex, France\\ 
$^9$Nikhef, Science Park,  Amsterdam, The Netherlands\\ 
% ...

}
\newpage

\vspace*{0.4cm}
\begin{center}
{\LARGE \bf Table of Contents}
\end{center}
{\normalsize
\begin{enumerate}
\vspace{1cm}
\item Aart Heijboer, {\it "Recent results from the ANTARES deep sea neutrino telescope"} \\
\item Claudio Bogazzi, {\it "Searching for point sources of high energy cosmic neutrinos with the ANTARES telescope"} \\
\item Juan Pablo G\'omez-Gonz\'alez, {\it "Search for point sources with the ANTARES neutrino telescope using the EM algorithm"} \\
\item Fabian Sch\"ussler, {\it "Autocorrelation analysis of ANTARES data"} \\
\item Fabian Sch\"ussler, {\it "Search for a diffuse flux of high-energy muon neutrinos with the ANTARES neutrino telescope"} \\
\item Dimitirs Palioselitis, {\it "Muon energy reconstruction and atmospheric neutrino spectum unfolding with the ANTARES detector"} \\
\item Corey Reed, Mieke Bouwhuis, Eleonora Presani, {\it "Searches for neutrinos from GRBs using the ANTARES telescope"} \\
\item Damien Dornic, {\it "Search for neutrino emission of gamma-ray flaring blazars with the ANTARES telescope"} \\
\item Damien Dornic {\it et al.}, {\it "Search for neutrinos from transient sources with the ANTARES telescope and optical follow-up observations"} \\
\item Vladimir Kulikovskiy, {\it "SN neutrino detection in the ANTARES neutrino telescope"} \\
\item Jelena Petrovic, {\it "Study on possible correlations between events observed by the ANTARES neutrino telescope and the Pierre Auger cosmic ray observatory"} \\
\item V\'eronique Van Elewyck, {\it "Searches for high-energy neutrinos in coincidence with gravitational waves with the ANTARES and VIRGO/LIGO detectors"} \\
\item Guillaume Lambard, {\it "Indirect dark matter search in the Sun direction using the ANTARES dta 2007-2008 for the two common theoretical frameworks (CMSSM, mUED)"} \\
\item Goulven Guillard, J\"urgen Brunner, {\it "On neutrino oscillations searches with ANTARES"} \\
\item Nicolas Picot-Clemente, {\it "Search for magnetic monopoles with the ANTARES underwater neutrino telescope"} \\
\item Vlad Popa, {\it "Nuclearite search with the ANTARES neutrino telescope"} \\
\item Ching-Cheng Hsu, {\it "Studying Cosmic Ray Composition around the knee region with the ANTARES Telescope"} \\
\item Goulven Guillard, {\it "ANTARES sensitivity to steady cosmic ray sources"} \\
\item Salvatore Mangano, {\it "Muon induced electromagnetic shower reconstruction in the ANTARES neutrino telescope"} \\
\item Salvatore Mangano, {\it "Optical properties in deep sea water at the site of the ANTARES detector"} \\
\item Colas Rivi\`ere, Carla Distefano, {\it "Moon shadow observation with the ANTARES neutrino telescope"} \\
\item Robert Lahmann, {\it "Status and recent results of the acoustic neutrino detection test system AMADEUS"} 
\end{enumerate}
}

\end{onecolumn}
\newpage
\mbox{}

\begin{twocolumn}

{\normalsize



\title{Recent Results from the ANTARES Deep-sea Neutrino Telescope}

\shorttitle{ANTARES collaboration results}

\authors{Aart Heijboer$^{1}$, ANTARES collaboration }
\afiliations{$^1$Nikhef, Amsterdam}
\email{aart.heijboer@nikhef.nl}

\maketitle
\begin{abstract}

  The ANTARES observatory is currently the largest neutrino 
  telescope in the Northern Hemisphere. Located at a depth 
  of 2.5 km in the Mediterranean Sea, it aims to detect high 
  energy neutrinos that are expected from cosmic ray 
  acceleration sites. The status of the experiment will be 
  discussed, including a broad target-of-opportunity program. 
  The latest results will be presented, including searches
  for a diffuse high-energy cosmic neutrino flux, neutrinos 
  from Gamma Ray Bursts, and for (galactic) point-like sources.
\end{abstract}


%
%
%

\section{The ANTARES Neutrino Detector}

 Cosmic Rays are thought to originate in Galactic and extra-
 Galactic sources that accelerate protons and other nuclei up 
 to high energies. Identification of the responsible objects
 could be achieved by detecting the distinct signatures of these
 cosmic accelerators, which are high energy neutrinos and gamma 
 rays produced through hadronic interactions with ambient gas 
 or photoproduction on
 intense photon fields near the source. While gamma rays can be produced
 also by directly accelerated electrons, the detection of high-energy 
 neutrinos from these objects would provide
 unambiguous and unique information on the sites of the cosmic
 accelerators and hadronic nature of the accelerated particles.

 The ANTARES Collaboration has constructed a neutrino telescope 
 \cite{detpaper} at a depth 
 of about 2475 meters, offshore Toulon, France. Neutrinos 
 are detected by photomultiplier tubes (PMTs), housed
 in pressure resistant glass spheres, which are regularly 
 arranged on 12 detection lines. Each line accommodates up 
 to 25 triplets  of PMTs, located between 100 and 450 m
 above the sea bed. The lines are connected to the shore via
 a junction box and a single, 40 km electro-optical cable, 
 which provides both power and an optical data link.
 On shore, a computer farm runs a set of trigger algorithms to 
 identify events containing Cherenkov light from high energy 
 muons within the data stream, which otherwise consists mostly 
 of signals from radioactive decay and bioluminescence. The selected 
 events are stored for offline reconstruction. In 2007, the first 5 
 detector lines became operational, followed, in May 2008, by the
 completion of the full 12-line detector.

 The reconstruction of muon tracks is based on the arrival time
 of the Cherenkov photons on the PMTs. For high energy neutrinos, the
 angular resolution is determined by the timing accuracy, which is 
 limited by the transit time spread of the PMTs (1.3ns). Time calibration
 is performed by a number of independent systems, including LED and
 laser beacons \cite{timing} located throughout the detector. 
 The relative inter-line timing has been calibrated using 
 the time residuals measured in a large number of down-going 
 reconstructed muon events, in addition to the optical beacon
 systems.
 The positions of the PMTs vary with time because of the 
 sea currents. Using an acoustic positioning system, combined
 with information from internal compasses and tiltmeters, the positions
 of the PMTs are determined every 2 minutes with an accuracy of 
 $\sim10$ cm. 
 
 Most of the analyses described here use a muon track 
 reconstruction algorithm (based on \cite{thesis}) that 
 consists of multiple fitting steps. The final step is based on a 
 full likelihood description of the arrival times of the detected
 Cherenkov photons, which also accounts for background light. 
 The achieved angular resolution is, by necessity, 
 determined from simulations. However, several aspects of the simulations
 were confronted with data in order to constrain the possible systematic
 effects in the timing resolution that would result in a deteriorated 
 angular resolution. The angular resolution (i.e. the median angle between
 the neutrino and the reconstructed muon) was found to be $0.4 \pm 0.1 \rm$ (sys)
 degrees for the detector with all 12 lines operational. Studies of
 the detector and the optical water properties \cite{salvi} are ongoing
 and may help to further improve and constrain the angular resolution
 in the near future. Moreover, a study to observe the shadow of the moon
 using down-going muons might in the future provide additional information
 on the (absolute) pointing accuracy \cite{riviere}.

 In the following, a selection of results recently obtained by
 the ANTARES experiment will be summarized; many of them are
 discussed in more detail in dedicated contributions to this 
 conference.

\section{Searches for high energy cosmic sources }

 Searches for cosmic neutrinos and their sources comprise
 a main goal of the ANTARES experiment. Various searches 
 for high energy cosmic neutrinos have been performed using
 the first years of data.

\subsection{Search for a diffuse neutrino flux}

 A search for a diffuse cosmic neutrino flux has been  
 conducted using 334 live-days of data collected in 
 2008 and 2009 \cite{diffuse}. Such a flux would result in an
 excess of high energy events over the irreducible background
 of atmospheric neutrinos. A measure of the energy is provided
 by an observable $R$, which measures the number of PMTs that
 detect multiple photons separated in time. The distribution
 of the $R$ variable agrees well with the background-only 
 simulations and shows no evidence for a contribution from a 
 cosmic diffuse $E^{-2}$ flux, which would result in an excess
 of high-$R$ events. Consequently, a 90\% C.L. limit on such a flux is
 obtained  in the energy range  20 TeV - 2.5 PeV. The limit is 
 shown in figure \ref{fig_difflims} together with previously 
 published limits from other experiments.

 \begin{figure}[!t]
  \vspace{5mm}
  \centering
  \includegraphics[width=3.3in]{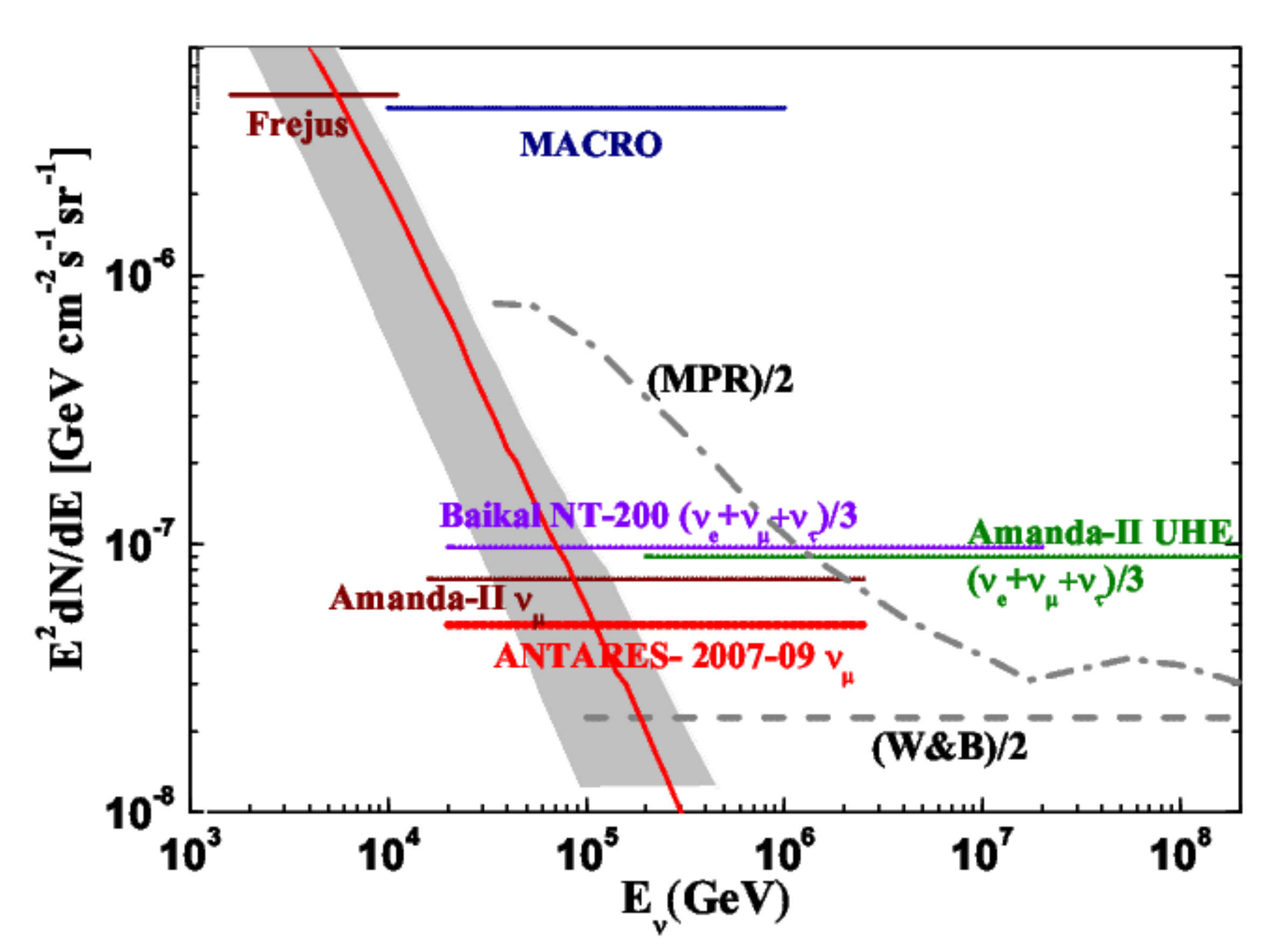}
  \caption{Upper limit on the diffuse neutrino flux of HE neutrinos 
           obtained from the 2007-2009 ANTARES data, compared to 
           theoretical predictions 6 and to limits set by other neutrino
           telescopes. (see \cite{diffuse} for references). }
  \label{fig_difflims}
 \end{figure}

 \subsection{Point source search}
 
 Cosmic point-like source of neutrinos have been searched for
 using 813 live-days of data from 2007 up to and including 2010 
 \cite{claudio}. An earlier version of the analysis is described
 in \cite{pntsrcpaper}.
 Event selection criteria have been applied which optimize both
 the sensitivity and the discovery potential. Events are required 
 to be reconstructed as upward-going and to have a good 
 reconstruction quality, quantified by a 
 variable based on the reduced log-likelihood of the track fit, 
 and an angular error estimate better than $1^{\circ}$.
 The resulting event sample consists of 3058 neutrino candidates, 
 of which $\sim 84(16)\%$ is expected to be atmospheric neutrinos
 (muons misreconstructed as upward-going).
 To search for point sources, the analysis uses an unbinned 
 maximum likelihood method, which exploits the knowledge on the
 angular resolution\footnote{Since part of the data in this analysis 
 was taken by a 5-line detector, the resolution is slightly worse than
 the $0.4^{\circ}$ mentioned earlier for the full detector.} 
 of $0.5^{\circ}$ and the rate of background events as a 
 function of the declination. 

 Two different versions of the search were conducted: in the 'full-sky'
 search, the full visible sky is searched for point sources. In the
 'candidate search', neutrinos are searched for only in the 
 direction of 24 a-priori selected candidate source-locations, corresponding
 to known gamma ray objects of interest.
 Neither search yields a significant excess of events over the background:
 the post-trial $p$-values are $2.5\%$ (for a cluster of events 
 at $\alpha,\delta = (-46.5^{\circ}, -65.0^{\circ})$
 for the full sky search and $41\%$ for 
 the most signal-like source in the candidate source list (HESS J1023-575).
 Limits have been extracted on the intensity\footnote{The limits are on $\phi$, which is defined by the
 the following expression for the neutrino flux:
 $dN/dE = \phi \times (E / \rm {GeV} )^{-2} ~ GeV^{-1} cm^{-2} s^{-1}$.} of 
 an assumed $E^{-2}$ neutrino flux from the candidate sources. 
 They are shown in Figure \ref{fig_pntlims}.
 The limit computation is based a large number of generated 
 pseudo experiments in which systematic uncertainties on the 
 angular resolution and acceptance are taken into account.

 These limits are more stringent than those from previous experiments in
 the Northern hemisphere (also indicated in the figure) and
 competitive with those set by the IceCube 
 observatory \cite{ic40:2010rd} for declinations $ <-30^{\circ}$.
 The various experiments are sensitive in 
 different energy ranges, even though they all set limits 
 on $E_{\nu}^{-2}$ spectra.
 For this spectrum, ANTARES detects most events at
 energies in a broad range around 10 TeV, which is
 a relevant energy range for several galactic source 
 candidates. 

 An independent point source analysis was performed using 
 a different search method based on the 'EM-algorithm' \cite{jp}.
 This cross-check yielded similar results as the 
 likelihood based analysis described above.

 The sample of neutrino candidates from the previous
 search \cite{pntsrcpaper} has been used for additional 
 studies, which are also reported on at this conference:
\begin{itemize}
\item An auto-correlation analysis \cite{fabian} was performed  in
       order to test for unexpected (larger scale) structures in the
       neutrino candidate sample. No such structures were found.
\item An analysis has been performed to search for directional 
       correlations between the neutrino candidate events and 69 published 
       ultra-high energy cosmic rays events detector by the Pierre 
       Auger Observatory \cite{petrovic}. The data were found to contain
       no such correlations.
\end{itemize}

 \begin{figure}[!t]
  \vspace{5mm}
  \centering
  \includegraphics[width=3.3in]{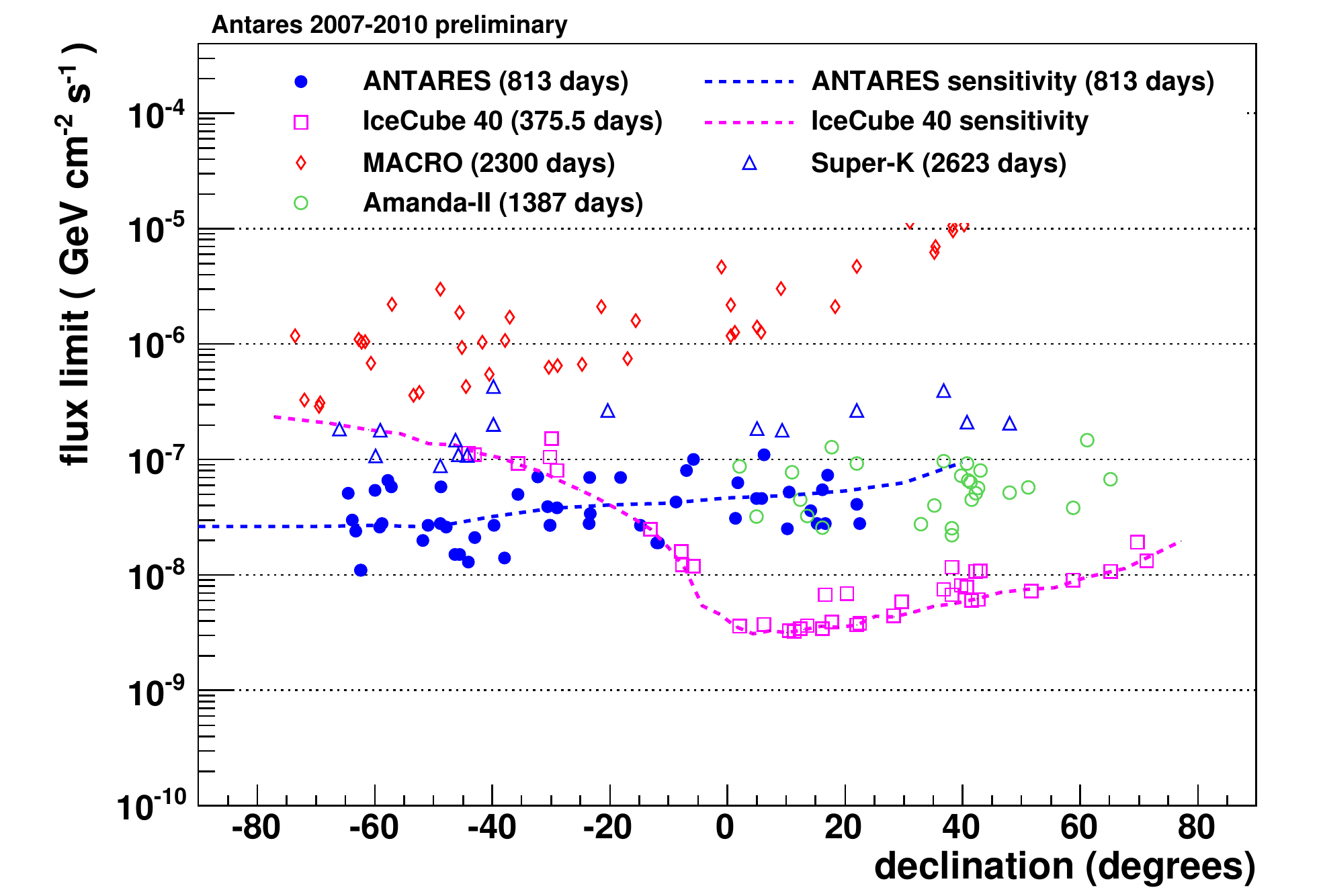}
  \caption{Limits set on the normalisation of an $E^{-2}_{\nu}$ 
           spectrum of high energy neutrinos from selected candidates. 
           Also shown is the 
           sensitivity, which is defined as the median expected limit.
           In addition to the present result, several previously published 
           limits on sources in both the Southern and Northern sky are 
           also shown (see \cite{claudio} for references). }
  \label{fig_pntlims}
 \end{figure}

\section{Multi-messenger Astronomy}

 Several analyses are performed in ANTARES, which focus on coincident
 measurement of neutrinos with a variety of external measurements. A
 selection is described below.

\subsection{Neutrinos from flaring blazars}

 In addition to the time-integrated searches described above, 
 a time-dependent point source search 
 has been conducted to look for neutrinos in 
 correlation to the variable gamma-ray emission 
 from blazars measured by the LAT instrument on-board the Fermi 
 satellite. By restricting the search to the 'high state' (typically
 1-20 days) of the
 gamma emission, the background is reduced compared to the time-integrated
 point source search. An analysis using 60 days of live time 
 collected during 2008 is presented in \cite{dornic}; no significant
 excess above the expected background was observed.

\subsection{Neutrinos from GRBs}

 Various models predict high energy neutrinos to be emitted
 by Gamma Ray Burst events.  Restricting the neutrino
 search to the duration (i.e. $T_{90}$) of the GRB virtually 
 eliminates all background events. Hence, the detection
 of only a few events could already constitute a discovery. 
 Two of such searches \cite{corey} have been performed. The first one
 uses the muon-neutrino channel, exploiting the good
 angular resolution of the detector to demand directional correlation
 in addition to the time. This search has so far been performed
 using 37 GRBs and ANTARES data from 2007 (5 detector lines). 
 No neutrino events were found in the a-priori-defined search cone and limits
 on the neutrino flux were obtained; see figure \ref{fig_grb}.

 The second search is ongoing and searches for 'shower' events, 
 which are the result of a localized energy deposition in the 
 detector. These events can be produced by e.g. electron neutrinos 
 which produce an electromagnetic shower, or by neutral current 
 interactions of all neutrino flavours. A reconstruction algorithm
 for these events has been developed. The sensitivity of this analysis
 to GRB neutrinos of all flavours is presented in \cite{corey}.

\begin{figure}[!t]
  \vspace{5mm}
  \centering
  \includegraphics[width=3.3in]{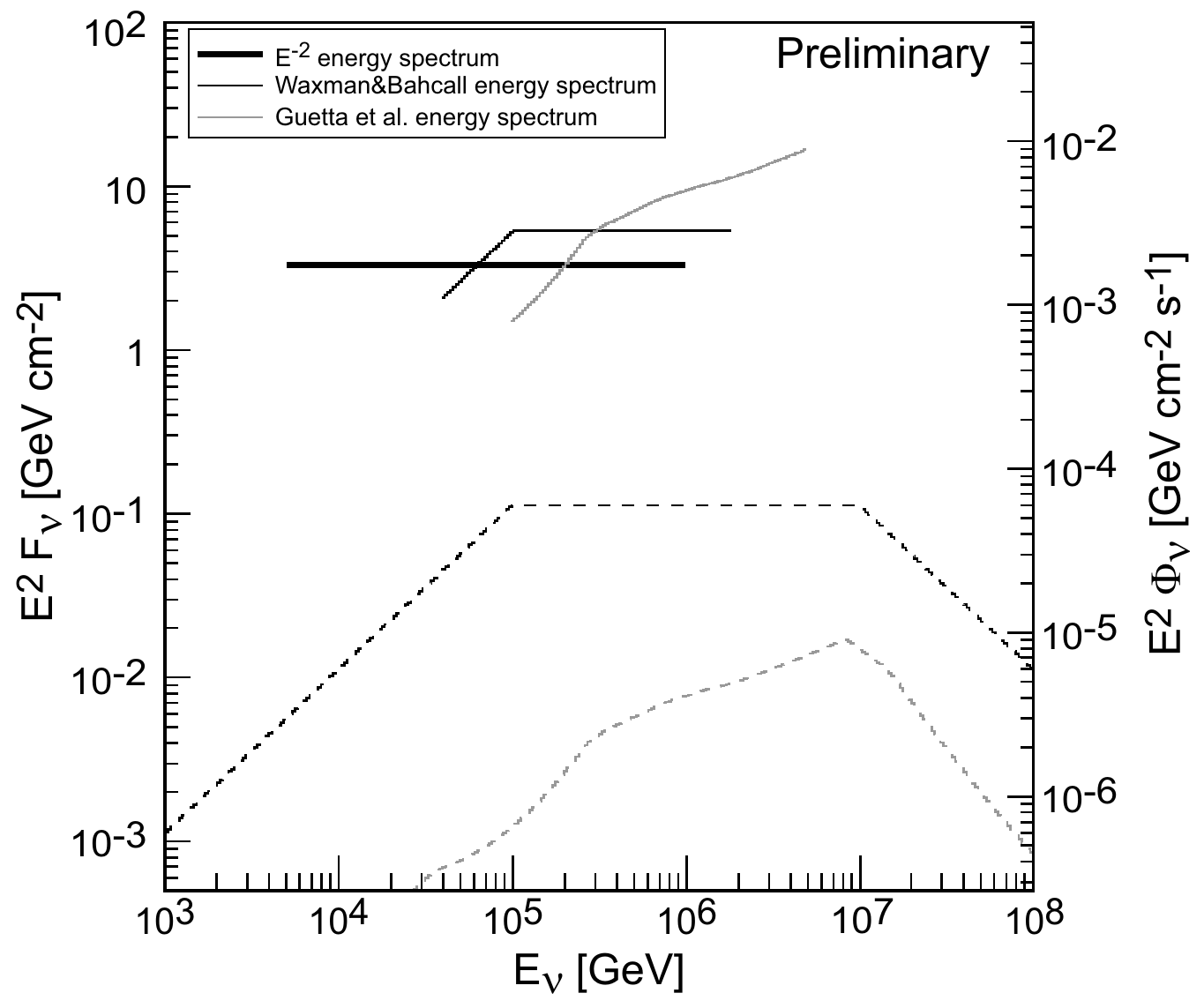}
  \caption{The upper limits (solid lines) for 37 GRBs obtained 
           by the muon track search for the specified neutrino
           flux models (dashed lines) of gamma-ray bursts. The 
           limits were placed using data taken during 2007, 
           when the telescope consisted of 5 detector lines.}
  \label{fig_grb}
\end{figure}

\subsection{Optical follow-up of ANTARES events}
 
 To search for transient sources of neutrinos with an optical counterpart,
 a system has been setup to enable fast
 optical observations in the direction of detected neutrino events.
 A reconstruction algorithm that does not require full alignment
 information \cite{bbfit} is run online and alerts are produced for
 network of small automatic optical telescopes. Such alerts are 
 produced for very high energy neutrinos or for multiple neutrinos
 that coincide in time and direction.
 Since February 2009, ANTARES has 
 sent 37 alert triggers to the TAROT and ROTSE telescope networks, 
 27 of them have been followed. First results on the analysis of the
 resulting optical images to search for GRB and core-collapse SNe will be
 shown at the conference \cite{dornic}.

 Another combination of measurements consists of correlating
 neutrino events with the signals from the gravitational wave
 detectors LIGO and VIRGO. A joint analysis is being performed
 that searches for a gravitational wave signal in coincidence with
 a sample of neutrino candidate events detected by ANTARES in 
 2007 \cite{elewyk}.

\section{Searches for Exotic physics}

 ANTARES is also searching for signatures of physics beyond the
 Standard Model. An analysis is performed that looks for neutrinos 
 produced by Dark Matter particles annihilating in the Sun and 
 the Galactic center \cite{lambard}. 

 Magnetic Monopoles with masses between $10^{10}$ and $10^{14}$
 traversing the detector volume would be detected as a very 
 bright track. A search for this signature has been conducted
 \cite{pimente} and a limit on the flux of monopoles with $\beta>0.55$
 has been obtained; see Figure \ref{fig_monopoles}. This limit is
 more stringent than those from previous experiments.

 \begin{figure}[!t]
 \vspace{-5mm}
  \centering
  \includegraphics[width=3.3in, height = 2.3in]{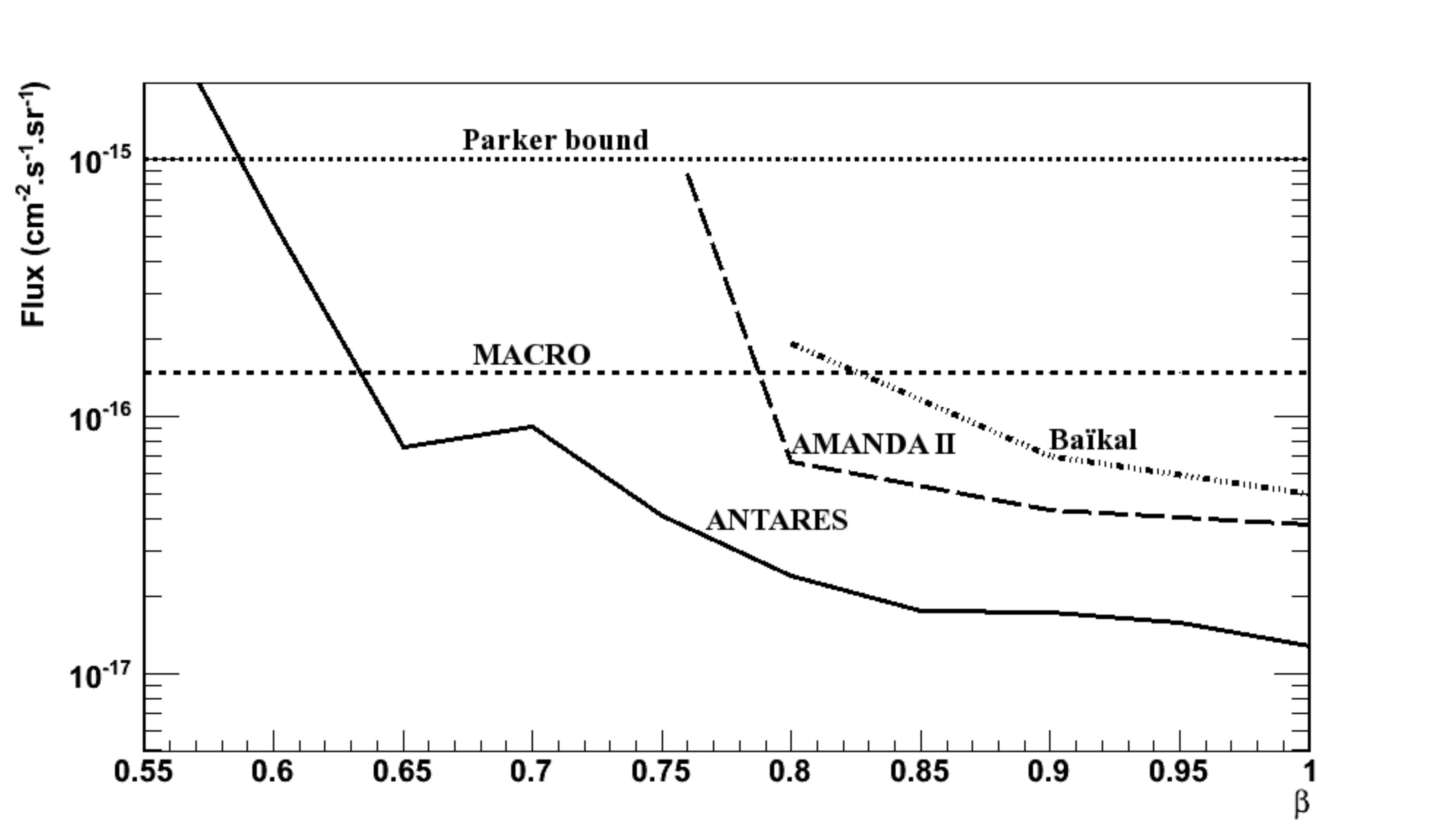}
  \caption{Flux upper limit (90 \% C.L.) on up-going magnetic monopole 
           for relativistic velocities ($0.55 \leq \beta \leq 0.995$) 
           is compared to the published upper limits
           set by other experiments are also shown (see \cite{pimente} for references)
           as well as the theoretical Parker bound.}
  \label{fig_monopoles}
 \end{figure}

 Another hypothetical form of matter is formed by Nuclearites: particles 
 composed of strange quark matter. The signature is a
 slow-moving (e.g. $10^{-3} c$) bright point traversing the detector. 
 The searches are described in \cite{poppa}.

\section{Conclusion}
 
 The first deep-sea neutrino telescope, ANTARES, has been taking data
 for four and a half years now. A large number of analysis are being
 performed, looking for astrophysical signals of neutrinos, either
 stand-alone or by looking for coincident observations with a variety
 of other experiments. The geographical position, combined with the
 good angular resolution allow ANTARES to explore, in particular, Galactic
 neutrino sources in the relevant energy range. In addition, several 
 analyses are aimed at detecting signals from non-standard model 
 particles.
 
 The successful operation of ANTARES, and analysis
 of its data, is an important step towards KM3NET \cite{km3net}
 a future km3-scale high-energy neutrino observatory and
 marine sciences infrastructure planned for construction
 in the Mediterranean Sea.

\setcounter{figure}{0}
\setcounter{table}{0}
\setcounter{footnote}{0}
\setcounter{section}{0}
\newpage



\title{Searching for Point Sources of High Energy Cosmic Neutrinos with the ANTARES telescope}

\shorttitle{Claudio Bogazzi on behalf of the ANTARES Collaboration}

\authors{Claudio Bogazzi$^{1}$ on behalf of the ANTARES Collaboration }
\afiliations{$^1$FOM Instituut voor Subatomaire Fysica Nikhef, Science Park 105, 1098 XG Amsterdam, The Netherlands}
\email{claudiob@nikhef.nl}

\maketitle
\begin{abstract}
ANTARES is currently the largest neutrino detector on the Northern Hemisphere. It consists of a tri-dimensional array of 885 photomultipliers arranged on 12 vertical lines, placed at a depth of 2475 meters in the Mediterranean Sea near Toulon, France. The telescope, completed in 2008, detects the Cherenkov radiation of muons produced by high energy neutrinos interacting in or around the detector. Muon tracks are then reconstructed using a likelihood-based algorithm. 
One of the main goals of the experiment is the search for high-energy neutrinos from astrophysical point-like sources. Due to its location, ANTARES is sensitive to up-going neutrinos from many potential galactic sources in the TeV to PeV energy regime. New results from an unbinned method as well as the sensitivity of the detector are presented.
\end{abstract}


\section{Introduction}
One of the main questions in astroparticle physics is the origin of high energy Cosmic Rays (CRs). In the last decade progress has been made related to energy spectrum and composition \cite{Waxmann}. However, the origin of CRs remains unknown. Many acceleration sites have been suggested, such as supernova remnants, microquasars and active galactic nuclei \cite{Becker}. The final signature of these cosmic accelerators are gamma rays and high energy neutrinos produced through hadronic interactions. The observation of a point-like source of neutrinos would then offer a unique occasion to study the mechanism of CRs acceleration.
\subsection{The ANTARES detector}
The ANTARES detector is located at a depth of 2475 m in the Mediterranean Sea, 42 km from Toulon in the south of France ($42^{\circ}48 N, 6^{\circ} 10 E$). It consists of a tri-dimensional array of 885 optical sensors arranged on 12 vertical lines. Each line comprises up to 25 detection storeys each equipped with 3 downward-looking 10-inch photo-multipliers (PMTs), oriented $45^{\circ}$ to the line axis. The spacing between storeys is 14.5 m while the lines are spaced by 60-70 m. A buoy at the top of the line keep them to stay vertical.

The telescope operates by detecting Cherenkov light emitted by charged particles that result from neutrino interactions in or around the detector. The arrival time and amplitude of the Cherenkov light on the PMTs are digitized into 'hits' \cite{Electronics} and transmitted to shore.

\section{Data Selection}\label{sec:selection}
The data analysed in this work were collected between January 31st 2007 and December 30th 2010. The total livetime of the analysis is 813
days of which 183 days were with 5 lines, while for the remaining 630 days the detector consisted of 9, 10 and 12  lines

The reconstruction of the muon track is achieved using the time and position information of the hits. The algorithm is based on a maximum likelihood method \cite{Aart} where a multi-stage fitting procedure is applied in order to maximise the likelihood of the observed hit times as a function of the muon direction and position. The quality of the reconstruction is defined by the variable $\Lambda$, which is based on the maximisation of the log-likelihood \cite{Aart}. Figure \ref{fig:lambda} shows the cumulative distribution of $\Lambda$ for upward-going events with the simulated contributions of atmospheric muons and neutrinos. Atmospheric muons are simulated with the MUPAGE package \cite{MUPAGE}; neutrinos are instead generated with the GENNEU \cite{genneu} package and the Bartol model \cite{Bartol}.

 \begin{figure}[!t]
  \vspace{5mm}
  \centering
  \includegraphics[width=3.in]{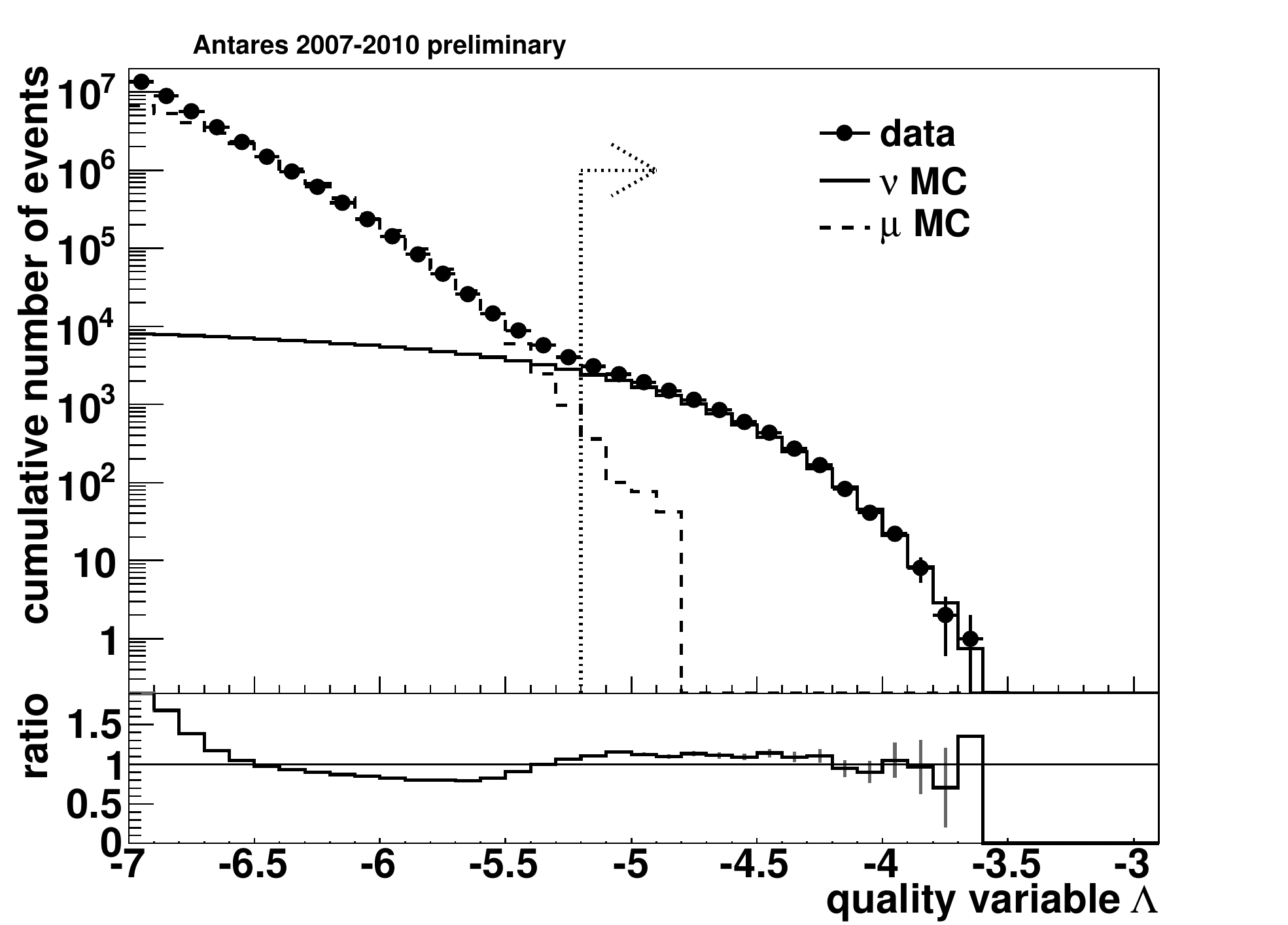}
  \caption{Cumulative distribution of the reconstructed quality variable $\Lambda$. The dashed line is for simulated atmospheric muons while the solid line corresponds to simulated atmospheric neutrinos. The bottom plot shows the ratio between data and Monte Carlo. }
  \label{fig:lambda}
 \end{figure}

Neutrino candidates events are selected requiring an upward going track, i.e. zenith angle $< 90^{\circ}$, and a value for the lambda variable $\Lambda > -5.2$. The latter is obtained by optimizing the background reduction and the signal efficiency, in terms of the discovery potential. Another cut is then applied in order to reject mis-reconstructed atmospheric downward going muons using the information of the uncertainty on the reconstructed muon track direction obtained from the fit. This value is required to be $\le 1^{\circ}$. The final sample consists of 3058 events; from the simulations 84\% of them are estimated to be neutrinos, while the rest are mis-reconstructed atmospheric muons.

\section{Detector Performance}
The angular resolution and the acceptance of the detector have been obtained from simulation. The systematic uncertainty on the angular resolution has been computed following the procedure described in \cite{aart2007} by smearing the hit times according to a Gaussian with a width of $\sigma_{t}$ in order to artificially deteriorate the simulated timing accuracy. The study leads to exclude an additional smearing of 3 ns which was found to be incompatible with data at the 2$\sigma$ level where $\sigma$ is the uncertainty on the flux model \cite{Barr}. The best agreement between data and Monte Carlo is obtained for $\sigma_{t} = 2 \pm 0.5$ ns. This value is indeed used for the simulations presented in this paper.



\subsection{Angular resolution}
Figure \ref{fig:reco_angle} shows the cumulative distribution of the angle between the reconstructed muon direction and the generated neutrino direction for neutrino events where we assumed an energy spectrum proportional to $E_{\nu}^{-2}$ with $E_{\nu}$ the neutrino energy.
 \begin{figure}[!t]
 \vspace{5mm}
 \centering
 \includegraphics[width=5.5cm]{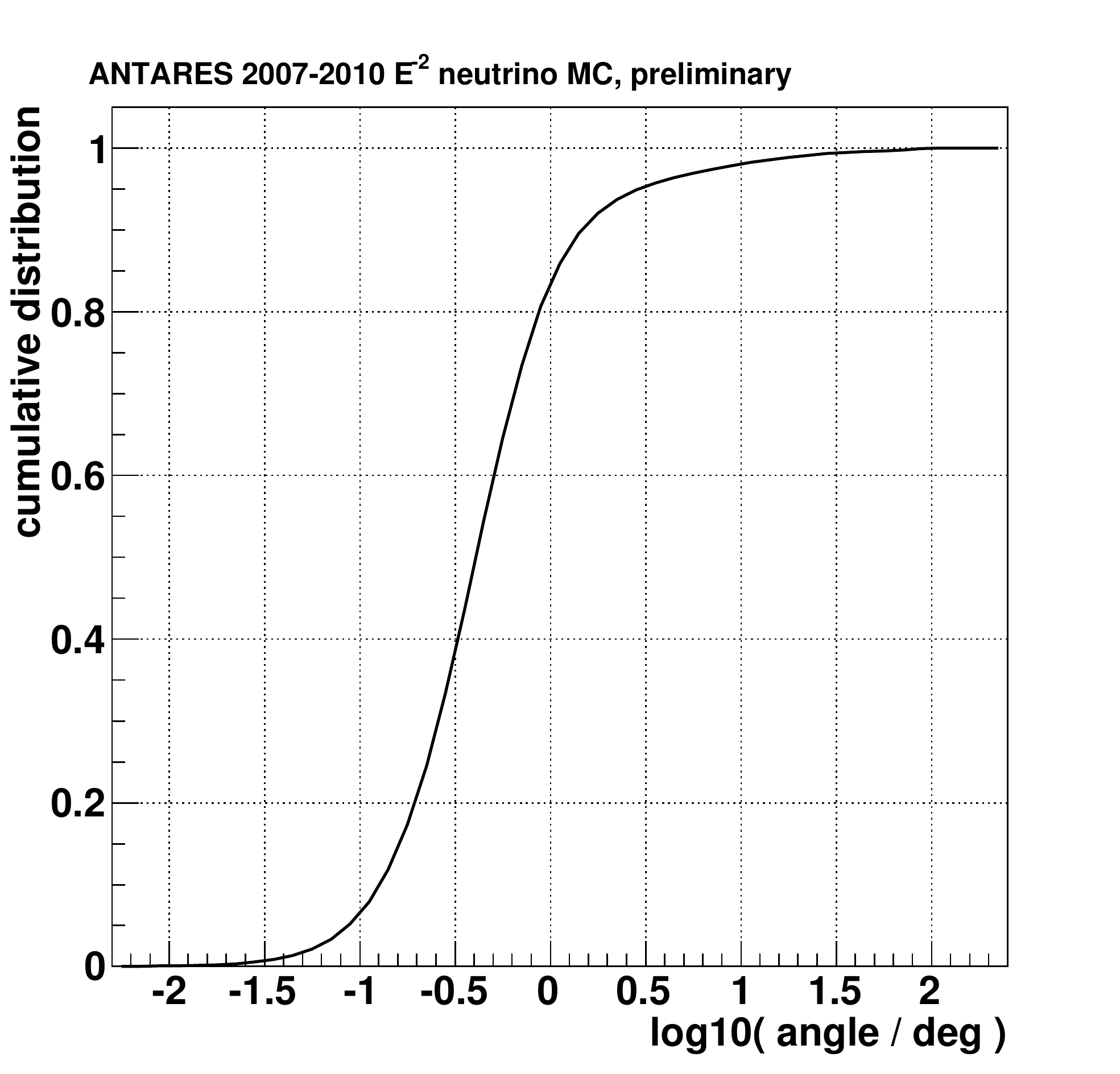}
 \caption{Cumulative distribution of the angle between the reconstructed muon direction and the true neutrino direction for simulated upward going neutrinos that pass the cuts described in Section \ref{sec:selection} assuming a $E_{\nu}^{-2}$ neutrino spectrum.}
 \label{fig:reco_angle}
 \end{figure}
 The median of this angular error is $0.46$ degrees.
  
\subsection{Acceptance}
The acceptance for signal neutrinos is also estimated using simulations. In the search, we deal with fluxes of the form of
\begin{equation}
\frac{dN}{dE} = \phi(\frac{E_{\nu}}{GeV})^{-2} GeV^{-1}s^{-1}cm^{-2},
\end{equation}
where $\phi$ is the flux normalisation. The acceptance, A, is defined as the constant of proportionality between $\phi$ and the number of selected events. Figure \ref{fig:acceptance} shows exactly this proportionality: for a source at a declination of -90 (0) degrees, $A=8.8(4.8)\times 10^{7}$ GeV cm$^{2}$ s. Systematic uncertainties on the acceptance are constrained by the agreement between the simulated atmospheric neutrino sample and data. For the computation of the flux limits an uncertainty of 15\% is assumed.
\begin{figure}[!t]
\vspace{5mm}
 \centering
 \includegraphics[width=5.5cm]{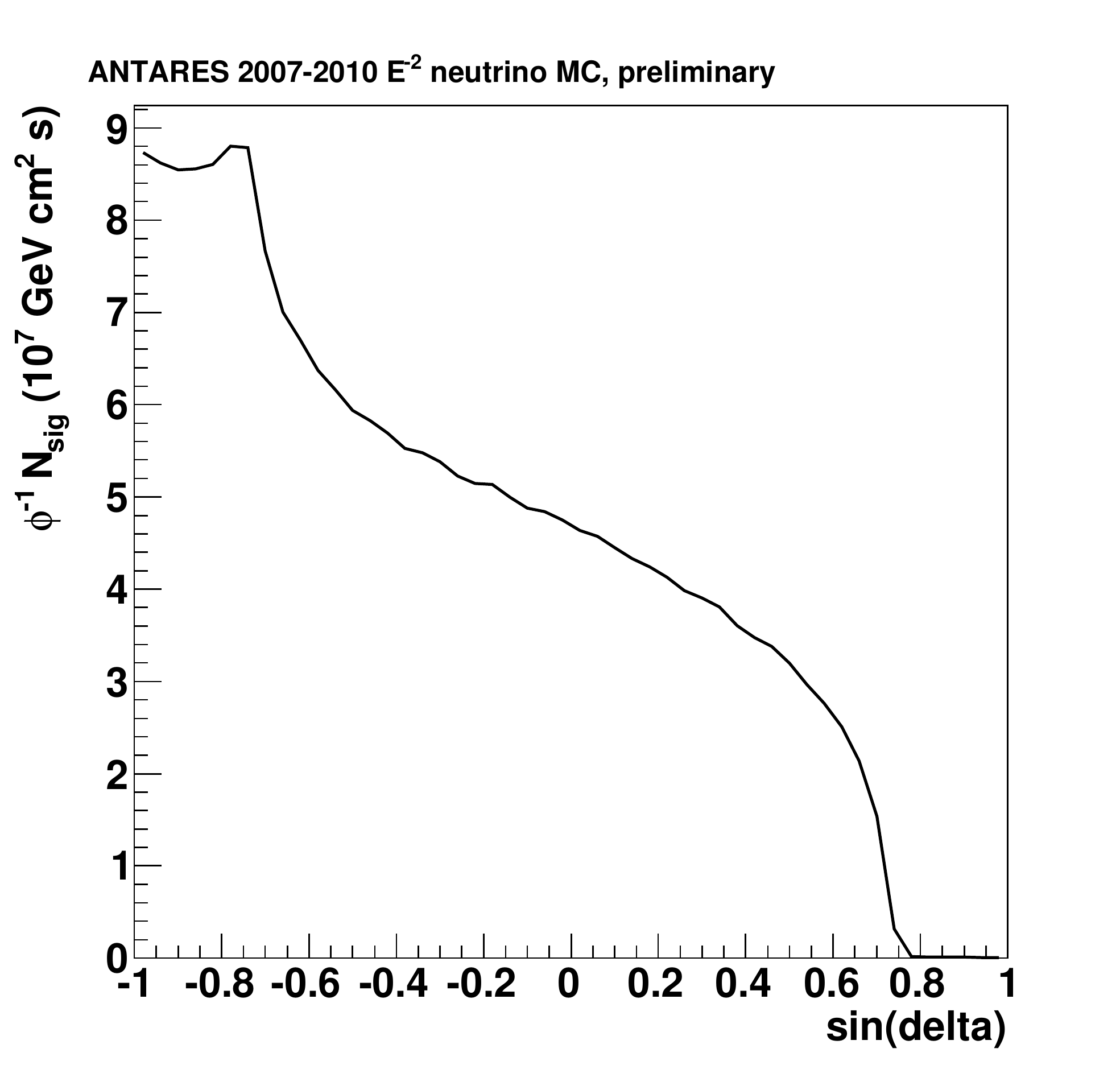}
 \caption{Acceptance, i.e. the constant of proportionality between the normalisation factor for an $E^{-2}$ flux and the selected number of events. }
 \label{fig:acceptance}
 \end{figure}
\section{Search method}
The algorithm used on the analysis is based on the likelihood of the observed events which is defined :
\begin{IEEEeqnarray*}{rCl}
\log {\cal L}_{\rm s+b} &=& \sum_i \log [ \mu_{\rm sig}\times {\cal F}(\beta_i(\delta_s,\alpha_s))\times {\cal N}(N_{hits}^{i,sig})\\
&+& {\cal B}_i \times{\cal N}(N_{hits}^{i,bkg})] + \mu_{\rm tot},
\label{eq:lik}
\end{IEEEeqnarray*}
where the sum is over the events, $\cal F$ is the point spread function, i.e. the probability density of reconstructing an event $i$ at a distance $\beta_i$ from the true source declination and right ascension $\delta_s, \alpha_s$; $\cal B$ is a parametrization of the background rate, obtained from the observed declination distribution of the events; $\mu_{sig}$ is the mean number of detected events that the source produces and $\mu_{tot}$ represents the total number of expected events and ${\cal N}(N_{hits}^{i})$ is the probability for an event i to be reconstructed with $N_{hits}$ number of hits (this probability was not included in the analysis with 2007 and 2008 data \cite{aart2007}).

In order to compute the test statistic the free parameters of the likelihood are maximized. We have now to distinguish between the two different analysis presented in this paper: in the candidate list search only the $\mu_{sig}$ are fitted while in the full sky search we have in addition the source coordinates ($\delta_s , \alpha_s$) to fit. In both cases the results of the fit are the maximum likelihood value ${\cal L}_{s+b}^{max}$ and the estimates of the free parameters. The test statistic is then defined as:

\begin{equation}
{\cal Q} = {\cal L}_{s+b}^{max} - { \cal L}_{b}
\end{equation}
where ${\cal L}_{b}$ is the likelihood computed for the background only case. The higher ${\cal Q}$ the more the data are compatible with signal.

Just using the number of hits information in the likelihood let us to gain a 25\% (22\%) factor for the 3 (5) $\sigma$ discovery probability as shown in Figure \ref{fig:disco} for the full sky search. 
 \begin{figure}[!t]
 \vspace{5mm}
 \centering
 \includegraphics[width=6cm]{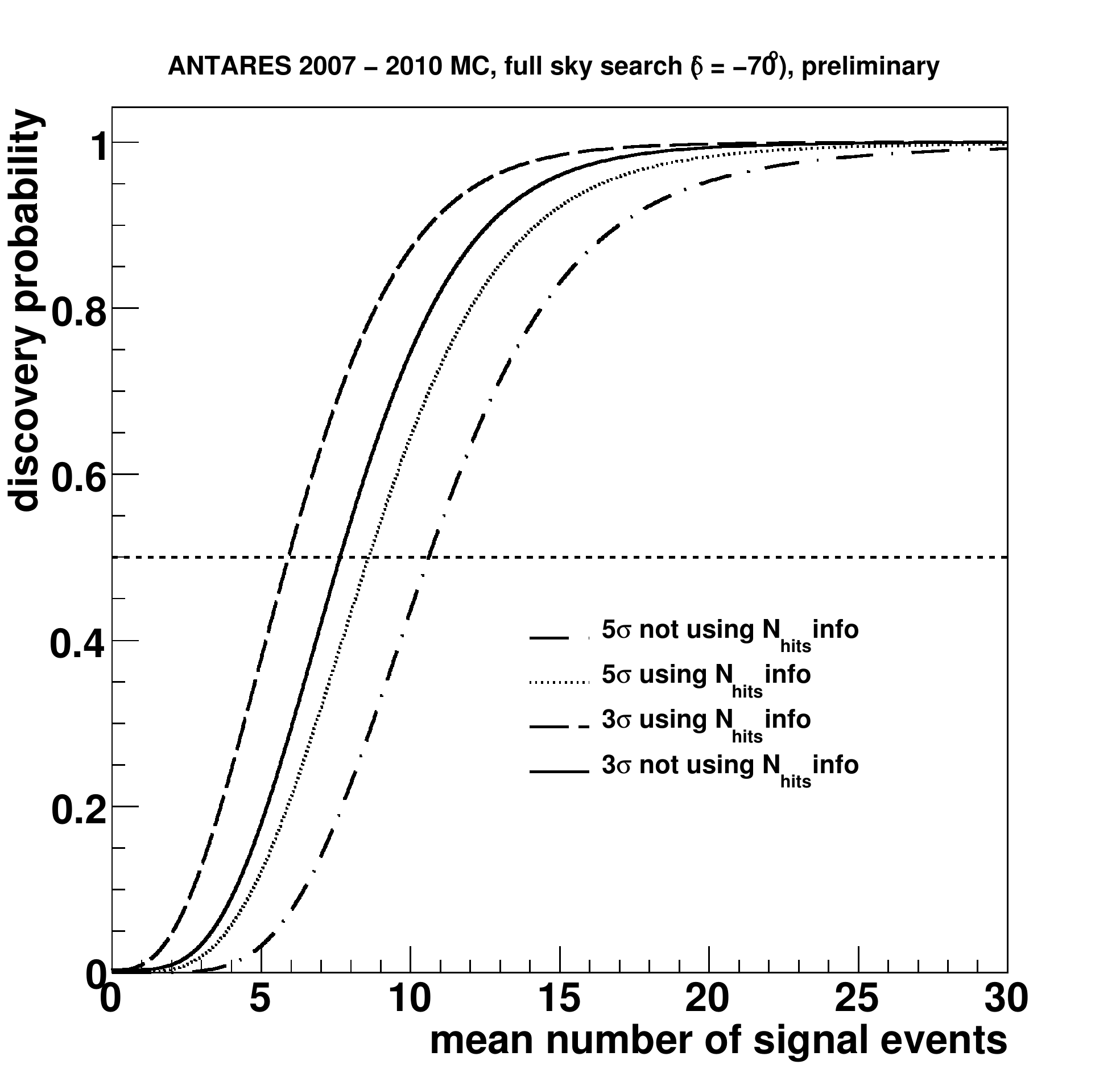}
 \caption{Probability for a 3$\sigma$ (dashed and solid lines) and 5$\sigma$ (dotted and dashed-dotted lines) discovery as a function of the mean number of signal events for the case where we use the number of hits information in the likelihood (dashed and dotted lines) and for the case where we do not use it (solid and dashed-dotted lines) for the full sky search. In this case the signal was added at a declination of -70$^{\circ}$.}
 \label{fig:disco}
 \end{figure}

\section{Results}
As mentioned above two different analyses have been done. The first one is a full sky survey with no assumptions about the source position. In the second analysis, we made a search for a signal excess in an \textit{a priori} defined spot in the sky corresponding to the position of some interesting astrophysical objects.
\subsection{Full sky search}
In the full sky search, no significant clusters of neutrino candidates were found. The most signal-like cluster is located at $\alpha_s , \delta_s = ( -46.5^{\circ}, -65.0^{\circ})$. The fit assigns 5 events above the background. The value of the test statistic for this cluster is 13.0 which yields to a p-value of 2.6\%. Figure \ref{fig:skymap} shows a sky map of the selected events in galactic coordinates with the location of the most signal like cluster. 
 \begin{figure}[!t]
 \vspace{5mm}
 \centering
 \includegraphics[width=9cm]{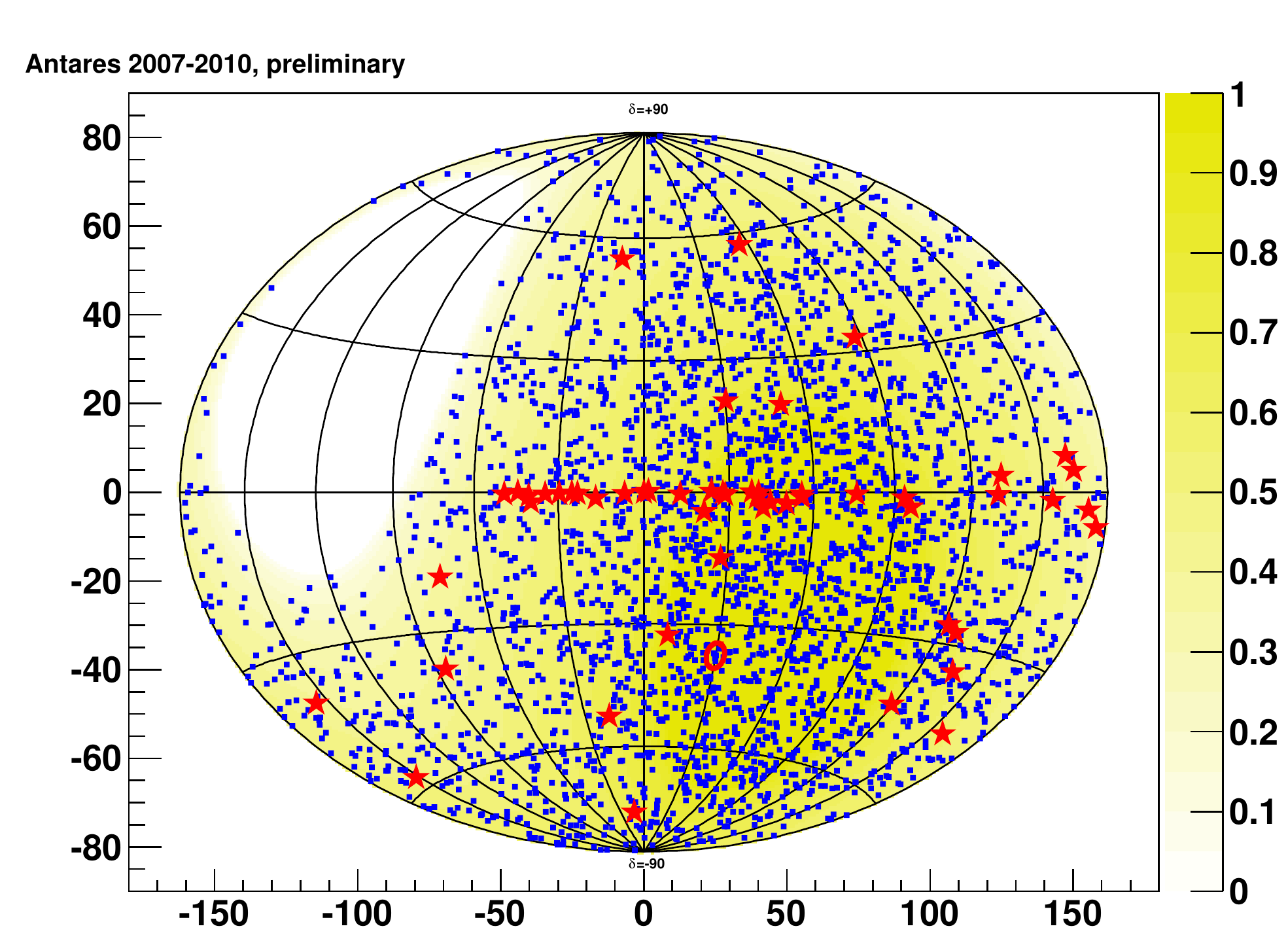}
 \caption{Galactic skymap showing the 3058 data events. The position of the most signal-like cluster is indicated by the circle. The stars denote the position of the 51 candidate sources.}
 \label{fig:skymap}
 \end{figure}
\subsection{Candidate list search}
The results of the search in the direction of 51 pre-defined candidate sources are shown in Table \ref{tab:lims}. None of the sources have a significant excess of events. The most signal-like candidate source is HESS J1023-575 where the post-trial p-value is of 41\%.
\begin{table}
\begin{tabular}{lr@{.}lr@{.}l@{\hspace{0.18cm}}ll}
\hline
         source
         & \multicolumn{2}{c}{$\alpha_s(^{\circ})$ }
         & \multicolumn{2}{c}{$\delta_s(^{\circ})$ }
         & \multicolumn{1}{c}{$p$}
         & \multicolumn{1}{c}{$\phi^{90\% \rm CL}$} \\
\hline
HESS J1023-575  & 155&83 & -57&76  &  0.41  & 6.6  \\
3C 279     &     -165&95 & -5&79  &  0.48   &  10.1   \\
GX 339-4    &    -104&30 & -48&79 & 0.72   &    5.8\\
Cir X-1     &    -129&83 & -57&17 &   0.79    & 5.8 \\
MGRO J1908+06 &  -73&01 & 6&27   &   0.82    &  10.1  \\
ESO 139-G12  &   -95&59  & -59&94 &   0.94   &  5.4  \\
HESS J1356-645 & -151&00 & -64&50 &  0.98   &    5.1 \\
PKS 0548-322  &  87&67  & -32&27 &   0.99    &   7.1  \\
HESS J1837-069  & -80&59 &  -6&95  & 0.99   &    8.0  \\  
PKS 0454-234  &  74&27   & -23&43 & 1.00    &   7.0    \\
IceCube hotspot  &       75&45 &  -18&15 &   1.00   &   7.0   \\
PKS 1454-354   &  -135&64 & -35&67 &  1.00    &    5.0   \\
RGB J0152+017 &  28&17  & 1&79  &  1.00   &   6.3\\
Geminga   &      98&31  & 17&01  & 1.00  &   7.3 \\
PSR B1259-63  &  -164&30 & -63&83 & 1.00    &   3.0    \\
PKS 2005-489  &  -57&63  & -48&82 &  1.00    &  2.8  \\
HESS J1616-508 &  -116&03 & -50&97 & 1.00   &   2.7  \\
HESS J1503-582 &  -133&54 & -58&74 &  1.00  &   2.8  \\
HESS J1632-478 &  -111&96 & -47&82 & 1.00 &   2.6  \\
H 2356-309   &   -0&22   & -30&63 & 1.00   &  3.9  \\
MSH 15-52  &     -131&47 & -59&16 & 1.00    &  2.6  \\
Galactic Center & -93&58  & -29&01 &  1.00  &  3.8  \\
HESS J1303-631  & -164&23 & -63&20 &  1.00  &  2.4  \\
HESS J1834-087 &  -81&31 & -8&76 &  1.00  &  4.3  \\
PKS 1502+106 &   -133&90 & 10&52 & 1.00  &  5.2  \\
SS 433      &    -72&04 & 4&98  &  1.00   &  4.6 \\
HESS J1614-518 & -116&42 & -51&82 &  1.00 &  2.0   \\
RX J1713.7-3946 & -101&75 & -39&75 & 1.00  &   2.7  \\
3C454.3    &     -16&50 & 16&15 & 1.00 &     5.5  \\
W28     &        -89&57 & -23&34 &  1.00  &   3.4 \\
HESS J0632+057 &  98&24  & 5&81 &  1.00 &   4.6 \\
PKS 2155-304   & -30&28  & -30&22 &  1.00   &    2.7 \\
HESS J1741-302 & -94&75 & -30&20 & 1.00  &  2.7 \\
Centaurus\ A   &  -158&64 & -43&02 & 1.00  &   2.1  \\
RX J0852.0-4622& 133&00 & -46&37 & 1.00  &  1.5 \\
1ES 1101-232   & 165&91  & -23&49 &  1.00    &  2.8  \\
Vela X       &   128&75  & -45&60 &  1.00 &  1.5\\
W51C        &    -69&25 & 14&19 &  1.00  &  3.6\\
PKS 0426-380 &   67&17  & -37&93 & 1.00  &  1.4 \\
LS 5039    &     -83&44  & -14&83 & 1.00  &   2.7\\  
W44    &         -75&96 & 1&38   & 1.00 &    3.1\\
RCW 86      &    -139&32 & -62&48 &  1.00  &     1.1\\  
Crab    &        83&63  & 22&01 & 1.00  &   4.1\\
HESS J1507-622  &-133&28 &  -62&34 & 1.00  &   1.1\\  
1ES 0347-121 &   57&35  &  -11&99 &  1.00 &  1.9 \\
VER J0648+152 &  102&20  & 15&27  & 1.00 &  2.8 \\ 
PKS 0537-441 &   84&71  &  -44&08 & 1.00  &  1.3  \\
HESS J1912+101  &-71&79  & 10&15  &  1.00 &     2.5  \\
PKS 0235+164  &  39&66  & 16&61 & 1.00   &    2.8 \\
IC443    &       94&21  & 22&51  & 1.00  &   2.8  \\
PKS 0727-11   &  112&58  -& 11&70 & 1.00 &  1.9\\
\hline
\end{tabular}
\caption{Results of the candidate source search. The source coordinates and the p-values ($p$) are shown as well as the
         limits on the flux intensity $\phi^{90\% \rm CL}$; the latter has units $10^{-8} \rm GeV^{-1} cm^{-2} s^{-1}$.}
\label{tab:lims}
\end{table}
Figure \ref{fig:limits} shows the 90\% confidence level limits on $\phi$ using the Feldman-Cousins prescription \cite{Feldman} and assuming an $E^{-2}$ neutrino spectrum for each of the source candidates as a function of the source declination. The sensitivity of this analysis is also presented, defined as the median expected limit and resulting in a factor 2.7 better than the one obtained with data collected during 2007 and 2008 only \cite{aart2007}. Limits from other experiments are also shown. However, it should be noted that for this spectrum, ANTARES detects most events at energies around 10 TeV while the limits in the Southern Hemisphere published by the IceCube Collaboration \cite{Icecube} apply to the PeV region.

 \begin{figure}[!t]
 \vspace{5mm}
 \centering
 \includegraphics[width=7cm]{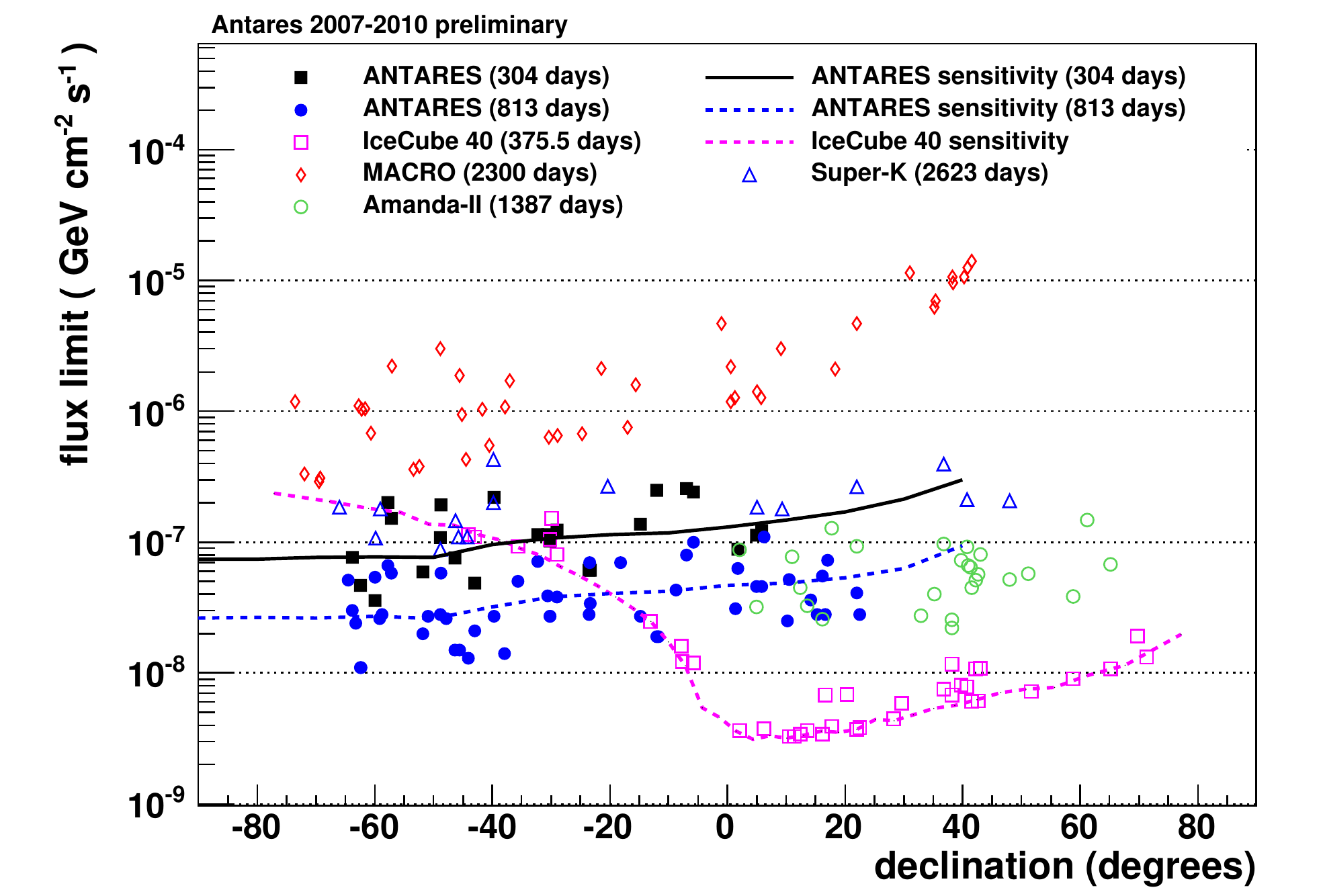}
 \caption{Limits set on the normalisation $\phi$ of an $E^{-2}_{\nu}$ spectrum of high energy
         neutrinos from selected candidates (see Table \ref{tab:lims}). Also shown is the
         sensitivity, which is defined as the median expected limit.
         In addition to the present result, several previously published limits on sources in both
         the Southern and Northern sky are also shown.}
 \label{fig:limits}
 \end{figure}
 
 \section{Conclusions}
A search of high energy cosmic neutrinos has been performed. Data were taken during the first four years of operation when ANTARES consists of 5 line for most of the first year considered and 9, 10 and 12 for the rest. A measurement of the angular resolution using MonteCarlo data yields to $0.46$ degrees.
 Neither the full sky search nor the candidate list search show a significant excess of events therefore limits have been obtained on the neutrino flux.  


\clearpage

\setcounter{figure}{0}
\setcounter{table}{0}
\setcounter{footnote}{0}
\setcounter{section}{0}
\newpage




\title{Search for point sources with the ANTARES neutrino telescope using the EM algorithm}

\shorttitle{Juan Pablo G\'omez Gonz\'alez, Search for point sources with ANTARES}

\authors{Juan Pablo G\'omez-Gonz\'alez$^{1}$, on behalf of the ANTARES Collaboration. }

\afiliations{$^1$ IFIC - Instituto de F\'isica Corpuscular, Edificios de Investigaci\'on de Paterna, CSIC - Universitat de Val\`encia, 
Apdo. de Correos 22085, 46071 Valencia, Spain.}
\email{jpablo@ific.uv.es}

\maketitle

\begin{abstract}
The ANTARES detector, currently the largest deep-sea neutrino telescope in the
Northern Hemisphere, consists of a three-dimensional array of 885 optical modules
arranged over 12 detection lines anchored at a depth of 2475 m in the Mediterranean
Sea, 40 km offshore from Toulon (France). The photomultiplier tubes detect the
Cherenkov light induced by the charged particles produced in the interaction of
cosmic neutrinos with the matter surrounding the detector. The trajectories of the
resulting muons are reconstructed with high precision, revealing the direction of the
incoming neutrinos.
The main scientific goal of ANTARES is the search for high energy neutrinos coming
from astrophysical sources. This contribution describes a point source search using a
dedicated clustering algorithm, based on the analytical maximization of the
likelihood. The results of shuch analysis using four years of data will be
presented.
\end{abstract}



\section{Introduction}


The ANTARES neutrino telescope \cite{antares} started data taking in 2007 and is fully operational since 2008. Located at 40 km off the 
coast of Toulon it consists of 12 detection lines anchored to the seabed at a depth of 2475 m 
and sustained vertically by means of buoys. Each line has 25 floors (or storeys) composed by 
a triplet of photomultiplier tubes (PMTs) housed in pressure resistant glass spheres called optical modules (OMs). 
The OMs are facing downward at $45^\circ$ from the vertical for an increased detection- efficiency for up-going neutrinos.

This three-dimensional photo-detector array detects the Cherenkov light emitted by the charged leptons originated in the 
interaction of high energy neutrinos with the matter surrounding the detector. The tracks of the produced muons 
can be reconstructed using the position and timing information of the hits arriving to the PMTs.
An accurate timing and position calibration \cite{calib} of the detector OMs is necessary in order to achieve the best attainable angular
resolution. 

The main goal of the experiment is the detection of high energy neutrinos from extraterrestial origin, and one of the most promising ways of establishing 
their existence is the search for point sources. Here, we present  such a search  using data collected between years 2007 and 2010 for a total of 813 days 
of livetime. In section 2 the track reconstruction method and data selection criteria are described. The detector performance is reviewed in section 3. 
The clustering algorithm applied in this analysis is explained in section 3. Finally the search results are presented in section 4.

\section{Data selection and track reconstruction}
Data runs used in this analysis were recorded in the first four years of detector operation.
Taking into account the time spent on sea operations (like the deployment of new lines) 
and sporadic data taking problems of the detector, the total livetime of the analysis is 813 days; about 77$\%$ of this data were
collected using 9, 10 and 12 detection lines, while the remaining 183 days correspond to data gathered with the initial 5-lines configuration.
 
The reconstruction method \cite{treco} is based on the maximization of the likelihood function describing the probability
density function (PDF) for the residuals, defined as the difference between the measured hit time
and the expected arrival time of the hits. 
The goodness of the track reconstruction is described by the $\Lambda$ parameter, which is basically the log-likelihood
of the fitted track. This parameter can be used to eliminate badly reconstructed tracks by selecting an appropriate cut on the $\Lambda$ value.
The corresponding cumulative $\Lambda$ distribution for events reconstructed as upgoing is shown on Figure \ref{lambda}. 
The contribution from the different components of the expected background is also included. The simulation reproduces well the data.

For this analysis atmospheric muons were simulated using the MUPAGE package \cite{mupage}, while the atmospheric neutrinos were generated with the GENNEU \cite{genneu} 
package according to the Bartol model \cite{bartol}.

 \begin{figure}[!t]
  \vspace{5mm}
  \centering
  \includegraphics[width=3.0in]{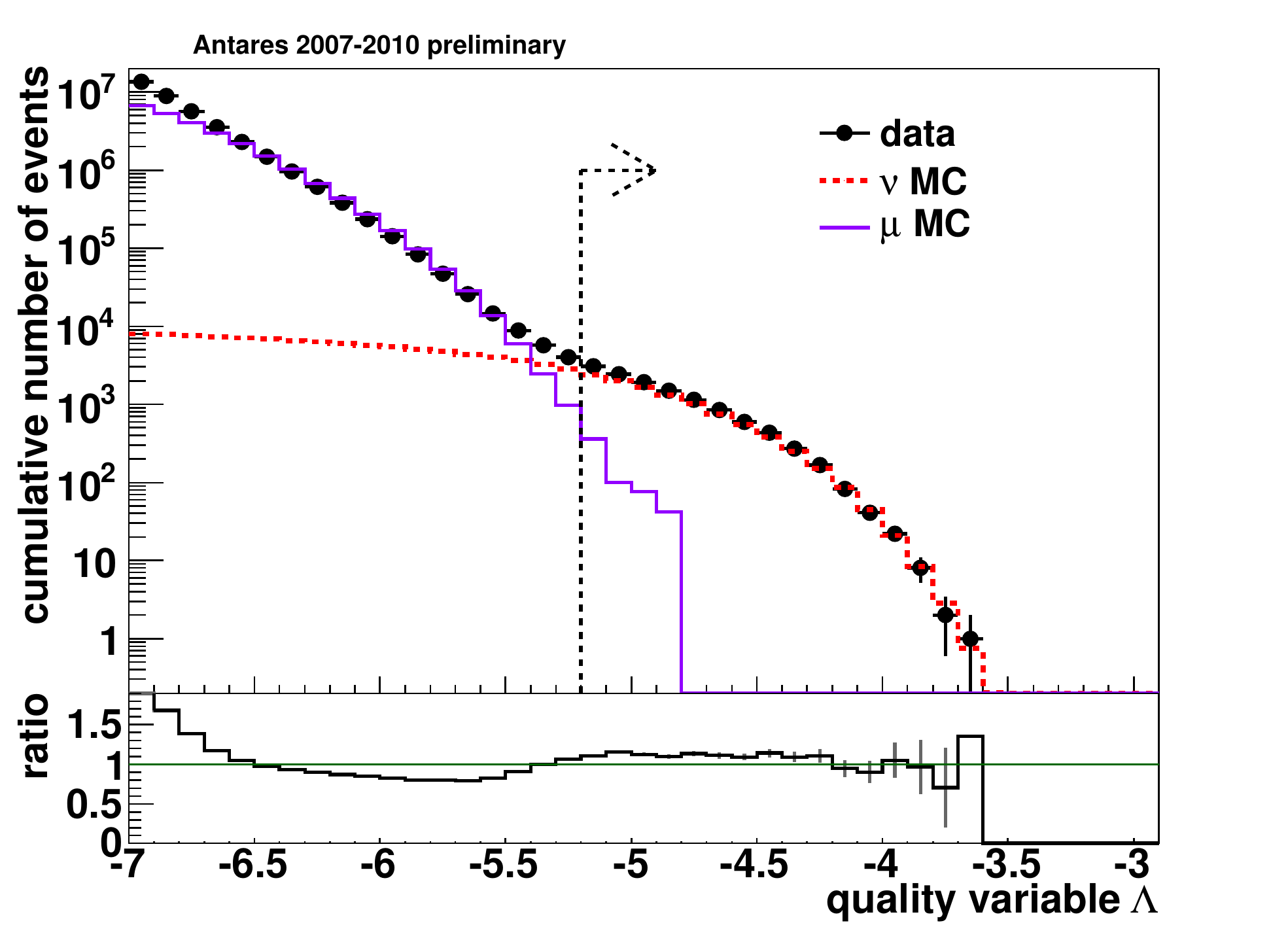}
  \caption{Cumulative distribution of the quality of the reconstruction parameter for data and MC upgoing events.}
  \label{lambda}
 \end{figure}

A cut on the quality of the reconstruction at $\Lambda \geq$ -5.2 was found to be the optimal for the search discovery potential using 
muon tracks recostructed as upward going ($\theta < 90^{\circ}$). Additionally, the uncertainty on the muon direction estimated from 
the fit is required to be $\leq$ $1^{\circ}$. The selected sample contains 3058 events, out of which it is estimated from MC simulations that about 84$\%$ are 
neutrino events and only 16$\%$ downgoing atmospheric muons mis-reconstructed 
as upgoing.

\section{Detector performance}

The two main parameters describing the performance of a neutrino telescope are the angular resolution and
the acceptance. Both parameters are estimated from simulations.
Figure \ref{angres} shows the cumulative angular resolution for upgoing neutrino events following an $E^{-2}$ spectrum
and complying the selection criteria described in Section 2. The plot shows that roughly 80$\%$ of the signal events are 
reconstructed with an angular error less than $1^\circ$, being the median value of the reconstruction error equal to 0.46 $\pm$ $0.1^\circ$. 
The uncertainty on this value has been computed considering all the effects leading to a deterioration of the detector timing resolution \cite{ps_paper}. 
In addition, the uncertainty on the absolute orientation of the detector, which is estimated to be of the order of $0.1^\circ$, is also taken into 
account in the limits computation (see section 4).\\

The acceptance allows us to relate the detected event-rate with the neutrino flux of the source at the Earth, and it is shown on Figure \ref{effarea} as
a function of the sinus of the declination considering a flux normalization $\phi=10^{-8}GeV^{-1}cm^{-2}s^{-1}$. 
Based on the agreement between data and simulations a 15$\%$ systematic error on the detection efficiency 
has been considered for the limits calculation.

 \begin{figure}[!t]
  \vspace{5mm}
  \centering
  \includegraphics[width=2.2in]{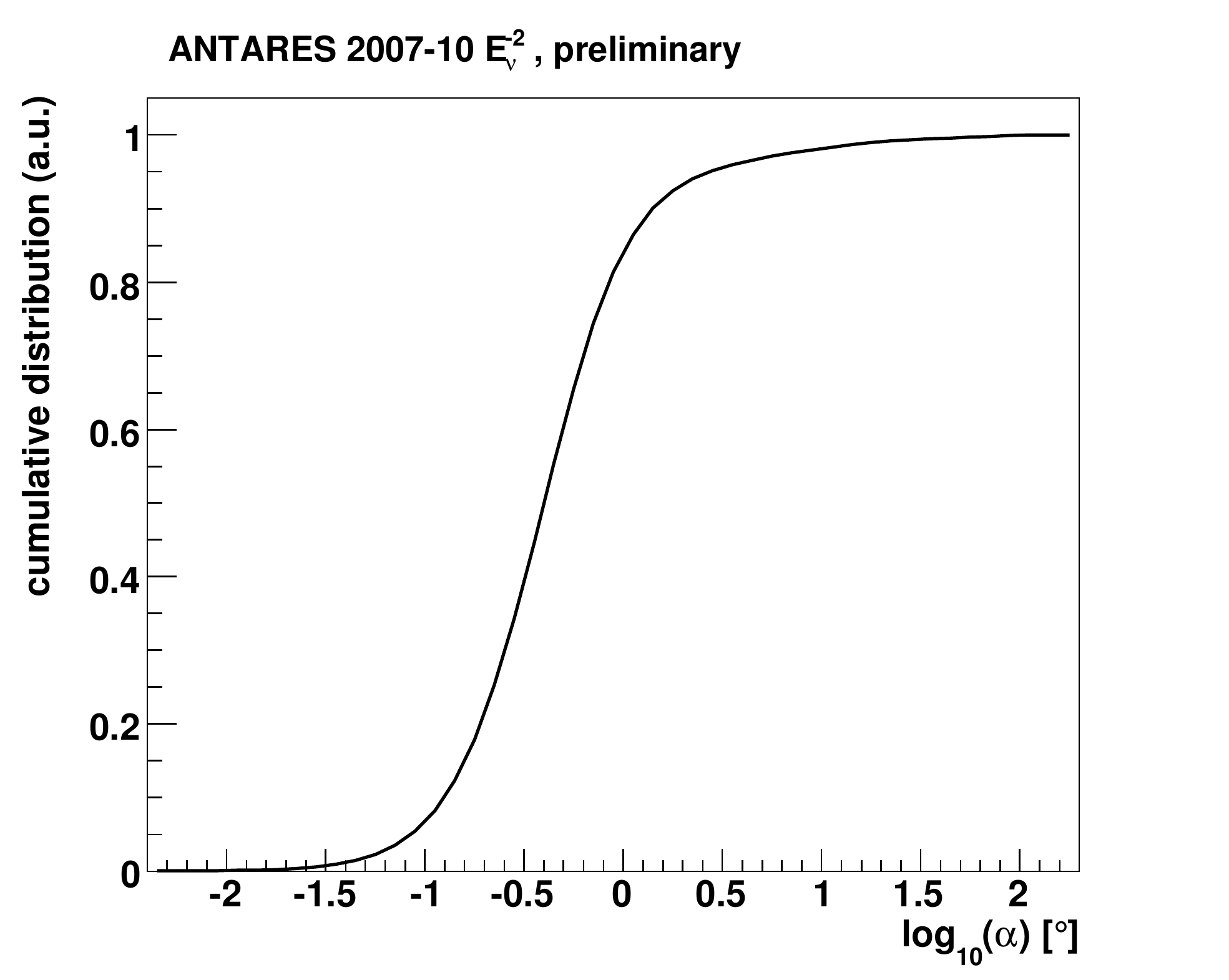}
  \caption{Cumulative angular resolution for $E^{-2}$ upgoing neutrinos selected for this analysis.}
  \label{angres}
 \end{figure}

 \begin{figure}[!t]
  \vspace{5mm}
  \centering
  \includegraphics[width=2.2in]{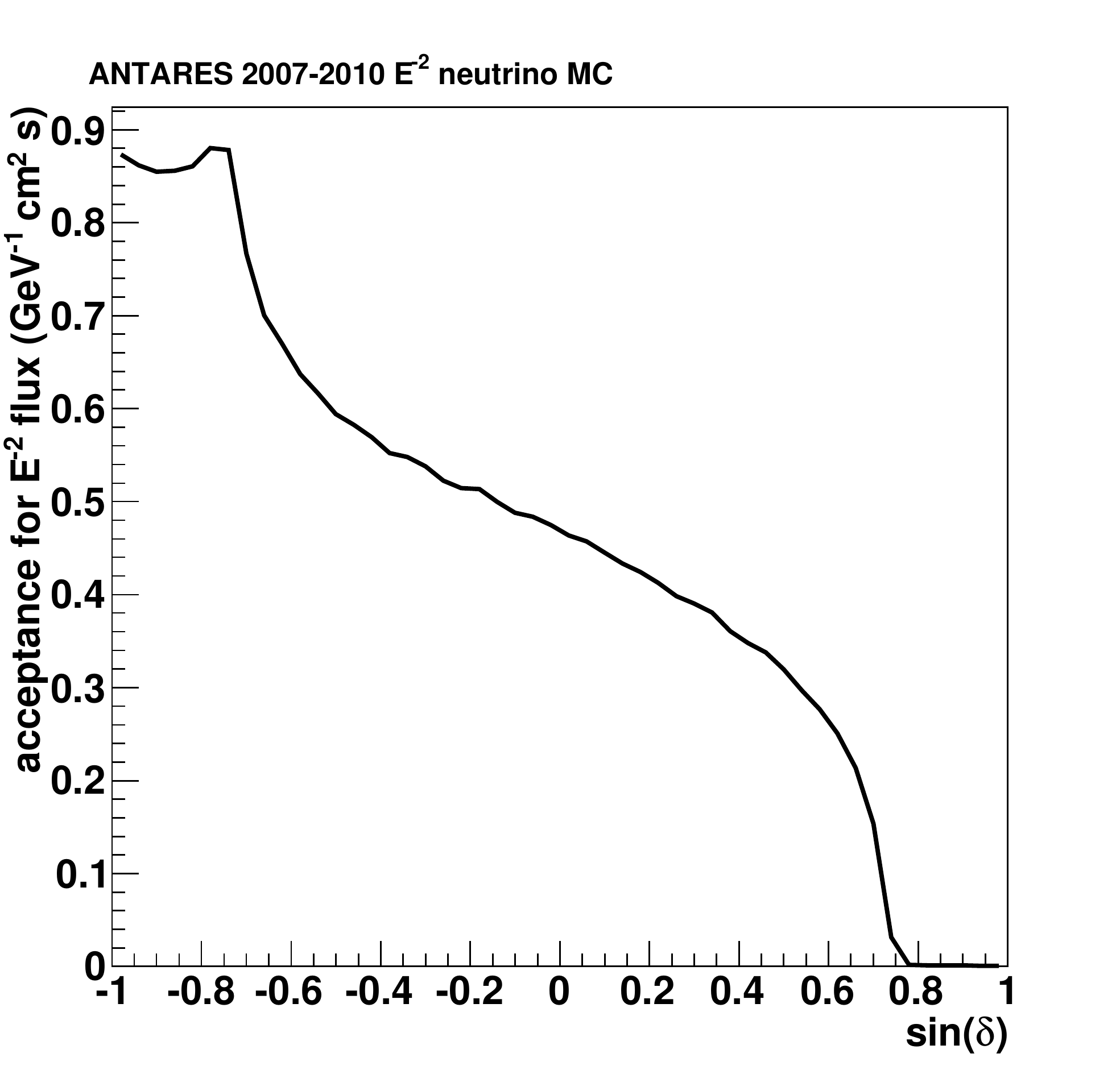}
  \caption{Detector acceptance as a function of the sinus of the declination.}
  \label{effarea}
 \end{figure}

\section{Clustering method}
Point source clustering techniques try to identify and separate events coming from real sources from background
events. The Expectation-Maximization (EM) algorithm \cite{em_paper} used in this analysis is a pattern recognition algorithm that 
analytically maximizes the likelihood in finite mixture problems. These mixture models are different groups
of data described by different density components. In the case of a search for neutrino point sources \cite{em_ap} the mixture 
problem can be expressed as:

\begin{equation}
\label{eq_02}
p(x) = \pi_{bg}P_{bg}(\delta) + \pi_{sg}P_{sg}(x;\mu;\Sigma)\cdot(P^{nhits}_{sg}/P^{nhits}_{bg}) 
\end{equation}
\noindent
where $\pi_{bg}$ and $\pi_{sg}$ are the so-called mixing proportions, $x =(\alpha,\delta)$ is the position of the signal 
event in equatorial coordinates, $\mu=(\mu_{\alpha},\mu_{\delta})$ and $\Sigma=(\sigma_{\alpha},\sigma_{\delta})$ are, respectivly,
the mean and the covariance vector of the Gaussian distribution, and $P^{nhits}$ is the probability for an event to be reconstructed
using $nhits$ number of hits. 

In this analysis the expected density distributions of background and signal events are parametrized. The pdf describing the background
is obtained from the declination distribution of data events, while the source signals are supposed to follow a two-dimensional Gaussian distribution. 

The EM algorithm works in two steps. In the first step called ``Expectation'' the expected value
of the complete data log-likelihood is computed for a given set of parameters. In the ``Maximization'' step a new set of parameters
that maximizes the likelihood is found.
In our case, the parameters to be maximized are the two components of the Gaussian width, the expected number of signal events and, 
in the full sky search (see next section), the coordinates of the signal source. 

After likelihood maximization the so called test-statistic, defined as the likelihood ratio of the two mixture
models, is computed. Lower values of this quantity indicate that data is more likely to be produced by the background,
while larger values are more likely to be produced by the presence of the searched signal. 

 \begin{figure}[!t]
  \vspace{5mm}
  \centering
  \includegraphics[width=3.5in]{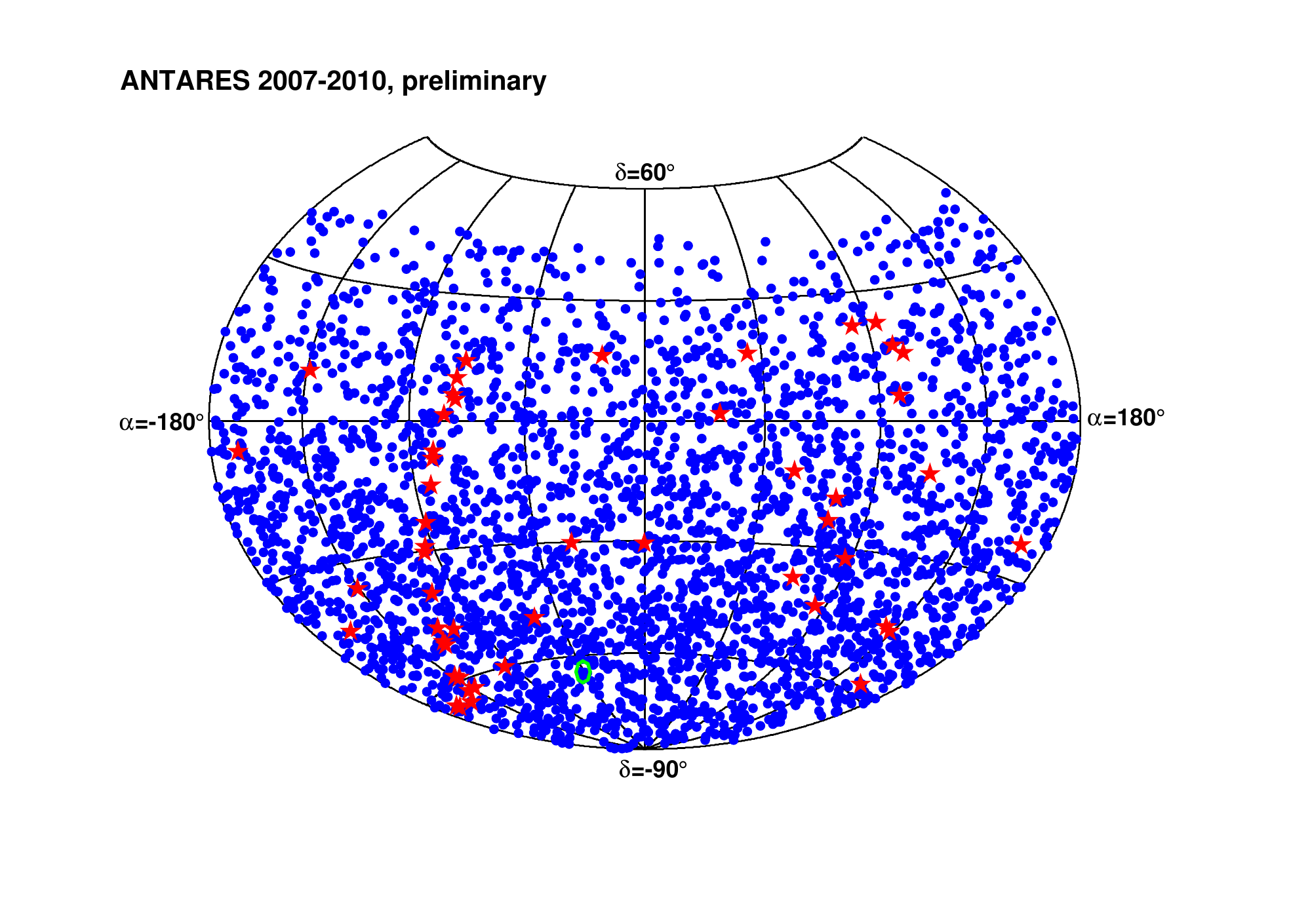}
  \caption{Skymap with the equatorial coordinates of the 3058 events selected. The position of the most significant cluster is denoted by the green circle and the
coordinates of the candidate list sources are indicated with the stars.}
  \label{skymap}
 \end{figure}

\section{Results}
Two different searches for point sources have been considered in this analysis.
In the first approach a blind survey is done looking everywhere in the whole ANTARES visible sky.
The second search used a catalog of candidate sources to look for presence of signal at particular locations
in the sky. 
The candidate list of sources includes both galactic and extra-galactic sources known to be gamma-ray emitters. 
The detector visibility and PSF was taking into account when defining
the list. 

No significant excess of events was found neither in the full sky search, nor in the candidate list search. 
The most signal-like cluster was found at ($\delta=-64.87^{\circ}$, $\alpha=-46.49^{\circ}$) in the all sky survey.
For this cluster a 2.6$\%$ excess p-value is found. 

The locations (in equatorial coordinates) of the most significant cluster, the 3058 events selected and the 51 candidate sources 
are shown on figure \ref{skymap}.
Upper limits \footnote{Here we follow the Neyman prescription.} on the $E^{-2}$ neutrino flux spectrum are reported in table \ref{table_results} and in figure \ref{limits} 
as a function of the declination for the sources in the candidate list. The ANTARES sensitivity (defined as the median value of the expected limit) it is shown, as well as 
limits reported by other neutrino experiments included for comparison.

 \begin{figure}[!t]
  \vspace{5mm}
  \centering
  \includegraphics[width=3.2in]{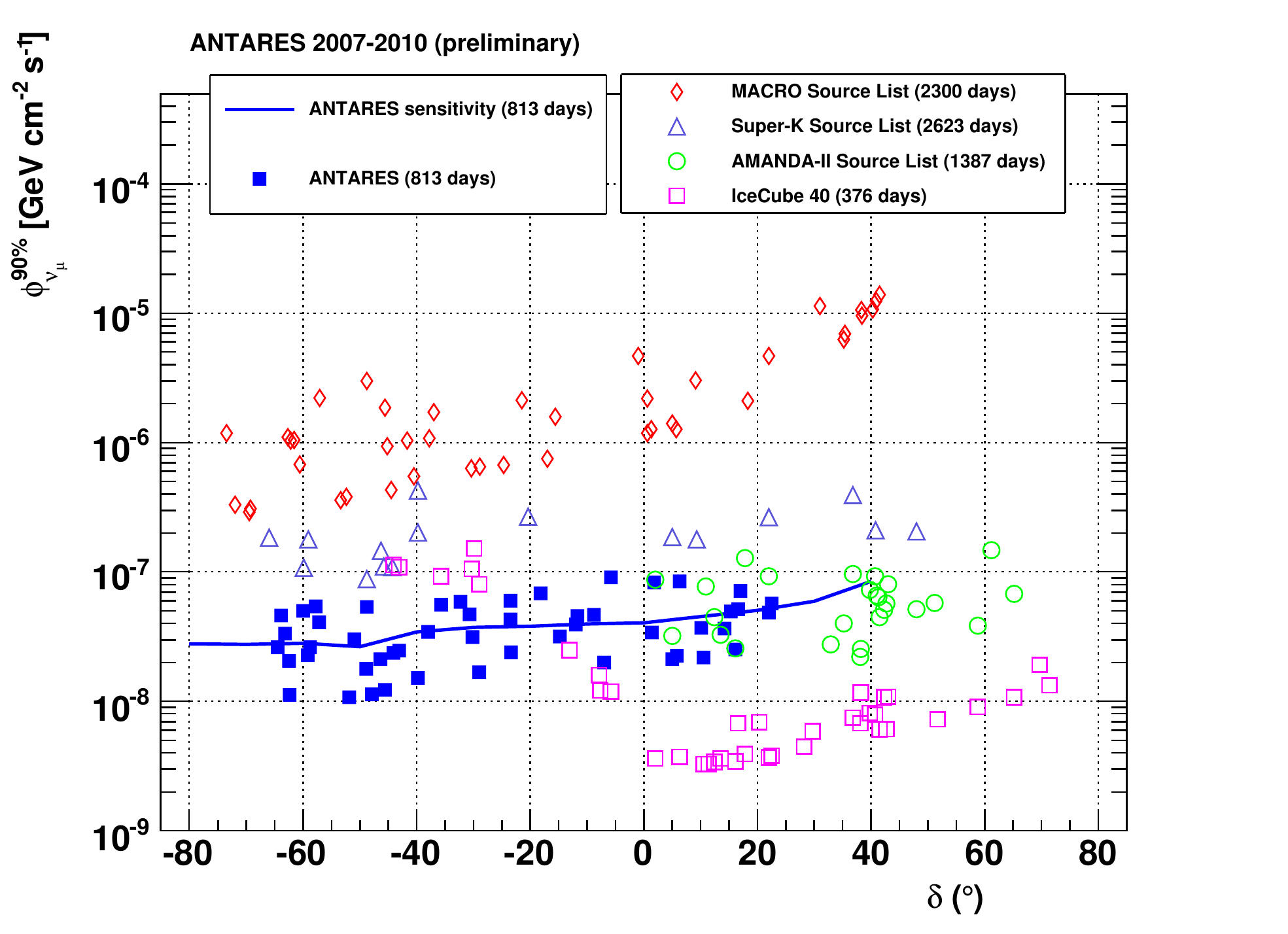}
  \caption{Limits on the $E^{-2}_{\nu}$ flux for the 51 sources in the candidate list search. Upper limits previously reported by other neutrino experiments 
    for both Northern and Southern sky are shown. The ANTARES sensitivity computed in this analysis is also included.}
  \label{limits}
 \end{figure}

\begin{table}
\begin{scriptsize}
\begin{tabular}{l l l l l l}
\hline
Source & ra $(^{\circ})$ & $\delta$ $(^{\circ})$ & Q & p-value & $\phi^{90CL}$ \\
\hline
3C 279 & -165.95 & -5.79 & 1.97 & 0.04 & 9.1\\ 
GX 339-4 & -104.30 & -48.79 & 1.62 & 0.06 & 5.4 \\  
HESS J1023-575 & 155.83 & -57.76 & 1.49 & 0.07 & 5.4\\ 
MGRO J1908+06 & -73.01 & 6.27 & 1.12 & 0.09 & 8.5\\ 
RGB J0152+017 & 28.17 & 1.79 & 1.16 & 0.09 & 8.3\\ 
ESO 139-G12 & -95.59 & -59.94 & 1.02 & 0.11 & 5.0\\ 
ICECUBE & 75.45 & -18.15 & 0.74 & 0.14 & 6.8\\ 
PSR B1259-63 & -164.30 & -63.83 & 0.63 & 0.16 & 4.6\\ 
PKS 0548-322 & 87.67 & -32.27 & 0.41 & 0.19 & 5.9\\ 
PKS 1454-354 & -135.64 & -35.67 & 0.39 & 0.20 & 5.6\\ 
1ES 1101-232 & 165.91 & -23.49 & 0.35 & 0.21 & 6.0\\ 
Cir X-1 & -129.83 & -57.17 & 0.30 & 0.22 & 4.1\\ 
Geminga & 98.31 & 17.01 & 0.25 & 0.22 & 7.1\\ 
H 2356-309 & -0.22 & -30.63 & 0.02 & 0.34 & 4.7\\ 
PKS 0454-234 & 74.27 & -23.43 & 0.00 & 1 & 4.3\\ 
HESS J1356-645 & -151.00 & -64.50 & 0.00 & 1 & 2.6\\ 
HESS J1837-069 & -80.59 & -6.95 & 0.00 & 1 & 2.0\\ 
PKS 2005-489 & -57.63 & -48.82 & 0.00 & 1 & 1.8\\  
HESS J1616-508 & -116.03 & -50.97 & 0.00 & 1 & 3.0\\ 
HESS J1503-582 & -133.54 & -58.74 & 0.00 & 1 & 2.6\\ 
HESS J1632-478 & -111.96 & -47.82 & 0.00 & 1 & 1.1\\ 
MSH 15-52 & -131.47 & -59.16 & 0.00 & 1 & 2.3\\ 
Galactic Center & -93.58 & -29.01 & 0.00 & 1 & 1.7\\  
HESS J1303-631 & -164.23 & -63.20 & 0.00 & 1 & 3.3\\ 
HESS J1834-087 & -81.31 & -8.76 & 0.00 & 1 & 4.7\\ 
PKS 1502+106 & -133.90 & 10.52 & 0.00 & 1 & 2.2\\ 
SS 433 & -72.04 & 4.98 & 0.00 & 1 & 2.1\\ 
HESS J1614-518 & -116.42 & -51.82 & 0.00 & 1 & 1.1\\  
RX J1713.7-3946 & -101.75 & -39.75 & 0.00 & 1 & 1.5\\ 
3C454.3 & -16.50 & 16.15 & 0.00 & 1 & 2.5\\ 
W28 & -89.57 & -23.34 & 0.00 & 1 & 2.4\\ 
HESS J0632+057 & 98.24 & 1.81 & 0.00 & 1 & 2.3\\ 
PKS 2155-304 & -30.28 & -30.22 & 0.00 & 1 & 3.2\\ 
HESS J1741-302 & -94.75 & -30.20 & 0.00 & 1 & 3.2\\ 
Centaurus A & -158.64 & -43.02 & 0.00 & 1 & 2.5\\ 
RX J0852.0-4622 & 133.00 & -46.37 & 0.00 & 1 & 2.1\\ 
Vela X & 128.75 & -45.60 & 0.00 & 1 & 1.2\\ 
W51C & -69.25 & 14.19 & 0.00 & 1 & 3.6\\ 
PKS 0426-380 & 67.17 & -37.93 & 0.00 & 1 & 3.4\\ 
LS 5039 & -83.44 & -14.83 & 0.00 & 1 & 3.2\\ 
W44 & -75.96 & 1.38 & 0.00 & 1 & 3.4\\ 
RCW 86 & -139.32 & -62.48 & 0.00 & 1 & 2.1\\ 
Crab & 83.63 & 22.01 & 0.00 & 1 & 4.9\\ 
HESS J1507-622 & -133.28 & -62.34 & 0.00 & 1 & 1.1\\ 
1ES 0347-121 & 17.35 & -11.99 & 0.00 & 1 & 3.9\\ 
VER J0648+152 & 102.20 & 15.27 & 0.00 & 1 & 5.0\\ 
PKS 0537-441 & 84.71 & -44.08 & 0.00 & 1 & 2.4\\ 
HESS J1912+101 & -71.79 & 10.15 & 0.00 & 1 & 3.7\\ 
PKS 0235+164 & 39.66 & 16.61 & 0.00 & 1 & 1.2\\ 
IC443 & 94.21 & 22.51 & 0.00 & 1 & 1.7\\ 
PKS 0727-11 & 112.58 & -11.70 & 0.00 & 1 & 4.6\\ 
\hline
\end{tabular}
\caption{Results for the 51 sources used in the candidate list search. 
The equatorial coordinates of the sources, Q values, number of fitted
signal events, pre-trial p-values and upper limits on the $E^{-2}_{\nu}$ 
flux ($10^{-8} GeV^{-1} cm^{-2} s^{-1}$) are shown.}\label{table_results}
\end{scriptsize}
\end{table}

\section{Conclusions}
This contribution presented the analysis of 813 days of livetime using data collected in the first four years of the ANTARES neutrino 
telescope operation. 
No statistically significant excess of events has been found neither in the search using a candidate list of interesting sources, nor in the full sky search. 
The most significant cluster, with a post-trial probability of 2.6 $\%$ was found at coordinates $\delta = -64.87^{\circ}$, $\alpha = -46.49^{\circ}$.
Some of the most stringent limits to $E^{-2}_{\nu}$ flux were obtained for sources located in the ANTARES field of view.
Using a different search method the results presented here are consistent with the main analysis \cite{nikhef} reporting upper limits
for the 51 candidate sources using the Feldman-Cousins prescription.

I greatfully acknowledge the financial support of MICINN (FPA2009-13983-C02-01 and MultiDark CSD2009-00064) 
and of Generalitat Valenciana (Prometeo/2009/026).

\clearpage

\setcounter{figure}{0}
\setcounter{table}{0}
\setcounter{footnote}{0}
\setcounter{section}{0}
\newpage




\title{Autocorrelation analysis of ANTARES data}

\shorttitle{F. Sch\"ussler \etal Antares autocorrelation}

\authors{Fabian Sch\"ussler$^{1}$ on behalf of the ANTARES collaboration}
\afiliations{$^1$ Commissariat \`a l'\'energie atomique et aux \'energies alternatives\\ Institut de recherche sur les lois fondamentales de l'Univers\\
91191 Gif-sur-Yvette Cedex, France}
\email{fabian.schussler@cea.fr}

\maketitle
\begin{abstract}
Clustering of neutrino arrival directions would provide hints for their astrophysical origin. The two-point autocorrelation method is sensitive to a large variety of cluster morphologies and, due to its independence from Monte Carlo simulations, provides complementary information to searches for the astrophysical sources of high energy muon neutrinos. We present the analysis of the autocorrelation function as a function of the angular scale of data collected during 2007-08 with the ANTARES neutrino telescope.
\end{abstract}



\section{Introduction}
The key question to resolve the long standing mystery of the origin of cosmic rays is to locate the sources and study the acceleration mechanisms able to produce fundamental particles with energies orders of magnitude above man-made accelerators. Over the last years it has become more and more obvious that multiple messengers will be needed to achieve this task. Fundamental particle physics processes like the production and subsequent decay of pions in interactions of high energy particles predict that the acceleration sites of high energy cosmic rays are also sources of high energy gamma rays and neutrinos. The detection of astrophysical neutrinos and the identification of their sources is one of the main aims of large neutrino telescopes operated at the South Pole (IceCube), in Lake Baikal and in the Mediterranean Sea (ANTARES). 
 
\subsection{The ANTARES neutrino telescope}
Whereas physics data taking started already during the deployment phase, the ANTARES detector~\cite{Antares_DetectorPaper} became fully equipped and operational in 2008. The detector is composed of 12 detection lines placed at a depth of 2475m off the French coast near Toulon. The detector lines are about 450m long and hold a total of 885 optical modules (OMs), 17'' glass spheres housing each a 10'' photomultiplier tube. The OMs look downward at $45^\circ$ in order to optimize the detection of upgoing, i.e. neutrino induced, tracks. The geometry and size of the detector makes it sensitive to neutrinos in the TeV-PeV energy range. A schematic layout is shown in Fig.~\ref{fig:layout}.\\

\begin{figure}[!t]
  \vspace{5mm}
  \centering
  \includegraphics[width=2.7in]{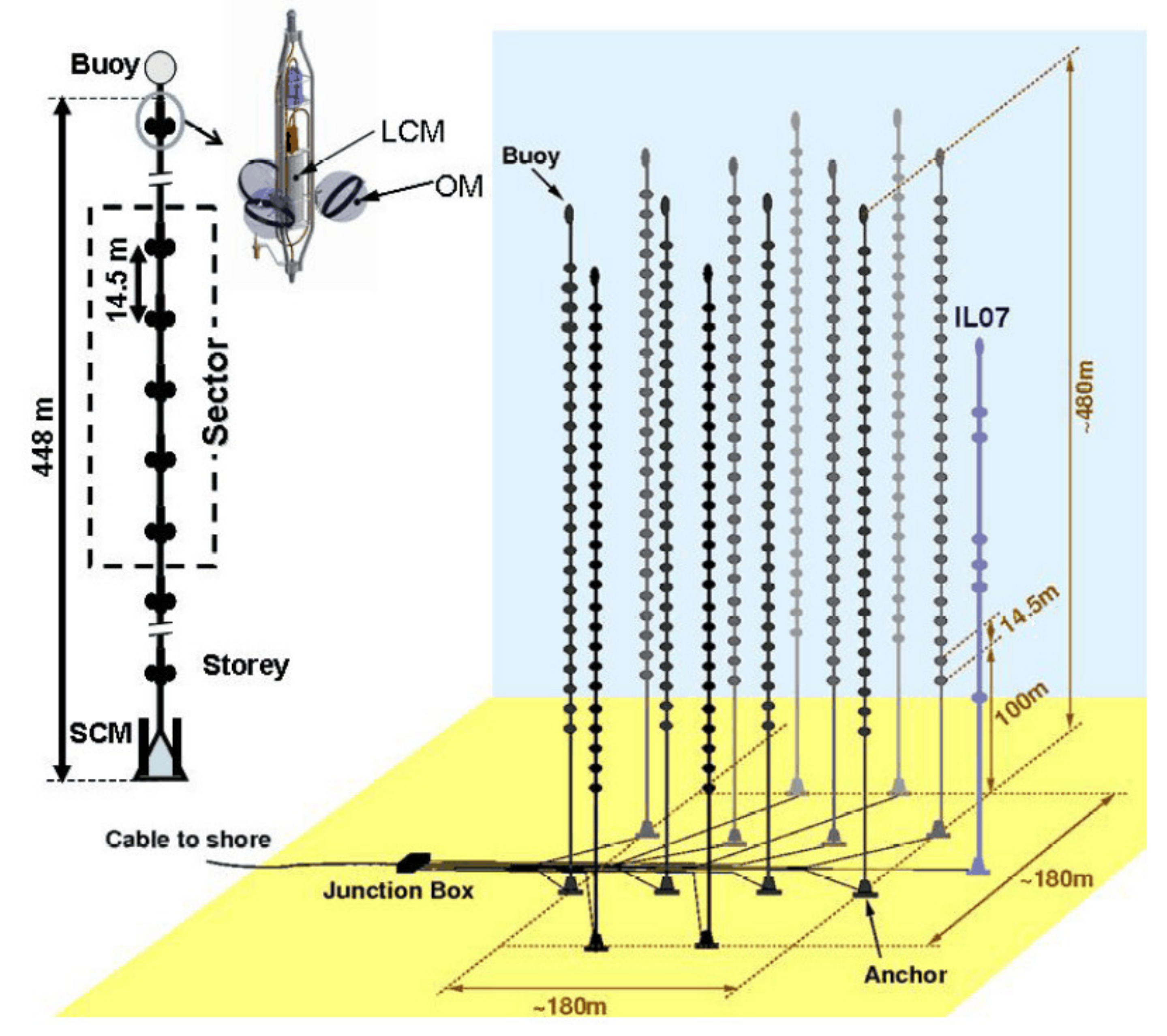}
  \caption{Schematic view of the ANTARES detector.}
\label{fig:layout}	
 \end{figure}
 
\begin{figure*}[!t]
   \centerline{\includegraphics[width=3.4in]{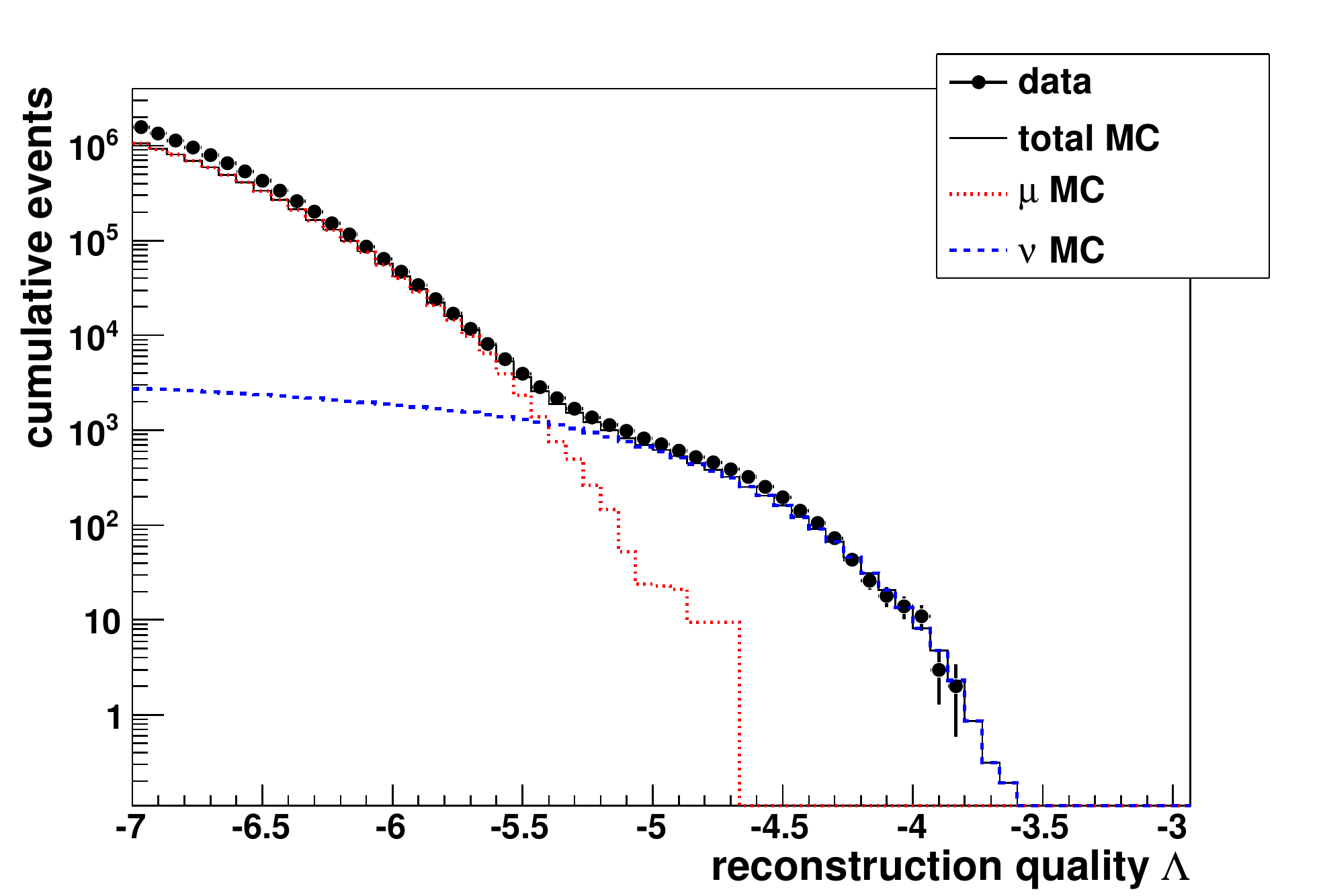}
              \hfill
              \includegraphics[width=3.4in]{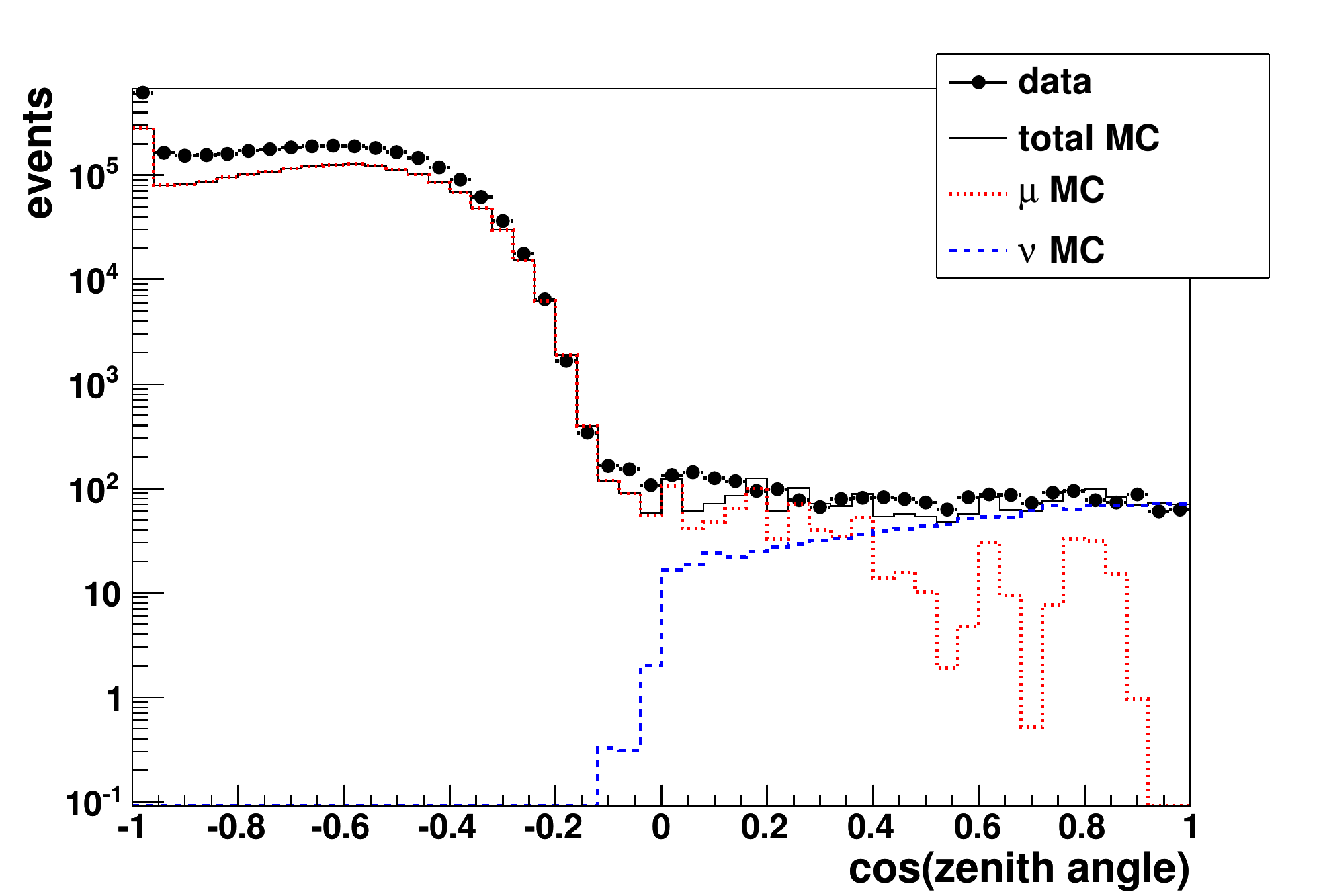} 
             }
   \caption{Distribution of the quality parameter $\Lambda$ (left plot) and the cosine of the reconstructed zenith angle (right plot) of the selected data events (markers) compared to Monte Carlo simulations of atmospheric muons (red, dotted line) and atmospheric neutrinos (blue, dashed line).}
   \label{fig:eventSelection}
 \end{figure*}
 
\subsection{Neutrino detection}
The neutrino detection relies on the emission of Cherenkov light by high energy muons originating from charged current neutrino interactions inside or near the instrumented volume. All detected signals are transmitted via an optical cable to a shore station, where a farm of CPUs filters the data for coincident signals or {\it hits} in several adjacent OMs. The muon direction is then determined by maximising a likelihood which compares the times of the hits with the expectation from the Cherenkov signal of a muon track.\\

\subsection{Astrophysical neutrinos}
Two main backgrounds for the search for astrophysical neutrinos can be identified: downgoing atmospheric muons which have been mis-reconstructed as upgoing and atmospheric neutrinos originating in cosmic ray induced air showers at the opposite side of the Earth. Depending on the requirements of the analysis both backgrounds can at least partially be discriminated using various parameters like the quality of the event reconstruction or an estimator for the deposited energy~\cite{ICRC11_EnergyEstimator}.\\

In addition, analysing the reconstructed arrival directions of the events allows to search for an excess over the uniform atmospheric backgrounds. Despite important efforts, no clear signature for point sources of astrophysical neutrinos has been found so far~\cite{Amanda_PointSources2009, IceCube_PointSources2011, Antares_PointSources2011, ICRC11_PointSources}. Both the distribution and morphologies of sources potentially emitting neutrinos in the TeV energy range are yet unknown but are possibly very inhomogeneous with most of them being located in the Galactic disk and spatially extended (e.g. shell type supernova remnants like RXJ1713~\cite{HESS_RXJ1713_2006}). It seems therefore interesting to study the intrinsic clustering of the arrival directions of neutrinos. Possible analysis biases are naturally avoided as no prior information about the potential sources is required. Covering a large angular range, i.e. neutrino emission regions of very different sizes, this study complements the searches for point like sources and, if successful, would provide hints for underlying, yet unresolved, source morphologies and source distributions.  

 
\section{Autocorrelation analysis}
The most commonly used method to detect intrinsic clusters within a set of $N$ events is the standard 2-point autocorrelation distribution. It is defined as the differential distribution of the number of observed event pairs $N_\mathrm{p}$ in the data set as a function their mutual angular distance $\Delta \Omega$. To suppress statistical fluctuations that would reduce the sensitivity of the method, we analyse here the cumulative autocorrelation distribution defined as \begin{equation}
\mathcal{N}_\mathrm{p} (\Delta \Omega) = \sum\limits_{i=1}^{N} \sum\limits_{j=i+1}^{N} H(\Delta \Omega_{ij} - \Delta \Omega), \label{equ:autocor}
\end{equation}
where $H$ is the Heaviside step function.

\begin{figure*}[!t]
   \centerline{\includegraphics[width=3.4in]{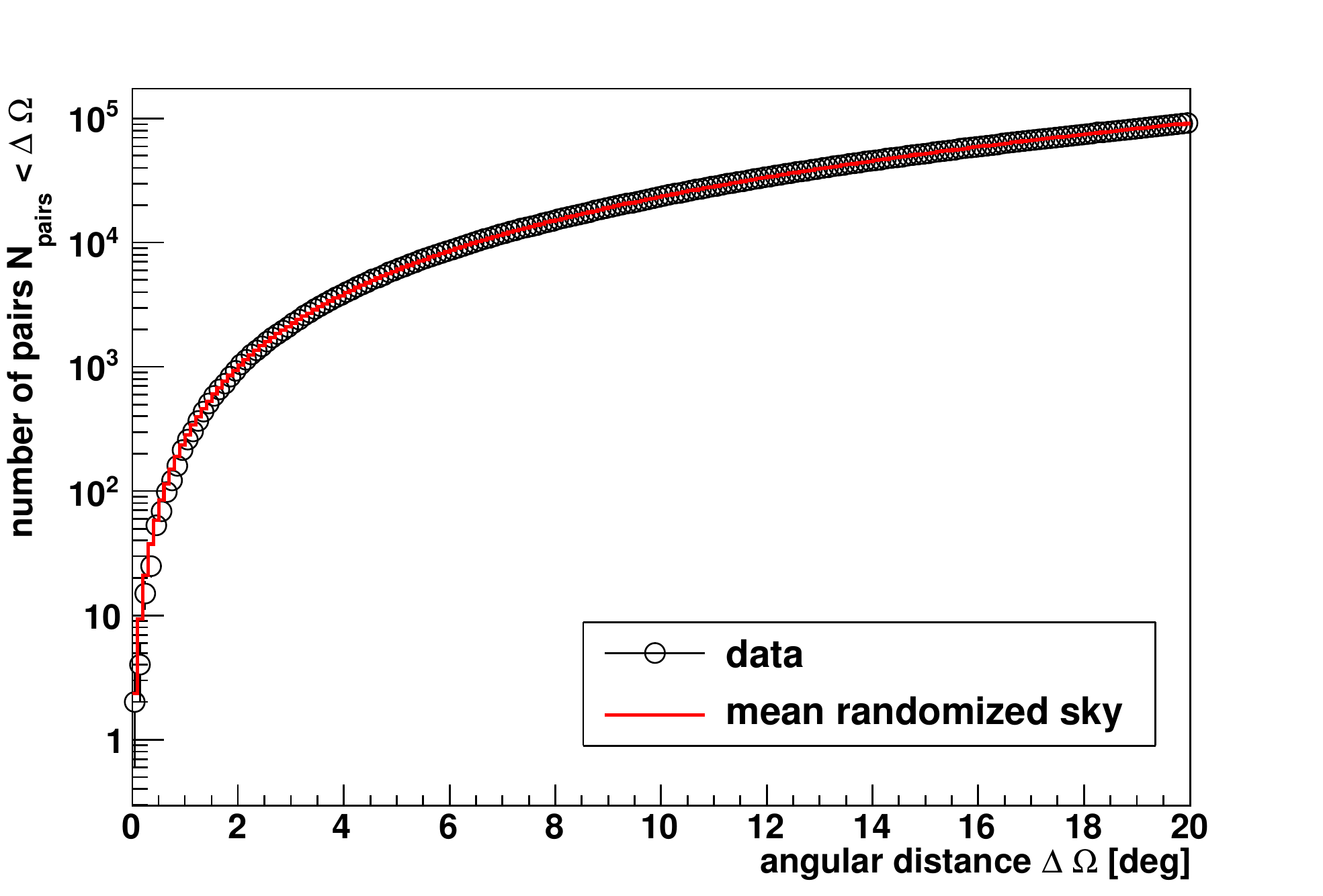}
              \hfill
               \includegraphics[width=3.4in]{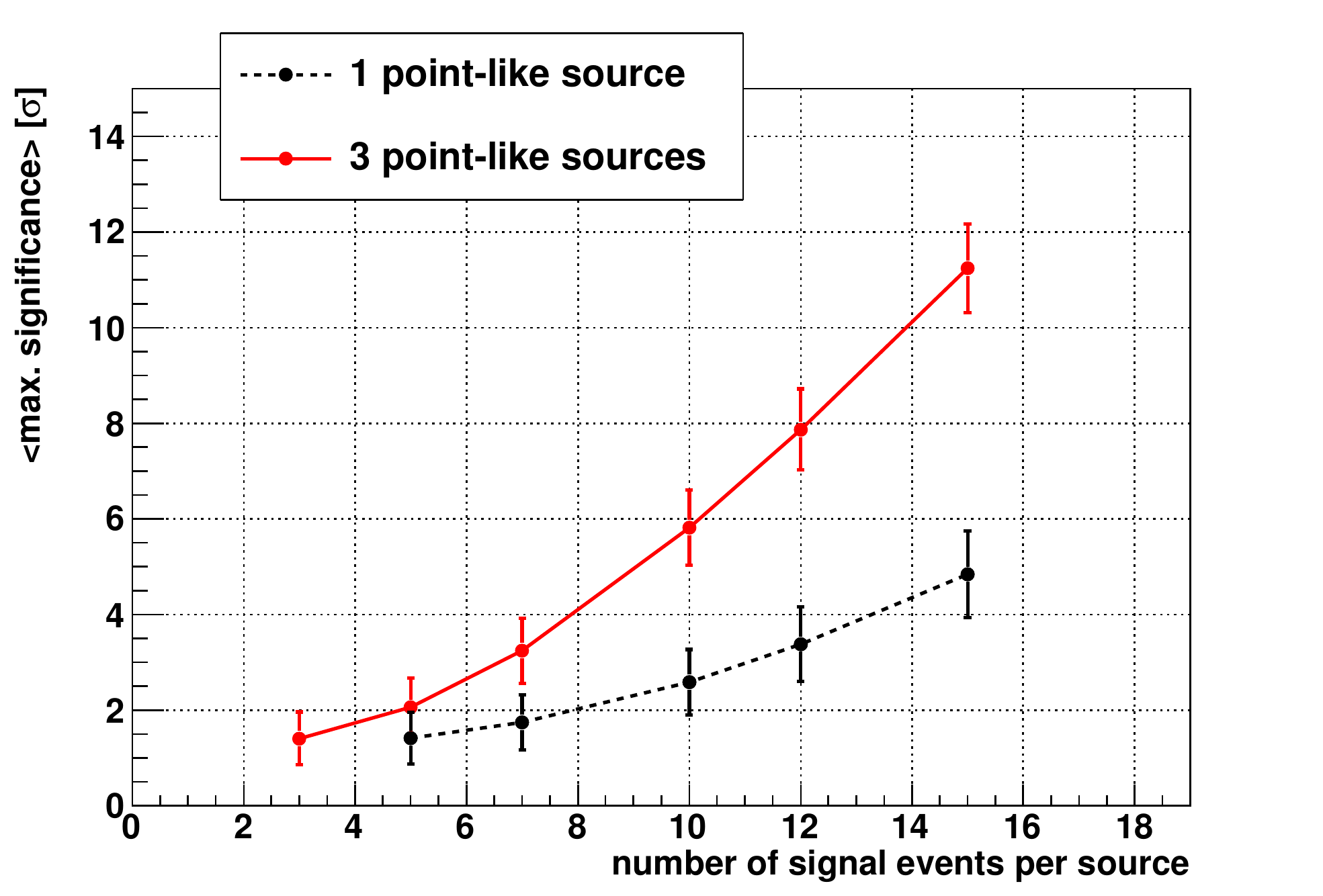}
             }
   \caption{Left: Cumulative autocorrelation distribution of the selected dataset (open markers) shown in comparison with the reference distribution (red solid line). Right: Performance of the algorithm to detect a single or multiple point-like sources.}
   \label{fig:autocorrelation}
 \end{figure*}
  
\subsection{Data set}
The analysed data set has been recorded by the ANTARES neutrino telescope in 2007 and 2008. During this period the detector was in its construction phase and has been operated in various setups ranging from 5 active lines at the beginning of 2007 to a fully operational detector of 12 lines since mid 2008. After applying a run selection removing for example periods without precise detector alignment information, the dataset corresponds to about 300 effective days. Comprising mainly atmospheric muons, about 100 million events were reconstructed with the standard ANTARES reconstruction algorithm. Basic selection criteria include a cut on the reconstructed zenith angle $\theta$ to ensure that only upgoing muon tracks are selected ($\cos(\theta) > 0$) and a cut on the angular uncertainty $\beta$ given by the covariance matrix of the final likelihood fit ($\beta < 1^\circ$). The final selection criteria is a cut on the fit quality parameter $\Lambda$, which is derived from the value of the maximal likelihood itself. Before unblinding the data, this cut has been optimized by means of MC simulations to yield the best average upper limit on the neutrino flux in the search for point like sources~\cite{Antares_PointSources2011, ICRC11_PointSources}. 2190 events pass the final criterion $\Lambda > -5.4$. The $\Lambda$ and zenith angle distributions of events passing all quality criteria (except the ones shown in the plot) are shown in figure~\ref{fig:eventSelection}.
 
Following eq.~\ref{equ:autocor}, the cumulative autocorrelation distribution of the selected events has been determined. It is shown in the left plot of figure~\ref{fig:autocorrelation}.

\subsection{Reference autocorrelation distribution}
To detect structures in the sky distribution of the analysed events we need a reference autocorrelation distribution to compare with. This reference has been determined by scrambling the data themselves, a method which allows to avoid systematic uncertainties introduced by the use of Monte Carlo simulations. The scrambling method uses the local coordinates (zenith and azimuth) and the detection time $T_{i}$ of all selected data events. While keeping the pairs of zenith/azimuth for all events in order to avoid losing information about possible correlations between them, the detection time is drawn randomly from another event within the same detector configuration in order to keep track of the changing asymmetry of the detector. Using all selected events, a randomized sky map with the same number of events as in the data and naturally the same sky coverage is constructed. This randomized sky is then analysed in exactly the same way as the data to derive the autocorrelation function. The randomization process is performed about $10^6$ times and the derived autocorrelation distributions are averaged in order to suppress statistical fluctuations. The resulting reference distribution is shown as a red dotted line in the left plot of figure~\ref{fig:autocorrelation}.

\subsection{Comparison between data and reference}
Structures in the sky distribution of our data will show up as differences between the autocorrelation distributions of the data and the reference distribution. The comparison between them is performed by using the formalism introduced by Li\&Ma~\cite{LiMa}. This formalism results in the raw significances of the differences as a function of the cumulative angular scale which is shown in figure~\ref{fig:result}. As the comparison is performed bin-by-bin and as we scan over different angular scales, this result has to be corrected for the corresponding trial factor. We apply the method proposed by Finley and Westerhoff~\cite{FinleyWesterhoff2004} and perform about $10^5$ pseudo experiments in which the autocorrelation distributions of randomized sky maps are compared with the reference distribution. The probability to obtain the same or higher significance as the maximum deviation observed in the data is calculated and given as final p-value of the analysis.

\begin{figure*}[!th]
  \centering
  \includegraphics[width=5in]{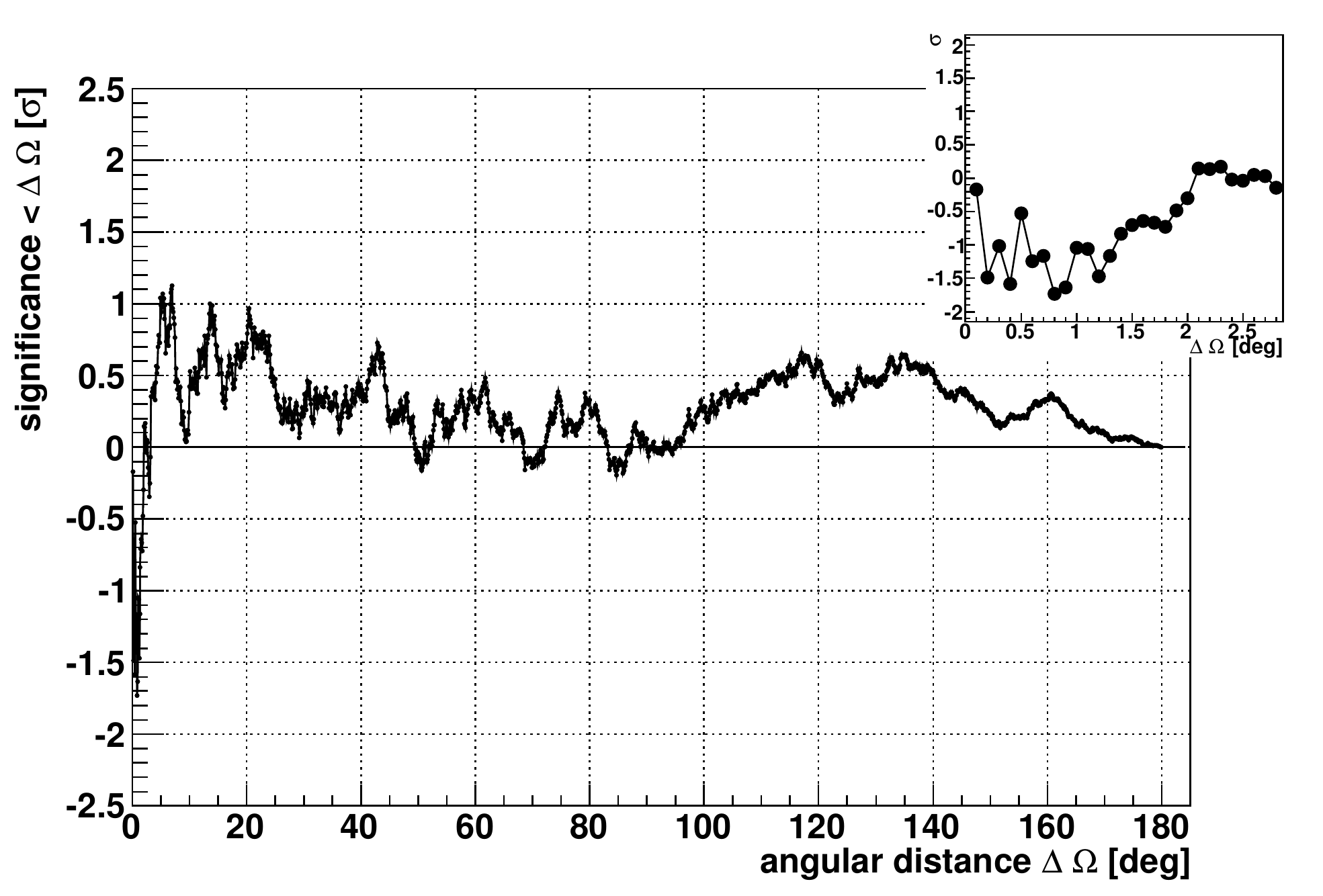}
  \caption{The significance of intrinsic clustering of data taken with the ANTARES neutrino telescope in 2007/2008. The trial factor is not corrected for. The inset shows an enlarged view of the significance for small angular distances.}
  \label{fig:result}
 \end{figure*}
 
\subsection{Performance and sensitivity}
The performance of the algorithm has been determined using mock data sets for which we scrambled the selected data events as described above. While keeping the total number of events in the toy model constant and taking into account the angular resolution, we added predefined source structures with various sizes and source luminosities. These mock data sets where then analysed in exactly the same way as described above. As can be seen in the right plot of figure~\ref{fig:autocorrelation}, the algorithm is sensitive enough to obtain a $3\sigma$ evidence in the exemplary case of 3 point like sources emitting each about 6 events. This source luminosity is at the detection threshold of the dedicated search for a point like excess in the same dataset~\cite{Antares_PointSources2011, ICRC11_PointSources}, which underlines the complementarity of the two methods. 

An important free parameter of the analysis is the binning of the autocorrelation distribution as it will determine the sensitivity to certain angular scales and which is connected to the angular resolution. For the used quality selection an average angular resolution of $0.5^\circ$ has been determined from Monte Carlo simulations. Using toy simulations with various source scenarios an optimal binning of $0.1^\circ$ has been determined.

\subsection{Results and discussion}
The described analysis has been applied to the 2090 selected data events recorded by the ANTARES neutrino telescope between 2007 and 2008. The uncorrected significance as a function of the cumulative angular scale is shown in figure~\ref{fig:result}. A maximum deviation between the data and the reference distribution of $1.1~\sigma$ is found for an angular scale $<7^\circ$. Correcting for the scanning trial factor this corresponds to a p-value of $55~\%$ and is therefore not significant.

In the search for the sources of high energy cosmic rays, the detection of astrophysical sources of neutrinos may play a crucial role. Various experiments are currently taking data or are in a preparatory phase to achieve this goal and the recorded data is scrutinized in numerous ways in order to extract a maximum of information. We presented here the first search for intrinsic clustering of data recorded with the ANTARES neutrino telescope. The data, taken during the deployment phase of the detector, do not show evidence for deviations from the isotropic arrival direction distribution expected for the background of atmospheric neutrinos and contamination by mis-reconstructed atmospheric muons.

\vspace{\baselineskip}

\clearpage

\setcounter{figure}{0}
\setcounter{table}{0}
\setcounter{footnote}{0}
\setcounter{section}{0}
\newpage




\title{Search for a diffuse flux of high-energy muon neutrinos with
   the ANTARES neutrino telescope}

\shorttitle{F. Sch\"ussler \etal Antares diffuse flux}

\authors{Fabian Sch\"ussler$^{1}$ on behalf of the ANTARES collaboration}
\afiliations{$^1$ Commissariat \`a l'\'energie atomique et aux \'energies alternatives\\ Institut de recherche sur les lois fondamentales de l'Univers\\
91191 Gif-sur-Yvette Cedex, France}
\email{fabian.schussler@cea.fr}

\maketitle
\begin{abstract}
We present the search for the diffuse flux of astrophysical muon neutrinos using data collected by the ANTARES neutrino telescope. We introduce a novel method to estimate the energy of high-energy muons traversing the ANTARES detector and discuss detailed comparisons between data and Monte Carlo simulations. Using data recorded in 2008 and 2009 a search for a high-energy excess over the expected atmospheric neutrino background is presented and stringent limits on the diffuse flux of astrophysical muon neutrinos in the energy range 20 TeV - 2.5 PeV are derived.
\end{abstract}



\section{Introduction}
Despite enormous efforts throughout the last century, the mystery of the origin of high-energy cosmic rays remains unsolved. Over the last years it became more and more obvious that multiple messengers will be needed to achieve this task. Fortunately fundamental particle physics processes like the production and subsequent decay of pions in interactions of high-energy particles predict clear links between high-energy cosmic rays and high-energy neutrinos as well as gamma rays. The detection of astrophysical neutrinos and the identification of their sources is one of the main aims of large neutrino telescopes operated in ice at the South Pole (IceCube) and in water at Lake Baikal and in the Mediterranean Sea (ANTARES). 
 
\subsection{The ANTARES neutrino telescope}
The ANTARES detector~\cite{Antares_DetectorPaper} became fully equipped and operational in 2008. The detector is composed of 12 detection lines placed at a depth of 2475m off the French coast near Toulon. The detector lines are arranged on the seabed in an octagonal configuration, covering a base of $180 \times 180~\mathrm{m}^2$ and are about 450m high. They hold a total of 885 optical modules (OM), 17'' glass spheres housing each a 10'' photomultiplier tube. The OMs look downward at $45^\circ$ in order to optimize the detection of upgoing, i.e. neutrino induced, tracks. The geometry and size of the detector make it sensitive to neutrinos in the TeV-PeV energy range. A schematic layout is shown in Fig.~\ref{fig:layout}.\\

\begin{figure}[!t]
  \vspace{5mm}
  \centering
  \includegraphics[width=2.7in]{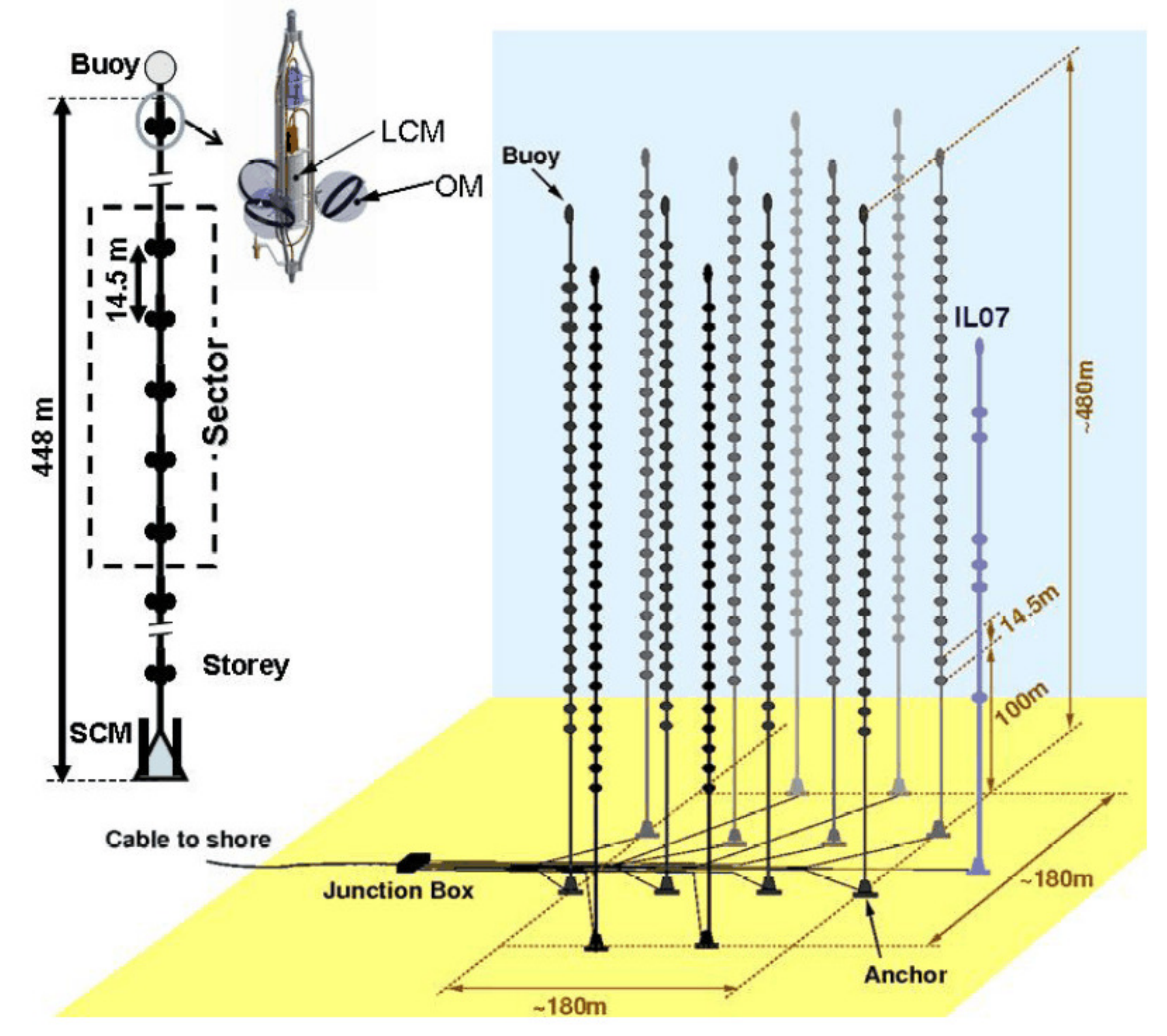}
  \caption{Schematic view of the ANTARES detector.}
\label{fig:layout}	
 \end{figure}
  \begin{figure*}[!t]
   \centerline{\includegraphics[width=3.in]{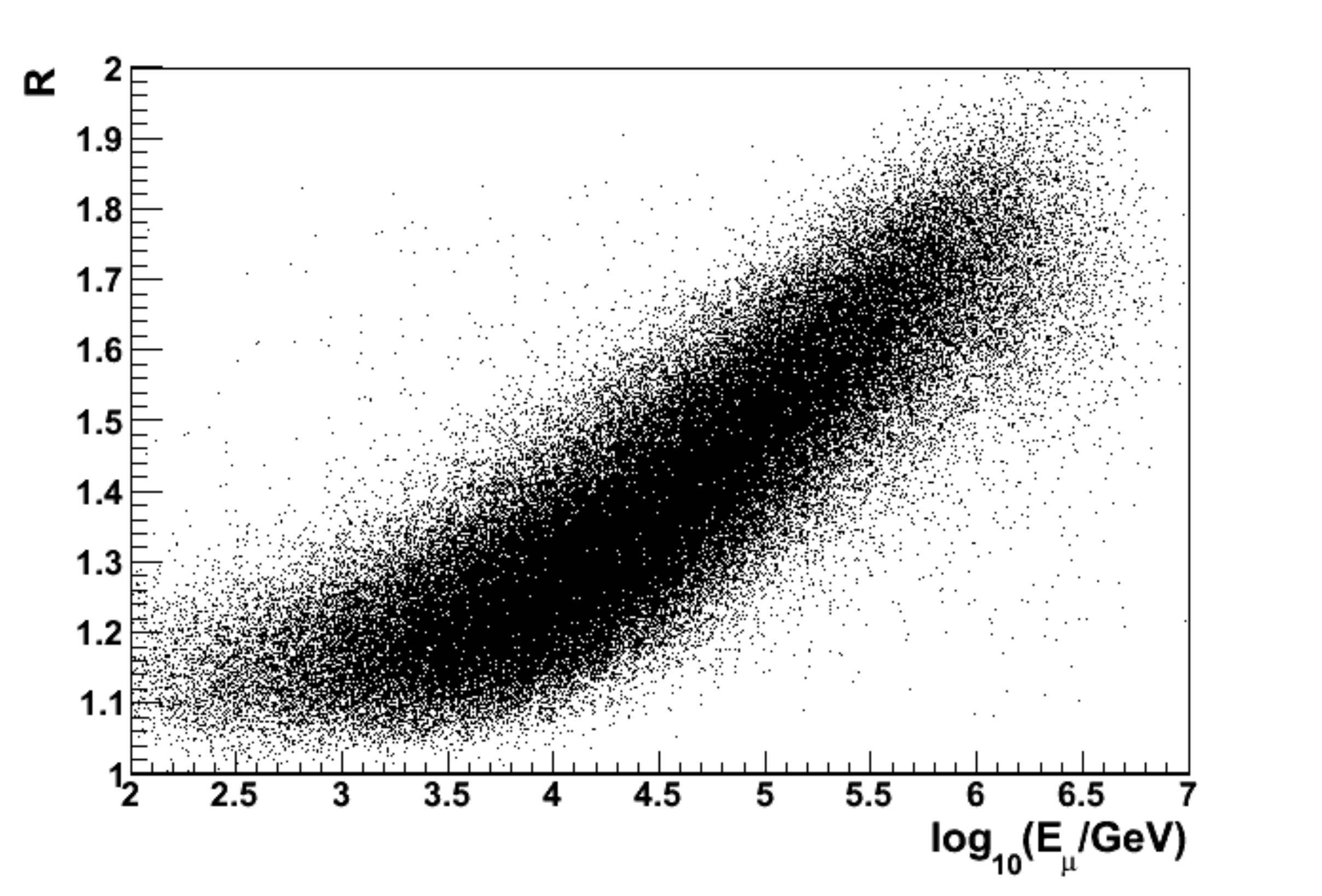} 
            \hspace*{1cm}
           \includegraphics[width=3.in]{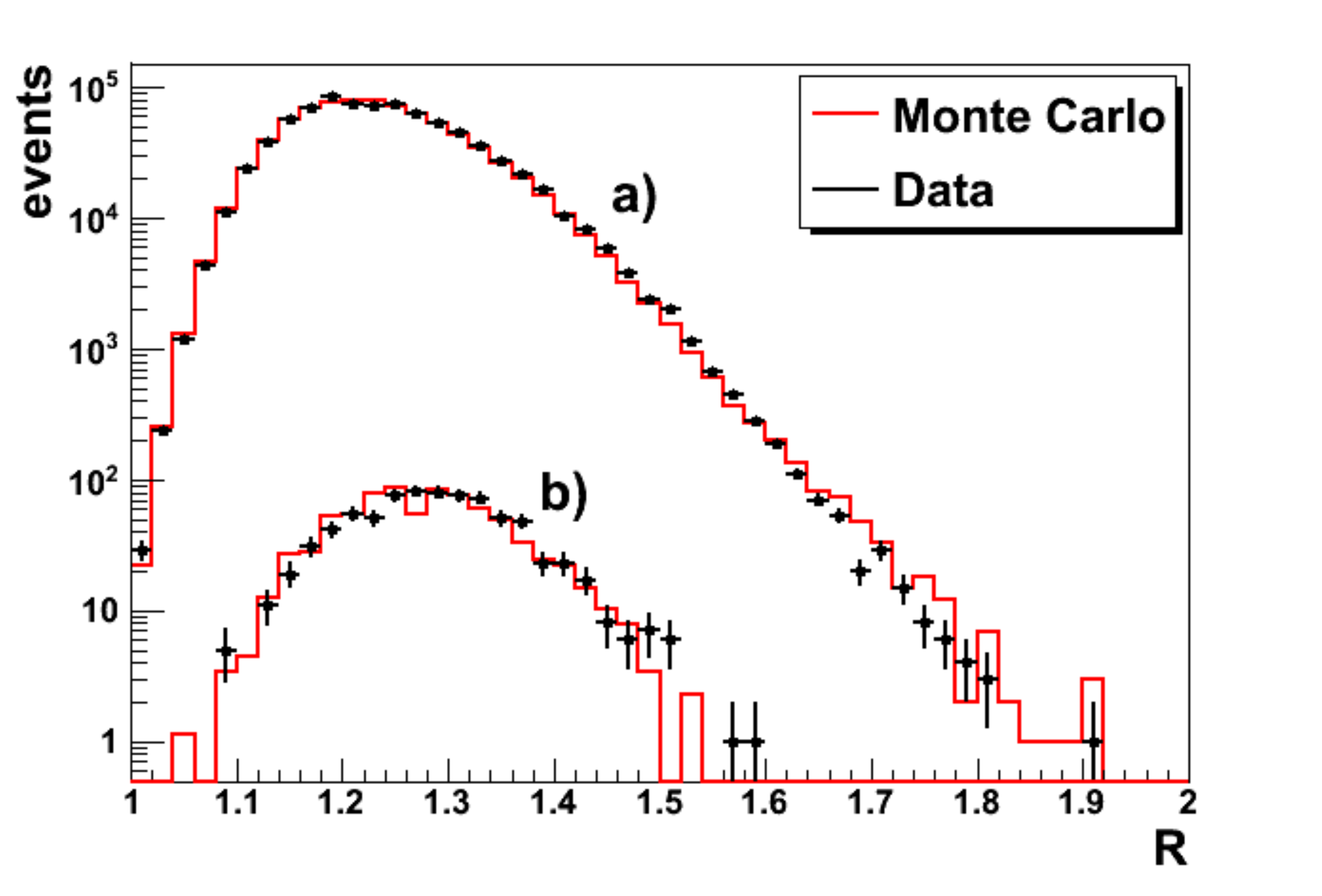}
             }
   \caption{Left plot: Correlation between the energy estimator $R$ and the generated muon energy after all quality cuts. Right plot: Distribution of the R energy estimator of data (markers) and simulated atmospheric muons normalized to the data (histograms) applying only loose quality cuts. Good agreement of the shape is found for both downgoing muons (a) and events mis-reconstructed as upgoing (b).}\label{fig:EnergyEstimator}
 \end{figure*}
 
\subsection{Neutrino detection and dataset}
The neutrino detection relies on the emission of Cherenkov light by high-energy muons originating from charged current neutrino interactions near and inside the instrumented volume. All detected signals are transmitted via an optical cable to a shore station, where a farm of CPUs filters the data for coincident signals or {\it hits} in several adjacent OMs. The muon direction is then determined by maximising a likelihood which compares the times of the hits with the expectation from the Cherenkov signal of a muon track.\\

We analyse here data taken with the ANTARES detector between 12/2007 and 12/2009. Related to the construction and maintenance efforts, this period includes data from a detector comprised of 9, 10 and 12 active detection lines. Data runs were selected according to a set of basic quality criteria, which require for example low environmental background noise. The selection corresponds to a total live time of 334 days (136 days with 9 lines, 128 days with 10 lines and 70 days with 12 lines).

\section{Diffuse astrophysical neutrino flux}
The measured flux of high-energy cosmic rays has been used to derive upper bounds for the expected diffuse neutrino flux~\cite{WaxmannBahcall, MPR2000}. For the TeV to PeV energy range considered here, this flux is typically assumed to originate from particle interactions at or close to the cosmic ray acceleration sites. Although only weakly constrained, the neutrino energy spectrum is typically modelled by a simple $E^{-2}$ power law.

\subsection{Background discrimination}
Two main backgrounds for the measurement of the flux of these astrophysical neutrinos can be identified: downgoing atmospheric muons which have been mis-reconstructed as upgoing and atmospheric neutrinos originating in cosmic ray induced air showers at the opposite side of the Earth. Both backgrounds can at least partially be discriminated using various parameters like the quality of the event reconstruction or an estimator for the energy of the muon. To optimize the selection criteria detailed Monte Carlo (MC) simulations have been used. The atmospheric muon flux has been simulated with the MUPAGE package~\cite{MUPAGE}. Generated atmospheric neutrinos are weighted corresponding to the 'Bartol' parametrisation~\cite{BartolFlux}. Due to the lack of information on the production of charm mesons in high-energy hadronic interactions, the presence of an additional component at high energies (above $\sim 10~\mathrm{TeV}$) is possible. Among the models considered in~\cite{prompt} the Recombination Quark Parton Model (RQPM) was used. It gives the largest 'prompt' contribution to the atmospheric neutrino flux. Both event types, atmospheric muons and neutrinos, are processed with the full ANTARES detector simulation and reconstruction chain. Special care has been taken to reproduce the changing detector configuration during the analysed data taking period and the details of the data acquisition by including for example afterpulses in the PMT simulations. The simulated instrumental and environmental background noise  has been extracted for each of the detector configurations from a representative real data-taking run.

  \begin{figure*}[!t]
   \centerline{\includegraphics[width=3.in]{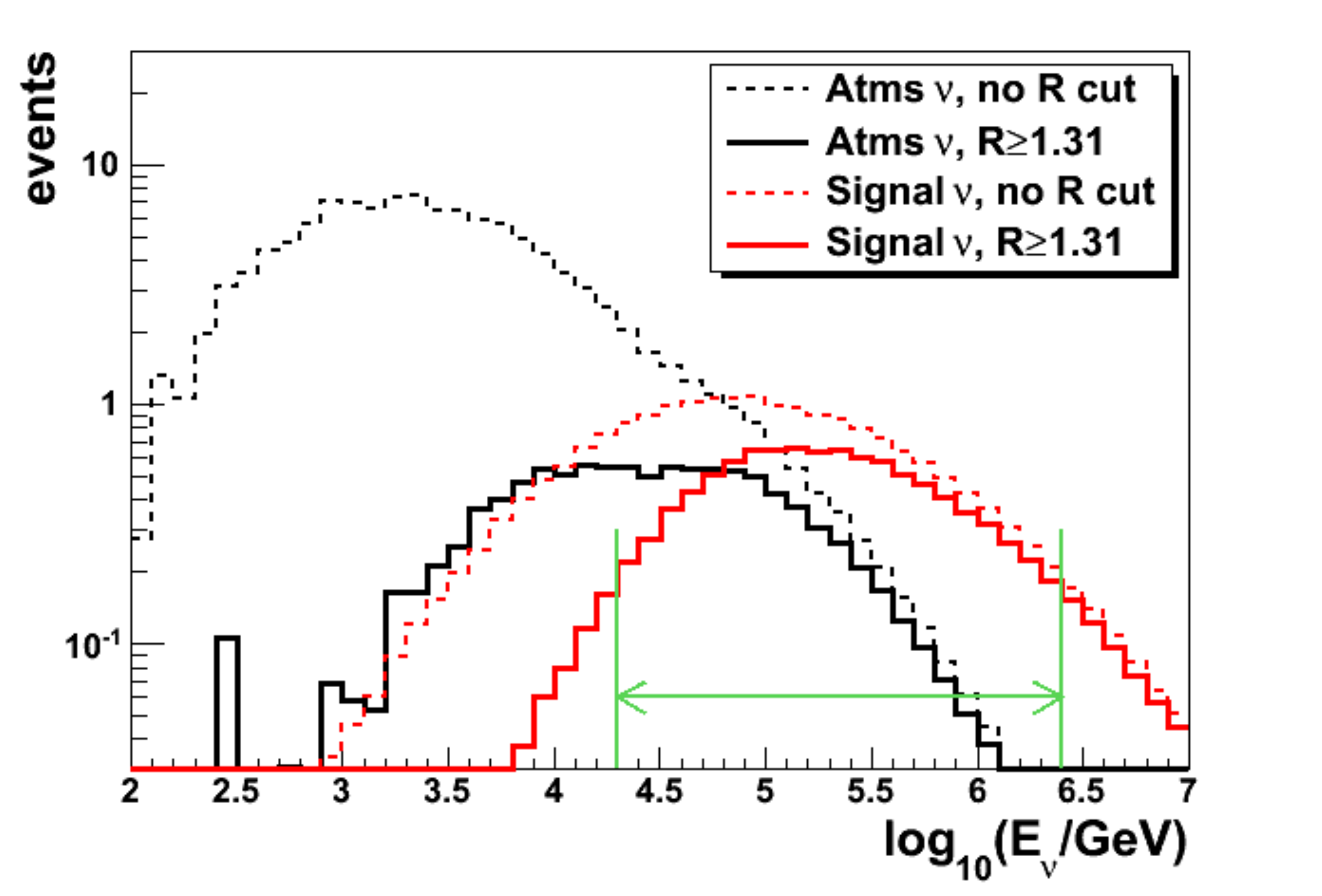}
            \hspace*{1cm}
              \includegraphics[width=3.in]{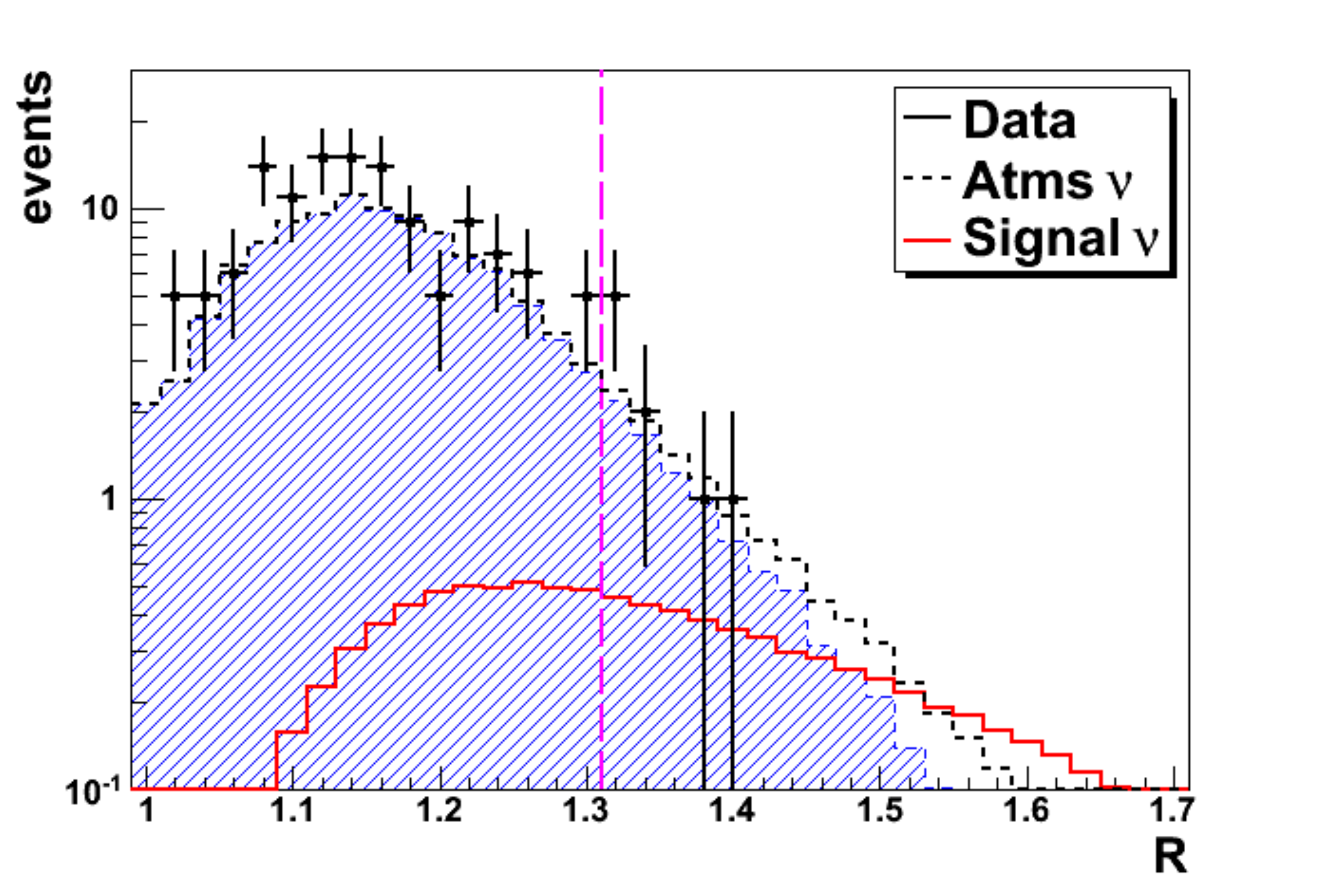} 
             }
   \caption{Left plot: Energy distribution for the Bartol+RQPM atmospheric neutrino flux and a $E^{-2}$ astrophysical signal with arbitrary normalisation after all quality cuts. The horizontal arrow denotes the interval in which $90\%$ of the signal events is expected. Right plot: Distribution of the energy estimator $R$ for data (markers), the Bartol atmospheric neutrino flux (filled histogram) and the 'prompt' contribution (RQPM model, dashed histogram). The signal at the level of the derived upper limit is shown as the full, red line together with the cut at $R>1.31$ (vertical, dashed line).}\label{fig:REstimator}
   
 \end{figure*}
\subsection{Atmospheric muon rejection}
Atmospheric muons are recorded with the ANTARES detector at a rate of several Hz and dominate the detector trigger rate. To remove a large majority of them from the dataset the selected events have to fulfil the following basic quality criteria:
\begin{itemize}
\item detection with at least two detector lines
\item more than 60 hits available for the reconstruction
\item reconstructed zenith angle $\theta < 80^\circ$, i.e. upgoing tracks
\end{itemize}
To fully suppress mis-reconstructed atmospheric muons, an additional 2-dimensional cut has been derived. It combines the quality parameter $\Lambda$ which is derived from the likelihood value of the track fitting algorithm and the number of hits used in the fitting procedure $N_{\mathrm{hit}}$. The events have to pass the selection
\begin{displaymath}
\Lambda > \left\lbrace
\begin{array}{ll}
-4.59 - 5.88 \cdot  10^{-3} \; N_\mathrm{hit} \quad \mathrm{for} \quad N_\mathrm{hit} \leq 172 \\
- 5.60 \qquad \qquad \quad \quad \quad \quad \;\;\;\; \mathrm{for}\quad N_\mathrm{hit} > 172
\end{array}
\right.
\end{displaymath}

Applied to Monte Carlo simulations, these cuts completely remove the $2\cdot 10^{8}$ reconstructed tracks induced by atmospheric muons and reduce the contribution from atmospheric neutrinos (Bartol + RQPM) from $7\cdot 10^{3}$ to 116 events~\cite{SimoneBiagi_CRIS2010}. 

\subsection{Energy estimator}
As the flux of astrophysical neutrinos is expected to follow a harder spectrum ($\propto E^{-2}$) than that of the atmospheric neutrino background it should become visible as an excess of high-energy events. This discrimination requires an estimation of the neutrino, or as best approximation, the muon energy. Various energy estimators are under study within the ANTARES collaboration~\cite{ICRC11_EnergyEstimator}. Here we exploit the structure of the arrival times of photons created along the muon track at the OMs. This time structure is sensitive to the energy as higher energy muons have a higher probability to create electromagnetic showers along the track. The light emitted by these showers is responsible for delayed hits in the OMs with respect to the detected Cherenkov photons. A robust parameter sensitive to delayed photons is the mean number of hit repetitions $R$ within an event. It is calculated by averaging the number of hits $R_i$ recorded by an OM within a $500~\mathrm{ns}$ time window over all OMs contributing to the reconstructed muon track. The clear correlation with muon energy is shown for simulated events in Fig.~\ref{fig:EnergyEstimator}, left plot. The estimator has been extensively studied both on MC and on (atmospheric muon dominated) data (e.g. Fig.~\ref{fig:EnergyEstimator}, right plot) and an average HWHM resolution after the discussed quality criteria of $\log(E_\mathrm{rec}/E_\mathrm{true}) = 0.4$ could be determined.

\begin{figure*}[!th]
  \centering
  \includegraphics[width=5in]{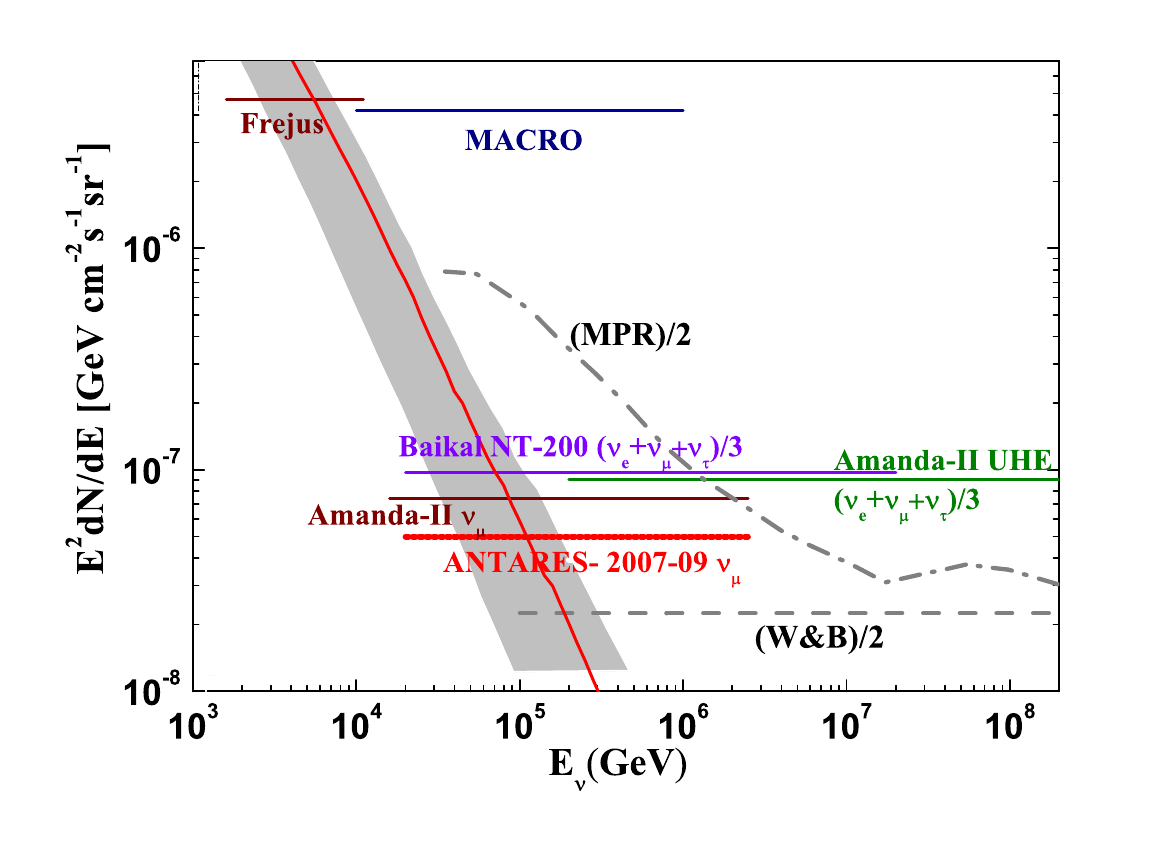}
  \caption{The upper limit for a $E^{-2}$ diffuse high-energy $\nu_\mu + \bar{\nu}_\mu$ flux derived from data of the ANTARES neutrino telescope compared to previous measurements and phenomenological upper bound predictions.}
  \label{fig:result}
 \end{figure*}
 
\subsection{Optimisation of the event selection}

The final discrimination between the atmospheric neutrino background and the astrophysical signal is achieved by a cut in the $R$ variable. This cut has been optimized before un-blinding the data by minimizing the Model Rejection Factor (MRF)~\cite{MRF}. Taking into account all possible fluctuations of the number of background events $n_\mathrm{B}$, the average upper limit $\bar{\mu}_{90\%} (n_\mathrm{B})$ of background-only pseudo-experiments is derived. The optimal value of the cut in $R$ is then determined by minimizing the MRF given as $\bar{\mu}_{90\%}(n_\mathrm{B})/n_\mathrm{S}$, where $n_\mathrm{S}$ is the number of expected signal events from an $E^{-2}$ test spectrum. The optimum has been found when selecting events above $R_\mathrm{cut}=1.31$. 

The determination of the energy estimator cut allows to define the energy range to which the analysis will be sensitive. We define this range as the interval containing $90\%$ of the signal events and obtain $20~\mathrm{TeV} < E_\nu < 2.5~\mathrm{PeV}$ (see Fig.~\ref{fig:REstimator}, left plot).

\subsection{Low energy region and systematic uncertainties}
Before un-blinding the data in the defined region $R\,>\,1.31$ the distribution of the selected data events below this cut has been compared to the corresponding Monte Carlo simulations. Whereas $125$ events are selected from data, the Bartol flux parametrisation combined with the RQPM model for the prompt contribution predicts $105$ events. The observed difference is well within the systematic uncertainty of the atmospheric neutrino flux given as $25-30\%$~\cite{AtmoNuUncertainty} and has been corrected before un-blinding the high-energy region by applying a scale factor $k=105/125= 1.19$ to the MC predictions. 

Further systematic uncertainties on the number of expected events in the high-energy signal region include 
\begin{itemize}
\item the contribution of prompt events derived as maximal deviation between the models discussed in~\cite{prompt} of ${}^{+1.7}_{-0.3}$ events. The maximal value ($1.7$ events) is used in the following.
\item uncertainties in the spectral shape of the atmospheric neutrino flux which is related to uncertainties in the primary cosmic ray energy spectrum. By varying the spectral index of the neutrino spectrum by $\pm 0.1$ independently below and above $\approx 10~\mathrm{TeV}$ we obtained a sys. uncertainty of $\pm 1.1$ events.
\item uncertainties in the details of the description of the detector and environmental parameters like the angular acceptance of the OMs, details of PMT afterpulses and water properties like absorption and scattering lengths which lead to a total uncertainty of $5\%$.
\end{itemize}

\section{Results}
Applying the discussed quality and energy selection criteria to the simulated dataset gives a background estimation of $n_\mathrm{B}=10.7 \pm 2$ events in the high-energy region. In the analysed dataset of the ANTARES detector $9$ events were selected. Including the systematic uncertainty on the background expectations following the method discussed in~\cite{Pole} we derive the $90\%$ c.l.~upper limit on the number of signal events as $\mu_{90\%}(n_\mathrm{B})=5.7$. The corresponding upper limit on the neutrino flux is given by $\phi_{90\%} = \phi \cdot \mu_{90\%}/n_\mathrm{S}$, where $n_\mathrm{S}$ is the number of events expected from the flux $\phi$. We obtain~\cite{DiffuseFluxPLB}:
\begin{equation}
E^{2} \phi_{90\%} = 5.3 \times 10^{-8}~\mathrm{GeV}^{-2}\mathrm{s}^{-1}\mathrm{sr}^{-1}
\end{equation}

As can be seen in Fig.~\ref{fig:result}, the derived limit is competitive with previous results and further constrains models of the diffuse flux high-energy $\nu_\mu + \bar{\nu}_\mu$ flux.

\vspace{\baselineskip}

\clearpage

\setcounter{figure}{0}
\setcounter{table}{0}
\setcounter{footnote}{0}
\setcounter{section}{0}
\newpage





\title{Muon energy reconstruction and atmospheric neutrino spectrum unfolding with the ANTARES detector}

\shorttitle{D. Palioselitis \etal Energy reconstruction and atmospheric neutrinos with ANTARES}

\authors{D. Palioselitis$^{1}$ on behalf of the ANTARES collaboration}
\afiliations{$^1$Nikhef (Nationaal instituut voor subatomaire fysica), Science Park 105, 1098XG Amsterdam, The Netherlands}
\email{dimp@nikhef.nl}

\maketitle
\begin{abstract}
Measurement of the atmospheric muon neutrino spectrum is important since it is a test on the expected conventional atmospheric neutrino fluxes as well as on the prompt contribution from charm decay. In addition to that, it forms the irreducible background for high energy cosmic neutrinos searches. A number of methods have been developed within the ANTARES collaboration for the estimation of the energy of muon tracks traversing the detector. These include a maximum likelihood method, an approach based on neural networks as well as an estimator based on the energy loss of the muon along its track. The reconstructed muon energy can be used to calculate the energy spectrum of the atmospheric neutrinos by using a singular value decomposition approach on regularized unfolding. The most recent results on the energy reconstruction and spectrum deconvolution will be presented.
\end{abstract}


\section{Introduction}

The ANTARES detector is a high energy neutrino telescope deployed in the Mediterranean sea, 40\,km off the coast of Toulon in France. It consists of a three dimensional array of photomultiplier tubes designed to detect light produced by charged particles crossing the instrumented volume. Its main scientific goal is the detection of astrophysical high energy neutrinos. 

When a neutrino interacts with a nucleon in the area close to or inside the instrumented volume, it produces a detectable signature. ANTARES is optimized to detect muons produced by the charged current neutrino interaction with nuclei in the surrounding medium. The detection principle is based on the \v{C}erenkov light  produced by charged particles when they traverse a medium with a velocity higher than the speed of light. This light is detected by the photomultiplier tubes, housed into glass spheres called optical modules (OMs). The OMs record information about the charge and time of each hit. The position of each OM is provided by an acoustic positioning system. This information allows for both directional \cite{lab1, lab2} and energy reconstruction. 

The majority of neutrino candidate events detected with ANTARES is expected to be due to atmospheric neutrinos, coming from the decay of pions and kaons produced by cosmic ray interactions in the atmosphere. Neutrinos produced by this mechanism constitute the so-called ``conventional'' component of the atmospheric neutrino spectrum. At higher energies, the decay lengths of these mesons are longer than their path lengths in the atmosphere thus leading to a suppression of produced neutrinos. The ``prompt'' decay of charmed mesons, produced at high energies, provides an additional contribution to the atmospheric neutrino flux at the higher energies end of the spectrum \cite{lab3}. The prompt neutrino spectrum is less steep than the conventional atmospheric neutrino spectrum. Measurement of the atmospheric neutrino spectrum can therefore provide valuable information on the prompt component which is expected to dominate the atmospheric neutrino flux at energies above $\sim10\,\mathrm{TeV}$. Downgoing atmospheric muons, produced by cosmic rays interactions in the atmosphere constitute an enormous background to neutrino candidate events. This background is reduced by selecting well-reconstructed upgoing events since the muons cannot cross the Earth. The atmospheric neutrinos cannot be rejected this way, resulting in an irreducible background for high energy neutrino searches of galactic or extragalactic origin. Fortunately, a cosmic diffuse neutrino flux is expected to fall as $E_{\nu}^{-2}$ while the atmospheric neutrino spectrum is steeper, proportional to $E_{\nu}^{-3.7}$. This allows the extraction of the cosmic neutrino flux statistically by measuring eventual  changes in the slope of the measured neutrino energy spectrum. The presence of a prompt component can be established in the same way.

In the present paper, three methods developed within the ANTARES collaboration aiming at the reconstruction of the muon energy are presented in section \ref{energy}. At the end of this section their performance is presented and discussed. Section \ref{unfolding} contains a description of the unfolding procedure that will be used to reconstruct the atmospheric neutrino energy spectrum. A test of the method on a neutrino Monte Carlo sample is performed.

\section{Energy reconstruction}\label{energy}

Relativistic muons passing through matter lose energy by means of various processes. The most common one is ionization of water molecules. Above 1\,TeV radiative processes start dominating, leading to an almost linear dependence of the energy loss per unit length on the energy of the muon. Pair production, bremsstrahlung radiation and photonuclear interactions are the processes that are responsible for the increase of energy losses at higher energies. The stochastic nature of these phenomena requires the use of a mean energy loss per unit length approximation described by
\begin{equation}\label{eq:1}
-\frac{dE}{dx}=a(E)+b(E)E.
\end{equation}
The first term in this equation is almost constant and accounts for the ionization energy losses while the second term describes the stochastic energy loss phenomena that dominate the energy losses above $1\,\mathrm{TeV}$. 

The three strategies presented in the following are attempting to reconstruct the muon energy by basically inverting numerically equation \ref{eq:1} and using the amount of light detected by the telescope as an estimate for the muon energy loss. Technically, the first strategy uses a maximum likelihood method, the second a neural network, whereas the third uses only analytical approximations. In addition to these three energy reconstruction strategies, an energy estimator based  on the average number of multiple hits recorded on the same OM during the event has been used for the  diffuse flux search by the ANTARES collaboration~\cite{lab4, lab4b}.
The results presented in this section are expectations from a realistic simulation of the detector, including detailed OM electronics response, photon tracking and optical  background generation.

\subsection{Maximum likelihood method}

The first method presented here is based on a maximum likelihood estimation of the energy of the muon. We define a function which gives the likelihood that the observed time $t$ and amplitude $A$ of each hit on the OMs are the result of a given muon track. The likelihood function is defined as
\begin{equation}\label{lik}
\mathcal{L}(E)=\prod_{i}^{N_{OM}}P_i(E),
\end{equation}
with the product running over all OMs within 300\,m from the track and $P_i(E)$ the probability for individual OMs to record a hit of a certain amplitude or not.
The form of $P_i(E)$ is given by
\begin{eqnarray}
P(A;\left<n\right>)&=&\sum_{\mathrm{n}=1}^{\mathrm{n}_{\mathrm{max}}}P_{p}(n;\left<n\right>)\cdot P_{g}(A;n),\label{hit}\\
P(0;\left<n\right>)&=&e^{-\left<n\right>}+P_{\mathrm{threshold}}(\left<n\right>).\label{nohit}
\end{eqnarray}

Equation \ref{hit} describes the probability of observing a certain amplitude $A$, given that the expected number of photoelectrons is $\left<n\right>$. $P_{p}(n;\left<n\right>)$ is the Poissonian probability of having $n$ photoelectrons given that the expectation is $\left<n\right>$, while the probability of $n$ photoelectrons in the photocathode producing an amplitude $A$ is given by $P_{g}(A;n)$ and is assumed to be a Gaussian. Equation \ref{nohit} represents the Poissonian probability of having no photoelectrons, when the expectation is $\left<n\right>$ photoelectrons, to which the probability $P_{th}$ that the produced photoelectrons give amplitudes below the threshold of the electronics is added. The best energy estimate $\hat{E}$ is found by minimizing the negative logarithm of the likelihood function $\mathcal{L}(E)$. The energy dependence of the likelihood function is included in the expected number of photoelectrons $\left<n\right>(E)$, which is calculated using the probability density functions of the photon arrival times on the OMs. In addition to the muon energy, these density functions depend on the track geometry i.e. distances of each OM to the track and the OM orientation compared to the track. A linear fit is applied to determine the relation between the energy estimate $\hat{E}$ and the true muon energy $E_\mu$. 

\subsection{Artificial neural networks}

A second method used to determine the muon energy is based on neural networks. An artificial neural network (ANN) is a simulated collection of interconnected nodes where each node produces a certain response to a set of input signals. A number of input parameters $x_i$ are given to the input layer of the network. The values of the input parameters are transmitted to the nodes of the next layer, where a connection weight $w_{ij}$ is assigned for each input $x_i$ connected to a node j. The method described here implements a feed-forward neural network i.e. the nodes are grouped in layers and information flows only in one direction, from the input layer to the output node. Three hidden layers are being used in the ANN energy reconstruction method.  The input to the j$^{th}$ node of a layer is  given by
\begin{equation}
 z_j=\sum_iw_{ij}x_i+w_0,
 \end{equation}
where $w_0$ is the bias of the node. The summation is performed on the nodes of the previous layer, and its output is given by an activation function $g_j(z_j)$. The back propagation algorithm is used to determine the weights $w_{ij}$ and the bias $w_0$ that optimize the performance of the neural networks method by minimizing an error function. The training is performed on Monte Carlo simulation samples. The network is a mapping from the space of input variables $x_i$ onto a space of output variables $y_i$. The only output in the present method is the energy of the muon. The space, time and charge distributions of the recorded hits, the number of hits per OM at different distances from the muon track as well as various geometrical parameters such as the track direction and the distance from the track to the center of gravity of the detector are used as input parameters. The optical background rate has also been proven to be a useful input parameter. The ANN energy reconstruction method uses  a total of 56 input parameters.

\subsection{$dE/dx$ estimator}

This method \cite{lab5} is based on the construction of an energy loss $dE/dx$ estimator $\rho$. The estimator is constructed as
\begin{equation}
\rho=\frac{\sum\limits_{i=1}^{N_\mathrm{hits}}A_i}{L_\mu\cdot\epsilon},
\end{equation}
where the sum of all hit amplitudes in the event is divided by the muon path length $L_\mu$ in the detector sensitive volume and the detector acceptance $\epsilon$. The sensitive volume extends 2.5 effective attenuation lengths $\lambda$ away from the instrumented volume. The acceptance $\epsilon$ is given by
\begin{equation}
\epsilon = \frac{1}{N_{OM}}\sum\limits_{j=1}^{N_{OM}}\frac{\alpha(\theta)}{r_j}e^{\frac{-r_j}{\lambda}},
\end{equation}
where $N_{OM}$ is the number of OMs in the detector and $\alpha(\theta)$ is the photomultiplier tube's angular efficiency. The term $\frac{1}{r_j}e^{\frac{-r_j}{\lambda}}$ describes the number of photons reaching the OM after traveling distance $r_j$ in water. The detector acceptance $\epsilon$ measures the fraction of light which can be seen by the OMs. An interpolation between tabulated values of the estimator $\rho$ and the true muon energy $E_\mu$ is used to determine the relation between them.

\subsection{Energy reconstruction performance}

In order to study the performance of an energy reconstruction algorithm, one has to look at the mean and standard deviation of the $\mathrm{log}_{10}\frac{E_{\mathrm{reco}}}{E_{\mathrm{true}}}$ distribution. Since the performance is dependent on the muon energy these distributions are examined as function of the true energy. For all three methods presented here, these distributions are well described by a Gaussian fit. The mean of the Gaussian fit shows how far away is the energy estimate from the true energy while the standard deviation represents the resolution. A mean of the fit very close to zero and a small standard deviation indicate a high performance of the method. The bias and the resolution of the three methods as a function of the muon energy are shown in figures~\ref{mean} and~\ref{sigma}. For this study the true Monte Carlo track has been used, in an attempt to decouple the performance of the energy reconstruction from the quality of the track fit. The overestimation of the true energy by the maximum likelihood and $dE/dx$ methods for lower energies, as shown in figure~\ref{mean}, is due to the fact that for muon energies lower than a few TeV the energy loss is almost constant, making it difficult to distinguish between e.g. a 100\,Gev and a 500\,GeV muon. The optical background from potassium decay in sea water and bioluminescence can affect the performance below the critical energy since these low energy events do not produce sufficient light and the effect of the optical background noise is more pronounced. The resolution of the energy reconstruction methods presented here is below 0.5 in $\mathrm{log}_{10}E\,[GeV]$ at energies above 5-10\,TeV. The maximum likelihood and artificial neural networks methods have a more stable behavior with the neural network approach reaching as low as 0.25 at higher energies. 
 \begin{figure}[!t]
  \vspace{5mm}
  \centering
  \includegraphics[width=0.47\textwidth]{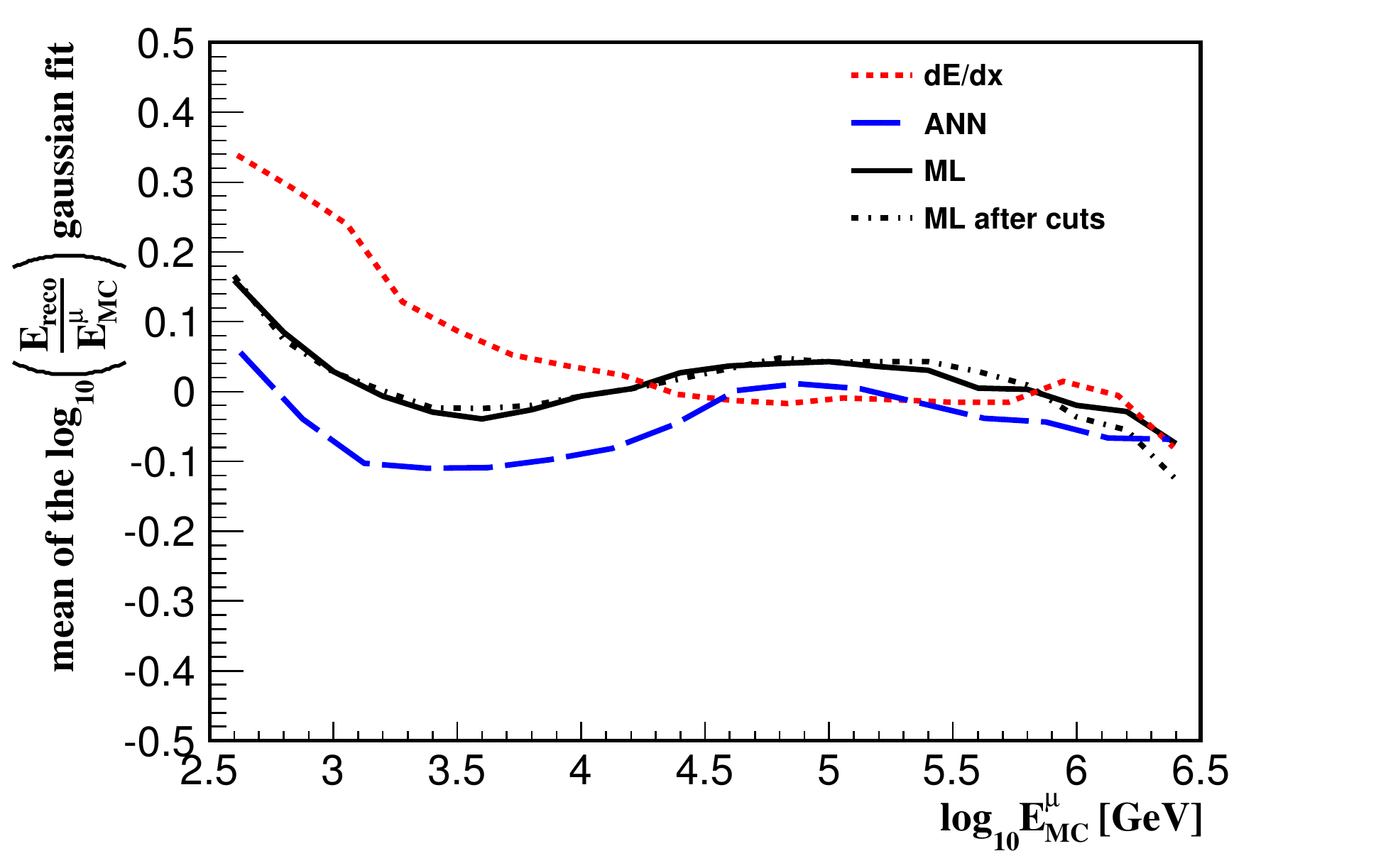}
  \caption{Mean of the gaussian fit on $\mathrm{log}_{10}\frac{E_{reco}}{E_{MC}}$ distribution as a function of the true muon energy for the different methods. A perfectly unbiased energy reconstruction is characterized by zero mean. }
  \label{mean}
 \end{figure}
 \begin{figure}[!t]
  \vspace{5mm}
  \centering
  \includegraphics[width=0.47\textwidth]{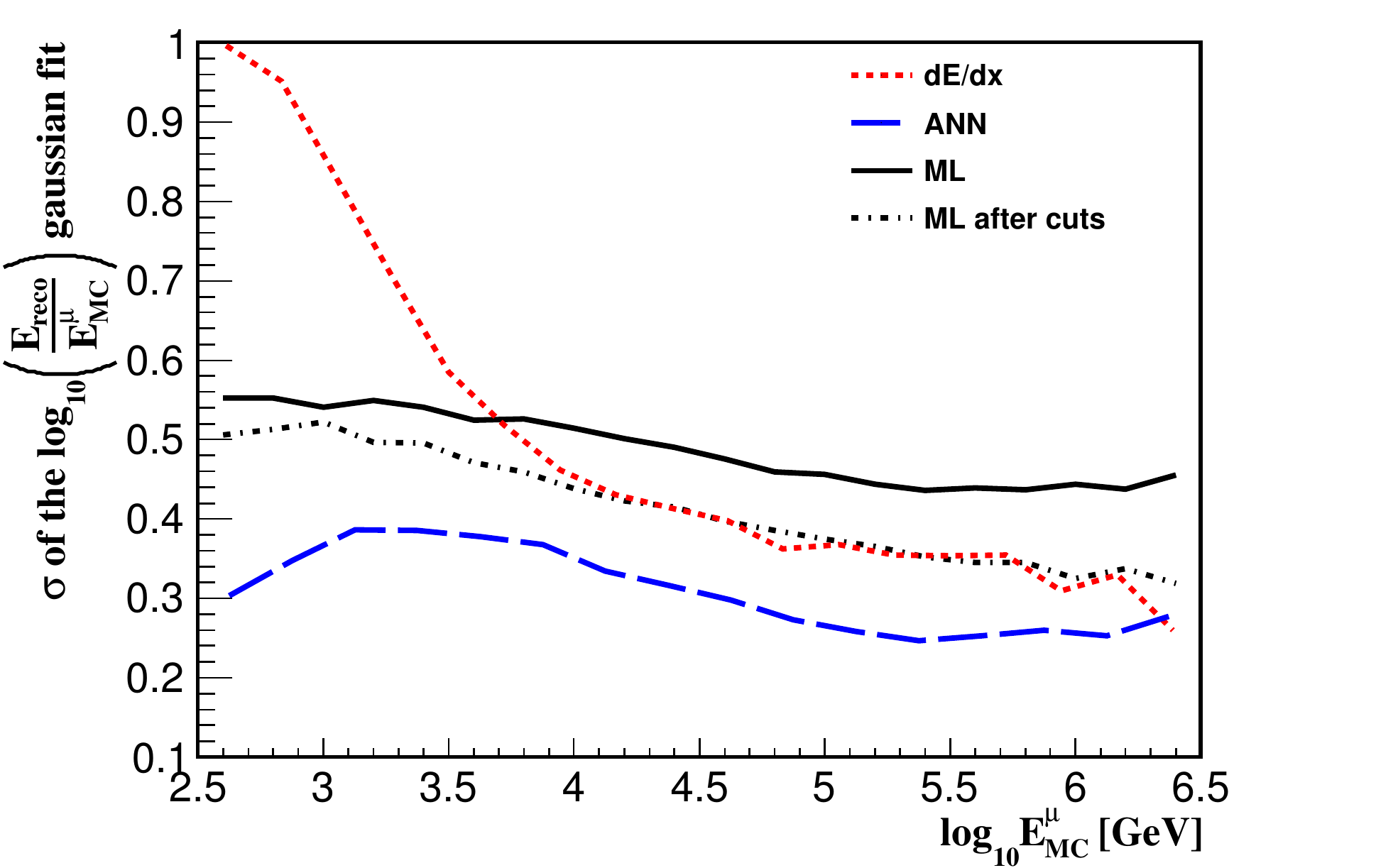}
  \caption{Standard deviation of the gaussian fit on $\mathrm{log}_{10}\frac{E_{reco}}{E_{MC}}$ distribution as a function of the true muon energy for the different methods. A small standard deviation corresponds to a good resolution.}
  \label{sigma}
 \end{figure}

The impact of using the reconstructed track was estimated for the maximum likelihood method, using tracks reconstructed with the standard ANTARES reconstruction strategy \cite{lab2} and selected to increase the population of well reconstructed tracks and minimize the atmospheric muon background events. The selection efficiency, estimated from simulation, varies  between 10\% at 1\,TeV and 20\% at 100\,TeV with almost 99\% purity. The median angular resolution is $0.25^\circ$ above 1\,TeV. The performance of the energy reconstruction using the reconstructed track is shown in figures \ref{mean} and \ref{sigma}, indicated as ``ML after cuts'' in the legend.
\section{Atmospheric neutrino spectrum deconvolution}\label{unfolding}
The atmospheric neutrino spectrum cannot be reconstructed by just assigning each event to the corresponding estimated energy bin. This is due to the limited resolution of the energy reconstruction. The atmospheric neutrino spectrum is steeply falling so events with overestimated energy will introduce a large distortion on the spectrum at high energies. This can be overcome by the use of unfolding techniques~\cite{lab6a}. The unfolding is performed following the method presented in \cite{lab6}, using  the package \textit{RooUnfold}~\cite{lab7}. 

The unfolding problem can be formulated as follows. The true neutrino spectrum $x(E)$ is distorted by a number of different factors. The energy that is measured is only part of the muon energy, depending on the part of its track that is visible by the detector. An additional distortion comes from the fact that the muon carries only a fraction of the parent neutrino energy. Finally, the limited resolution of the energy reconstruction itself contributes further to the distortion of the spectrum. Let $y(E)$ be the vector of the measured spectrum, then in matrix notation the problem is expressed as $y(E)=Rx(y)$. R is called the response matrix and describes the effect of the detector and the reconstruction on the true spectrum. The response matrix is constructed by means of Monte Carlo simulation. A simple matrix inversion turns out to be a naive attempt to solve the problem since statistical fluctuations in the data vector $x(E)$ will give a highly oscillating solution $y(E)$ with large errors. The problem can be located and addressed by performing a singular value decomposition on the matrix $R = USV^T$. This is equivalent to expressing the solution vector as a sum of terms weighted by the inverse singular values of the matrix $S$. The small singular values are responsible for enhancing the statistically insignificant coefficients in the solution. These coefficients can be damped out by imposing an additional constraint on the smoothness of the solution $y(E)$. Since there is no reason to expect abrupt irregularities in the spectrum, the solution is not allowed to exhibit significant bin to bin variations. Various methods exist to decide the amount of regularization one should impose on the solution in order to have the optimal trade off between bias towards the expected solution and the size of the covariance matrix.

The unfolding method was tested using the results from the maximum likelihood energy reconstruction on a realistic 12 line configuration Monte Carlo dataset, equivalent to 3 years of data taking livetime. The response matrix was constructed assuming the atmospheric neutrino flux parametrizations from~\cite{lab9, lab10} while the fluxes used to generate the test data are taken from~\cite{lab11, lab12}. After unfolding to the neutrino spectrum at the detector level, the atmospheric neutrino flux (figure~\ref{spectrum}) can be determined by including the effects of neutrino propagation through the Earth and detector efficiency. The method succeeds in reconstructing the ``true'' Monte Carlo neutrino spectrum.
\begin{figure}[!t]
  \vspace{5mm}
  \centering
  \includegraphics[width=0.48\textwidth]{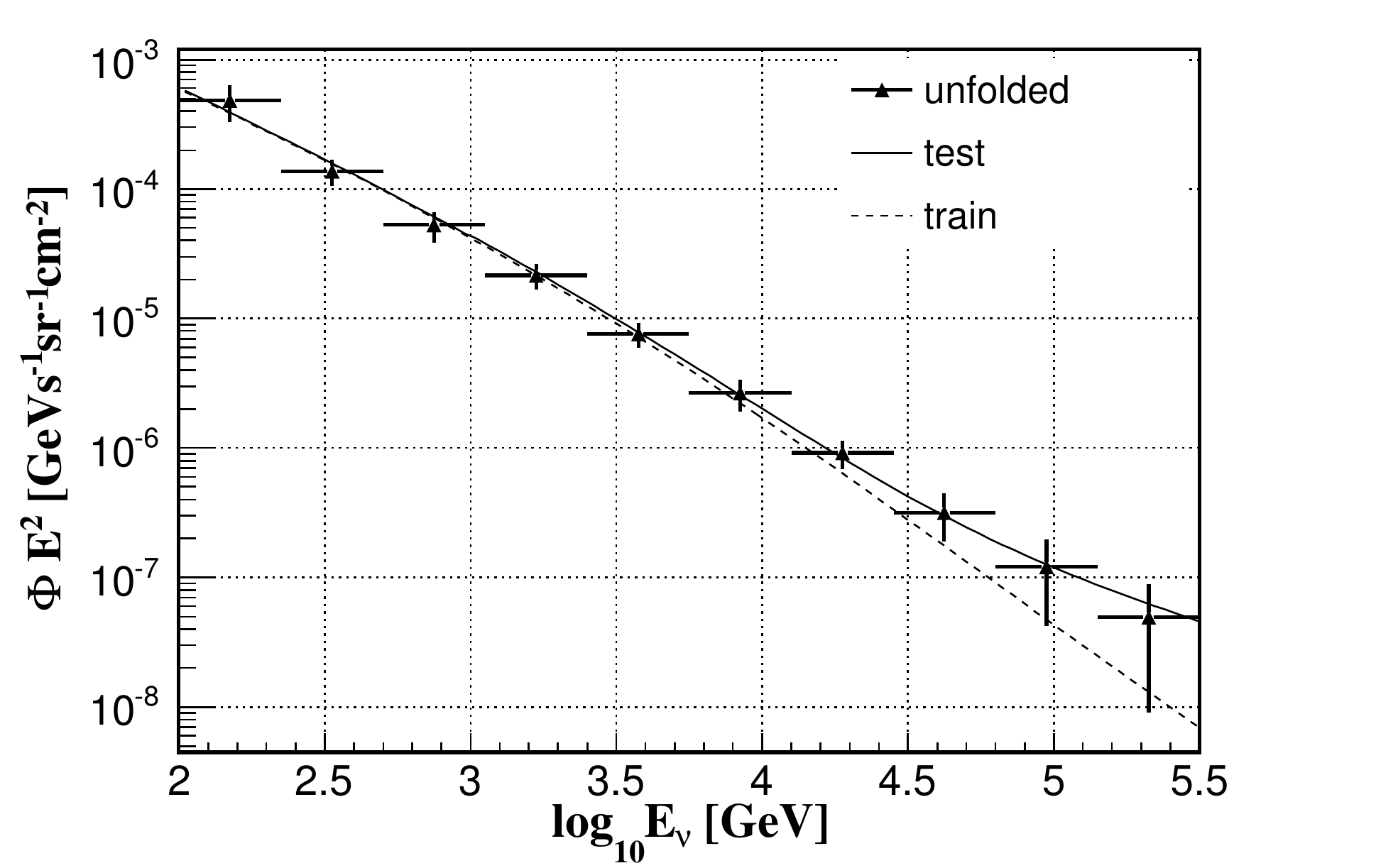}
  \caption{Application of the unfolding method on a realistic Monte Carlo dataset for the determination of the atmospheric neutrino flux weighted with $E_\nu^2$. Systematic effects are not included. See text for explanation of the fluxes.}
  \label{spectrum}
 \end{figure}
\section{Conclusions}
Three  different methods for muon energy reconstruction currently under evaluation in ANTARES have been described. Their precision varies from $0.25$ to $0.5$ in $\mathrm{log}_{10}E\,[GeV]$ above 5-10\,TeV when the true track geometry is used as an input. This resolution may be  improved for particular classes of events, especially in the case of the maximum likelihood and dE/dx methods. The results of the maximum likelihood reconstruction on a simulated dataset have been used to test the unfolding method. Systematic uncertainties are under study and have not been included in these proceedings. 


\clearpage

\setcounter{figure}{0}
\setcounter{table}{0}
\setcounter{footnote}{0}
\setcounter{section}{0}
\newpage




\title{Searches for Neutrinos from GRBs Using the ANTARES Telescope}


\shorttitle{Reed, C. \etal GRB Neutrinos at ANTARES}

\authors{Reed, C.$^{1}$, Bouwhuis, M.$^{1}$, Presani, E.$^{1}$\
 for the ANTARES Collaboration}
\afiliations{$^1$National Institute for Subatomic Physics (Nikhef), Amsterdam,\
The Netherlands}
\email{cjreed@nikhef.nl}

\maketitle
\label{icrc1085:begin}

\begin{abstract}
Multiple searches for neutrinos in correlation with gamma-ray
bursts (GRBs) using data from the ANTARES neutrino telescope have been
performed. One search uses data taken during 2007, at which time the
telescope consisted of 5 detector lines, to measure the number of
neutrino-induced muons in correlation with 37 GRBs that occurred during
the data-taking period. While no correlations are observed, upper limits
have been placed on the flux of neutrinos for different GRB models. A
second search uses an alternative method to identify the shower at the
neutrino-interaction vertex. This search is particularly sensitive to
electron-neutrinos, but is able to observe neutrinos of any flavor. The
sensitivity of this method to measure correlations between neutrinos and
prompt photons emitted by GRBs is presented for a typical neutrino flux
model.
\end{abstract}



\section{Introduction}


First discovered in the 1960's, a gammy-ray burst (GRB) is observed as
an extremely luminous flash of photons from a seemingly random point in
the sky. The energy spectrum of these photons can be parametrized by a
broken power law of the form $dN/dE{\propto}E^{\alpha}$, where
$\alpha\approx-1$ below the break energy (typically around an {\mev})
and $\alpha\approx-2$ above it~\cite{Band:1993eg}. The large energy
output of a typical GRB, around $10^{52}$~erg, combined with the GRB
occurrence rate, about one per galaxy per million years, makes gamma-ray
bursts a candidate source of high energy cosmic
rays~\cite{Waxman:1995vg}. If GRBs are indeed accelerating hadrons, then
these hadrons can be expected to suffer inelastic collisions with
particles found in the acceleration regions, typically shock waves in
relativistic jets~\cite{Rhoads:1997ps,Rees:1994nw,Meszaros:1993tv}. Such
hadronic interactions would ultimately yield mesons that decay to
produce high-energy neutrinos.

The ANTARES telescope seeks to observe such cosmic neutrinos. For this
purpose, the sea water of the Mediterranean Sea is used both as a target
and {\cheren} medium. Neutrinos interacting with a nucleus in or around
the instrumented volume will produce a hadronic shower at the collision
vertex and will also produce an energetic lepton in the case of a
charged-current interaction. These relativistic charged particles will
produce {\cheren} radiation which can then be measured by
photo-sensitive detection units in the ANTARES telescope. Neutrinos with
an energy of around 10~{\tev} or higher will yield leptons traveling
along the same trajectory as the neutrino. The source of such a neutrino
can then be identified through its correlation with the neutrino's
arrival time and/or direction.

To measure the light produced by charged particles generated in neutrino
interactions, the ANTARES telescope consists of 12~detector lines that
hold photo-sensitive units. Each line is anchored to the sea floor at a
depth of 2475~{\m} and supports 25~triplets of photo-multiplier tubes
(PMTs) spaced evenly along the line in 14.5~{\m} intervals. Further
details of the detector configuration can be found in
Ref.~\cite{Collaboration:2011ns}. The detector lines are not rigid and
are therefore affected by sea currents. The position and orientation of
each PMT is determined by multiple positioning systems, described in
reference~\cite{Brown:2009dp}.

While the sea water around the telescope is extremely dark, ANTARES can
and does detect photons that are not due to neutrino interactions. The
main sources of such background signals are (a)~down-going muons
produced by the interactions of cosmic rays with the Earth's atmosphere,
(b)~the decay of the radioactive isotope $^{40}$K present in the sea
water and (c)~light emitted by living organisms in the water. The latter
two sources produce random signals in each photo-detection unit at a
rate which varies with the sea current and is around 60-150~{\khz} per
PMT. This background is reduced through the application of triggering
algorithms that select potential neutrino events~\cite{Aguilar:2006pd}.
The so-called atmospheric muons form a background which is reduced after
the trigger as part of a neutrino analysis procedure, as described in
the following sections. With the search for transient sources such as
gamma-ray bursts, the vast majority of the atmospheric-muon background
can be reduced simply by demanding that a potential neutrino event
correlates in time to the a priori known time of the source flare.

\section{Muon Track Search}\label{muontrk}


Data taken by the ANTARES telescope during 2007 has been analyzed to
search for neutrino-induced muons arriving in correlation with prompt
photons from gamma-ray bursts. The prompt photon signals were detected
by external satellite experiments, and were mainly long-duration GRBs
detected by Swift~\cite{Gehrels:2004am,Burrows:2005gfa}. During this
data-taking period, the ANTARES telescope was still under construction
and consisted of only 5~detector lines. Muon candidates are extracted
from raw data using a trigger algorithm that searches for a collection
of causally-connected hits consistent with photons emitted by a muon
traversing the detector~\cite{Aguilar:2006pd}. The average trigger rate
during the data-taking period was 1~{\hz}.

The muon trajectory is determined by a reconstruction package that uses
probability density functions (PDFs) describing the expected photon
arrival times, of both signal and background photons, to find the local
muon position and direction that maximizes the likelihood that the
observed hit times correspond to the expected
times~\cite{Heijboer:2004gc,Heijboer:2009dq}. Potential neutrino-
induced muons are required to have an up-going trajectory (to reject
muons from atmospheric cosmic-ray showers) and to have a good fit
quality. The fit quality is estimated using two parameters, one based on
the (log of the) value of the likelihood function, $\Lambda$, and one
based on the angular precision calculated by the fit procedure,
$\sigma_f$.

The distribution of these parameters observed in the data is well
reproduced by simulations of the background due to atmospheric muons and
neutrinos. While signal neutrinos have been simulated for several
different spectra, a single set of cuts on the quality parameters is
found to provide a good separation of signal from background events. The
angular resolution for simulated muons passing these cuts is shown in
\fig{angular_resolution}.

\begin{figure}[t]
   \begin{center}
      \includegraphics[width=0.72\linewidth]{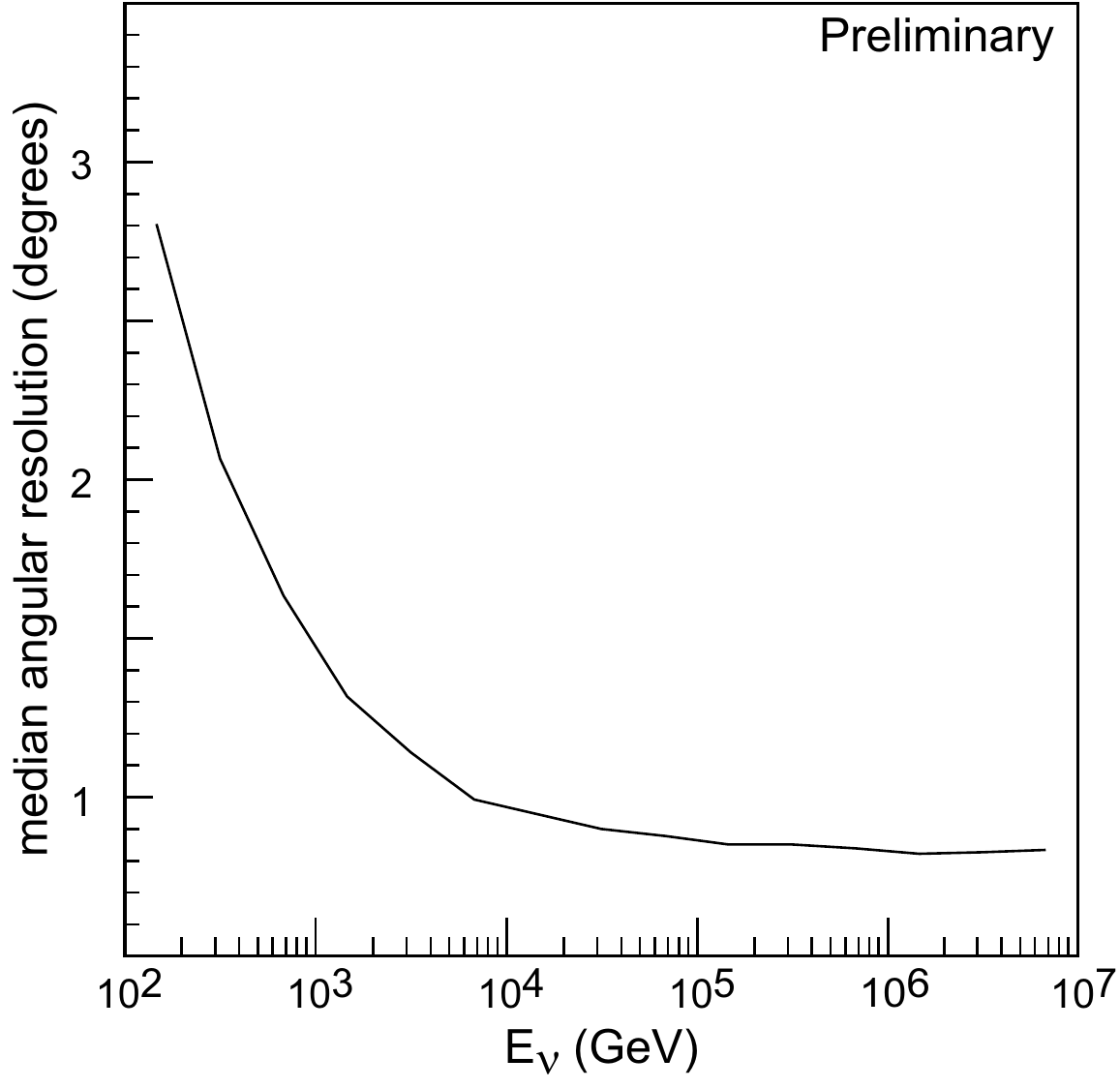}
   \end{center}
   \caption{\label{angular_resolution}
      The angular resolution for muon tracks passing the selection
      described in the text, as a function of the neutrino energy.}
\end{figure}

The good angular resolution for muon tracks allows the direction, in
addition to the time, of neutrino candidate events to be used in the
search for correlations with gamma-ray bursts. The candidate tracks are
required to point back to the GRB position to within 2{\dg}, and to
occur during the arrival of prompt photons, i.e.~during the T$_{90}$
observed by the satellite. With these criteria, the expected background
rate is found to be about $1.0\times10^{-7}$~{\hz}. Of the remaining
background events, 90\% are due to (misreconstructed) atmospheric muons
from showers above the detector and 10\% to atmospheric neutrinos
producing up-going muons.


This search for muons produced by neutrinos in correlation with
gamma-ray bursts has been applied to the data taken during the flares of
37~GRBs that occurred during 2007. No neutrino candidate events are
observed in correlation with the GRBs. The limits placed on the total
$\nu_\mu+\bar{\nu}_\mu$ flux of these bursts at the 90\% confidence
level, using the Feldman-Cousins recipe~\cite{Feldman:1997qc}, for three
different GRB models are presented in \fig{upper_limit90_GRBline5}. In
addition to a general E$^{-2}$ spectrum, two GRB neutrino spectra have
been explored, both based on the fireball model. The Waxman-Bahcall
model is that described in Ref.~\cite{Waxman:1998yy}, and all GRBs are
assumed to follow the same spectrum. The model of Guetta~{\etal} is
described in Ref.~\cite{Guetta:2003wi}. For this model, the neutrino
spectrum of each GRB has been calculated separately using the
(satellite) measured GRB parameters.

The reliability of the simulations used in this analysis has been
studied in depth. The uncertainty with the largest impact arises from
the efficiency and angular acceptance used for PMTs in the simulations.
Reducing the efficiency of the tubes by 15\% results in a 12\% reduction
of the acceptance of the telescope to E$^{-2}$ muon-neutrinos, without
affecting its angular resolution. A degradation of the PMT timing of
only a few nanoseconds in the simulations is found to be incompatible
with muon data. Such studies allow a conservative estimate of 15\% to be
placed on the uncertainty of the angular resolution of the telescope.
Uncertainties on PMT timing can also reduce the acceptance of the
telescope. The aggregate effect of the PMT timing and efficiency results
in an uncertainty of 15\% on the relative acceptance of ANTARES to
cosmic muon-neutrinos.

\begin{figure}[t]
   \begin{center}
      \includegraphics[width=0.85\linewidth]{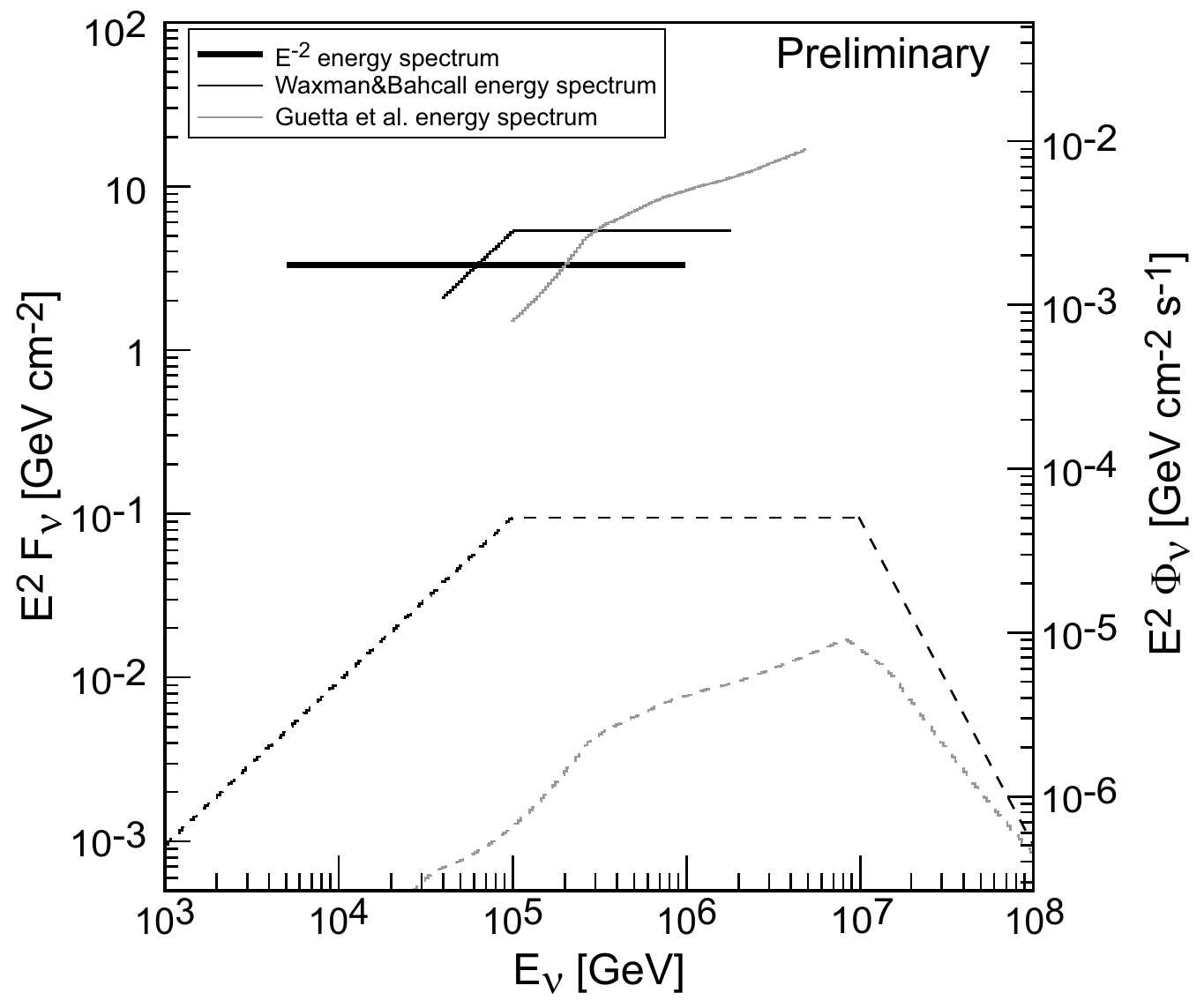}
   \end{center}
   \caption{\label{upper_limit90_GRBline5}
      The upper limits on the total $\nu_\mu+\bar{\nu}_\mu$ flux of 37
      GRBs, obtained by the muon track search, are shown by the solid
      lines for different flux models. Two of the flux models are shown
      by the dashed lines. The expected Waxman-Bahcall
      flux~\cite{Waxman:1998yy} is shown in black and the expected
      Guetta~{\etal} flux~\cite{Guetta:2003wi} is shown in gray. The
      limits have been placed using data taken during 2007, when the
      telescope consisted of 5~detector lines. }
\end{figure}

\section{Collision Vertex Shower Search}\label{shower}


An additional search method has been developed to search for neutrinos
of any flavor in correlation with gamma-ray bursts. This is done by
locating the shower(s) produced at the vertex of the collision between
the neutrino and a nucleus in or around the detector. Such a deep
inelastic scattering will produce a hadronic shower irrespective of the
neutrino flavor. In the case of electron neutrinos undergoing a charged
current interaction, an electromagnetic shower is also produced at the
vertex by the resulting electron or positron. Such showers occurring
during the prompt photon emission of a GRB are then sought. This
analysis is the first to seek to measure neutrinos of multiple flavors
in the ANTARES data.

As the extent of the vertex shower is small compared to the spacing
between detection units in the telescope, photons radiated by charged
particles in the shower appear to originate from the same point in space
and time. A simple procedure to reconstruct the position and time of the
vertex shower has been developed. This reconstruction is applied to
candidate neutrino events that have passed at least one trigger
criteria. In addition to the trigger algorithm described in
\sect{muontrk}, a second trigger condition in which clusters of
time-correlated hits in neighboring PMT triplets are sought has been
applied. This second trigger algorithm was implemented during the 2008
data taking period.

\begin{figure}[t]
   \begin{center}
      \includegraphics[width=0.75\linewidth]{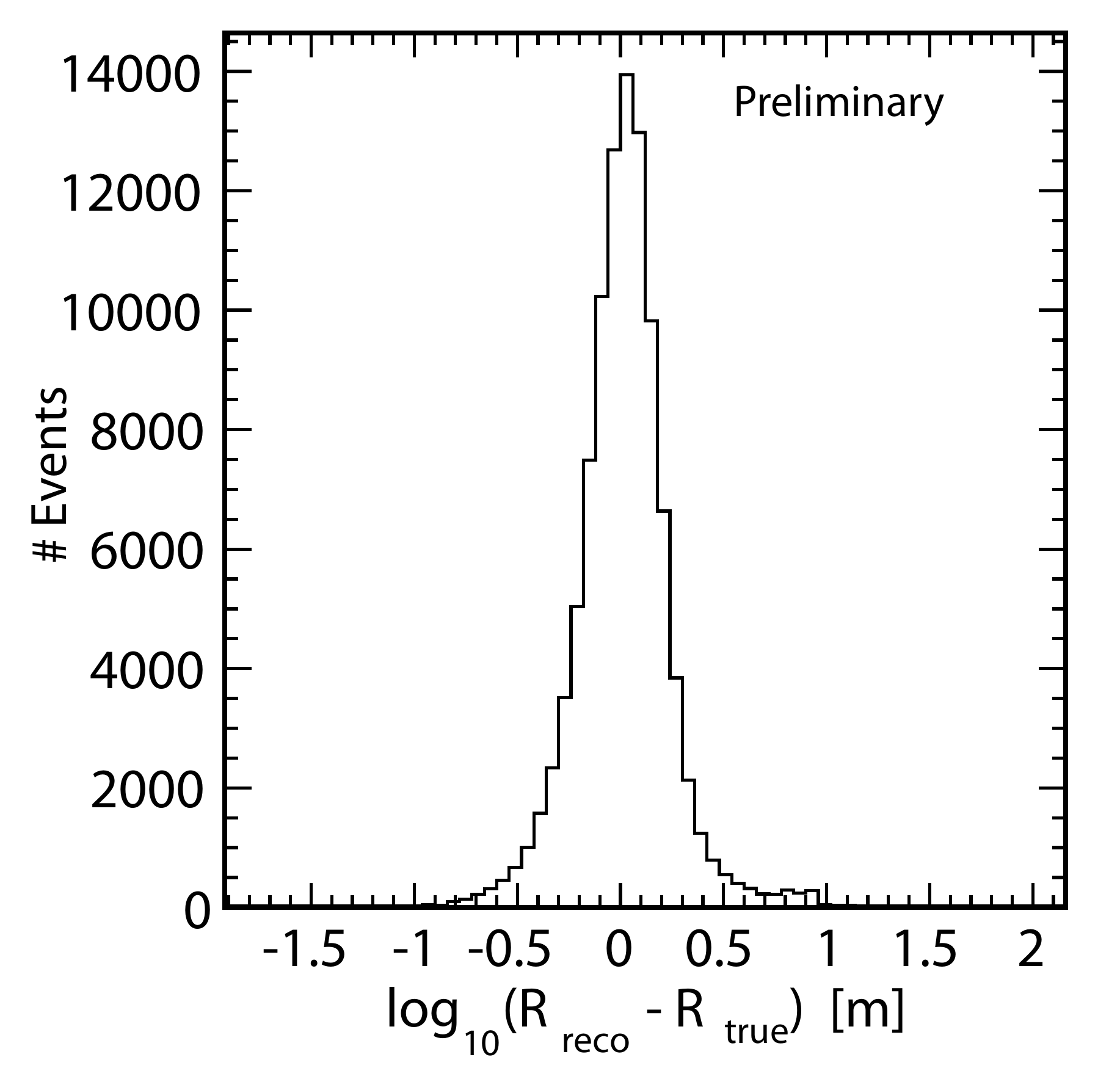}
   \end{center}
   \caption{\label{SpaceResolution}
      The spacial resolution of the shower reconstruction algorithm,
      determined using special calibration data in which LED beacons
      were flashed.}
\end{figure}

The shower vertex position and time is then determined by minimizing the
difference between the expected photon arrival times and the measured
hit times. The arrival time of photons is assumed to depend only on the
distance traveled from the shower vertex.
The resolution of this reconstruction
procedure has been studied using special calibration data taken by
ANTARES in which LED beacons were used to obtain a burst of photons
originating from a single point. The spacial resolution of the shower
reconstruction is found to be around 1~{\m}, as shown in
\fig{SpaceResolution}, and the timing resolution around 2.4~{\ns}.

A sample of reconstructed showers from triggered events will contain
many electromagnetic showers produced by the stochastic energy loss of
atmospheric muons traversing the telescope. Neutrino-induced showers are
extracted from this background by placing cuts on quality parameters.
The most discriminating variable is found to be the number of direct
hits, $N_{DirHits}$: the number of selected hits having a time residual
better than 15~{\ns}. In addition, requiring the number of detector
lines used in the fit to be greater than two is found to further reduce
the atmospheric muon background. The cut on $N_{DirHits}$ has been
determined by finding the value that minimizes the flux of a GRB
required to discover a source with T$_{90}=$100~s at the 5$\sigma$ level
in 50\% of test cases. The GRB flux used during cut optimization is
assumed to follow the Waxman-Bahcall flux~\cite{Waxman:1998yy}.

While a rare occurrence, a sparking PMT could produce a signal with a
similar profile to that of a neutrino-induced shower. To eliminate such
events, showers are rejected if they are reconstructed too close to any
PMT. This cut effectively reduces the fiducial volume of the telescope
by about 2\%, but since the detector is capable of measuring high-energy
neutrino-induced showers from far outside the instrumented volume, the
ultimate effect on the sensitivity is much less.

The expected rate of background events passing the full set of quality
cuts has been estimated from data taken during 2007-2008 in which no GRB
was observed by a satellite experiment. During this time, the ANTARES
telescope was under construction and data was taken while the telescope
consisted of 5, 9, 10 and 12 detector lines. The average rate of surviving
background events over the 2007 (5 line) and 2008 (9-12 lines)
periods are found to be $4.4\times10^{-5}$~{\hz} and
$6.8\times10^{-5}$~{\hz}, respectively. Note that the background is much
larger than that obtained by the muon track search method described in
\sect{muontrk}, due to the lack of directional information. With the
simple shower reconstruction, the field of view of the telescope cannot
be restricted to the region of sky around the GRB.

The (Neyman) sensitivity of the vertex shower search method to observe
the total flux of neutrinos and anti-neutrinos of all three flavors from
GRBs~\cite{Waxman:1998yy} has been calculated.
The sensitivity obtained for the 2007 and 2008 data taking periods,
averaged over the viewable sky, is shown in \fig{myAmandaComp2}. For the
2007 period, the detector consisted of 5 lines and the sensitivity is
shown for the same 37 GRBs studied in \sect{muontrk}. For the 2008 period,
the sensitivity is shown for 65 GRBs and averaged over the 9, 10 and 12 line
detector configurations.

The impact of systematic uncertainties in the detector simulations are
under study. A significant effect comes from the possibility that the
efficiency of each PMT is not accurately reproduced in the simulations.
This is being studied by reducing the efficiency with which signal
photons are converted to hits. Other simulation properties have also
been studied by varying the {\cheren} light propagation, altering the
high energy shower generation and using different models of the
neutrino-nucleus cross sections. These effects lead to a reduction of
the sensitivity of the shower analysis that is currently estimated to
be 10-20\%.

\begin{figure}[t]
   \begin{center}
      \includegraphics[width=0.9\linewidth]{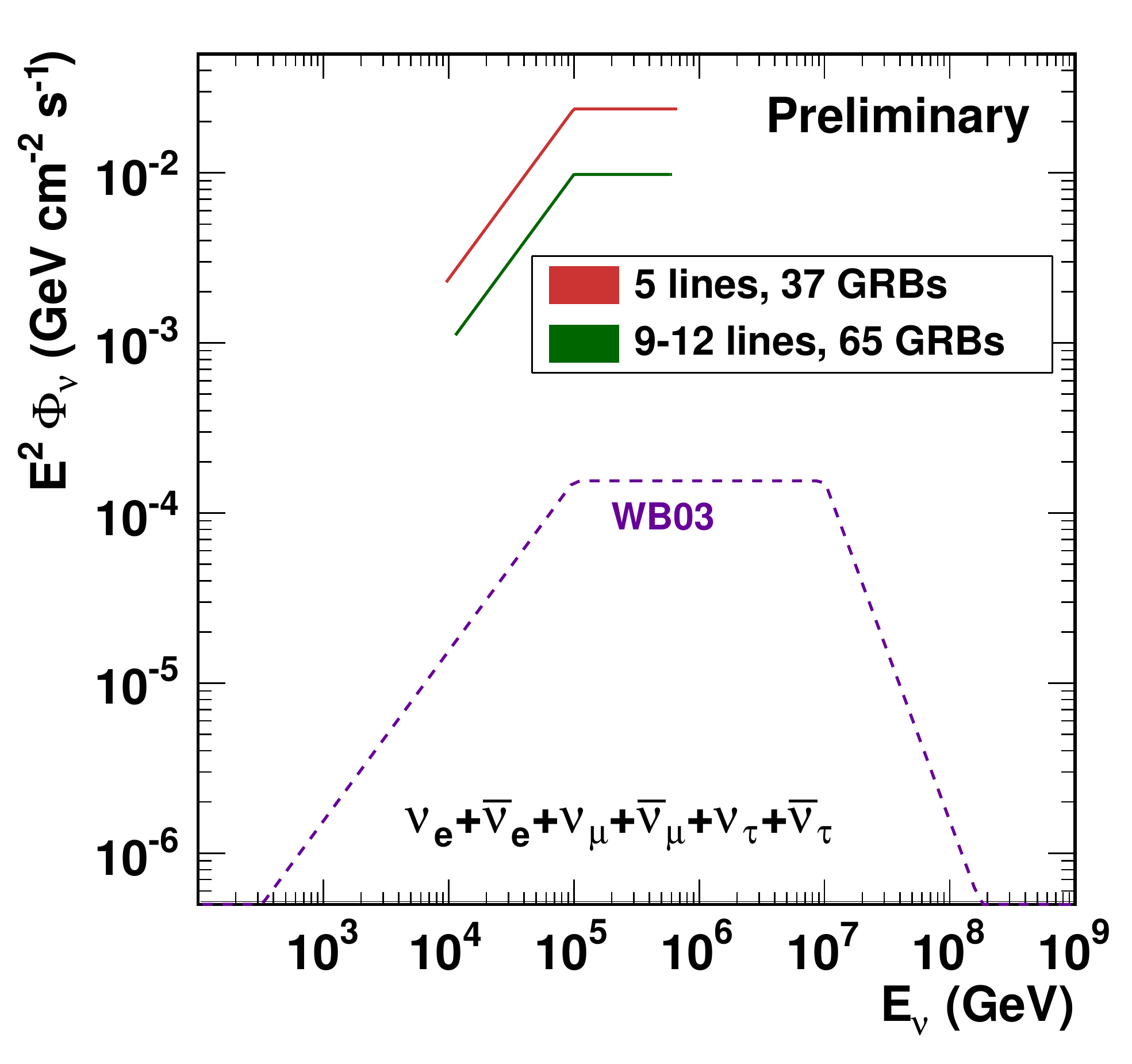}
   \end{center}
   \caption{\label{myAmandaComp2}
      The sensitivity of the shower vertex search method applied to the
      average ANTARES detector as it was during the 2007 and 2008 data taking
      periods to the total neutrino and anti-neutrino flux of GRBs.
      The neutrino spectra are assumed to follow the
      Waxman-Bahcall spectrum~\cite{Waxman:1998yy}.}
\end{figure}

\section{Summary}

Two methods have been developed that can be used to search for a
detectable flux of neutrinos emitted by gamma-ray bursts. Both methods
attempt to measure neutrinos arriving at the Earth in conjunction with
the prompt photons emitted by GRBs (as observed by external satellite
experiments). For both methods, muons and neutrinos generated by the
interaction of cosmic-rays with the Earth's atmosphere form a
substantial background that must be reduced.

The first method reduces this background by measuring the trajectory of
the muon produced by muon-neutrino interactions. The muon is demanded to
be well reconstructed and to come from within 2{\dg} of the GRB. It has
been applied to data taken by the ANTARES telescope during 2007, when
the experiment consisted of 5~detector lines. No neutrino events are
observed in correlation with 37~selected GRBs of 2007. Limits on the
maximum flux that would have only a 10\% chance to produce no events in
the telescope have been calculated.

The second method seeks to measure the shower generated at the neutrino
collision vertex and is the first analysis of ANTARES to be sensitive to
neutrinos of any flavor. Showers produced by the stochastic energy loss
of atmospheric muons are rejected by studying the topology of the shower
event. Due to the lack of directional information available from the
shower reconstruction, the background rate of this method is higher than
that of the muon analysis. The sensitivity of the ANTARES telescope as
it was during 2007-2008 to showers from a flux of neutrinos from GRBs
has been determined.

\label{icrc1085:end}


\setcounter{figure}{0}
\setcounter{table}{0}
\setcounter{footnote}{0}
\setcounter{section}{0}
\newpage




\title{Search for neutrino emission of gamma-ray flaring blazars with the ANTARES telescope}

\shorttitle{Damien Dornic \etal Time-dependent neutrino point source search}

\authors{Dornic D.$^{1}$ ANTARES Collaboration }
\afiliations{$^1$ IFIC - Instituto de Fisica Corpuscular, Edificios Investigaci�on de Paterna, CSIC - Universitat de Valencia, Apdo. de Correos 22085, 46071 Valencia, Spain	    	     
	     }
\email{dornic@ific.uv.es}
\maketitle

\begin{abstract}
The ANTARES telescope is well suited to detect neutrinos produced in astrophysical
transient sources as it can observe a full hemisphere of the sky at all the times with a
duty cycle close to unity. The background and point source sensitivity can be drastically
reduced by selecting a narrow time window around the assumed neutrino production
period. Radio-loud active galactic nuclei with their jets pointing almost directly towards
the observer, the so-called blazars, are particularly attractive potential neutrino point
sources, since they are among the most likely sources of the observed ultra high energy
cosmic rays and therefore, neutrinos and gamma-rays may be produced in hadronic
interactions with the surrounding medium. The gamma-ray light curves of blazars
measured by the LAT instrument on-board the Fermi satellite reveal important time
variability information. A strong correlation between the gamma-ray and the neutrino
fluxes is expected in this scenario.

An unbinned method based on the minimization of a likelihood ratio was applied to a
subsample data collected in 2008 (61 days live time). By looking for neutrinos
detected in the high state periods of the AGN light curve, the sensitivity to these sources
has been improved by about a factor 2 with respect to a standard time-integrated point
source search. First results on the search for ten bright and variable Fermi
sources are presented.
\end{abstract}


\section{Introduction}

The production of high-energy neutrinos has been proposed for several kinds
of astrophysical sources, such as active galactic nuclei, gamma-ray bursters, supernova
remnants and microquasars, in which the acceleration of hadrons may occur. 
Neutrinos are unique messengers to study the high-energy universe as there are neutral and 
stable, interact weakly and travel directly from their point of creation in the source without absorption. 
Neutrinos could play an important role in understanding the mechanisms of cosmic ray acceleration and their 
detection from a source would be a direct evidence of the presence of hadronic acceleration in that source.

Radio-loud active galactic nuclei with their jets pointing almost directly towards
the observer, the so-called blazars, are particularly attractive potential neutrino point
sources, since they are among the most likely sources of the observed ultra high energy
cosmic rays and therefore, neutrinos and gamma-rays may be produced in hadronic
interactions with the surrounding medium~\cite{bib:AGNhadronic}. The gamma-ray light curves of blazars
measured by the LAT instrument on-board the Fermi satellite reveal important time
variability information on timescale of hours to several weeks, with intensities
always several times larger than the typical flux of the source in its quiescent state~\cite{bib:FermiLATAGNvariability}. 
A strong correlation between the gamma-ray and the neutrino fluxes is expected in this scenario. 

In this paper, the results of the first time-dependent search for cosmic neutrino sources in
the sky visible to the ANTARES telescope are presented. The data sample used in
this analysis is described in Section 2,
together with a discussion on the systematic uncertainties. The point source search
algorithm used in this time-dependent analysis is explained in Section 3. The results are presented in
Section 4 for a search on a list of ten selected candidate sources.

\section{ANTARES}

The ANTARES collaboration has completed the construction of a neutrino
telescope in the Mediterranean Sea with the connection of its twelfth detector line
in 2008~\cite{bib:Antares}. The telescope is located 40 km on the southern coast of France
(42$^{o}$48'N, 6$^{o}$10'E) at a depth of 2475 m. It comprises a three-dimensional array of
photomultipliers housed in glass spheres (optical modules), distributed along twelve
slender lines anchored at the sea bottom and kept taut by a buoy at the top. 
Each line comprises up to 25 storeys of triplets of optical modules (OMs), each housing a single 10" PMT. 
Since lines are subject to the sea current and can change shape and orientation, a positioning system 
comprising hydrophones and compass-tiltmeters is used to monitor the detector geometry. The main goal of the experiment 
is to search for neutrinos of astrophysical origin 
by detecting high energy muons ($>$100~GeV) induced by their neutrino charged current interaction in the vicinity 
of the detector. 

The arrival time and intensity of the Cherenkov light on the OMs are digitized into hits and transmitted to shore, 
where events containing muons are separated from the optical backgrounds due to natural radioactive decays and bioluminescence, 
and stored on disk. A detailed description of the detector and the data acquisition is given in~\cite{bib:Antares}~\cite{bib:antaresdaq}. The arrival times 
of the hits are calibrated as described in reference~\cite{bib:TimeCalib}. The online event 
selection identifies triplets of OMs that detect multiple photons. At least 5 of these are required throughout the detector, 
with the relative photon arrival times being compatible with the light coming from a relativistic particle. Independently, 
events were also selected which exhibit multiple photons on two sets of adjacent, or next to adjacent floors. 

The data used in this analysis corresponds to the period from September 6th to
December 31st, 2008 (54720-54831 modified Julian day), taken with the full detector. Some filtering has been applied 
in order to exclude periods in which the bioluminescence-induced optical background was high. The resulting effective life time is 60.8 days.
Atmospheric neutrinos are the main source of background in the search for astrophysical neutrinos. These neutrinos 
are produced from the interaction of cosmic rays in the Earth's atmosphere.Only charged 
current interactions of neutrinos and antineutrinos were considered. An additional source of background 
is due to the mis-reconstructed atmospheric muons. 
The track reconstruction algorithm derives the muon track parameters that maximize a likelihood function built from the 
difference between the expected and the measured arrival time of the hits from the Cherenkov photons emitted along the muon 
track. This maximization takes into account the Cherenkov photons that scatter in the water and the additional photons 
that are generated by secondary particles (e.g. electromagnetic showers created along the muon trajectory). The algorithm used is outlined in~\cite{bib:AAfit}.
The value of the log-likelihood per degree of freedom ($\Lambda$) from the track reconstruction fit is a measure of the track fit 
quality and is used to reject badly reconstructed events, such as atmospheric muons that are mis-reconstructed as upgoing tracks. 
Neutrino events are selected by requiring that tracks are reconstructed as upgoing and have a good reconstruction quality. In addition, the error 
estimate on the reconstructed muon track direction obtained from the fit is required to be less than 1$^{o}$.

The angular resolution can not be determined directly in the data and has to be estimated from simulation. However, the comparison of 
the data and MonteCarlo simulations from which the time accuracy of the hits has been degraded has yielded a constrain on the 
uncertainty of the angular resolution of the order of 0.1$^{o}$~\cite{bib:AAfitps}. Figure~\ref{fig:Angres} shows the cumulative distribution of the angular difference between 
the reconstructed muon direction and the neutrino direction with an assumed spectrum proportional to $E_{\nu}^{-2}$, where $E_{\nu}$ is 
the neutrino energy. For this period, the median resolution is estimated to be 0.4 +/- 0.1 degree. 

\begin{figure}[ht!]
\centering
\includegraphics[width=0.4\textwidth]{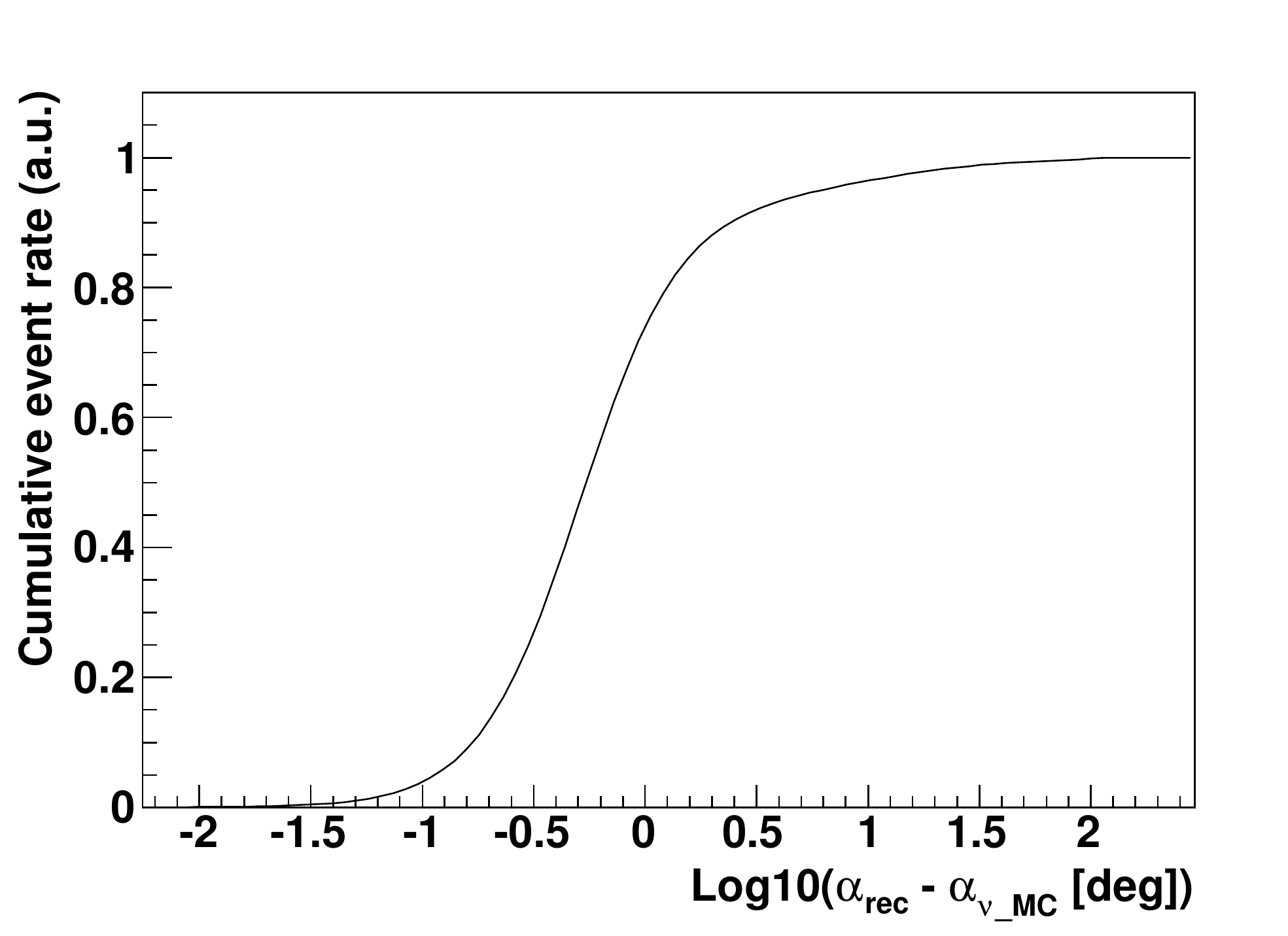}
\caption{Cumulative plot of the distribution of the angle between the true Monte Carlo neutrino direction and the reconstructed
muon direction for E$^{-2}$ upgoing neutrino events selected for this analysis.}
\label{fig:Angres}
\end{figure}

\section{Time-dependent search algorithm}
This time dependant point source analysis is done using an unbinned method based on a
likelihood ratio maximization. The data is parameterized as a two components
mixture of signal and background. The goal is to determine, at a given point in the sky and at a given
time, the relative contribution of each component and to calculate the probability to have a signal
above a given background model. The likelihood ratio $\lambda$ is the ratio of the probability density for the hypothesis 
of background and signal ($H_{sig+bkg}$) over the probability density of only background ($H_{bkg}$):

\begin{equation}
\lambda=\sum_{i=1}^{N} log\frac{P(x_{i}|H_{sig+bkg})}{P(x_{i}|H_{bkg})} 
\label{eq:EQ_likelihood}
\end{equation}
\begin{equation}
\lambda=\sum_{i=1}^{N} log\frac{\frac{n_{sig}}{N}.P_{sig}(\alpha_{i},t_{i}) + (1-\frac{n_{sig}}{N}).P_{bkg}(\alpha_{i},t_{i})}{P_{bkg}(\alpha_{i},t_{i})}   
\label{eq:EQ_likelihood2}
\end{equation}

where $n_{sig}$ and N are respectively the unknown number of signal events and the total number of events in the considered data sample. 
$P_{sig}(\alpha_{i},t_{i})$ and  $P_{bkg}(\alpha_{i},t_{i})$ are the probability density function (PDF) for signal and background respectively. 
For a given event i, $t_{i}$ and $\alpha_{i}$ represent the time of the event and the angular difference between the coordinate of this 
event and the studied source. 

The probability densities $P_{sig}$ and $P_{bkg}$ are described by the product of two 
components: one for the direction and one for the timing. The shape of the time PDF for the signal event is extracted directly from the gamma-ray light curve assuming 
 the proportionality between the gamma-ray and the neutrino fluxes. For the signal event, this directional PDF is described by the one 
dimension point spread function, which is the probability density of reconstructing an event at an angular distance from the true source position. 
The directional and time PDF for the background are derived from the data using respectively the observed declination distribution of selected events 
in the sample and the observed time distribution of all the reconstructed muons. 
Figure~\ref{fig:TimeDistri} shows the time distribution of all the reconstructed events and the selected upgoing events for this analysis. Once normalized to an integral equal to 1, the distribution 
for all reconstructed events is used directly as the time PDF for the background.
When data is at 0, it means that there are no data taken during these periods (ie detector in maintenance) or data with a very poor quality (high bioluminescence or bad calibration).

\begin{figure}[ht!]
\centering
\includegraphics[width=0.4\textwidth]{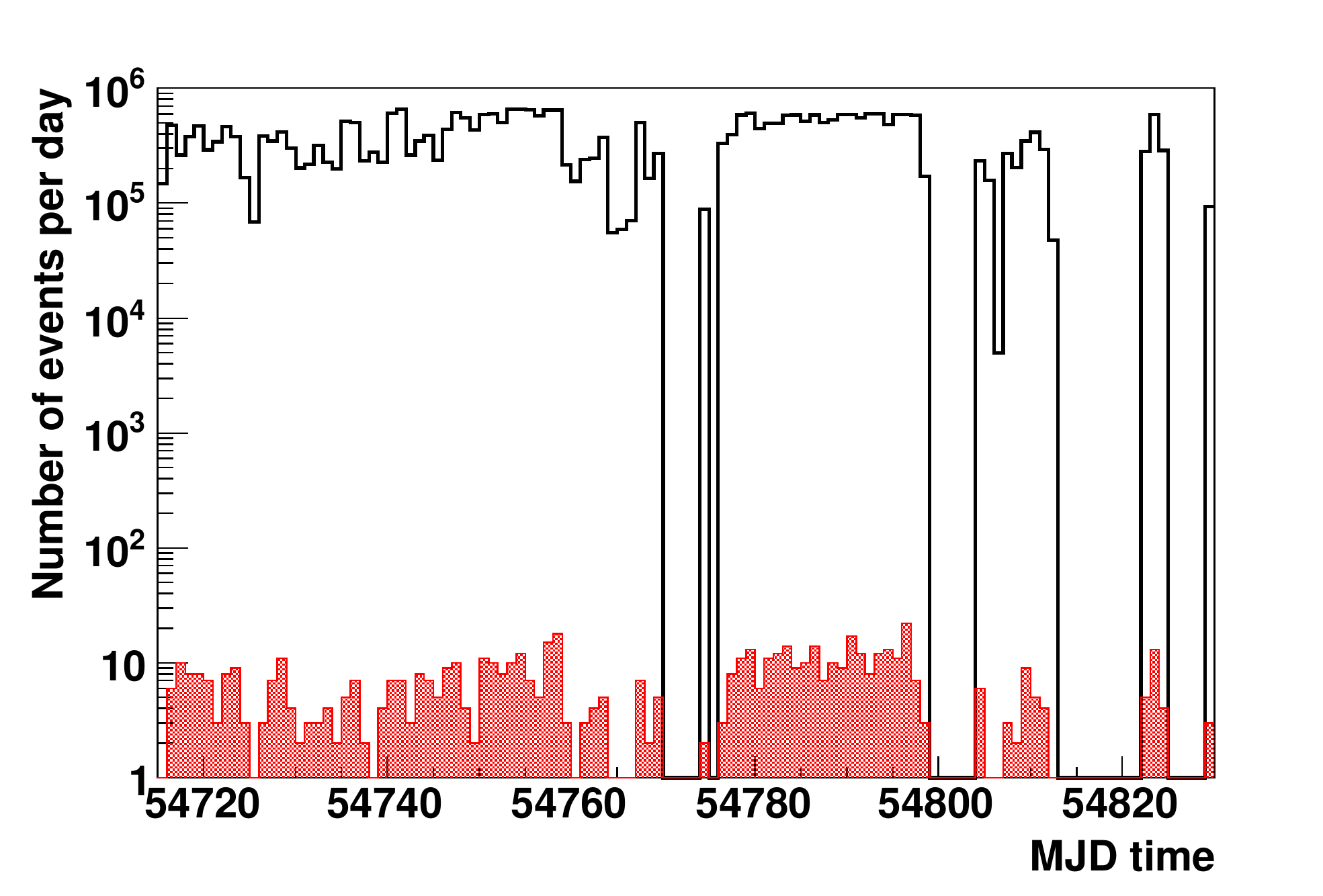}
\caption{Time distribution of the reconstructed events. Black: distribution for all reconstructed events. Red filled: distribution of selected upgoing events ($\Lambda>-5.4$ and $\beta<1^{o}$).
}
\label{fig:TimeDistri}
\end{figure}

The null hypothesis is given with $n_{sig}=0$. The obtained value of $\lambda_{data}$ on the data is then 
compared to the distribution of $\lambda$ given the null hypothesis. Large values of $\lambda_{data}$ compared to the distribution of $\lambda$ for the 
background only reject the null hypothesis with a confident level equal to the fraction of the scrambled trials above $\lambda_{data}$. This fraction of 
trials above $\lambda_{data}$ is referred as the p-value. The discovery potential is then defined as the average number of signal events required to 
achieve a p-value lower than 5$\sigma$ in 50~$\%$ of trials. Figure~\ref{fig:Nev5sigma} shows the average number of events required for a 5$\sigma$ discovery (50~\% C.L.) produced in one source located at a declination of -40$^{o}$ as 
a function of the total width of the flare periods. These numbers are compared to the one obtained without using the timing information. Using the timing information yields 
to an improvement of the discovery potential by about a factor 2-3 with respect to a standard time-integrated point source search~\cite{bib:AAfitps}.

\begin{figure}[ht!]
\centering
\includegraphics[width=0.4\textwidth]{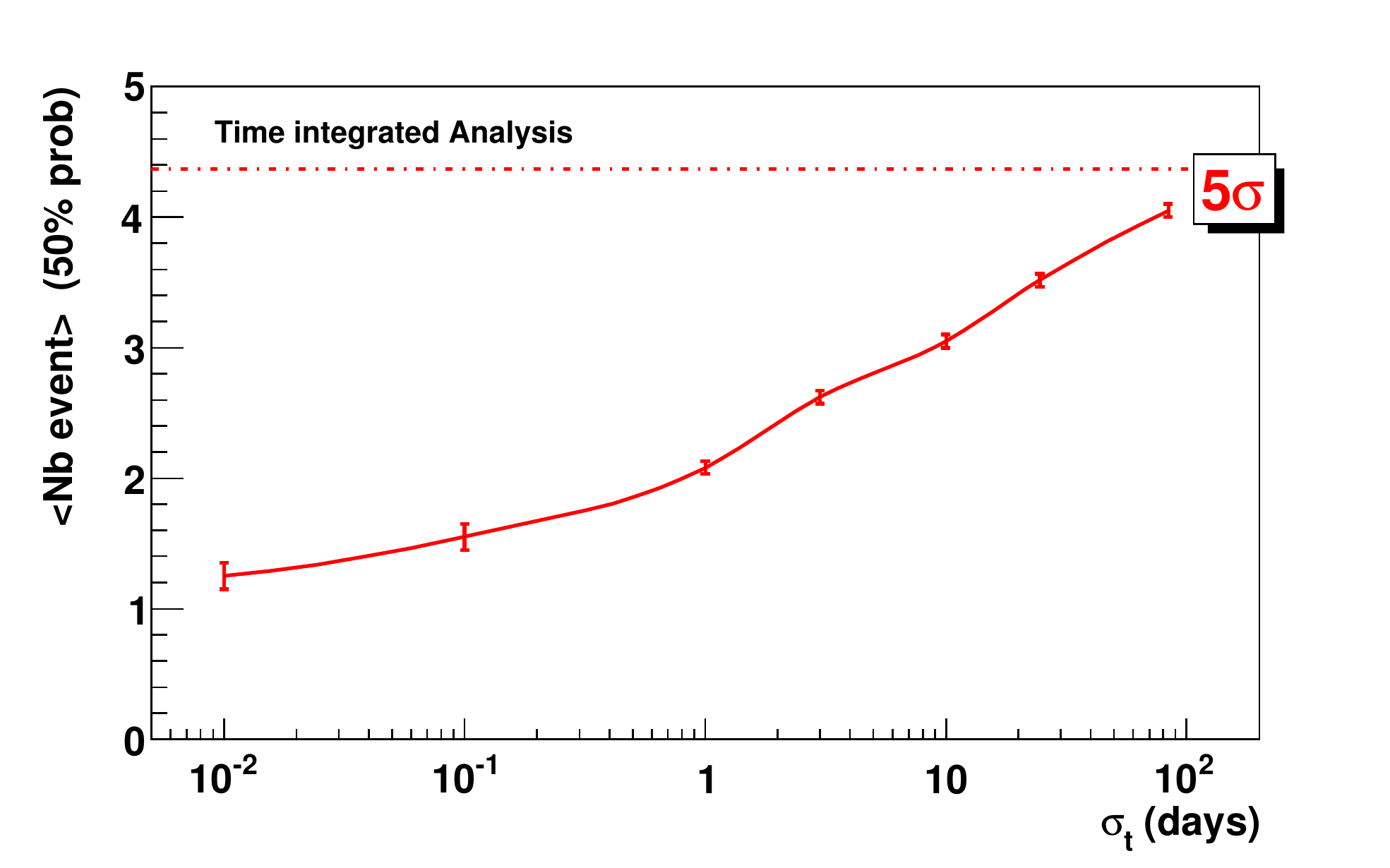}
\caption{Average number of events required for a 5$\sigma$ discovery (50~$\%$ C.L.) produced in one source located at a
declination of -40$^{o}$ as a function of the width of the flare period. These numbers are compared to the one obtained without
using the timing information.}
\label{fig:Nev5sigma}
\end{figure}

\section{Search for neutrino emission from gamma-ray flare}

This time-dependent analysis has been applied to bright and variable Fermi blazar sources reported in the first year Fermi LAT catalogue~\cite{bib:Fermicatalogue} and in the LBAS catalogue 
(LAT Bright AGN sample~\cite{bib:FermicatalogueAGN}). The sources located in the visible part of the sky by Antares from which the averaged 1 day-binned flux 
in the high state is greater than 20 10$^{-8}$ photons.cm$^{-2}$.s$^{-1}$ above 300~MeV in the studied time period and with a significant time variability are selected. 
This list includes six flat spectrum radio quasars (FSRQ) and four BLlacs. Table~\ref{tab:Sources} lists the characteristics of the ten selected sources.

\begin{table}[ht!]
\begin{center}
\begin{small}
\begin{tabular}{|c|c|c|c|c|}
\hline
Name & {OFGL name} & Class & redshift & {$F_{300}$} \\
\hline
{PKS0208-512} & {J0210.8-5100} & FSRQ & 1.003 & 4.43 \\
\hline
{AO0235+164} & {J0238.6+1636} & BLLac & 0.940 & 13.19 \\
\hline
{PKS0454-234} & {J457.1-2325} & FSRQ & 1.003 & 13.56 \\
\hline
{OJ287} & {J0855.4+2009} & BLLac & 0.306 & 2.48 \\
\hline
{WComae} & {J1221.7+28.14} & BLLAc & 0.102 & 2.58 \\
\hline
{3C273} & {J1229.1+0202} & FSRQ & 0.158 & 8.68 \\
\hline
{3C279} & {J1256.1-0548} & FSRQ & 0.536 & 15.69 \\
\hline
{PKS1510-089} & {J1512.7-0905} & FSRQ & 0.36 & 28.67 \\
\hline
{3c454.3} & {J2254.0+1609} & FSRQ & 0.859 & 24.58 \\
\hline
{PKS2155-304} & {J2158.8-3014} & BLLac & 0.116 & 7.89 \\
\hline
\end{tabular}
\caption{List of bright variable Fermi blazars selected for this analysis. $F_{300}$ is the gamma-ray flux above 300~MeV (10$^{-8}$ photons.cm$^{-2}$.s$^{-1}$).}
\label{tab:Sources}
\end{small}
\end{center}
\end{table}

The light curves published in Fermi web page for the monitored sources~\cite{bib:Fermimonitored} are used for this analysis. These light
curves correspond to 
the one-day binned time evolution of the average gamma-ray flux above a threshold of 100~MeV from August 2008 to August 2010. The high state periods are defined 
using a simple and robust method based on three main steps. First, the baseline is determined with an iterative linear fit. After each fit, the points where the 
flux is above a given threshold are suppressed. When the baseline is computed, all the points (green dots) where the flux minus its error are above the baseline 
plus two times its fluctuation and the flux is above the baseline plus three times its fluctuation are used as priors from which the flares are defined. The last 
step consists on, for each selected point, adding the adjacent points for which the emission is compatible with the flare. Finally, an additional delay of 0.5 day 
is added before and after the flare in order to take into account that the precise time of the flare is not known (1-day binned LC). With this definition, a flare 
has a width of at least two days. Figure~\ref{fig:3C454} shows the time distribution of the Fermi LAT gamma-ray light curve of 3C454 for almost 2 years of data 
and the determined high state periods (blue histogram). With the hypothesis that the neutrino emission follows the gamma-ray emission, the signal time PDF is 
simply the normalized de-noise light curve. 

\begin{figure}[ht!]
\centering
\includegraphics[width=0.4\textwidth]{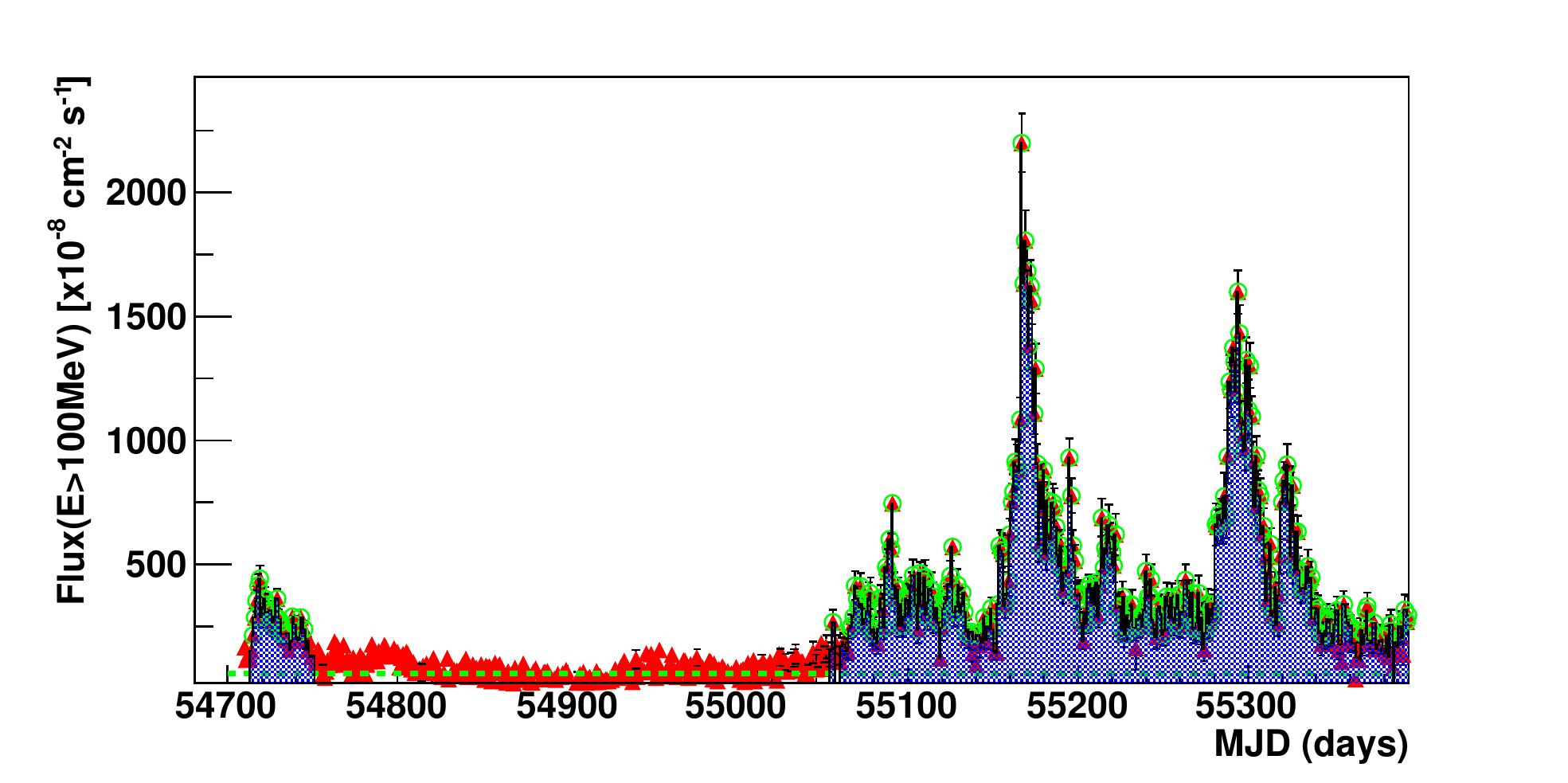}
\caption{Gamma-ray light curve (red dots) of the blazar 3C454.3 measured by the LAT instrument onboard the Fermi satellite above 100~MeV for almost 2 years 
of data. Blue histogram: high state periods. Green line and dots: baseline and significant dots above this baseline used for the determination of the flare 
periods. 
}
\label{fig:3C454}
\end{figure}

The most significant source is 3C279, which has a pre-trial p-value of 1.03~\%. 
The unbinned method finds one high-energy neutrino event located at 0.56$^{o}$ from the source location during a large flare in November 2008. 
Figure~\ref{fig:Result_3C279} shows the time distribution of the Fermi gamma-ray light curve of 3C279 and the time of the coincident neutrino event. 
This event has been reconstructed with 89 hits spread on 10 lines with a track fit quality $\Lambda=-4.4$ and an error estimate $\beta=0.3^{o}$. The post-trial 
probability is computed taking into account the ten searches. The final probability, 10~\% is compatible with background fluctuations.  

\begin{figure}[ht!]
\centering
\includegraphics[width=0.4\textwidth]{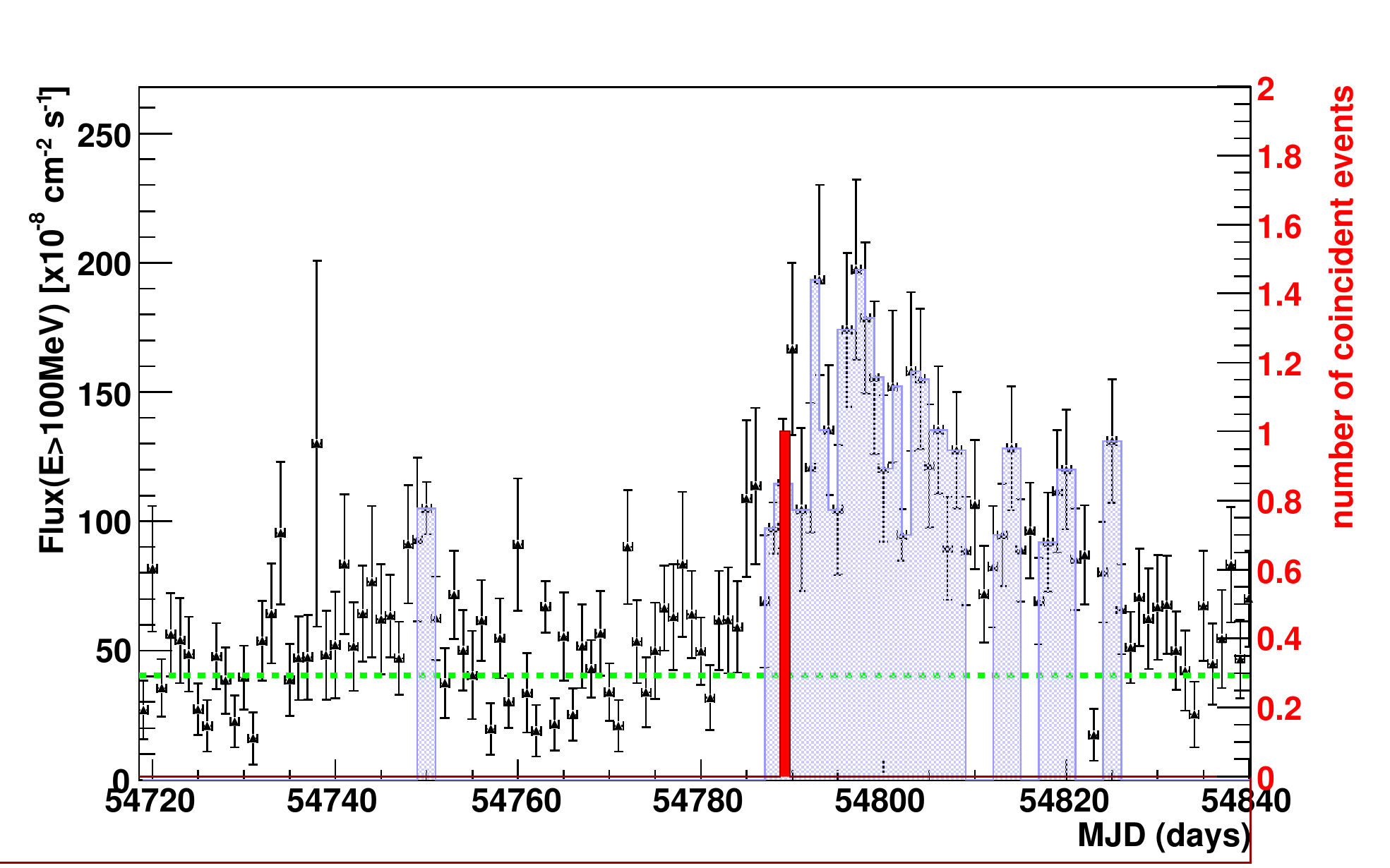}
\caption{Gamma-ray light curve (black dots) of the blazar 3C279 measured by the LAT instrument onboard the Fermi satellite above 100 MeV. Blue histogram: 
high state periods. Green dashed line: fit of a baseline. Red histogram: time of the ANTARES neutrino event in coincidence with 3C279.
}
\label{fig:Result_3C279}
\end{figure}

\section{Summary}
This paper discusses the first time-dependent search for cosmic neutrinos using the data taken with the full 12 lines ANTARES detector during the last four months of 2008. 
Time-dependent searches are significantly more sensitive than standard point-source search to variable sources thanks to the large reduction of the background of 
atmospheric muons and neutrinos over short time scales. This search has been applied to ten very bright and variable Fermi LAT blazars. The most significant observation 
of a flare is 3C279 with a p-value of about $10~\%$ after trials for which one neutrino event has been detected in time/space coincidence with the gamma-ray emission. Limits have been obtained on 
the neutrino fluence for the ten selected sources.
The most recent measurements of Fermi in 2009-11 show very large flares yielding a more promising search of neutrinos~\cite{bib:Fermi3C454}.


\section{Acknowledgments}
I greatfully acknowledge the financial support of MICINN (FPA2009-13983-C02-01 and MultiDark 
CSD2009-00064) and of Generalitat Valenciana (Prometeo/2009/026).

\clearpage

\setcounter{figure}{0}
\setcounter{table}{0}
\setcounter{footnote}{0}
\setcounter{section}{0}
\newpage




\title{Search for neutrinos from transient sources with the ANTARES telescope and optical follow-up observations}

\shorttitle{Damien Dornic \etal The TAToO project}

\authors{Ageron M.$^{1}$, Akerlof C.$^{2}$, Al Samarai I.$^{1}$, Basa S.$^{3}$, Bertin V.$^{1}$, Boer M.$^{4}$, Brunner J.$^{1}$, Busto J.$^{1}$, Dornic
D.$^{5}$, Klotz A.$^{4,6}$, Sch\"ussler F.$^{7}$, Vallage B.$^{7}$, Vecchi M.$^{1}$ and Zheng W.$^{2}$ on behalf the ANTARES, TAROT and ROTSE Collaborations }
\afiliations{$^1$ CPPM, CNRS/IN2P3 - Universit\'e de M\'editerran\'ee, 163 avenue de Luminy, 13288 Marseille Cedex 09, France
             $^2$ Randall Laboratory of Physics, Univ. of Michigan, 450 Church Street, Ann Arbor, MI, 48109-1040, USA
             $^3$ LAM, BP8, Traverse du siphon, 13376 Marseille Cedex 12, France
	     $^4$ OHP, 04870 Saint Michel de l'Observatoire, France
	     $^5$ IFIC - Instituto de F�\u0131sica Corpuscular, Edificios Investigaci�on de Paterna, CSIC - Universitat de Val`encia, Apdo. de Correos 22085, 46071 Valencia, Spain
	     $^6$ IRAP, 9, avenue du Colonel Roche, BP44346, 31028 Toulouse Cedex 4, France
             $^7$ CEA-IRFU, centre de Saclay, 91191 Gif-sur-Yvette, France	    	     
	     }
\email{dornic@ific.uv.es}

\maketitle

\begin{abstract}
The ANTARES telescope is well suited to detect neutrinos produced in astrophysical transient sources as it can observe a
full hemisphere of the sky at all the times with a duty cycle close to unity and an angular resolution better than 0.5
degrees. Potential sources include gamma-ray bursts (GRBs), core collapse supernovae (SNe), and flaring active galactid
nuclei (AGNs). To enhance the sensitivity of ANTARES to such sources, a new detection method based on coincident
observations of neutrinos and optical signals has been developed. A fast online muon track reconstruction is used to trigger
a network of small automatic optical telescopes. Such alerts are generated one or two times per month for special events
such as two or more neutrinos coincident in time and direction or single neutrinos of very high energy. Since February 2009,
ANTARES has sent 37 alert triggers to the TAROT and ROTSE telescope networks, 27 of them have been followed. First results
on the optical images analysis to search for GRB and core-collapse SNe will be presented. 
\end{abstract}


\section{Introduction}

The detection of high energy cosmic neutrino from a source would be a 
direct evidence of the presence of hadronic acceleration within the source and provide
important information on the origin of the high energy cosmic rays. Powerful
sources of transient nature, such as gamma ray bursts or core collapse supernovae, offer one of the
most promising perspectives for the detection of cosmic neutrinos as, due to their
short duration, the background from atmospheric neutrinos and muons is strongly reduced.
For example, several authors predict the emission of neutrinos in correlation with
multi-wavelength signals, e.g. the Fireball model of GRBs~\cite{bib:FireballRef}.
As neutrino telescopes observe a full hemisphere of the sky
(even the whole sky if downgoing events are considered) at all times,
they are particularly well suited for the detection of transient phenomena.

The ANTARES neutrino telescope~\cite{bib:Antares} is located in the
Mediterranean sea, 40~km South of the french coast of Toulon, at a depth of
about 2500~m below sea level. The detector is an array of photomultipliers
tubes (PMTs) arranged on 12 slender vertical detection lines.
Each string comprises up to 25 floors, i.e. triplets of optical modules (OMs)
housing one PMT each. Data taking started in 2006 with the operation of the
first line of the detector. The construction of the 12 line detector was completed in 2008.
The main goal of the experiment is to search for neutrinos of astrophysical
origin by detecting high energy muons ($\geq$100~GeV) induced by their
neutrino charged current interaction in the vicinity of the detector.
Due to the large background from downgoing cosmic ray induced muons, the
detector is optimised for the detection of upgoing neutrino induced muon tracks.

In this paper, the implementation and the first results of a strategy for the detection of transient sources is presented.
This method, earlier proposed in~\cite{bib:Marek}, is based on the optical follow-up of selected neutrino events
very shortly after their detection by the ANTARES neutrino telescope.
The alert system is known as ``TAToO'' (Telescopes and ANTARES Target of Opportunity)~\cite{bib:vlvnt09},

\section{ANTARES neutrino alerts}

The criteria for the TAToO trigger are based on the features of the
neutrino signal produced by the expected sources.
Several models predict the production of high energy neutrinos greater than
1 TeV from GRBs~\cite{bib:GRB} and from Core Collapse
Supernovae~\cite{bib:CCSN}. Under certain conditions, multiplet of
neutrinos can be expected~\cite{bib:CCSN1}.

Two online neutrino trigger criteria are currently implemented in the TAToO
alert system:
\begin{itemize} 
\item the detection of at least two neutrino induced muons coming from
similar directions within a predefined time window;
\item the detection of a single high energy neutrino induced muon.
\end{itemize}

A basic requirement for the coincident observation of a neutrino and an optical
counterpart is that the pointing accuracy of the neutrino telescope should be
at least comparable to the field of view of the TAROT~\cite{bib:Tarot} and ROTSE~\cite{bib:Rotse} telescopes ($\approx 2^\circ \times 2^\circ$).  

To select the events which might trigger an alert, a fast and robust algorithm
is used to reconstruct the calibrated data. This algorithm uses an idealized detector
geometry and is independent of the dynamical positioning calibration. As a result, the hits of the three 
OMs of a storey are grouped and their location assigned to the barycenter of the storey.
The storey orientations as well as the line-shape deviations from straight lines are
not considered in the online reconstruction.
A detailed description of this algorithm and its performances
can be found in~\cite{bib:BBfit}.
The principle is to minimize a $\chi^2$ which compares the times of selected
hits with the expectation from a Cherenkov signal of a muon track.
The resulting direction of the reconstructed muon track is available within
about $10~\mathrm{ms}$ and the obtained minimal $\chi^2$ is used as fit quality
parameter to remove miss-reconstructed tracks.

Atmospheric muons, whose abundance at the ANTARES detector~\cite{bib:atmu}
is roughly six orders of magnitude larger than the one
of muons induced by atmospheric
neutrinos, are the main background for the alerts and have to be
efficiently suppressed. Among the surviving events, neutrino candidates
with an increased probability to be of cosmic origin are
selected~\cite{bib:difflux}.

In order to establish the criteria used for our neutrino selection,
we have analysed a subsample of data taken by ANTARES after the completion of the
12-line detector, corresponding to a livetime of 70.3 days.
During this period, around 350 upgoing neutrino candidates were reconstructed
and have been compared to a Monte Carlo (MC) simulation of atmospheric muons
and neutrinos using the same lifetime.

Figure~\ref{fig:AngularResolutionAAFitBBFit} shows the angular
resolution of the online algorithm as a function of the neutrino energy.
This resolution is defined as the median of
the space angular difference between the direction of the incoming neutrino and the
reconstructed neutrino-induced muon.
For neutrinos with an energy higher than a few tens of TeV, an angular
resolution of 0.4 degree is achieved,  
despite of the approximations related to the detector geometry.
For example, the inclination of the ANTARES line for a typical sea current
of 5~cm/s induces a systematic angular deviation of less than 0.2 degree.

\begin{figure}[ht!]
\centering
\includegraphics[width=0.45\textwidth]{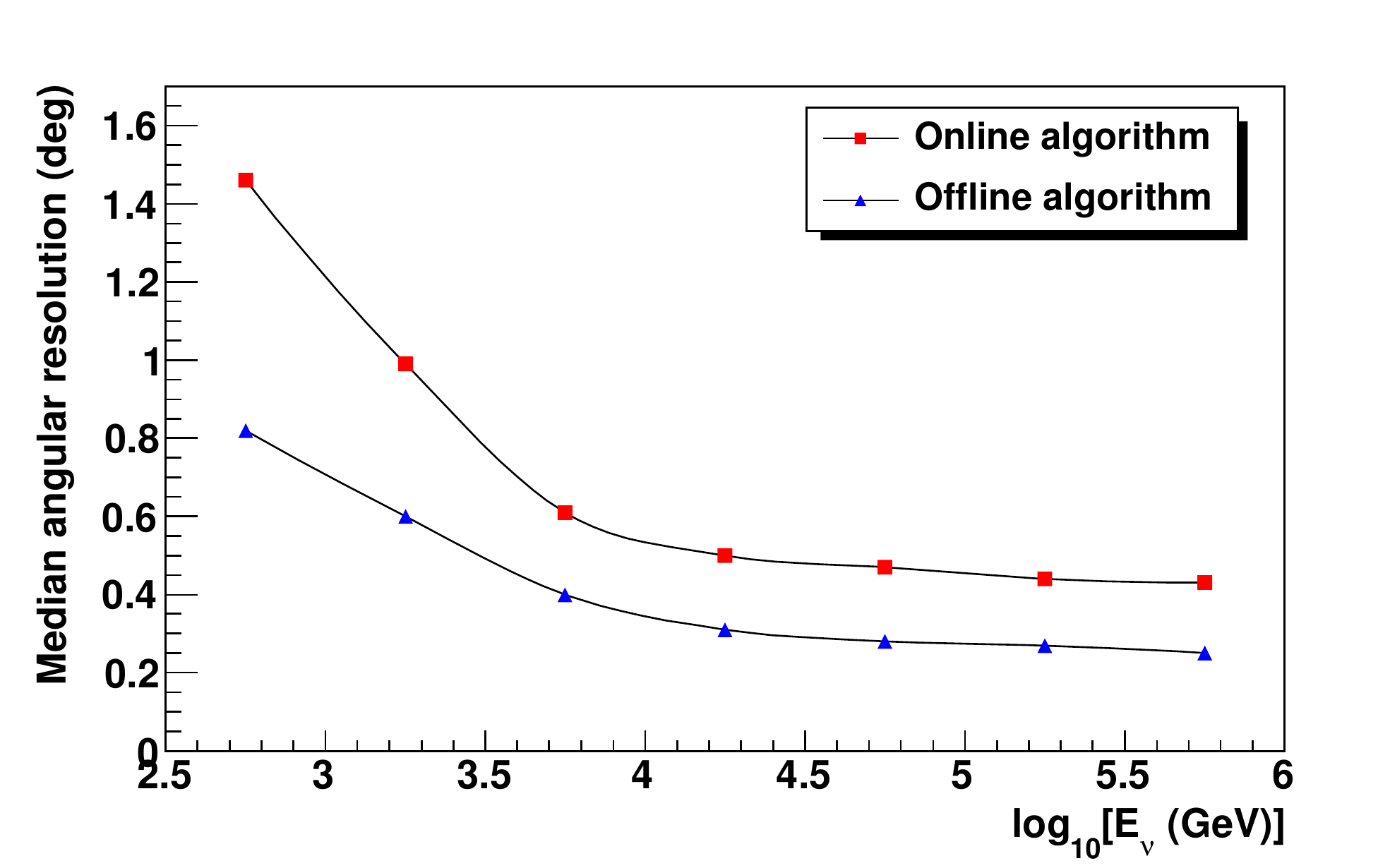}
\caption{Angular resolution obtained for both online and offline
  reconstructions as a function of the neutrino energy.}
\label{fig:AngularResolutionAAFitBBFit}
\end{figure}

\subsection{Multi-neutrino trigger}
The typical signature of the transient emission of high energy
neutrinos is a neutrino burst, i.e. a multiplet of neutrino
events originating from the source in a short time window.
A trigger for this event type is implemented as the detection
of two upgoing events reconstructed with at least two lines
in a 15~minutes time window
with a maximum angular difference of 3$^\circ$.
The time window was optimized to include most predictions 
of the neutrino emission by various models for transient sources.
The 3$^\circ$ angular window was selected to match
the convolution of the track reconstruction angular resolution and
the field of view of the robotic optical
telescopes ($\approx 2^\circ \times 2^\circ$).
The accidental coincidence rate due to background events, from two
uncorrelated upgoing atmospheric neutrinos,
is estimated to be $7\times10^{-3}$
coincidences per year with the full ANTARES detector.
With such a small background, the detection of a doublet (triplet) in
ANTARES would have a significance of about 3 (5) sigma.

\subsection{High energy event trigger}
Since the neutrino energy spectrum for signal events
is expected to be harder than
for atmospheric neutrinos, a cut on the reconstructed energy
efficiently reduces the atmospheric neutrino background
while most of the signal events are kept. The selection of the alert candidates is based on two simple energy estimators:
the number of storeys used in the track fit and the total amplitude (in photoelectrons) of the hits
in the storeys. 

The event selection for the high energy trigger has been tuned on
atmospheric neutrinos in order to obtain a false alarm rate
of about 25 alerts per year. This rate was agreed between ANTARES and the optical telescope collaborations.
A requirement of at least 20 storeys on at least three lines and an
amplitude greater than 180 photoelectrons will select around 25 
high energy events per year with the full 12 line configuration
of the ANTARES detector. The TAToO alert criteria select neutrinos of energies above 10 TeV for the single high
energy trigger (calculated with a neutrino Monte Carlo generated with an $E^{-2}$ energy spectrum).

Figure~\ref{fig:ZenithAzimuth} shows the estimation of the point spread function for a typical high 
energy neutrino alert. Around 70\% of the events are contained in the field of view of a typical 
robotical telescope ($\approx 2^\circ \times 2^\circ$). With a larger delay (few tens of minutes 
after the time of the burst), we are able to run the standard reconstruction tool which provides a much 
better angular resolution using the dynamical positioning of the detector lines~\cite{bib:AAFit}. Simulations indicate 
that, with this algorithm, ANTARES reaches an angular resolution smaller than about 0.3-0.4$^\circ$ for 
neutrino energies above 10~TeV (curve labeled 'offline algorithm' in Figure~\ref{fig:AngularResolutionAAFitBBFit}).

\begin{figure}[ht!]
\centering
\includegraphics[width=0.45\textwidth]{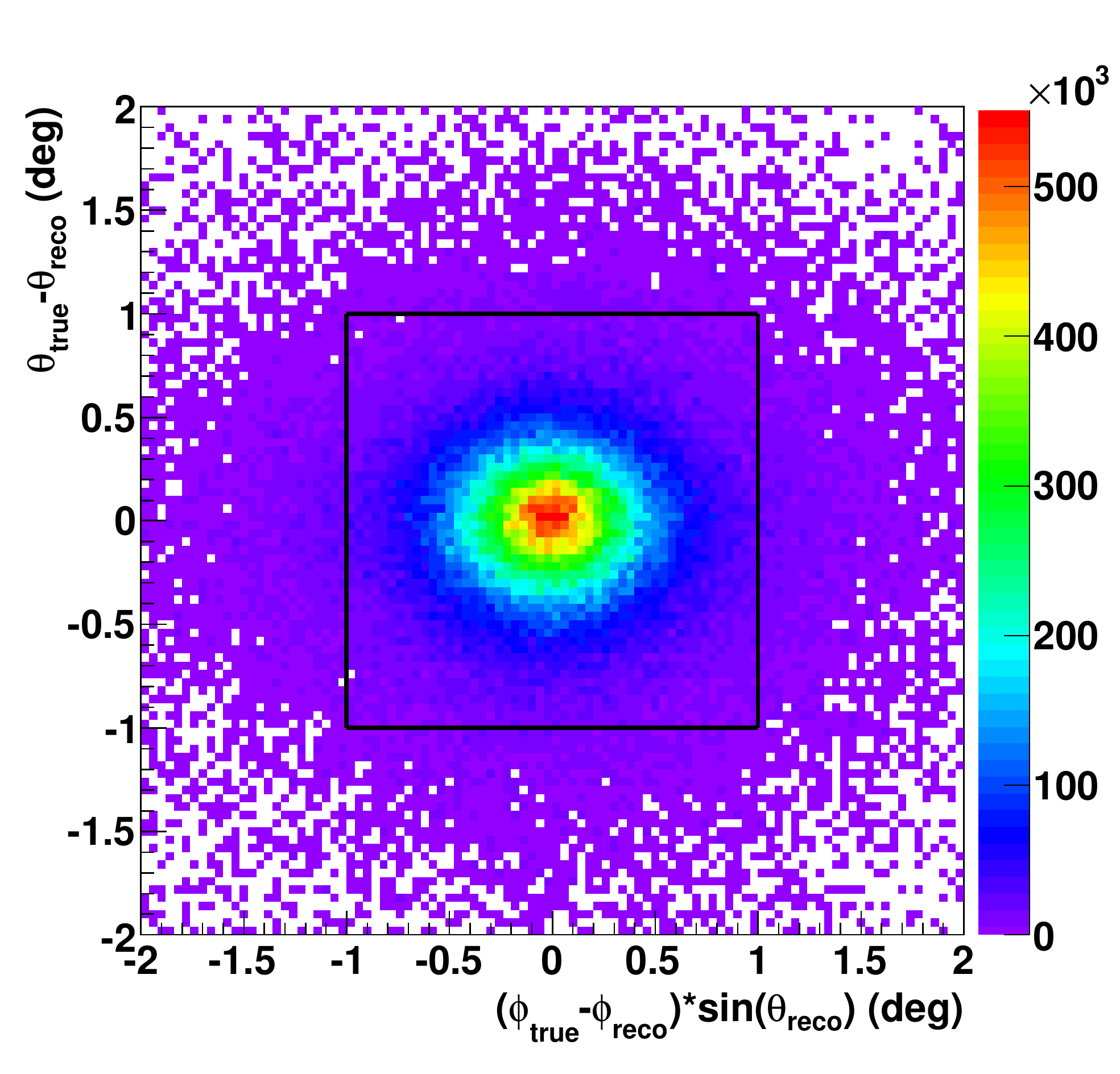}
\caption{Bi-dimensional angular resolution. The black square
corresponds to the TAROT telescope field of view ($\approx 2^\circ \times 2^\circ$).}
\label{fig:ZenithAzimuth}
\end{figure}

\section{Observation strategy of the robotical telescopes}

ANTARES is organizing a follow-up program in collaboration with the TAROT and ROTSE telescopes The 
TAROT~\cite{bib:Tarot} network is composed of two 25~cm optical robotic telescopes located at Calern 
(France) and La Silla (Chile). The ROTSE~\cite{bib:Rotse} network is composed of four 45~cm optical 
robotic telescopes located at Coonabarabran (Australia), Fort Davis (USA), Windhoek (Namibia) and Antalya 
(Turkey). The main advantages of these instruments are the large field of view of about 2 x 2 square degrees 
and their very fast positioning time (less than 10s). These telescopes are perfectly tailored for such a program. 
Thanks to the location of the ANTARES telescope in the Northern hemisphere (42.79 degrees latitude), all the six 
telescopes are used for the optical follow-up program. Depending on the neutrino trigger settings, the alert are sent 
at a rate of abour one or two times per month. With the current settings, the connected telescopes can start taking 
images with a latency of the order of one minute with respect to the neutrino event (T0).

As it was said before, the rolling search method is sensitive to all transient sources producing high energy 
neutrinos. For example, a GRB afterglow requires a very fast observation strategy in contrary to a core collapse 
supernovae for which the optical signal will appear several days after the neutrino signal. To be sensitive 
to all these astrophysical sources, the observational strategy is composed of a real time observation 
followed by few observations during the following month. For the prompt observation, 6 images 
with an exposure of 3 minutes and 30 images with an exposure of 1~min are taken respectively by the first 
available TAROT and ROTSE telescopes. The integrated time has been defined in order to reach an average magnitude of about 19.
For each delayed observation, six images are taken at T0+1,+2, +3, +4, +5, +6, +7, +9, +15, +27~days after the 
trigger for TAROT (8 images for ROTSE the same days plus T0+16 and T0+28~days).

\section{Optical image analysis}

Once the images are taken, they are automatically dark subtracted and flat-fielded at the telescope site. 
Once the data are copied from the telescopes, an offline analysis is performed combining the images
from all sites. This off-line program is composed by three main steps: astrometric and photometric 
calibration, subtraction between each image and a reference one and light curve determination for each variable 
candidates. 

Curently, two offline analysis pipelines are used: the ROTSE automated pipeline and one specially adapted to the TAROT and ROTSE image quality based on a
program originally developed 
for the supernova search in the SuperNova Legacy Survey (SNLS) project. Cases like variable PSF due to the atmospheric conditions or the lower quality images on the CCD 
edges have to be optimized in order not to loose any optical information. The choice of the reference is based on 
quality criteria such as the limiting magnitude and the seeing. For the GRB search, the reference is picked among 
the follow-up observations (few days after the alert) where no GRB signal is expected anymore while for SN 
search, either we consider the first night observation or we order it few months later to have a better quality 
reference in absence of a SN signal. It is also planned that the image analysis step will be included at the end of the automatic detection chain.

The ROTSE pipeline has been applied to five alerts from which optical images have been recorded during the first 24 hours after the neutrino alert sending. The
minimum delay between the neutrino detection and the first image is around 70~s. No object has been found for which the light curve is compatible with a fast
time decreasing signal.

\section{Conclusion}

The method used by the ANTARES collaboration
to implement the search for coincidence between high energy neutrinos and
transient sources followed by small robotic telescopes has been presented. 
Of particular importance for this alert system are the ability
to reconstruct online the neutrino direction and to reject efficiently the
background. With the described ANTARES alert sending capability,
the connected optical telescopes can start taking images with a latency of the order of one minute,
which will be reduced to about 15~s in the near future. 
The precision of the direction of the alert is much better than one degree.
The quasi-online availability of a refined direction obtained using the
measured geometry of the ANTARES detector further improves the
quality and efficiency of the alert system.

The alert system is operational since February 2009, and as of May 2011,
37 alerts have been sent, all of them triggered by the high energy selection
criterium. No doublet trigger has been recorded yet. After a commissioning phase in 2009, almost all alerts had an optical
follow-up in 2010, and the live time of the system over this year
is strictly equal to the one of the ANTARES telescope, namely 87\%.
These numbers are consistent with the expected trigger rate, after accounting for the duty cycle of the neutrino telescope.
The image analysis of the 'prompt' images has not permitted to discover a GRB afterglow associated to the high energy neutrino. 
The analysis of the rest of the images to look for the light curve of a core collapse SN is still on-going.

The optical follow-up of neutrino events significantly improves the perspective
for the detection of transient sources. A confirmation by an optical telescope
of a neutrino alert will not only provide information on the nature of
the source but also improve the precision of the source direction determination in order to
trigger other observatories (for example very large telescopes for
redshift measurement). The program for the follow-up of ANTARES neutrino
events is already operational with the TAROT and ROTSE telescopes
and results based on analysis of the optical images will be
presented in a forthcoming paper. This technique could be extended to observations in other wavelength
regimes such as X-ray or radio.


\section{Acknowledgments}
This work has been financially supported by the GdR PCHE in France. I greatfully 
acknowledge the financial support of MICINN (FPA2009-13983-C02-01 and MultiDark 
CSD2009-00064) and of Generalitat Valenciana (Prometeo/2009/026).

\clearpage

\setcounter{figure}{0}
\setcounter{table}{0}
\setcounter{footnote}{0}
\setcounter{section}{0}
\newpage




\title{SN neutrino detection in the ANTARES neutrino telescope}

\shorttitle{V.Kulikovskiy \etal SN neutrino detection in the ANTARES}

\authors{V.Kulikovskiy$^{1,2}$, on behalf of the ANTARES collaboration}
\afiliations{$^1$INFN - Sezione di Genova, Via Dodecaneso 33, 16146 Genova\\ $^2$Lomonosov Moscow State University Skobeltsyn Institute of Nuclear Physics (MSU SINP), 1(2), Leninskie gory, GSP-1, Moscow 119991, Russian Federation }
\email{vladimir.kulikovskiy@ge.infn.it}

\maketitle
\begin{abstract}
Neutrinos of all flavours are produced during a Supernova (SN) explosion. The antineutrinos can interact in the medium surrounding the underwater ANTARES telescope and produce positrons emitting Cerenkov radiation which is recorded by the optical modules (OMs) of the detector. The signature of this event is a simultaneous increase of the counting rate in the detector. The aim of this work is to develop several methods to detect the SN signal in ANTARES detector. In particular this analysis, using the specific geometrical configuration of the OMs, minimizes the noise effect introduced by the bioluminescence during the measurement.  The significance as a function of the number of active OMs and of the distance to SN is also evaluated.
\end{abstract}


\section{Introduction.}
The ANTARES neutrino telescope~\cite{general} is a large water Cerenkov detector that could provide the opportunity to detect the Cerenkov light produced by positrons from SN antineutrinos via reaction $\nu_{e}+p\to n+e^+$ .

To investigate this possibility we evaluate the response of the ANTARES optical modules (OM) to SN neutrinos with a help of Geant4 simulations and compare the possible SN signal rates to the ANTARES OMs rates, which are recorded during the normal run operations. 

The ANTARES detector is made of detection lines that consist of 25 storeys, a group of a 3 OMs that are ar-ranged close to each other (distance is less than 1m). This configuration provides an opportunity to reduce the background using only hits that are in coincidence be-tween OMs in the same storey.

The methods to minimize the effect of the bioluminescence burst are discussed in the Section 3 while the statistical analysis based on singles rates, double coincidences and triple coincidences in one storey is presented in Section 4, 5 and 6. The performances of the three different meth-ods are also compared, and significance is evaluated.

\section{Simulations.}
We performed the simulations using Geant4 with the detailed OM description. In these simulations we included the full 3 OMs ANTARES storey configuration while for the positrons we used the energy spectrum and the flux obtained averaging in time the distributions from the model 57 of A.Burrows~\cite{burrows} for a SN1987A like event at a 10 kpc distance.

As a result we have obtained an excess of 14 hits in a 105ms time interval on each OM. The time interval of 105ms was chosen because it both corresponds to the standard time slice of the data flow in ANTARES detector and to the period of SN explosion where the highest neutrino flux is produced according to~\cite{burrows}. To fully exploit the geometry of ANTARES detector, we have also obtained the number of coincidences produced by the positrons light between two or three OMs of the same storey in a 25ns time window. The possibility to use these multi coincidences will be described in the next sections. To evaluate the number of hits from an event at a distance R different from the 10 kpc we used a simple proportion, knowing that $\Phi*4\pi R=const$ .
\section{The bioluminescence filter.}
In addition to the Cerenkov light produced by muons, neutrino telescopes also detect Cerenkov light from the decay of radioactive elements and light from luminescent organisms. Fig.\ref{bg_time} represents the time variation of the optical background rate detected in one OM. One of the main contributions comes from the radioactive decay of $^{40}{\mbox K}$. The frequency of this noise is evaluated by MC calculations to be around 40kHz and does not change since the $^{40}{\mbox K}$ concentration depends on the salinity that is almost constant on time and sea depth. However in fig.\ref{bg_time} the mean rate is about 80kHz, the additional contribution being due to a constant bioluminescent activity.

 \begin{figure}[!t]
  \centering
  \includegraphics[width=0.8\columnwidth]{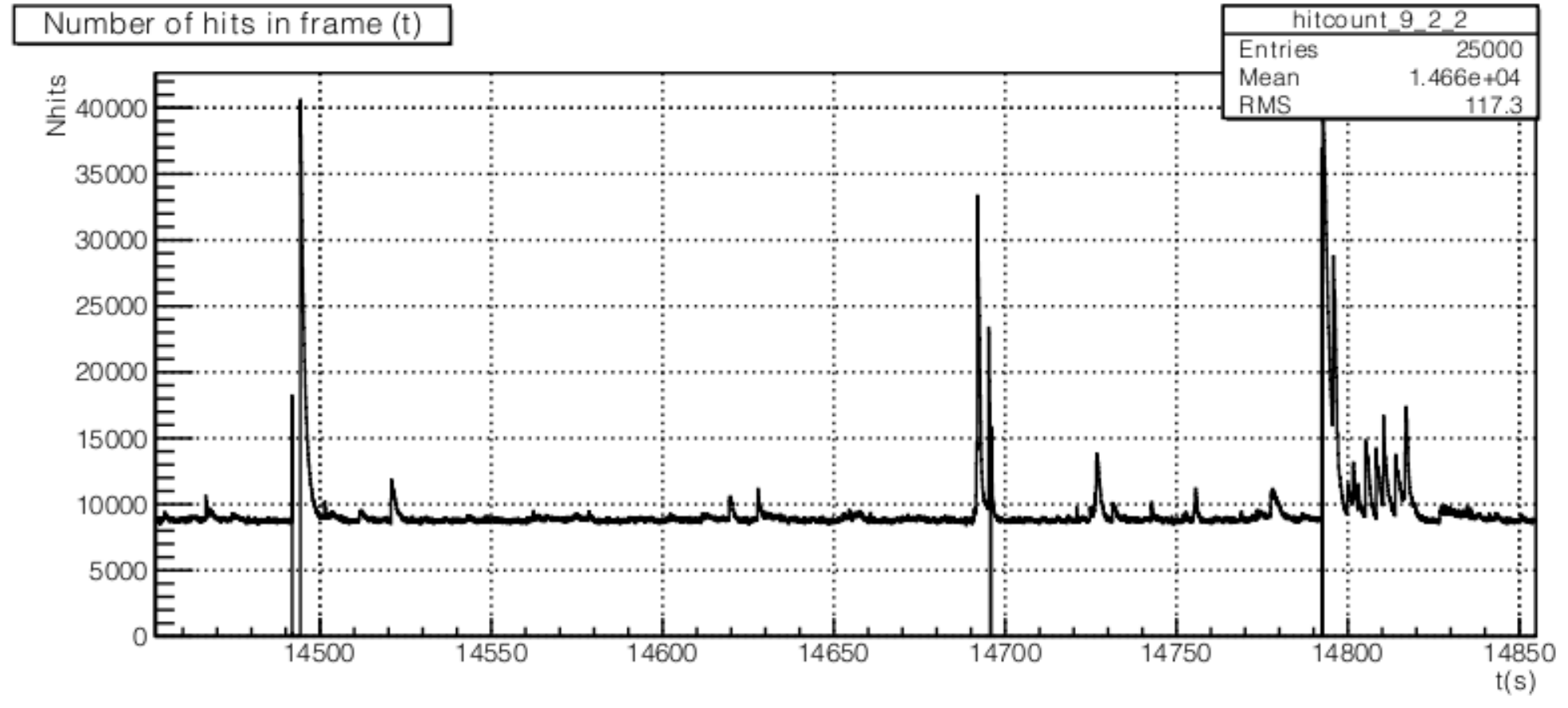}
  \caption{Example of singles rate (0.3pe threshold) events in one OM of the detector.}
  \label{bg_time}
 \end{figure}
Above the slow variable noise one can see high sharp peaks due to the emission of light bursts from macro organisms in transit close to the OM. Later in this note we will term these peaks as "bioluminescent burst" to distinguish them from the continuous or slowly variable bioluminescence noise.

To exclude the bioluminescence bursts on one OM we have fitted the distribution of the number of hits in a 105ms period. The fit is performed with a Poisson function. There are three parameters in the fit: $\lambda$ is the mean value of the Poisson distribution, $j^{max}$ is the right margin of the fit and A is the scaling factor (height of the distribution).

The filter for the bursts is very simple: if in one time frame the number of hits for a particular OM is higher than $j_i^{max}$, then the OM is considered as affected by a bioluminescence burst and for this time frame OMi is not used in the calculation of the total detector hit counts.
\section{4	Supernova signal with the single OMs rates.}
Detection significance can be defined as the ratio between the measured signal and the uncertainty of the measurement \begin{equation}
S=\frac{Signal}{\sigma_{measurement}}
\end{equation}In case of 14 signal hits ($h_{signal}$) and of 80kHz of back-ground for 900 active OMs ($N_{OM}$) the significance can be estimated by the formula:\begin{equation}
S=\frac{h_{signal}N_{OM}}{\sqrt{h_{background}N_{OM}}}
\end{equation}
The main idea of all methods described here being to measure the increase of the total hits in the detector in one time slice and to compare it with the previous measurements. For single hits in the specific case of ANTARES detector, this cannot be directly applied due to the bioluminescence bursts that have been therefore excluded using the filter as described above. The procedure we have used is the following:

a) The data collected in a 45min time interval, sufficiently large to have good statistics, were used to determine the fit $h_i^{exp}$ of the hit counts distribution for each $i^{th}$ OM in the detector. $h_i^{exp}$ can be seen as the probability density function (PDF) of the hits in OM$_{i}$. The results from the fit were also used to define quality cuts. In particular only OMs for which the hit distribution follows the next criteria where accepted for the analysis: the fit converges, $\lambda\ge$2000, the area of the Poisson component is larger than 30\% of the total area (in other words: there is less than 70\% of the bio-burst hits).

b) The data are then analysed time slice by time slice. At each time slice the hits of all the OMs that have passed the bioluminescence cut are summed together to determine the total hit count of the detector in that particular time slice.

c) The total hit count of the detector should be described by the "detector PDF" distribution, which is different for every time slice because OMs which have passed the bioluminescence cut are different. The detector PDF distribution can be evaluated using the PDFs for every OM, but this is a very time consuming procedure. To simplify we have made a simulation using all the parameters from fits of the single PDFs which proved that detector PDF is still a Gaussian with an average $M=\Sigma m_i$  where $m_i$ is mean of PDF for $i^{th}$ working OM and with a sigma $S=\sqrt{\Sigma\sigma^2_i}$  where $\sigma_i$ is a variance for $i^{th}$ PDF.

To estimate the significance two runs were selected. One of them was \#39656, recorded on 16 May 2009, where bioluminescence conditions were good (rate: 80kHz for the bottom, 60kHz - middle, 60kHz - top), another one was \#40154 recorded on 13 April 2009 when bioluminescence was higher (rate 200-160-140kHz). 

The parameters of the fit ($\lambda$, A, $j^{max}$) were then extracted from the experimental distributions as deduced from the analysis of the 2 runs. In order to compare with the other methods described in the note, we have calculated the significance assuming 900 working PMTs and keeping the same mean characteristics of the background of the two previous runs.

The results of these simulations are shown in fig.\ref{fig:2} where the significance is plotted as a function of the distance of the SN. 900 OMs in the same conditions of the runs 39656 and 40154 were simulated. The decrease of the significance with the increase of the SN signal (from distance 3.5 kpc and closer) is due to the bioluminescence filter: when the SN signal is very high a lot of OMs are excluded from the analysis. The dotted lines with a 1.7$\sigma$ label in the same figure are obtained when the effect due to the long-term fluctuations is included in the calculation. This effect is probably ascribed to the baseline rate which is slightly changing during the analysis time of 45min and which increases the width of the distribution by a factor of about 1.7.

A maximum significance of 5 for SN at the 4 kpc is obtained at low bioluminescence level (run \#39656) for the full working detector when the long-term fluctuation is included.
 \begin{figure}[!t]
  \centering
  \includegraphics[width=0.9\columnwidth]{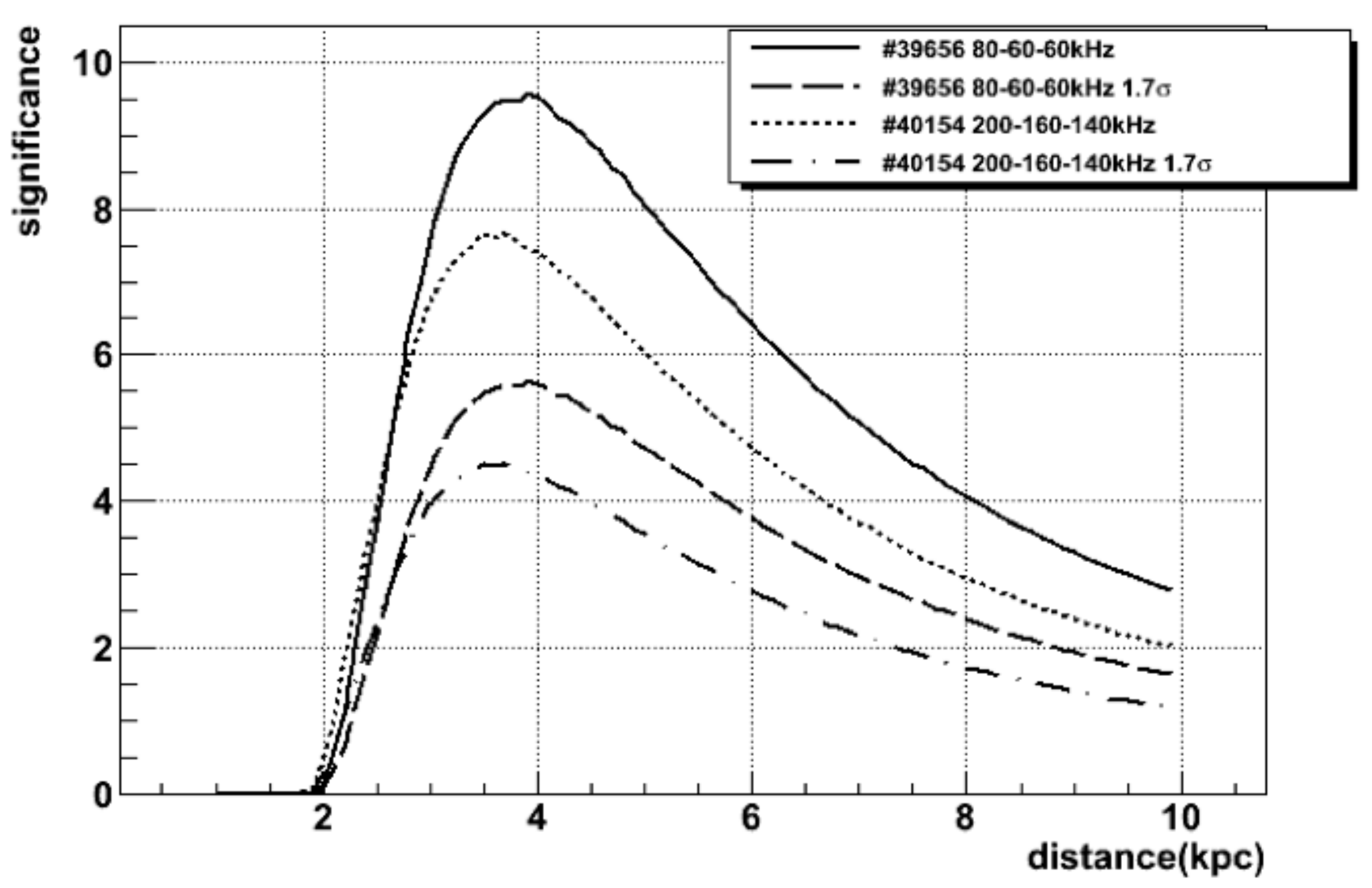}
  \caption{Simulated significance for 900 OMs detector in dependence of the distance from SN explosion. OMs signal is simulated according to run \#39656(solid) and \#40154(dotted) conditions. Dashed and dot dashed lines represent calculations with the 1.7 increase of sigma to account for long term fluctuations.}
  \label{fig:2}
 \end{figure} 

\section{Supernova signal for double coincidences between OMs}
The ANTARES configuration allows the light from $^{40}{\mbox K}$ decay to be simultaneously seen by a couple of OMs in a storey. This fact is rather well known and it is used in the detector calibrations [3]. Later on by the couple of OMs we assume only OMs from the same storey.

For every couple of OMs the plot of the time difference between every two hits from both OMs can be deter-mined. One example of this time distribution is shown in fig.\ref{fig:3}.

 \begin{figure}[!t]
  \centering
  \includegraphics[width=0.8\columnwidth]{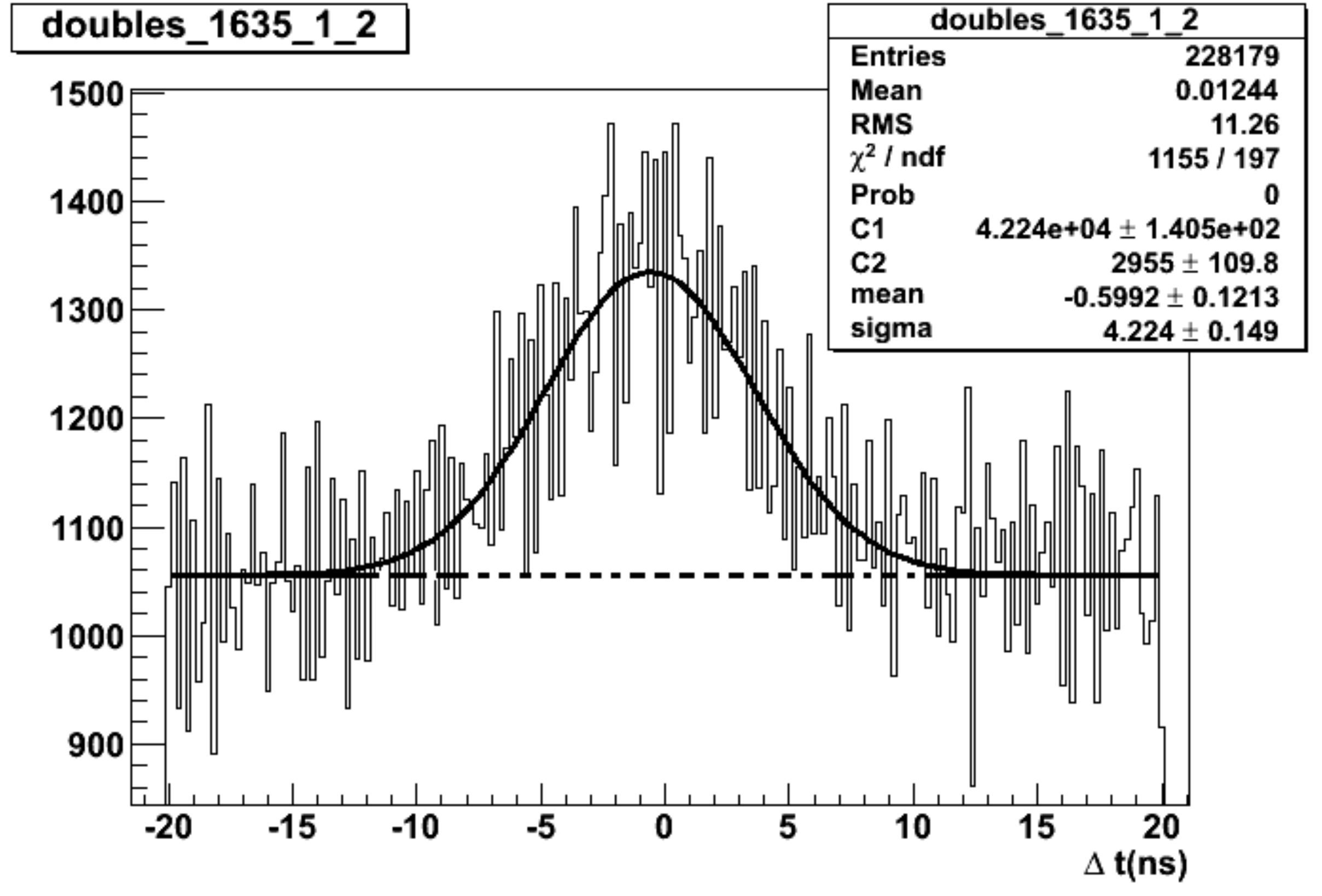}
  \caption{Example of the time coincidence distribution for storey 1635, OM 1 and 2. Dashed line represents random coincidence level. The area of the Gaussian ac-counts for the number of hits from $^{40}{\mbox K}$ coincidences.}
  \label{fig:3}
 \end{figure} 
We can distinguish a plateau due to the random coincidences in the time window $\tau$ of 20ns and a Gaussian shaped distribution due to $^{40}{\mbox K}$ decays seen by the both modules (later we call them �true coincidences� to distinguish from the random coincidences). To extract these two components the distribution was fitted with a Gaussian plus a constant. The rate of the true coincidences extracted from the fit is about 16Hz for a couple of OMs and it is independent from the bioluminescence activity. The rate is thus stable in time, but it depends from the couple of OMs since efficiency of the OMs is different.

To understand how the uncertainty on the true coincidence rate from the fit depends on the statistics i.e. on the number of the summed time slices, we performed a simple simulation of the time difference distribution for one couple of OMs assuming an 80kHz random background on each OM and a Gaussian distribution with 16Hz for the coincidences. These simulations gave the uncertainty on the extracted true coincidence rate as reported in fig.\ref{fig:4} for various number of time slices $N_{TS}$.

In this method we calculate the global rate of the coincidences in the detector at any time slice and compare it with the sum of the rates of each couples. The global detector rate was determined from the time differences between all coincidence hits of every OM couple that were collected in a common distribution. This distribution was then fitted with the same procedure described above and the error evaluated as in fig.\ref{fig:4} where the number of couples instead of the number of time slices was assumed. The true coincidence detector rate extracted from the fit is stable in time and equal to the sum of the every OM couple rates in case of normal work or higher in case of SN explosion. 
  \begin{figure}[!t]
  \centering
  \includegraphics[width=0.8\columnwidth]{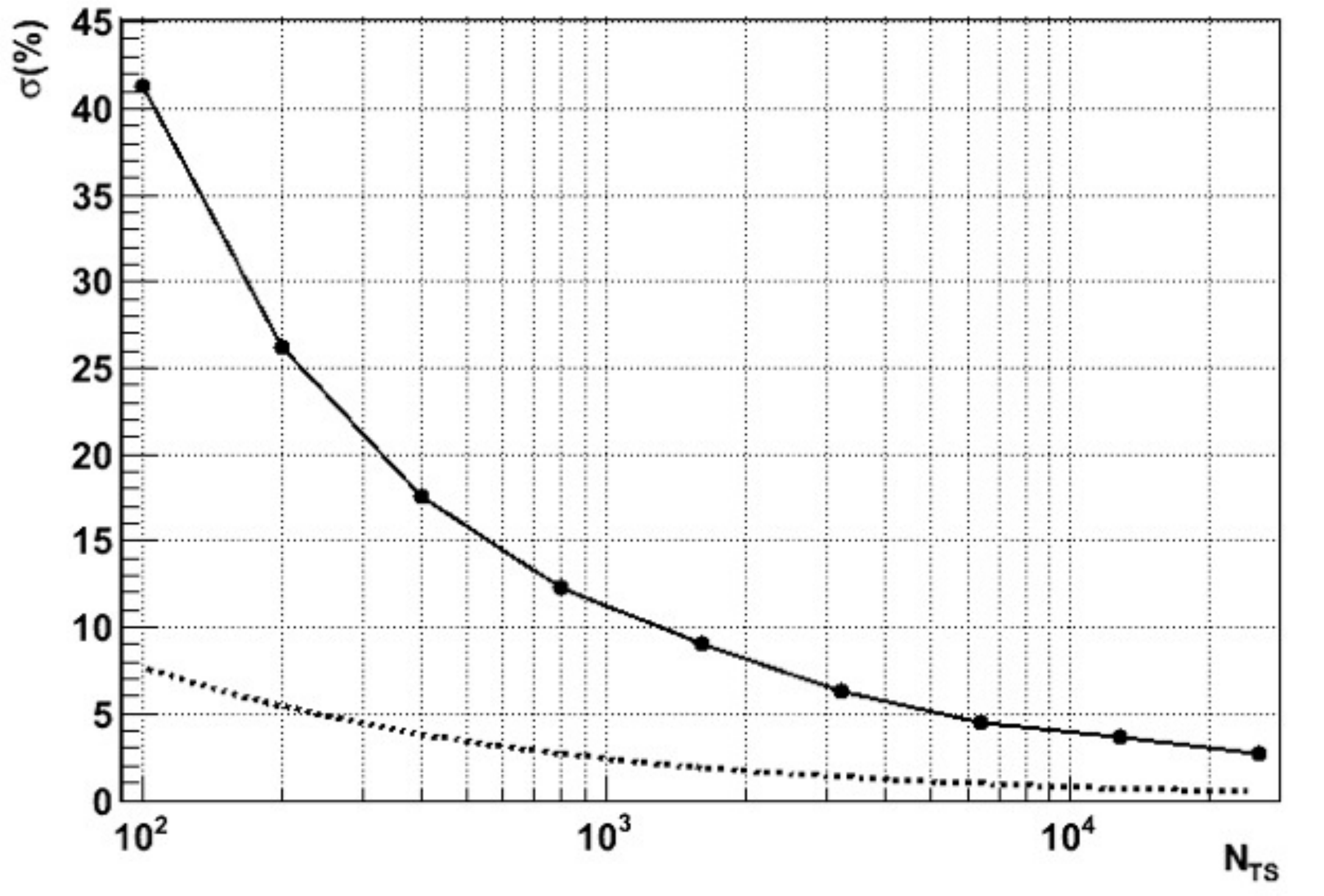}
  \caption{Uncertainty on the coincidence rate extracted from the fit of the simulated distribution for different time slice numbers (solid curve). Dotted curve represents error in absence of the random background.}
  \label{fig:4}
 \end{figure} 

We verified this procedure on real data. To this purpose one period with a 450 OM couples was analysed and the rate for every couple was calculated from 20000 time slices of the data to provide as seen from fig.\ref{fig:4} an uncertainty less than 5\%. The global detector true coincidence rate was then obtained from the fit for every time slice. The difference between this rate and the sum of the true coincidence rates of every couple was evaluated and normalized to the detector rate. As expected, the distribution of this normalized difference has a mean 0 and sigma 17\%, in good correspondence with the expected error from fig.\ref{fig:4} for 450 $N_{TS}$.
 
Simulations with Geant4 estimates a value of 22Hz for the true coincidence rate, the difference with respect to the measured 16Hz being ascribed mostly to the description of the lateral region of the OM. Geant4 simulations also show that the light produced by the positrons induced by SN neutrinos may simultaneously hit two and even three OMs in the same storey. These additional events increase the coincidence rate during the SN explosion. The Geant4 simulation of this effect gives a rate of about 2.8Hz in the first 100ms time window i.e. an increase of 0.28 events for each pair of OMs in the considered 105 ms time slice for a 10 kpc SN1987A. The measurement of this small increase, which was reduced down to 2Hz to take into account the difference between MC and data, can extend the significance to the SN explosion in ANTARES detector.

 \begin{figure}[!t]
  \centering
  \includegraphics[width=0.8\columnwidth]{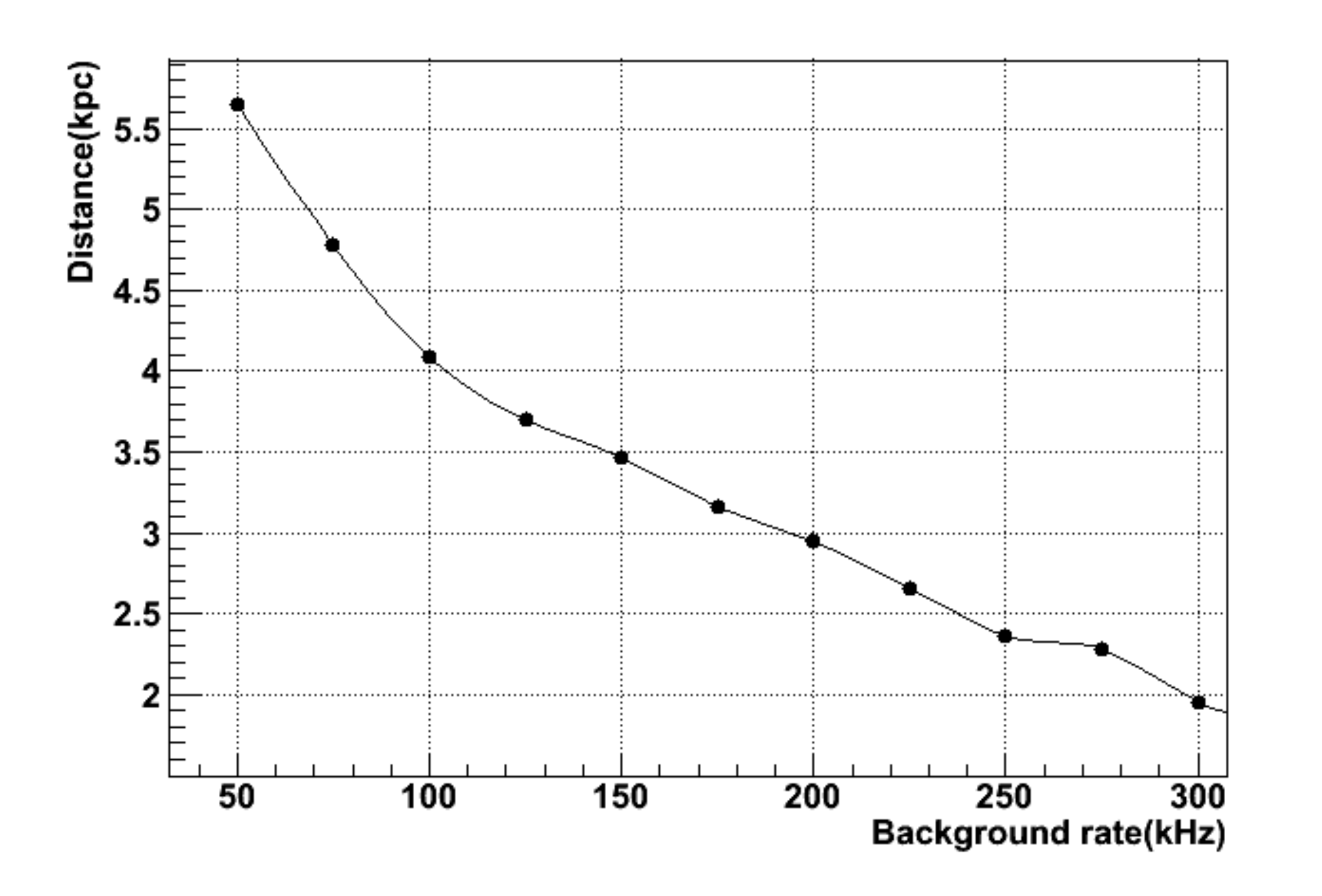} \\
  \includegraphics[width=0.8\columnwidth]{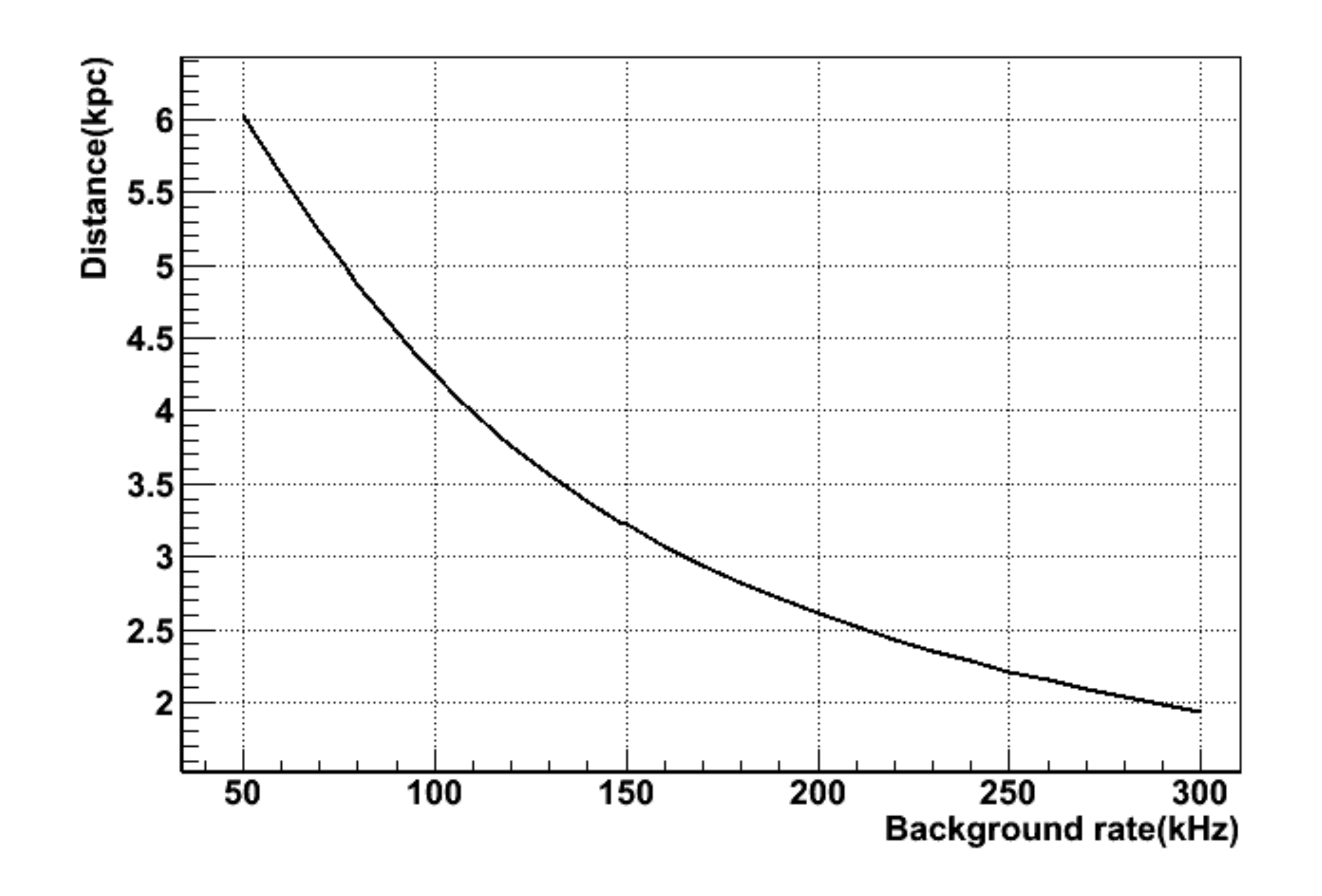} 

  \caption{Simulated 5$\sigma$ significance as a function of SN distance and background rate for the double (up) and a triple coincidence (bottom) methods.}
  \label{fig:5}
 \end{figure}

%

The 5$\sigma$ significance obtained from the simulations of 900 OMs detector is reported in fig.\ref{fig:5} as a function of the SN distance for various bioluminescence rates. For an event occurring at 4 kpc the significance is comparable to the previous method for a background rate of the 100kHz.
 
Given the ANTARES geometry, a triple coincidence in each storey can occur whenever a single event (light from a track, a $^{40}{\mbox K}$ decay) is simultaneously seen by the 3 OMs. In our approach we require that 3 hits, one for each OM, are detected during a 20ns time window. This means also that in every pair of the hits in the triple coincidence, the time difference is less than 20ns. In order to obtain the time difference distribution like for doubles it was decided to use three dependent distributions: $t_1-t_2$, $t_2-t_3$ and $t_3-t_2$, where ti is time of hits from OMi. In this case the random coincidences do not form a plateau in the time difference distribution. This is because, for example, the difference $t_1-t_2$ is constrained by the condition that both $t_1$ and $t_2$ must be close to $t_3$ by less than 20ns. Using Bayes' theorem one can evaluate the shape of the random coincidences part:\begin{equation}
f_{rand}(\Delta t=\frac{2\tau-|\Delta t|}{2\tau^2})
\label{form:4}
\end{equation}
The fit is thus done adding the Gaussian of the true coin-cidences to the random part (4). True coincidences rate for the storey is calculated as a mean value of the true coincidence rates extracted from the fit of the three dis-tributions $t_1-t_2$, $t_2-t_3$ and $t_3-t_2$.

To find the true triple coincidence rate, one month of data was analysed. It was found that the mean storey rate is about 0.065Hz and it is stable during 3 months. Simulations with Geant4 give a higher value: 0.2Hz. This large discrepancy can be explained considering the already observed behaviour in the double coincidence case where the simulations give a value around 22Hz against the observed 16Hz. In case of a triple coincidence this inefficiency is additionally amplified. The Geant4 simulation also estimates the total number of triple coincidences from SN to about 0.035 per storey in a time slice. Assuming the same damping in the detection efficiency as for $^{40}{\mbox K}$ rates (0.2Hz MC against 0.065Hz experimental), the SN triple coincidence rate is expected to be about 0.11Hz. 

Similarly to the double, the coincidences rate of the triple coincidences in the detector during one time slice is calculated. Unfortunately the number of triples is much less than the doubles, the fitting procedure is not applicable so that the significance is affected by the random coincidences~(\ref{form:4}). The simulated significance as a function of the SN distance and background is shown in the fig.\ref{fig:5}.

\section{Conclusion.}
We have investigated different possibilities for detecting SN neutrinos with the ANTARES telescope. The signifi-cances estimated for the three methods we have described are comparable (fig.\ref{fig:2}, fig.\ref{fig:5}). Double and triple coincidences seem preferable because significance increase with the SN signal and they do not strongly suffer from the bioluminescence activity. Of course the simultaneous acquisition of an external SN trigger like the one provided by the SNEWS collaboration will strength our analysis. The study to implement this possibility is now in progress.


\clearpage

\setcounter{figure}{0}
\setcounter{table}{0}
\setcounter{footnote}{0}
\setcounter{section}{0}
\newpage




\title{Study on possible correlation between events observed by the ANTARES neutrino telescope and the Pierre Auger cosmic ray observatory}

\shorttitle{J.Petrovic \etal ANTARES-PAO events correlation }

\authors{Jelena Petrovic$^{1}$ for the ANTARES collaboration }
\afiliations{$^1$Nikhef, The Netherlands}
\email{jelenap@nikhef.nl}

\maketitle
\begin{abstract}
According to the theory of hadronic acceleration, ultra high energy cosmic rays are expected to be accompanied by  gamma-rays and neutrinos from pion decays formed in the 
interactions of protons with photons. While gamma-rays have been linked to astrophysical sources by many experiments (H.E.S.S, MAGIC, Fermi), no point source of UHECRs or 
neutrinos have been found so far.
In this contribution, we present the results of multimessenger stacking sources analysis developed to investigate the correlation of arrival directions of neutrino candidate 
events and UHECRs. This analysis has been applied on neutrino candidate events detected during 2007-08 by the ANTARES telescope (ANTARES collaboration 2011), and 69 ultra-high 
energy cosmic rays observed by the PAO (The Pierre Auger Collaboration 2010).
\end{abstract} 


\section{Introduction}
Finding point sources of both UHECRs and neutrinos is a very challenging task, as both messengers have many down-sides.
UHECRs are rare and do not point back to their sources, since they are scrambled by galactic and intergalactic
magnetic fields \cite{lab1,lab2,lab3,lab4,lab5,lab6,lab7,lab8,lab9,lab10,lab11,lab12,lab13,lab14,lab15,lab16}.
Also, due to the interaction with the cosmic microwave background photons,
their range may be limited to the distance of about 100Mpc or less \cite{lab17,lab18,lab19,lab20,lab21}.
However, they are detectable with large shower arrays, like the Pierre Auger Observatory (PAO), which so far
reported the observation of few tens of events above 55EeV \cite{lab21,lab22}.
On the other side, cosmic neutrinos, as neutral and weak interacting particles, should point back to their sources
and their traveling distances should not be limited,
but at the same time they are very difficult to detect.
Currently operating neutrino telescopes ANTARES and IceCube
have not yet observed excess above the atmospheric neutrino flux coming from air showers
\cite{lab23,lab24}.

Previously, the Pierre Auger Observatory (PAO) reported an anisotropy in the arrival directions of UHECRs
\cite{lab21} and indicated a
correlation with Active Galactic Nuclei (AGN) from the VCV catalog \cite{lab25}. The correlation
was the most significant for 27 cosmic rays with energies higher than
57EeV and AGNs at distances less than $~$75Mpc.
The suggested correlation with the nearby AGN sources
mostly following the location of the supergalactic plane decreased in the subsequent analysis
\cite{lab22} with 69 events at energies above 55EeV, observed until 31st December 2009.

In this paper, we investigate the correlation of arrival directions of 2190 neutrino candidate events detected
by 5-12 line ANTARES neutrino telescope, and 69 UHECRs observed by the PAO.
This stacking sources analysis is developed for a blinded set of neutrino events. Blinding was performed by scrambling in right ascension.
At the end, the analysis was done with the unblinded neutrino dataset.

\section{The ANTARES telescope and data samples}

The ANTARES neutrino telescope is located in
the Mediterranean Sea, about 40km off the southern coast of France (42 48N, 6 10E), at a depth of 2475m.
It was completed in 2008, and its final configuration is
a three-dimensional array of 885 photomultipliers in glass spheres (optical modules),
distributed along twelve lines. These photomultipliers detect Cherenkov photons, from relativistic muons
produced in neutrino interactions nearby the detector.
The total instrumented volume of the ANTARES telescope is about 10$^7$m$^3$.
The data acquisition system of the detector is based on "all-data-to-shore" concept, in which signals from the
photomultipliers above a given threshold are digitized and sent to shore for processing.

The data used in this analysis were collected during 2007 and 2008, while the detector was operating with 
5 to 12 lines. For a part of that period, data acquisition was interrupted for the employment of new lines,
and in addition, some periods were excluded due to high bioluminescence-induced optical background. The resulting effective live time is
304 days. The final data sample consists of 2190 up-going (zenith angle up to 90$^{\circ}$) neutrino candidate
events. No selection was done based on the energy reconstruction. The angular resolution was estimated to be 0.5$\pm$0.1
degrees.
Details of data reconstruction, effective area, angular resolution and the dataset are given in \cite{lab24}.

This sample of neutrino candidate events was correlated with UHECR events recorded with the PAO surface detectors between January 1st 2004, and 31st December 2009.
Those events have zenith angles below 60 degrees, and reconstructed energy above 55EeV. 69 events satisfy these cuts,
and they are all in the ANTARES telescope field of view, as the field of view for the ANTARES telescope and the PAO greatly overlap.

\section{Background simulations}

A Monte Carlo set with 10$^{6}$ skies is generated, each with 2190 neutrinos and 69 cosmic rays. In each of the
million skies the position of the UHECRs is
fixed according to the PAO dataset, and the positions of 2190 neutrinos are obtained from the blinded 5-12 line dataset
scrambled in right ascension.
The numbers of neutrinos within bins of any size, centered on 69 UHECR events can be counted, probability density functions can be
calculated and fitted with Gaussian distributions, giving the mean neutrino count expected from the randomized background samples,
and the standard deviation of the neutrino count.

Figure \ref{prob} shows an example count of neutrinos within bins of 1-10 degree size. The count of events is done
by adding neutrinos for which the minimum angular distance to UHECRs is smaller than the bin size.
In this way, a double counting of neutrino events is avoided. After optimizing the bin, the significance of observed number of neutrino events within 69 bins
can be calculated by comparison with the distributions for the pure background MC samples.
After optimizing the bin, and unblinding the data, the significance of observed number of neutrino events within
69 bins can be calculated by comparison with the distributions for the pure background MC samples.

\section{Signal simulations}
Simulation of sources with E$^{-2}$ spectrum and equal flux strength from each 69 UHECR direction was performed.
Flux values from 0.5${\times}$10$^{-8}$GeV cm$^{-2}$ s$^{-1}$ to 10${\times}$10$^{-8}$GeV cm$^{-2}$ s$^{-1}$ are considered.
Further, for each flux,
the event rate per source is determined, using the effective area for 5-12 lines and the live time given in [24].
For every source, the amount of signal neutrinos is determined according to the Poisson distribution with the event rate per
source $R$ as mean value.

\begin{eqnarray*} \label{rate}
        R = t \int_{10^2 \textup{GeV}}^{10^7 \textup{GeV}} \Phi(E_{\nu}) A_{eff}(E_{\nu}) dE_{\nu}
\end{eqnarray*}

where $t$ is a live time of the ANTARES detector, $\Phi$ is the flux from each individual UHECR direction, and $A_{eff}$ is the ANTARES telescope effective area.

Signal neutrinos are randomly generated following a Gaussian distribution centered on the measured UHECR coordinates,
and the ANTARES telescope resolution \cite{lab24}. 

\begin{eqnarray*} \label{signal}
        \exp \left( -(x_1(\alpha,\delta) - \mu)^2 / (2\sigma_1^2) \right)  \to  \lefteqn{x_2(\alpha,\delta)} \\
        \exp \left( -(x_2(\alpha,\delta) - \mu)^2 / (2\sigma_2^2) \right)  \to  \lefteqn{x_{fin}(\alpha,\delta)}
\end{eqnarray*}

where $x_i$ = $x_i$ ($\alpha$, $\delta$) are neutrino coordinates, $x_{fin} (\alpha,\delta)$ are the final signal neutrino coordinates,  $\sigma_1$ is the tolerance value for 
the magnetic deflection,
and $\sigma_2$ is the ANTARES telescope angular resolution.
The width of the tolerance value accounting for a magnetic deflection Gaussian is
chosen to be 3 degrees, based on the PAO results [22,25].

The same amount of background neutrinos is removed from a declination band of 10 degrees centered on each UHECR to
ensure that every random sky has 2190 events, and to keep the neutrino declination distribution profile close to the observed profile.
The angular resolution of PAO
air shower reconstruction is about 0.9 degrees, less than the expected magnetic deflection, so it is not considered additionally.

\section{Sensitivity and bin optimization}
Although it is not possible to know the actual upper limit that will result from an experiment until looking into
unblinded data, Monte Carlo predictions can be used to calculate the average upper limit, or so called Feldman-Cousins sensitivity
\cite{lab26,lab27}, that would be observed after hypothetical repetition of the experiment with expected background $n_b$, and no true signal
$n_s = 0$.
Over an ensemble of experiments with no true signal, the background $n_b$ will fluctuate to different $n_{obs}$ values with different
Poisson probabilities, and upper limits ${\mu}_{90}$.

The "mean upper limit" is the sum of these expected upper limits, weighted by their Poisson probability of occurrence:

\begin{eqnarray*} \label{mu}
        \overline{\mu^{90}}(n_b) = \sum_{n_{obs} = 0}^{\infty} \mu^{90}(n_{obs},n_b) \left( n_b^{n_{obs}} / n_{obs}\! \right) e^{-n_b}
\end{eqnarray*}

Over an ensemble of identical experiments, the strongest constraint on the expected signal flux corresponds to a set of cuts that
minimizes the so called "model rejection factor" ${\mu}_{90}/n_s$ and at the same time minimizes the mean flux upper
limit that would be obtained over the hypothetical experimental ensemble.

The described Feldman-Cousin's approach with the Rolke extension \cite{lab28}
was used to calculate the mean upper limit on E$^{-2}$ flux per source,
for a 90\% confidence level, from 10$^6$ background samples, as a function of a search bin, as shown on a Figure \ref{sens}.

The search bin that minimizes the mean upper limit for 3 degrees
tolerance value accounting for a magnetic deflection is found to be
4.9 degrees. 

\section{Discovery potential}

With the bin size optimized and fixed, it is possible to estimate the probability of making a 3${\sigma}$ or a 5${\sigma}$  discovery given a certain signal flux.
This is done by taking the chosen significance from the background MC samples, and comparing this to MC samples with signal of a certain flux.
The number of skies with signal, that have more neutrinos in the given bin than the chosen significance from background only, is counted and gives a direct
measure of the discovery potential for that particular flux.

Figure \ref{discpot} shows the discovery potential for 3${\sigma}$ (dashed line) and 5${\sigma}$ (solid line) discovery, for an optimized bin of 4.9
degree bins, and a tolerance value accounting
for a magnetic deflection of 3 degrees. Around 125 signal events on the whole sky are needed for a 5${\sigma}$ discovery in 50\% of trials, and around 75 events are needed
for
a 3${\sigma}$ hint. Those values correspond to source flux values of
about  2.16${\times}$10$^{-8}$ GeV cm$^{-2}$ s$^{-1}$ and 1.29${\times}$10$^{-8}$  GeV cm$^{-2}$ s$^{-1}$ respectively.

\section{Results from the unblinded data}
To analyze the level of correlation between the distribution 69 UHECRs reported by the Pierre Auger Observatory, and ANTARES neutrino candidates, we unblinded
2190 neutrino events. The significance of observed correlation is determined with the help of randomized background samples, using the bin size of 4.9 degrees.

The most probable count for the optimized bin size of 4.9 degrees, or the
mean background expectation from the randomized samples is 310.49 events (in all 69 bins, i.e. on the
whole sky), with the standard deviation of 15.22 events.

After unblinding 2190 ANTARES neutrino candidate events, a count of 290 events within 69 bins is obtained. On Figure \ref{count}, neutrino candidates are represented 
with crosses and neutrino candidates correlating with observed UHECRs are highlighted as large triangles. The observed number of correlated neutrino events is beneath 
expected
(negative correlation), with the significance of about 1.35${\sigma}$. This result is compatible with a background fluctuation.
The corresponding upper flux limit, assuming the equal flux from all UHECR sources, is 4.96${\times}$10$^{-8}$ GeV cm$^{-2}$ s$^{-1}$.

 \begin{figure}[!t]
  \vspace{5mm}
  \centering
  \includegraphics[width=3.in]{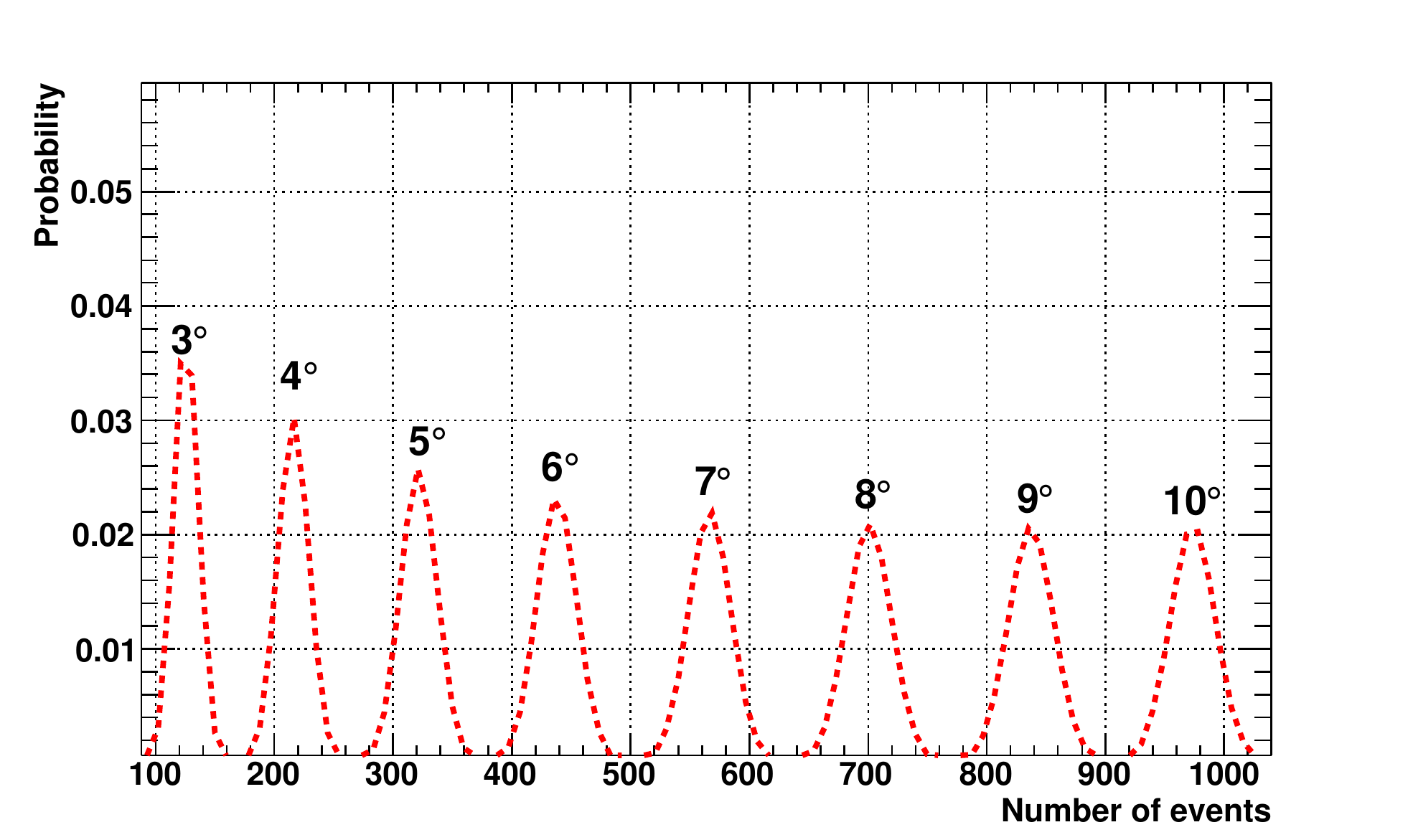}
  \caption{The probability density functions (PDFs) of the number of neutrino events in
1-10$^{\circ}$ bins centered on 69 UHECRs observed by the Pierre Auger Observatory. The counts were obtained from the background only
Monte Carlo simulations, each with 2190 neutrino events blinded by scrambling the observed events
in right ascension.}
  \label{prob}
 \end{figure}

\begin{figure}[!t]
  \vspace{5mm}
  \centering
  \includegraphics[width=3.in]{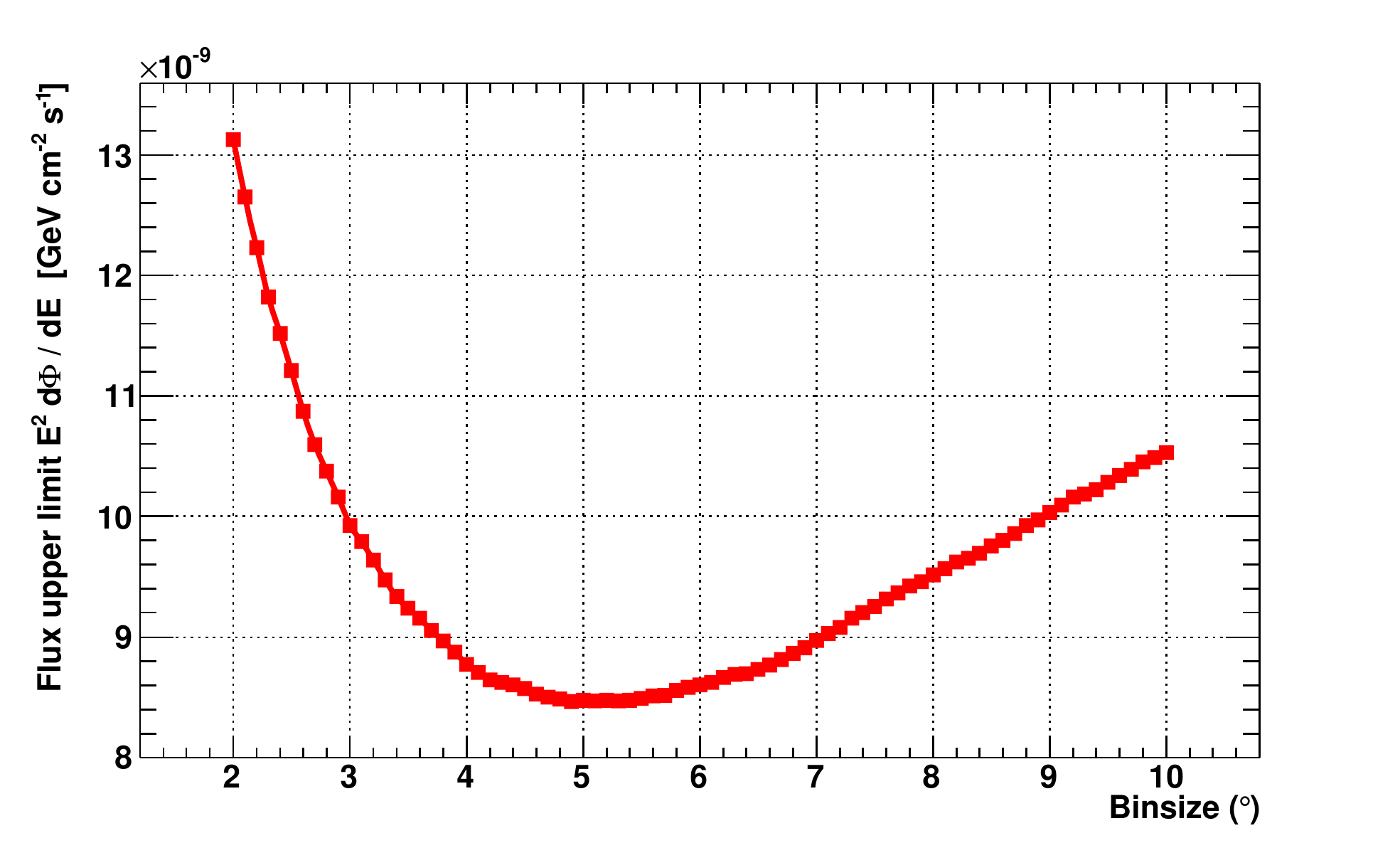}
  \caption{The mean upper flux limits as a function of the search bin size.
The minimum flux value for magnetic field deflection tolerance value of 3 degrees is obtained for 4.9$^{\circ}$ search bin.
}
  \label{sens}
 \end{figure}

\begin{figure}[!t]
  \vspace{5mm}
  \centering
  \includegraphics[width=3.in]{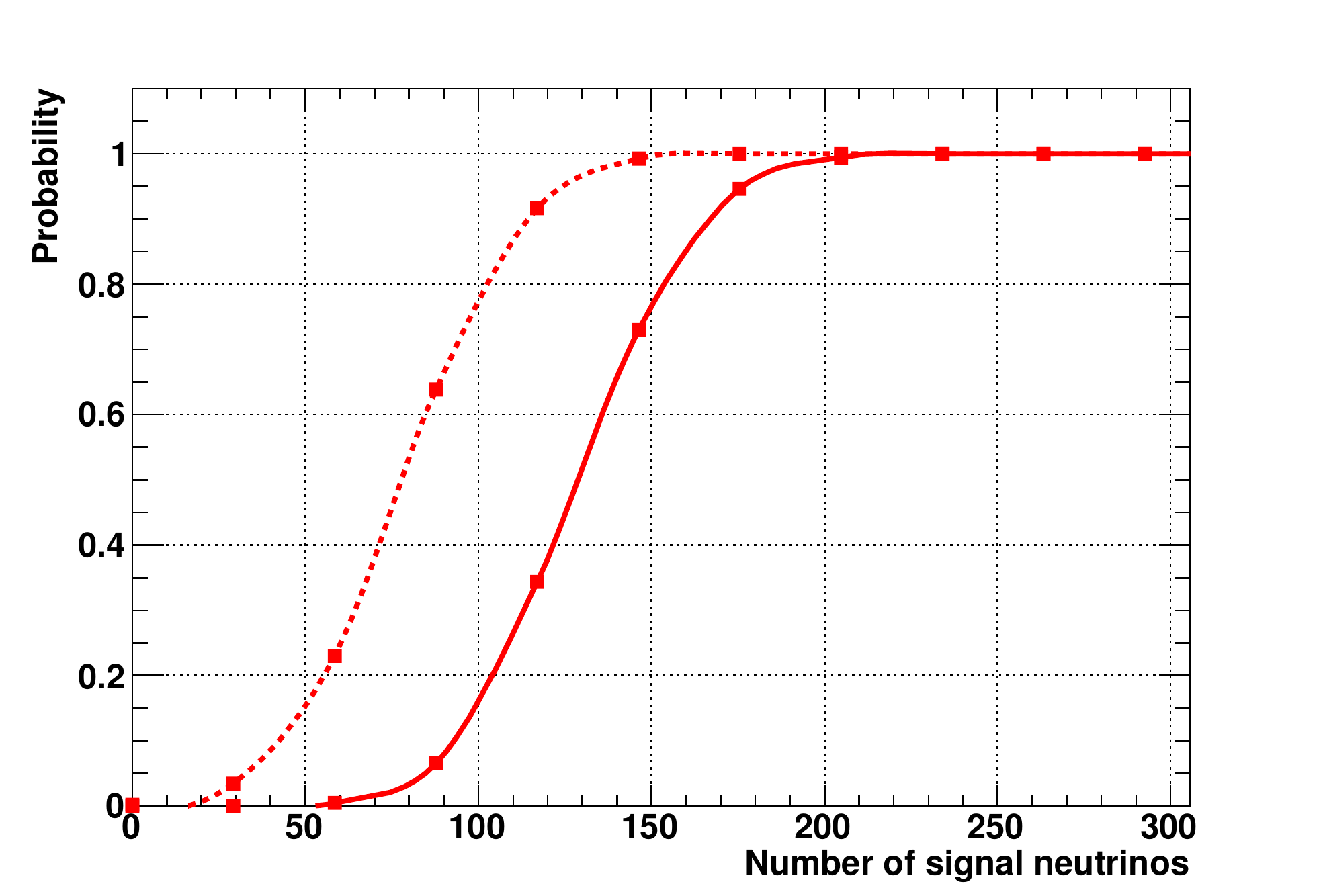}
  \caption{The discovery potential of 3${\sigma}$ (dashed line) and 5${\sigma}$ (solid line) as a 
  function of number of signal neutrino events on the whole sky derived from the signal source flux.}
  \label{discpot}
 \end{figure}


 \begin{figure*}[th]
  \centering
  \includegraphics[width=5in,height=3in]{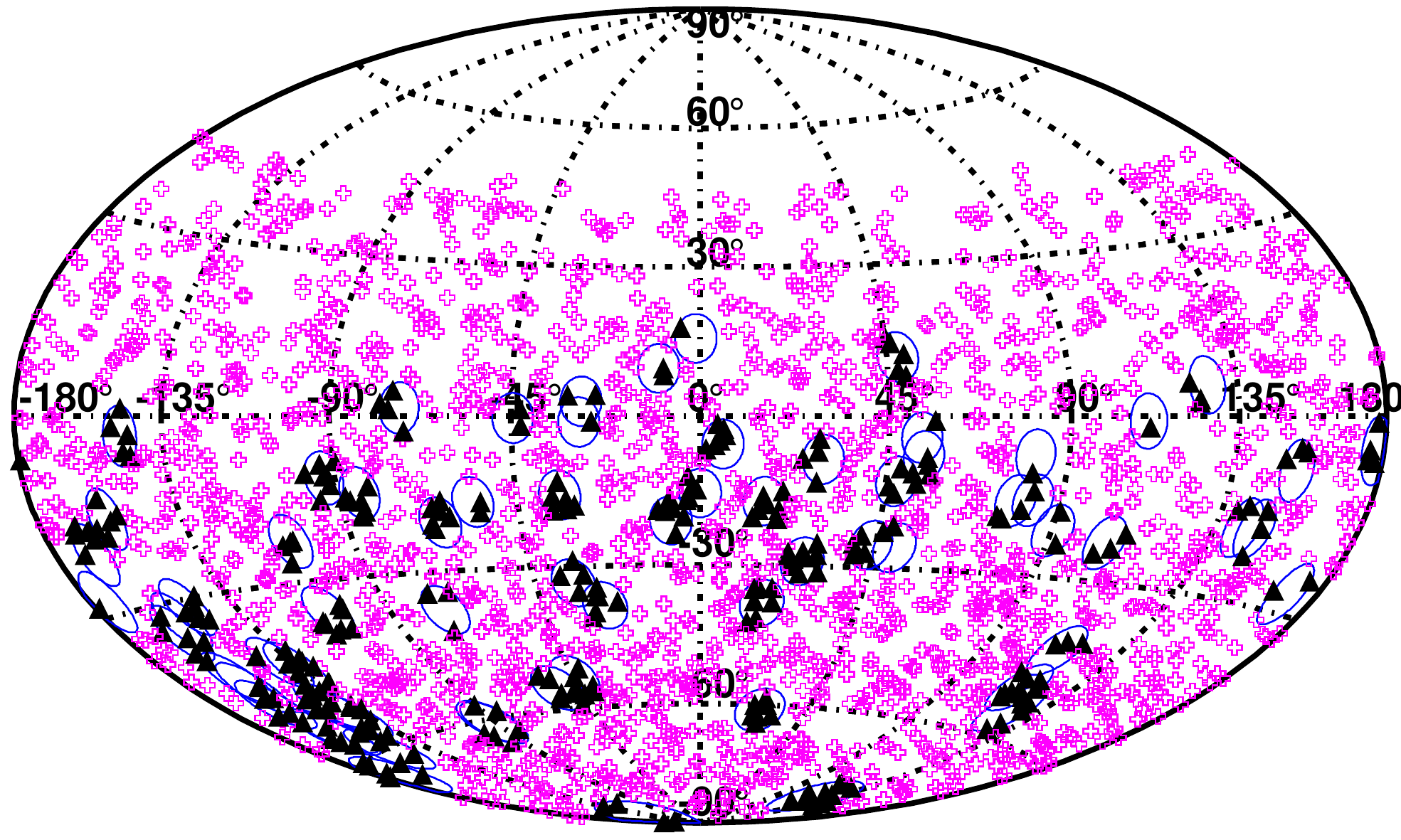}
  \caption{Crosses represent ANTARES neutrino events outside of 4.9 degree bins centered on UHECRs observed by the PAO,
and triangles represent ANTARES neutrino events correlating with observed UHECRs.}
  \label{count}
 \end{figure*}


\clearpage

\setcounter{figure}{0}
\setcounter{table}{0}
\setcounter{footnote}{0}
\setcounter{section}{0}
\newpage




\title{Searches for high-energy neutrinos in coincidence with gravitational waves with the ANTARES and VIRGO/LIGO detectors}

\shorttitle{V. Van Elewyck \etal Searches for GW+HEN coincidences with ANTARES and VIRGO/LIGO}

\authors{V\'eronique Van Elewyck$^{1}$, for the ANTARES Collaboration, the LIGO Scientific Collaboration and the VIRGO Collaboration}
\afiliations{$^1$AstroParticule et Cosmologie (APC), CNRS: UMR7164-IN2P3-Observatoire de Paris-Universit\'e Denis Diderot-Paris VII-CEA: DSM/IRFU, France
}
\email{elewyck@apc.univ-paris7.fr}

\maketitle
\begin{abstract}
Cataclysmic cosmic events can be plausible sources of both gravitational
waves (GW) and high-energy neutrinos (HEN). Both GW and HEN are alternative cosmic messengers that may traverse very dense media and travel unaffected over cosmological distances, carrying information from the innermost regions of the astrophysical engines. Such messengers could also reveal new, hidden sources that have not been observed by conventional photon astronomy.

A neutrino telescope such as ANTARES can determine accurately the time and direction of high energy neutrino events, and a network of gravitational wave detectors such as LIGO and VIRGO can also provide timing/directional information for gravitational wave bursts. Combining these informations obtained from totally independent detectors can provide original ways of constraining the processes at play in the sources, and also help confirming the astrophysical origin of a HEN/GW signal in case of concomitant observation. 
 
 This contribution describes the motivations and reach of a joint GW+HEN search using concomitant data taken with the ANTARES, VIRGO and LIGO detectors.  The specific strategies developed for the selection of HEN candidates in ANTARES and the combination of HEN and GW data are presented with a focus on the currently ongoing analysis of the concomitant datasets taken in 2007 during the VIRGO VSR1 and LIGO S5 science runs, while ANTARES was operating in a 5-line configuration. 
\end{abstract} 




\section{Introduction and motivations}

The recent years have witnessed the development and operation of a new generation of detectors offering unprecedented opportunities to observe the universe through all kind of cosmic radiations. In particular, both high-energy ($\gg$GeV) neutrinos (HENs) and gravitational waves (GWs), which have not yet been directly observed from astrophysical sources, are considered as promising tools for the development of a multi-messenger astronomy (see e.g.~\cite{becker} and~\cite{marka} for recent reviews). Contrary to high-energy photons (which are absorbed through interactions in the source and by the photon backgrounds) and charged cosmic rays (which are deflected by ambient magnetic fields), both HENs and GWs can escape from the core of the sources and travel over large distances through magnetic fields and matter without being altered. They are therefore expected to provide important information about the processes taking place in the core of the production sites and they could even reveal the existence of sources opaque to hadrons and photons, that would have remained undetected so far.

It is expected that many astrophysical phenomena produce both GWs, originating from the cataclysmic event responsible for the onset of the source, and HENs, as a byproduct of the interactions of accelerated protons (and heavier nuclei) with ambient matter and radiation in the source. The detection of coincident signals in both these channels would then be a landmark event and sign the first observational evidence that GW and HEN originate from a common source. The most plausible astrophysical emitters of GW+HEN are presented in Section~\ref{sec:sources}.

The feasibility of common GW and HEN astronomies is also related to the concomitant operation of both types of detectors, which is summarized in the time chart of Fig. \ref{fig:coinc}.  Section~\ref{sec:det} briefly describes the detection principles and the performances achieved by the ANTARES\cite{antares} neutrino telescope as well as by the GW inter\-fe\-ro\-meters VIRGO~\cite{virgo} and LIGO~\cite{ligo}, that are currently part of this joint search program\footnote{Similar studies involving the IceCube neutrino telescope\cite{ice3}, currently under deployment at the South Pole and looking for neutrinos from astrophysical sources in the Northern Hemisphere, are also under way\cite{aso}.}. 
As both types of detectors have completely independant sources of backgrounds,  the correlation between HEN and GW significances can also be exploited to enhance the sensitivity of the joint channel, even in absence of detection. The combined false alarm rate is indeed severely reduced by the requirement of space-time consistency between both channels.  In Section~\ref{sec:ana}, we describe in more detail the strategies being developed for a joint GW+HEN searches between ANTARES and the network of GW interferometers using the currently available datasets\footnote{More information can be obtained on the website of the first Workshop on Gravitational Waves and High-Energy Neutrinos: http://www.gwhen-2009.org .}. 

 \begin{figure}[!t]
  \vspace{5mm}
  \centering
  \includegraphics[width=3.in]{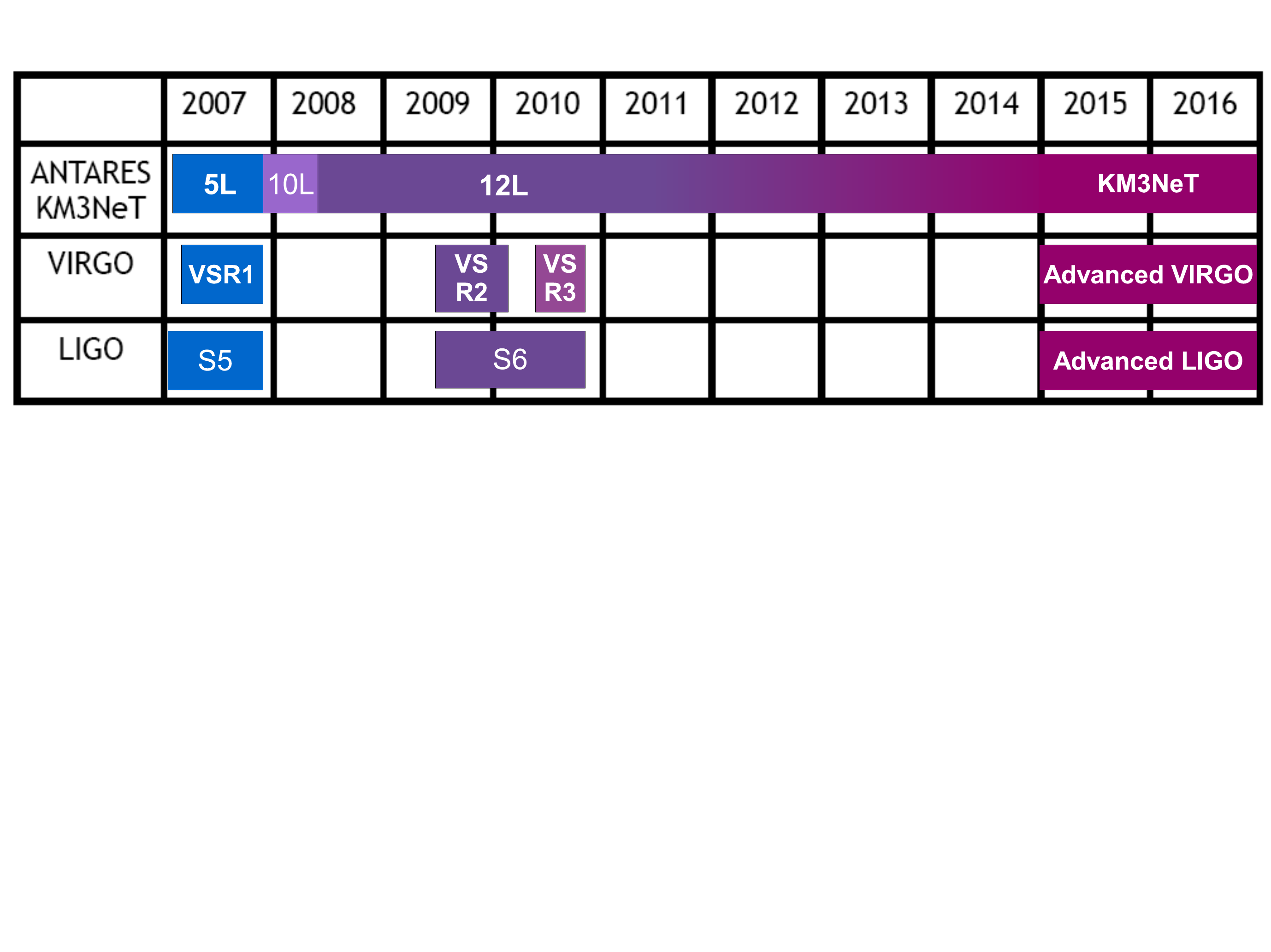}
  \vspace{-3cm}
  \caption{\footnotesize{Time chart of the data-taking periods for the ANTARES, VIRGO and LIGO experiments, indicating the respective upgrades of the detectors (as described in the text). The deployment of the KM3NeT neutrino telescope is expected to last three to four years, during which the detector will be taking data with an increasing number of PMTs before reaching its final configuration.\cite{km3}.}}
  \label{fig:coinc}
 \end{figure}

\section{Astrophysical emitters of gravitational waves and high-energy neutrinos}
\label{sec:sources}

Potential sources of GWs and HENs are likely to be very energetic and to exhibit bursting activity. Plausible GW+HEN emission mechanisms include two classes of galactic sources which could be accessible to the present generation of GW interferometers and HEN telescopes. 

{\bf Microquasars} are believed to be X-ray binaries involving a compact object that accretes matter from a companion star and re-emits it in relativistic jets associated with intense radio (and IR) flares. 
Such objects could emit GWs during both accretion and ejection phases~\cite{mqGW}; and the latter phase could be correlated with a HEN emission provided the jet has a hadronic component \cite{mq}. 
{\bf Soft Gamma Repeaters (SGRs)} are highly magnetized X-ray pulsars with a soft $\gamma$-ray bursting activity. In the popular magnetar model \cite{magnetar}, the outbursts are caused by star-quakes associated to large-scale rearrangements of the magnetic field. 
The deformation of the star during the outburst could lead to the emission of GWs 
within range of current interferometric GW detectors~\cite{Abadiemagnetar}. Sudden changes in the large magnetic fields would also accelerate protons or nuclei interacting with thermal radiation and generating a flux of HENs  \cite{ioka}. 

The most promising class of known extragalactic bursting sources are Gamma-Ray Bursts (GRBs). In the prompt and afterglow phases, HENs ($10^5 - 10^{10}$ GeV) are expected to be produced by accelerated protons in relativistic shocks and several models predict detectable fluxes in km$^3$-scale detectors~\cite{GRBnus}, although no evidence for GRB neutrinos has been observed yet by IceCube 40~\cite{ice3nu}.  While gamma-ray and HEN emission from GRBs are related to the mechanisms driving the relativistic outflow, GW emissions is closely connected to the central engine and hence to the GRB progenitor.

 {\bf Short-hard GRBs} are thought to originate from coalescing binaries involving  black holes and/or neutron stars; such mergers could emit GWs detectable from relatively large distances, with significant associated HEN fluxes \cite{GRBshort}. 
 {\bf Long-soft GRBs} are most probably  induced by  "collapsars",  i.e. collapses of a massive star into a black hole, with the formation of an accretion disk and a jet that emerges from the stellar envelope~\cite{GRBlong}. This model also involves the emission of a strong burst of GWs during the gravitational collapse of the (rapidly rotating) progenitor star and in the pre-GRB phase; however this population is distributed over cosmological distances so that the associated HEN signal is expected to be faint \cite{kotake}. 
  The subclass of {\bf low-luminosity GRBs}, with $\gamma$-ray luminosities a few orders of magnitude smaller, are believed to originate from a particularly energetic type Ibc  core-collapse supernovae, although the mechanism responsible for their onset is still debated~\cite{bromberg}. They could produce stronger GW signals together with significant high- and low-energy neutrino emission; moreover they are more frequent than typical long GRBs and often discovered at shorter distances \cite{razzaque}. 
  Finally, the {\bf choked GRBs} are thought to be associated with supernovae driven by mildly relativistic, baryon-rich and optically thick jets, so that no $\gamma$-rays escape~\cite{choked}. Such ``hidden sources'' could be among the most promising emitters of GWs and HENs, as current estimations predict a relatively high occurrence rate in the volume probed by current GW and HEN detectors \cite{ando}.

\section{The detectors}
 \label{sec:det}

  The {\bf ANTARES} detector \cite{antares} is the first undersea neutrino telescope;  its deployment at a depth of 2475m in the Mediterranean Sea near Toulon was completed in May 2008. It consists in a three-dimensional array of 884 photomultiplier tubes (PMTs) distributed on 12 lines anchored to the sea bed and connected to the shore through an electro-optical cable. Before reaching this final (12L) setup, ANTARES has been operating in various configurations with increasing number of lines, from one to five (5L)  and ten (10L).
  
   ANTARES detects the Cherenkov radiation emitted by charged leptons (mainly muons, but also electrons and taus) induced by cosmic neutrino interactions with matter inside or near the instrumented volume. The knowledge of the timing and amplitude of the light pulses recorded by the PMTs allows to reconstruct the trajectory of the muon and to infer the arrival direction of the incident neutrino. The current reconstruction algorithms achieve an angular resolution (defined as the median angle between the neutrino and the reconstructed muon) of about $0.4^\circ$ for neutrinos above 10 TeV\cite{aart}. The design of ANTARES is optimized for the detection of up-going muons produced by neutrinos which have traversed the Earth and interacted near the detector; its field of view is $\sim\, 2 \pi\, \mathrm{sr}$ for neutrino energies $100\ \mathrm{GeV} \lesssim E_\nu \lesssim 100\  \mathrm{TeV}$. 
  
  \begin{figure}[!t]
  \vspace{5mm}
  \centering
  \includegraphics[width=3.in]{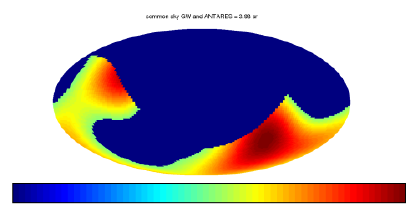}
  \caption{\footnotesize{Instantaneous common sky coverage for VIRGO + LIGO + ANTARES in geocentric coordinates. This map shows the combined antenna pattern for the gravitational wave detector network (above half-maximum), assuming that ANTARES has 100\% visibility in its antipodal hemisphere and 0\% elsewhere. The colour scale is from $0\%$ (left, blue) to $100\%$ (right, red).}}
  \label{fig:map}
 \end{figure}

  The data acquisition system of ANTARES is based on the "all-data-to-shore" concept, which allows to operate different physics triggers to the same data in parallel. Satellites looking for GRBs can also trigger the detector in real time via the GCN (Gamma-Ray Burst Coordinate Network) alert system.  ANTARES has also implemented the possibility to trigger an optical telescope network on the basis of "golden" neutrino events  selected by a fast, online reconstruction procedure \cite{tatoo}. All these characteristics make this detector especially suited for the search of astrophysical point sources, and transients in particular. ANTARES is intended as the first step towards a km$^3$-sized neutrino telescope in the Mediterranean Sea, currently under R\&D in the framework of the KM3NeT Consortium \cite{km3} and whose construction could start by 2014. 

The GW detectors {\bf VIRGO} \cite{virgo}, with one site in Italy, and {\bf LIGO} (see e.g. \cite{ligo}), with two sites in the United States, are Michelson-type laser interferometers. They consist of two light storage arms enclosed in vacuum tubes oriented at $90^\circ$ from each other. Suspended, highly reflective mirrors play the role of test masses. Current detectors are sensitive to relative displacements (hence GW amplitude) of the order of $10^{-20}$ to $10^{-22}$   Hz$^{-1/2}$. Their detection horizon is about 15 Mpc for standard binary sources.


The first concomitant data-taking phase with the whole VIRGO/LIGO network (VSR1/S5) was carried on in 2007, while ANTARES was operating in 5L configuration (see  Fig.~\ref{fig:coinc}). A second data-taking phase was conducted between mid-2009 and end 2010 with upgraded detectors (S6/VSR2 and VSR3), 
in coincidence with the operation of ANTARES 12L. Another major upgrade for both classes of detectors is scheduled for the upcoming decade: the Advanced VIRGO/Advanced LIGO and KM3NeT projects should gain a factor of 10 in sensitivities respect to the presently operating instruments.  As can be seen from Figure~\ref{fig:map}, the VIRGO/LIGO network monitors a good fraction of the sky in common with ANTARES: the instantaneous overlap of visibility maps is about 4~sr ($\sim 30\%$ of the sky).

\section{Joint GW+HEN searches using ANTARES data}
\label{sec:ana}

GW interferometers and HEN telescopes share the challenge to look for faint and rare signals on top of abundant noise or background events. The GW+HEN search methodology involves the combination of GW/HEN candidate event lists, with the advantage of significantly lowering the rate of accidental coincidences. Preliminary studies on the feasibility of such searches \cite{aso, pradier} indicate that, even if the constituent observatories provide several triggers a day, the false alarm rate for the combined detector network can be kept at a very low level ($\sim$1/(600 yr)).

An important ingredient of these searches is the definition of an appropriate coincidence time window between HEN and GW signals hypothetically arriving from the same astrophysical source. A case study that considered the duration of different emission processes in long GRBs, based on BATSE, Swift and Fermi observations, allowed to derive a conservative upper bound  $t_{GW} - t_{HEN} \in [-500s, +500s]$ on this time window~\cite{baret}. For short GRBs, this time-delay could be as small as a few seconds. 

Different strategies for a joint search are currently being developed. The first one consists in an event-per-event search for a GW signal  correlating in space and time with a given neutrino event considered as an external trigger. Such a search is rather straightforward to implement as it allows to make use of existing analysis pipelines developed e.g. for GRB searches. It is currently being applied to the concomitant set of data taken between January 27 and September 30,  2007 with ANTARES 5L/VSR1/S5. 

Fig. \ref{fig:flow} shows the schematic flowchart of such a HEN-triggered search. The ANTARES 5L data are filtered according to quality requirements similar to those selecting the well-reconstructed events that are used for the standalone searches for HEN point sources. The list of candidate HENs includes their arrival time, direction on the sky, and an event-by-event estimation of the angular accuracy,  which serves to define the angular search window for the GW. For the purpose of this joint search, the angular accuracy is defined as the 90\% quantile (and not the median) of the error distribution on the reconstructed neutrino direction, obtained from Monte Carlo studies. The on-source time window is taken to be $[-500s,+500s]$ around the neutrino arrival time. 

The list of HEN triggers is then passed to the X-pipeline~\cite{Xpipe}, an algorithm which performs coherent searches for unmodelled bursts of GWs on the combined stream of data coming from all active interferometers (ITFs). The background estimation and the optimization of the selection strategy are performed using time-shifted data from the off-source region in order to avoid contamination by a potential GW signal. Once the search parameters are tuned, the box is open and the analysis is applied to the on-source data set. If a coincident event is found, its significance is obtained by comparing to the distribution of accidental events obtained with Monte-Carlo simulations using time-shifted data streams from the off-source region. 
 
Alternatively, comprehensive searches for time-coincidences between independent 
lists of neutrino and GW events can also be performed, followed by a test of spatial correlation using the combined GW/HEN likelihood skymap. This second, more symmetrical, option requires the existence of two independent analysis chains scanning the whole phase space in search for interesting events. A  major issue for both types of analyses lies in the combined optimisation of the selection criteria for the different detection techniques. 

\begin{figure}[!t]
  \vspace{5mm}
  \centering
  \includegraphics[width=3.in]{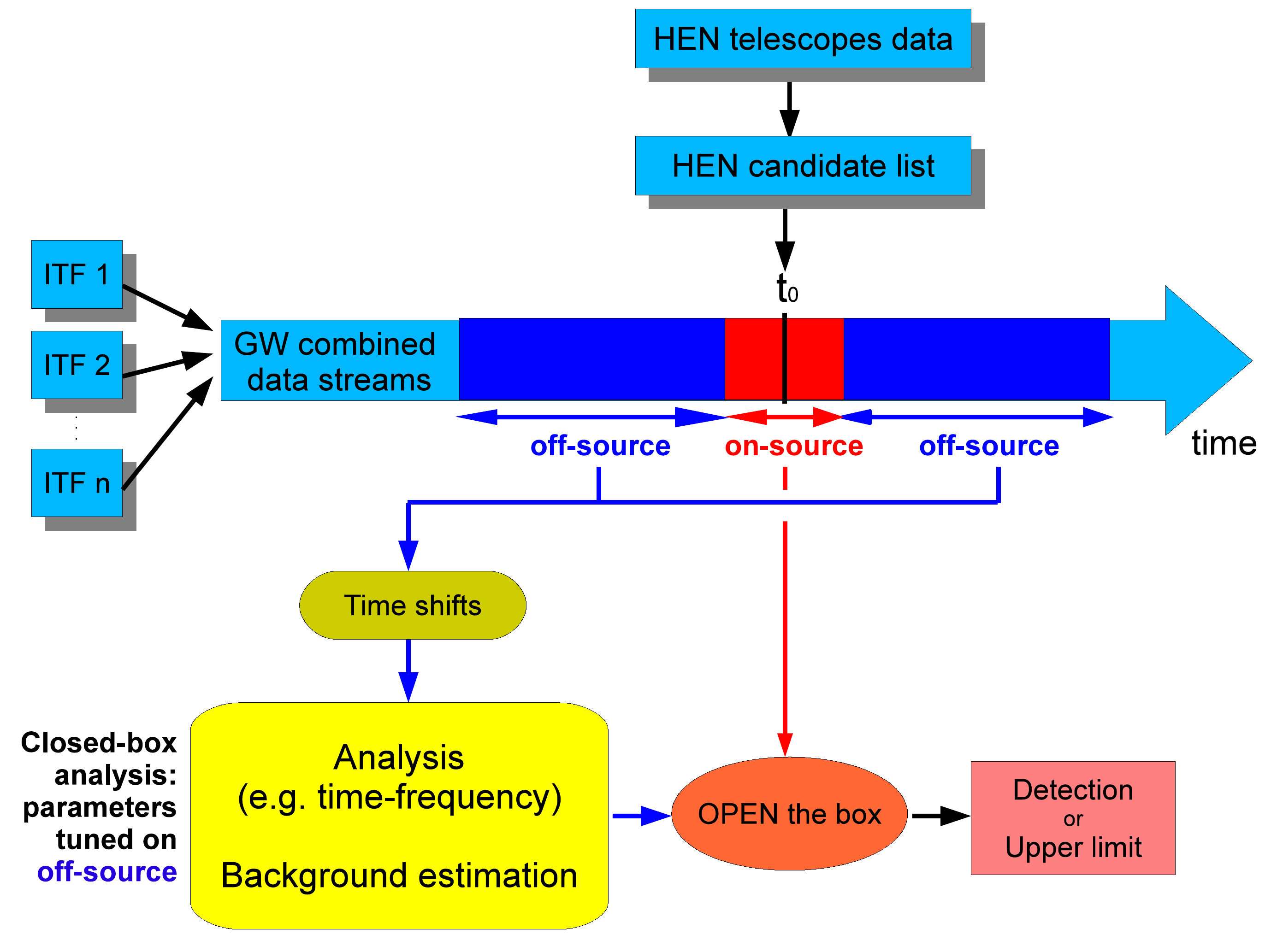}
  \caption{\footnotesize{Schematic flow diagram of a HEN-triggered search for GWs, as described in the text.}}
  \label{fig:flow}
 \end{figure}

\section{Conclusions}

The joint GW+HEN analysis program described here could significantly expand the scientific reach of both GW interferometers and HEN telescopes. 
Beyond the benefit of a potential high-confidence discovery, coincident GW/HEN (non-)ob\-servation shall play a crucial role in our understanding of the most energetic sources of cosmic radiation and in constraining existing models. A first search for GW signals in coincidence with HEN neutrinos detected by ANTARES 5L in 2007 is being performed, and optimisation studies are ongoing to exploit at best the concomitant datasets taken in 2009-2010 with upgraded detectors.

\newpage

\section*{Acknowledgements}
The author would like to thank the ICRC organizers for the interesting and fruitful conference. She acknowledges fi\-nancial support from the EC 7th FP (Marie Curie Reintegration Grant NEUTEL-APC 224898) and the Agence Nationale de la Recherche (ANR-08-JCJC-0061-01).


\clearpage

\setcounter{figure}{0}
\setcounter{table}{0}
\setcounter{footnote}{0}
\setcounter{section}{0}
\newpage




\title{Indirect Dark Matter search in the Sun direction using the ANTARES data 2007-2008 for the two common theoretical frameworks (CMSSM, mUED)}

\shorttitle{G. Lambard \etal Indirect Dark Matter search in the Sun with ANTARES}

\authors{Guillaume Lambard$^{1}$, on behalf of the ANTARES collaboration}
\afiliations{$^1$IFIC - Instituto de F\`{i}sica Corpuscular - Edificio Institutos de Investigaci\'{o}n, 
Apartado de Correos 22085 E-46071 Valencia, Spain}
\email{lambard@ific.uv.es}

\maketitle
\begin{abstract}
Using the ANTARES neutrino telescope, the largest neutrino telescope in the Northern 
hemisphere, with its first configuration with 5 lines of photodetectors to the actual nominal one corresponding to a total of 12 lines, we 
have studied our ability to search indirectly for an evidence of Dark Matter annihilations in heavy astrophysical objects as the Sun and the 
Galactic centre. First results have been obtained using the data recorded by ANTARES in 2007 and 2008, and compared with neutrino fluxes predicted 
within a minimal supersymmetric extension of the Standard Model with supersymmetry-breaking scalar and gaugino masses constrained to be universal 
at the GUT scale, the CMSSM, as well as a minimal Universal Extra-Dimensions 
scenario with one extra compact dimension where all the Standard Model fields propagate into the bulk, the mUED. 
The current limits over the neutrino/muon fluxes coming from Dark Matter annihilations, and the spin-dependent cross-section with protons, as well 
as the expected sensitivities predicted after several years of data taking with ANTARES will be presented for each source.
\end{abstract}


\section{Indirect dark matter search and the ANTARES neutrino telescope}
\label{indirect dark matter search and the antares neutrino telescope}

As the very old problem about the unseen planets, measuring a deviation from the known laws of gravitation and the theory of general relativity in large astrophysical 
systems, we assume the existence of a huge amount of unseen matter, the \textit{dark matter} \cite{darkmatter}, which represents $83$ $\%$ of the matter in the universe. Massives, 
and weakly interacting with the matter, the WIMPs (Weakly Interacting Massive Particles), defining the dark matter nature whatever a particular framework in Particle Physics, can be 
elastically scattered in a medium to be decelerated, and gravitationally attracted in heavy objects as the galactic center, or the stars like our Sun which is the point source 
of the present analysis. It's through the self-annihilation of WIMPs into the Sun's core, that muon neutrinos (independently of electron, and tau neutrinos) can be produced in an 
energy range reachable to a neutrino telescope as ANTARES.

Since the $29$th of May 2008, the neutrino telescope ANTARES has been completed at $40$ km offshore from Toulon at $2475$ m depth, making it the largest neutrino telescope 
in the northern hemisphere, and the first to operate in the deep sea \cite{antares}. It is made up of $12$ mooring detection lines holding ten-inch photomultiplier tubes, 
distributed into $25$ storeys on each line, designed for the measurement of neutrino induced charged particles based on the detection of \v{C}erenkov light emitted in water. 
With a low energy threshold of about $20$ GeV for well reconstructed muons, and an angular resolution of about $1^{\circ}$ at low energy ($E_{\nu} < 1$ TeV), ANTARES has a 
great potential to detect neutrinos induced by the self-annihilation of WIMPs into the Sun's core.  

\section{Dark Matter signal, and efficiency of ANTARES}
\label{dark matter signal and efficiency of antares}

Model-independently, the differential amount of neutrinos $dN_{\nu}/dz$ as a function of the ratio $z = E_{\nu}/M_{WIMP}$, where $M_{WIMP}$ is the mass of the WIMP, which can 
be created from the Sun's core, coming to the surface of the Earth, is computed using the software package WIMPSIM \cite{wimpsim}. With a high statistics of production 
($12\times10^{6}$ of generated events inside the Sun) for all possible self-annihilation channels which can be reached as $q\bar{q}$, $l\bar{l}$, $\phi\phi^{*}$, and $\nu\bar{\nu}$. 
Fig.~\ref{MainChannelsFig} gives a glimpse of the main channels in the two common frameworks studied here: CMSSM, and mUED. No branching ratios have been taken into account for 
the three first channels ($W\bar{W}$, $b\bar{b}$, and $\tau\bar{\tau}$), and all the contributions have been merged in the mUED case taking into account the following branching ratios
\cite{hooper}: $0.23$ for $\tau\bar{\tau}$, $0.077$ for $t\bar{t}$, $0.005$ for $b\bar{b}$, and $0.014$ for $\nu\bar{\nu}$. All of these are shown for $M_{WIMP} = 350$ GeV as an 
example. All of the physical contributions have been taken into account as the three-flavors oscillations, or the $\tau$ lepton regeneration through the Sun medium. 

\begin{figure}[!t]
\begin{minipage}[c]{.48\linewidth}
\includegraphics[width=\linewidth]{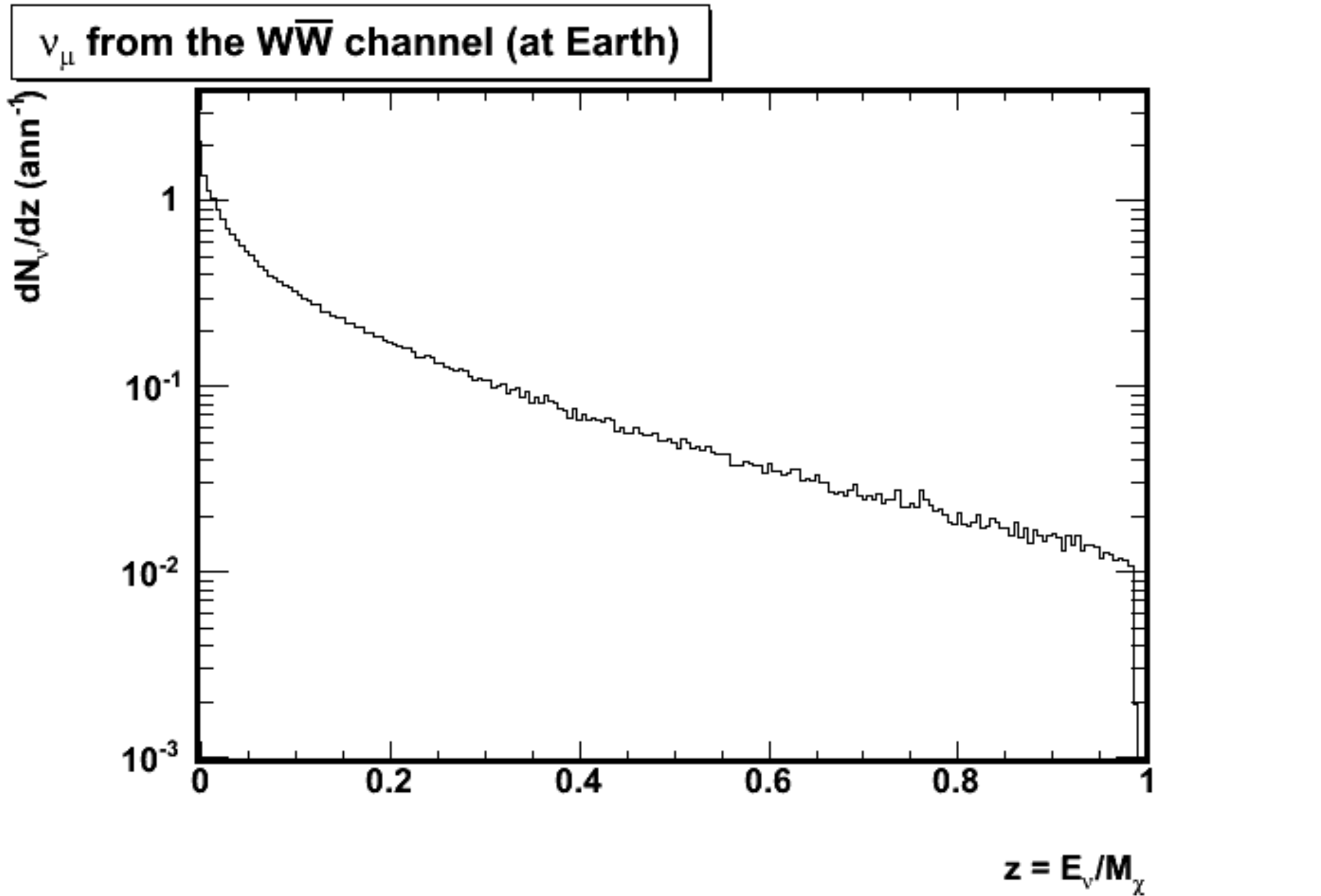} \\
\includegraphics[width=\linewidth]{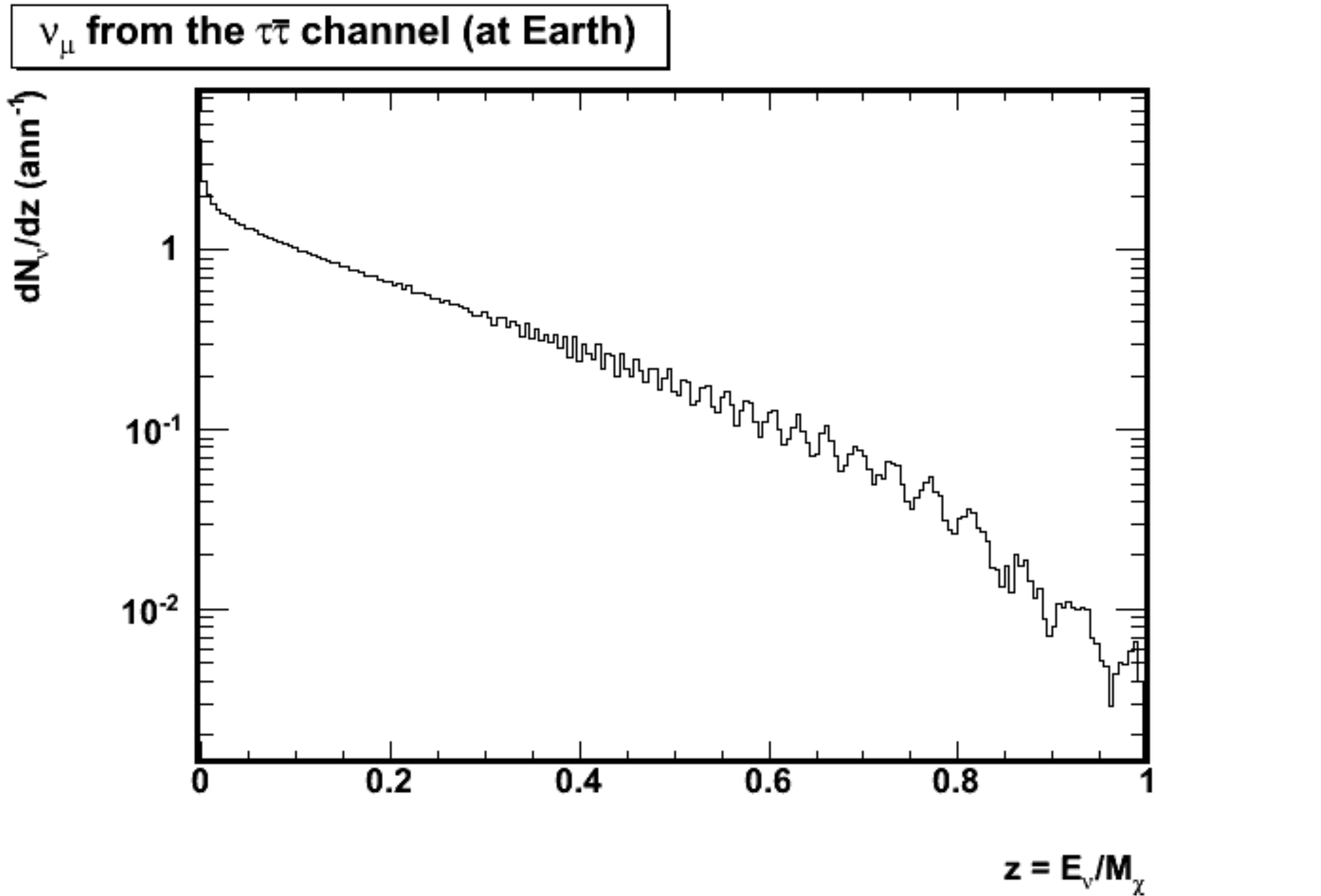}
\end{minipage}
\begin{minipage}[c]{.48\linewidth}
\includegraphics[width=\linewidth]{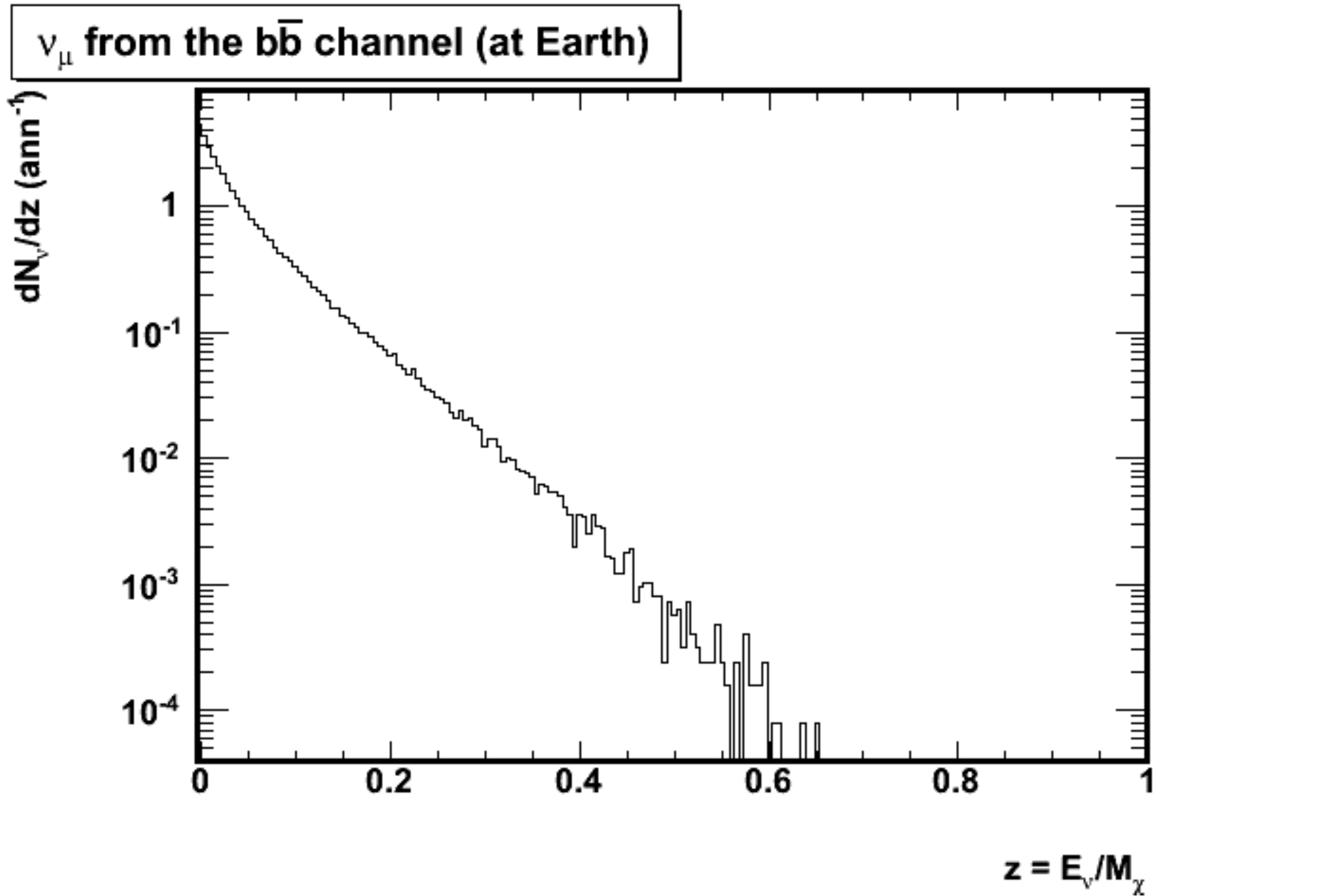} \\
\includegraphics[width=\linewidth]{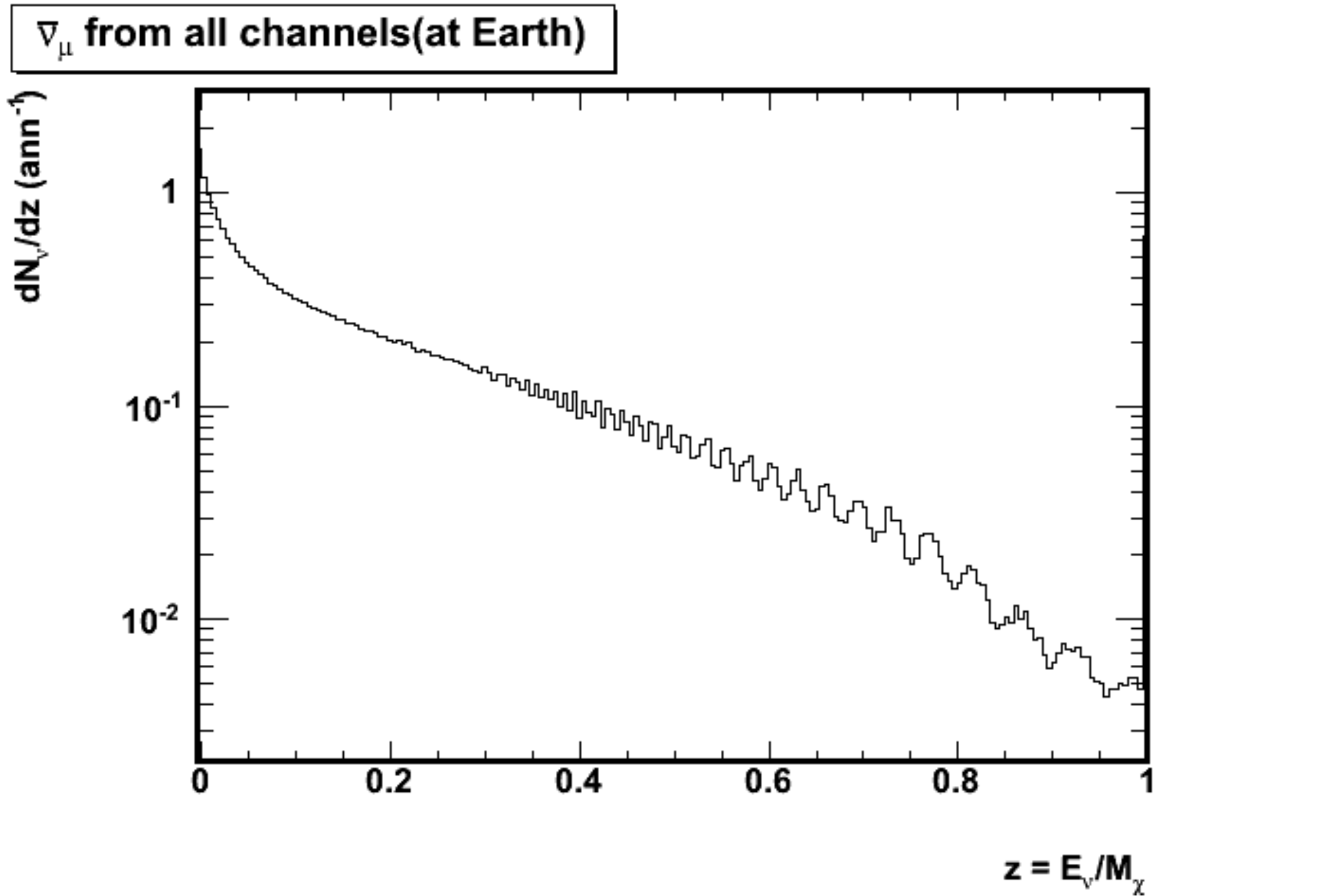}
\end{minipage}
\caption{Self-annihilation channels $W\bar{W}$, $b\bar{b}$, $\tau\bar{\tau}$, and the mUED particular case (from the top left to the bottom right).}
\label{MainChannelsFig}
\end{figure}

\bigskip
Then, an averaged effective area $A_{eff}(M_{WIMP})$, with the Sun as a point source, can be defined as a function of the WIMP masses, and a couple of cuts ($Q$,$\Psi$), where $Q$ 
is a track fit quality cut based on a fast reconstruction algorithm (see \cite{bbfitarticle} for more details), and $\Psi$ is the half-cone angle around the source. The following expression 
defines $A_{eff}(M_{WIMP})$ as: 

\begin{eqnarray}
A_{eff}(M_{WIMP}) & = & \int^{M_{WIMP}}_{0} \frac{N_{rec.}(Q,\Psi)}{N_{gen.}} \times A_{gen.} \nonumber \\ 
& & \times P_{Sun} \times P\left(\frac{dN_{\nu}}{dz}\right) \, dE_{\nu} ,
\end{eqnarray}

where $N_{rec.}(Q,\Psi)$ is the number of reconstructed neutrino events considering the cuts couple ($Q$,$\Psi$), $N_{gen.}$ is the number of isotropically generated events, $A_{gen.}$ is the 
surface used in the event generation, $P_{Sun}$ is the visibility of the Sun in the data taking period, and $P(dN_{\nu}/dz)$ is the dark matter muon neutrinos signal energy spectrum normalized to 
one. Fig.~\ref{WeightedAeffFig} shows (top) the simple multiplication, over the neutrino energy range $E_{\nu} \in$ [$10$ GeV;$M_{WIMP}$] ($M_{WIMP} = 350$ GeV here), of the effective 
area by the normalized differential amount of neutrino events at the surface of the Earth. As an example, a $Q < 1.4$ has been chosen with a half-cone angle $\Psi < 3^{\circ}$. Five different cases 
are presented here. First the effective area "alone" $A_{eff}^{\nu}$ in the direction of the Sun without any differential flux taken into account, then the effective area multiplied by 
the normalized differential amount of neutrinos for the $W^{+}W^{-}$ channel, for the $b\bar{b}$ channel, for the $\tau\bar{\tau}$, and for the mUED merged channels. Even if the contribution in 
number of neutrinos is very large at very low energy regime for all of these channels, the final number of events is going to be smoothed by the weakness of the effective area at very low energy. 
Then, the major channels, in sense of neutrinos production, are going to be the $W\bar{W}$, and the $\tau\bar{\tau}$, especially for the CMSSM framework. For the mUED, as this model is dominated by 
the production of leptons/anti-leptons due to a branching ratio closed to $0.23$~\cite{hooper}, the number of neutrinos produced is going to be very closed to the one from the $\tau\bar{\tau}$ 
channel, isolated in this plot for the CMSSM. Finally, a bump localized on the WIMP mass, caused by the direct production of neutrinos, is visible, direct production which is particular to the mUED 
framework.

\begin{figure}[!t]
\begin{center}
\begin{minipage}[c]{.8\linewidth}
\includegraphics[width=\linewidth]{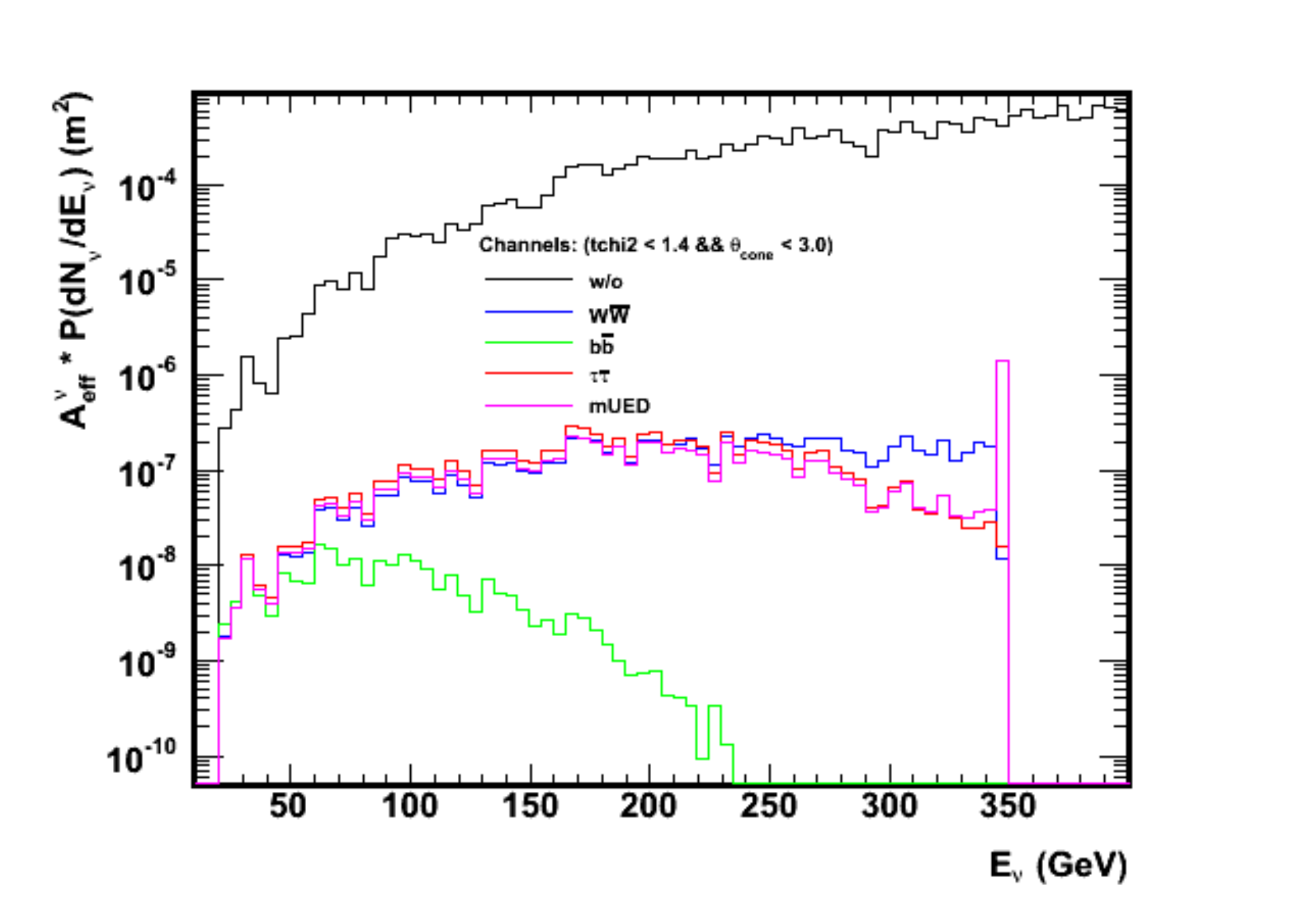} 
\end{minipage}
\begin{minipage}[c]{.8\linewidth}
\includegraphics[width=\linewidth]{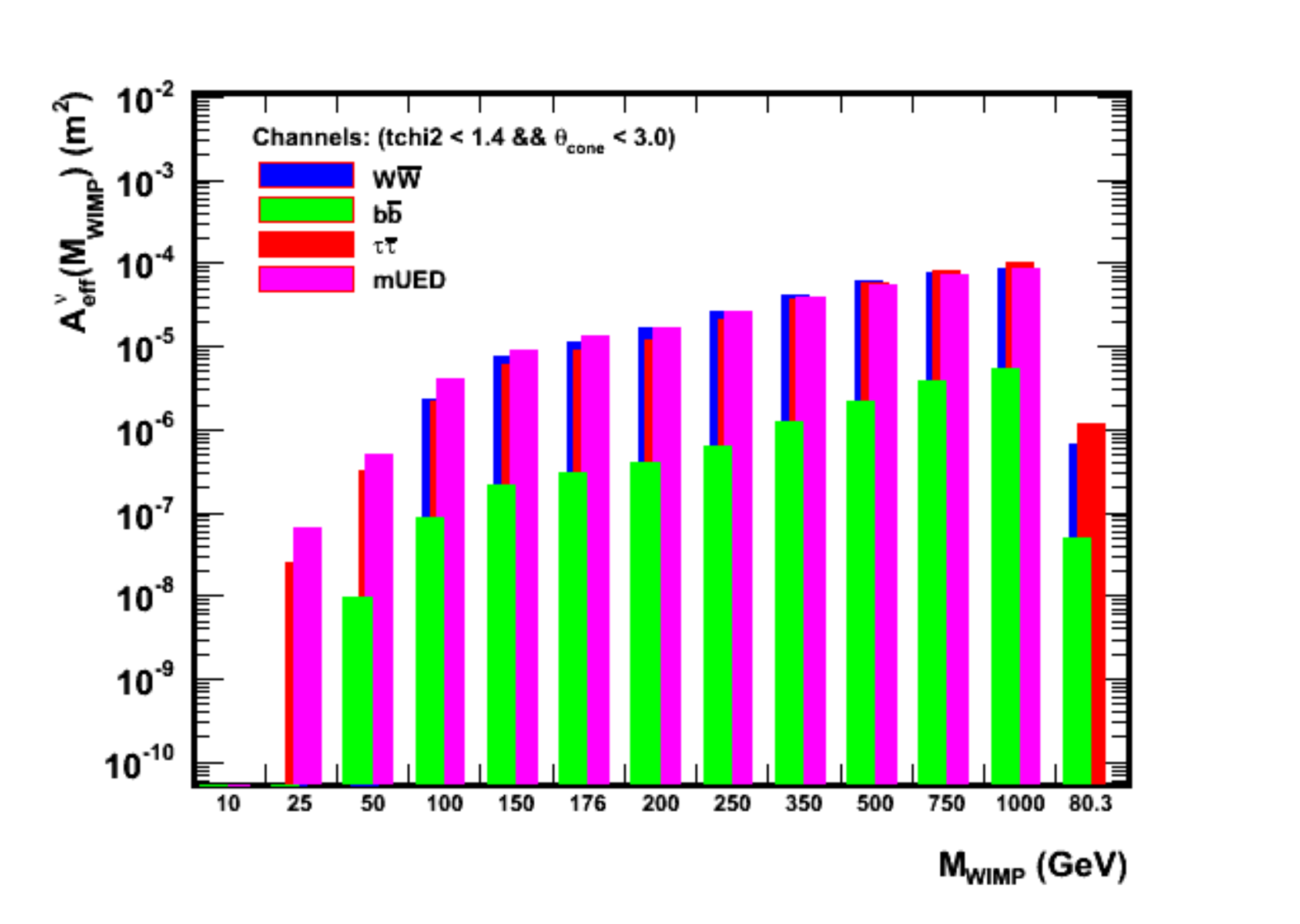}
\end{minipage}
\end{center}
\caption{Simple multiplication of the effective area $A_{eff}^{\nu}$ (m$^{2}$) by the normalized differential amount of neutrino events at the surface of the Earth, over the neutrino energy range 
$E_{\nu} \in$ [$10$ GeV;$M_{WIMP}$], on the top. And, normalized effective area $A_{eff}(M_{WIMP})$ ($m^{2}$) for a $Q < 1.4$, and a half-cone angle $\Psi < 3^{\circ}$ around the Sun, 
as a function of different WIMP masses, on the bottom.}
\label{WeightedAeffFig}
\end{figure}

The integration of the previous distributions $A_{eff} \times P(dN_{\nu}/dz)$ concludes on the effective areas dedicated to the WIMP masses $A_{eff}(M_{WIMP})$, and this for each channel considered 
here in fig.~\ref{WeightedAeffFig} (bottom). These $A_{eff}(M_{WIMP})$ have been computed for each ($Q$,$\Psi$), and for a set of thirteen different WIMP chosen masses ($10$, $25$, $50$, $80.3$, 
$100$, $150$, $200$, $250$, $350$, $500$, $750$, and $1000$ GeV). As an example, the same set of cuts and style of lines has been chosen than the previous plot, for the $W\bar{W}$, $b\bar{b}$, 
$\tau\bar{\tau}$, and mUED merged channels. The bin centered around a mass $M_{WIMP} = 80.3$ GeV has been isolated of the others because of its value of mass, closed to the W boson mass 
($M_{W} = 80.399 \pm 0.023$ GeV, PDG \cite{pdg}), corresponding to the beginning of the $W\bar{W}$ channel contribution (explaining the lacks in the mass bins centered on the $10$, $25$, and $50$ 
GeV), particular to a MSSM framework, and helicity vanished in the general UED framework. Main and equivalent contributions from the $W\bar{W}$ and the $\tau\bar{\tau}$ channels (usually called  
"Hard" channels) still appeared, and a lower contribution from the $b\bar{b}$ channel (usually called "Soft" channel). Concerning the mUED case, its integrated effective areas are very 
closed to the $\tau\bar{\tau}$ ones, main channel in this framework. Finally, a lack in statistics from the effective area and the dark matter simulations didn't allow to compute all the channels 
for the WIMP masses at $10$ (no channels), and $25$ GeV ($b\bar{b}$ is missing).

\section{Results and discussions}
\label{results and discussions}

\begin{figure}[!t]
\begin{center}
\includegraphics[width=0.8\linewidth]{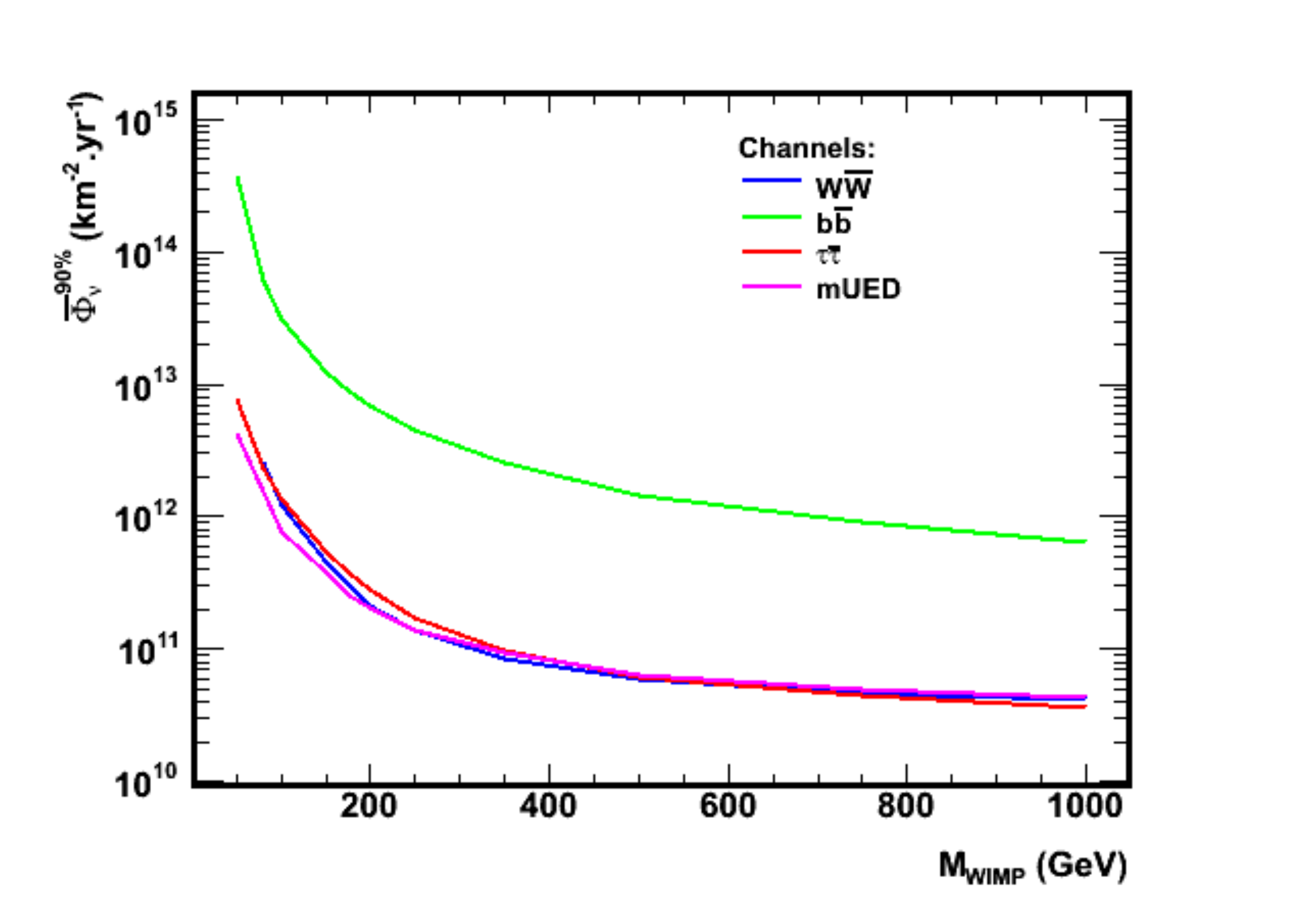}
\end{center}
\caption{Sentivity $\bar{\phi}_{\nu}^{90\%}$ in neutrino flux as a function of the WIMP masses, in the range $M_{WIMP}\in$[$10$ GeV;$1$ TeV] and from the data 2007-2008, 
for the channels $W^{+}W^{-}$, $b\bar{b}$, $\tau\bar{\tau}$, and mUED.}
\label{phinuaulFig}
\end{figure}

Following the Model Rejection Factor (MRF) technique~\cite{mrf} to optimize the ($Q$,$\Psi$) cuts couple in the Sun direction, the sensitivity in neutrinos $\bar{\phi}_{\nu}^{90\%}$ of ANTARES 
must be maximized for each value of WIMP mass, and each considered channel in the dark matter self-annihilation. For this, $\bar{\phi}_{\nu}^{90\%}$ can be expressed as:

\begin{equation}
\bar{\phi}_{\nu}^{90\%} = \frac{\bar{\mu}^{90\%}}{A_{eff}(M_{WIMP}) \times T_{eff}}, 
\label{phinuaulimiteqn}
\end{equation}

where $\bar{\mu}^{90\%}$ is the average upper limit considering a Poisson statistics in the Feldman-Cousins approach~\cite{feldmancousins}, which has been computed from a scrambled data set extracted 
from the $2007$-$2008$ data taking period of ANTARES, where the configuration of the detector has evolved from a $5$ detection lines structure to the final one at $12$ lines. Then, 
$T_{eff} \simeq 292.9$ days is the effective time of $2007$-$2008$ data taking period, and $A_{eff}(M_{WIMP})$ is the effective area defined in \ref{dark matter signal and efficiency of antares}.

As a direct consequence of the ($Q$,$\Psi$) couple optimization, a representation of the sensitivity $\bar{\phi}_{\nu}^{90\%}$ can be produced in fig.~\ref{phinuaulFig}. The most stringent 
sensitivities come from the "Hard" channels as $W^{+}W^{-}$ and $\tau\bar{\tau}$, for the Supersymmetry, explained by the amount of neutrinos produced through these channels (see in 
\ref{dark matter signal and efficiency of antares}), and by extension due to the effective area $A_{eff}(M_{WIMP})$ (fig.~\ref{WeightedAeffFig}, on the bottom) developped from them. For the 
particular case of the mUED framework, its sensitivity is very similar to the "Hard" channels previously mentioned because of the $\tau\bar{\tau}$ channel branching ratio supremacy (see in 
\ref{dark matter signal and efficiency of antares}).

\bigskip
After this, to compare properly the sensitivity in neutrino flux from ANTARES $2007$-$2008$ data to the other experiments, the sensitivity in muon flux must be produced, using the cross section 
of neutrinos with the Earth medium, the muon range, the nucleon density in the vicinity of the detector, and the neutrino transmission probability through the Earth, all derived from the atmospheric 
neutrinos (anti-neutrinos) Monte-Carlo. Fig.~\ref{CMSSMmUEDphimuaulFig} (top) shows it for the CMSSM framework.

\begin{figure}[!t]
\begin{center}
\begin{minipage}[c]{.8\linewidth}
\includegraphics[width=\linewidth]{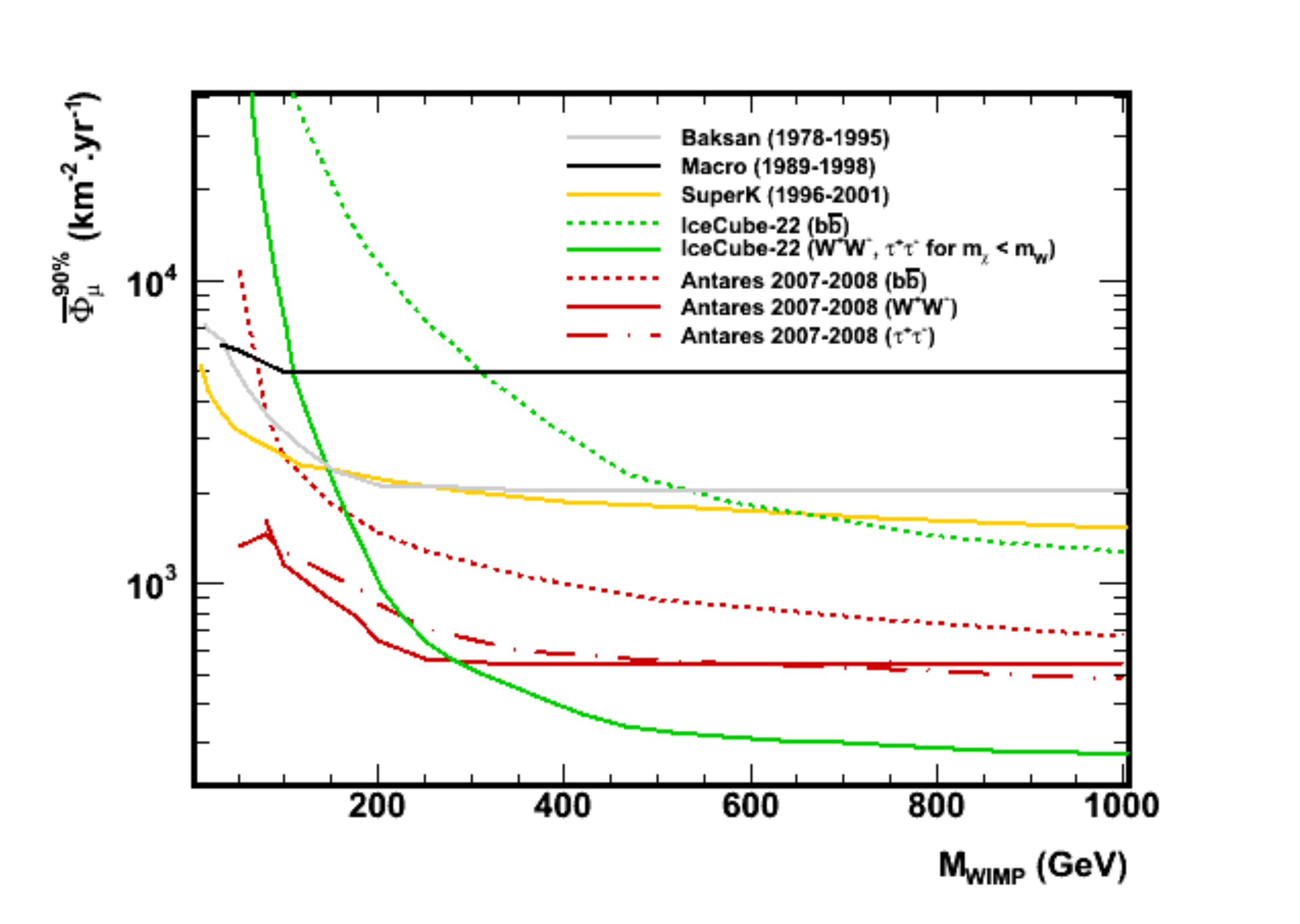}
\end{minipage}
\begin{minipage}[c]{.8\linewidth}
\includegraphics[width=\linewidth]{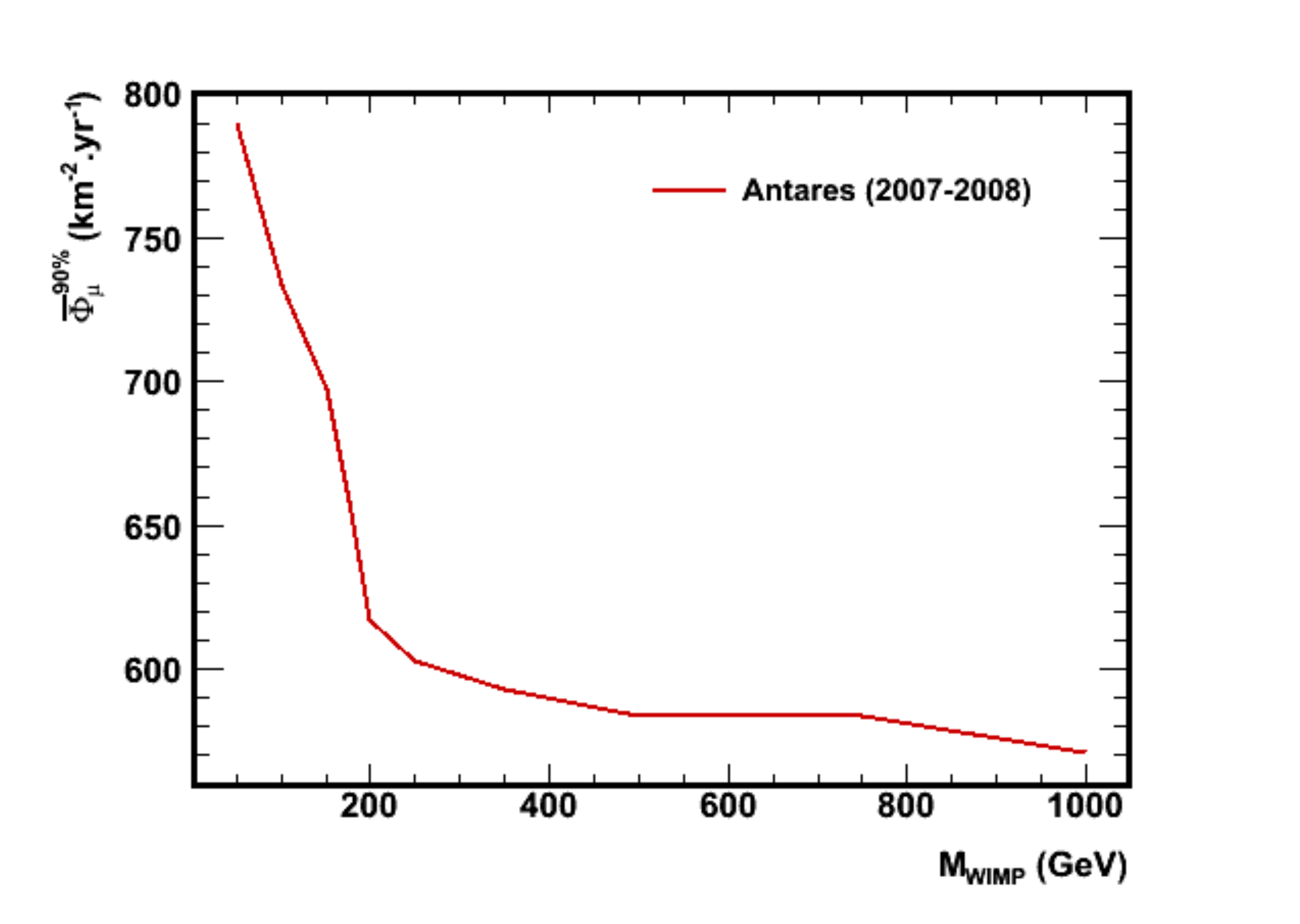}
\end{minipage}
\end{center}
\caption{Sentivity $\bar{\phi}_{\nu}^{90\%}$ in muon flux as a function of the WIMP masses, in the range $M_{WIMP}/in$[$10$ GeV;$1$ TeV] and from the data 2007-2008, 
for the CMSSM framework, on the left side, with Baksan $1978-1995$ (in grey), Macro $1989-1998$ (in black), SuperKamiokande $1996-2001$ (in yellow), IceCube-22 for the 
$b\bar{b}$ channel in $104$ days (in green dashed line), IceCube-22 for the $W^{+}W^{-}$ channel in $104$ days (in green solid line), ANTARES $2007-2008$ for the $b\bar{b}$ 
channel (in red dashed line), ANTARES $2007-2008$ for the $W^{+}W^{-}$ channel (in red solid line), and ANTARES $2007-2008$ for the $\tau\bar{\tau}$ channel 
(in red dot-dashed line). And the right side, the equivalent for the mUED framework.}
\label{CMSSMmUEDphimuaulFig}
\end{figure}

In the same way, the equivalent can be reached for the mUED framework in fig.~\ref{CMSSMmUEDphimuaulFig} (bottom). Unfortunately, any sensitivities from other experiments are not 
officially available in time for this proceeding. Concerning the present one for ANTARES, the shape is conformed to what is expected. This sensitivity in muon flux is very similar to the 
$W^{+}W^{-}$ one, or $\tau\bar{\tau}$ one, from the CMSSM framework due to the fact that the mUED neutrino flux construction is dominated by the $\tau\bar{\tau}$ channel through its branching ratio 
(see in \ref{dark matter signal and efficiency of antares}). More the effective area $A_{eff}(M_{WIMP})$ is large for high mass $M_{WIMP}$ values (see fig.~\ref{WeightedAeffFig}), more 
the resulting $\bar{\phi}_{\mu}^{90\%}$ is weak.

All of these sensitivities, for both frameworks studied in this paper, can be compared to the theoretical parameter spaces allowed by the experimental constraints, and derived from the SuperBayes 
simulation for the CMSSM framework \cite{austricmssm}, and for mUED \cite{austrimued}. In this way, the spin-dependent cross-section with protons $\bar{\sigma}_{H,SD}^{90\%}$ is 
developped from the sensitivity $\bar{\phi}_{\mu}^{90\%}$ using the method described in \cite{edsjosdcs}, as it appears in fig.~\ref{aulimitsdcsFig}, respectively for the CMSSM (top), and the mUED 
(bottom). For CMSSM and mUED, more the theoretical parameter space has a clear color more this one is allowed by the actual experimental contraints, with a comparison to the present direct detection 
experiment as KIMS~\cite{kims}, CDMS~\cite{cdms}, COUPP~\cite{coupp}, or Picasso~\cite{picasso}, and the usual indirect experiment like IceCube-22~\cite{icecube}, and SuperKamiokande~\cite{superk}. 
These curves show a $b\bar{b}$ channel sensitivity very closed to the SuperKamiokande one, and better than the IceCube-22 one for the same channel with more than one order of magnitude for the low 
energy regime ($m_{\chi} \leq 180$ GeV), and a $W^{+}W^{-}$ channel sensitivity better than the IceCube-22 one until $m_{\chi} \sim 440$ GeV, to be very closed to this last one for higher 
$m_{\chi}$ mass values. Concerning the $\tau\bar{\tau}$ channel, no equivalent has been found for IceCube or others, which places ANTARES as the actual best one for this channel in the CMSSM 
framework. This fact can be explained by the energy threshold of ANTARES around $20$ GeV, lower than for IceCube $\sim 50$ GeV, which has an effect over all the WIMP mass range due to the 
integration over the neutrino energy range, $E_{\nu}\in$[$10$ GeV;$M_{WIMP}$].

\begin{figure}[!t]
\begin{center}
\begin{minipage}[c]{.8\linewidth}
\includegraphics[width=\linewidth]{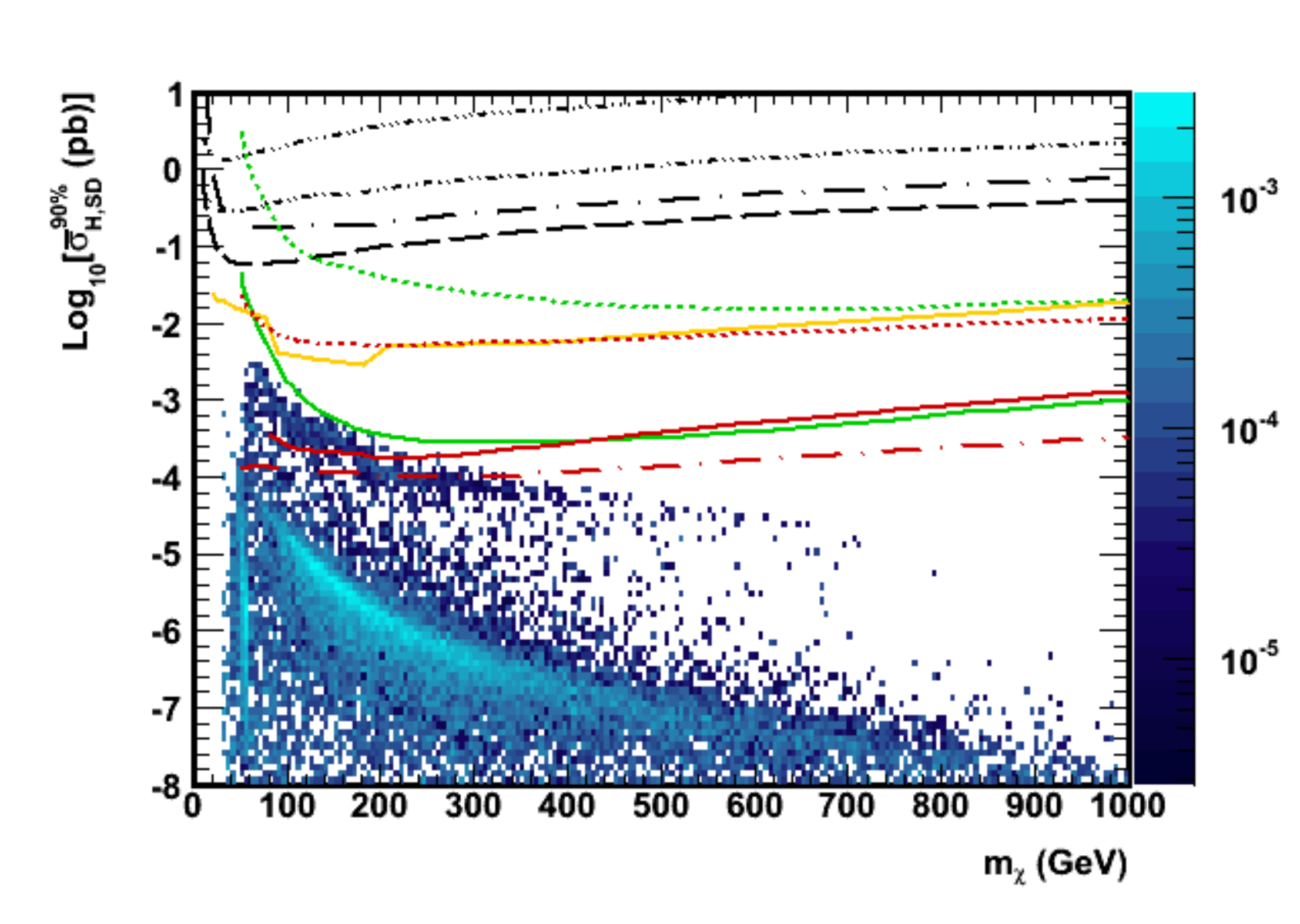} 
\end{minipage}
\begin{minipage}[c]{.8\linewidth}
\includegraphics[width=\linewidth]{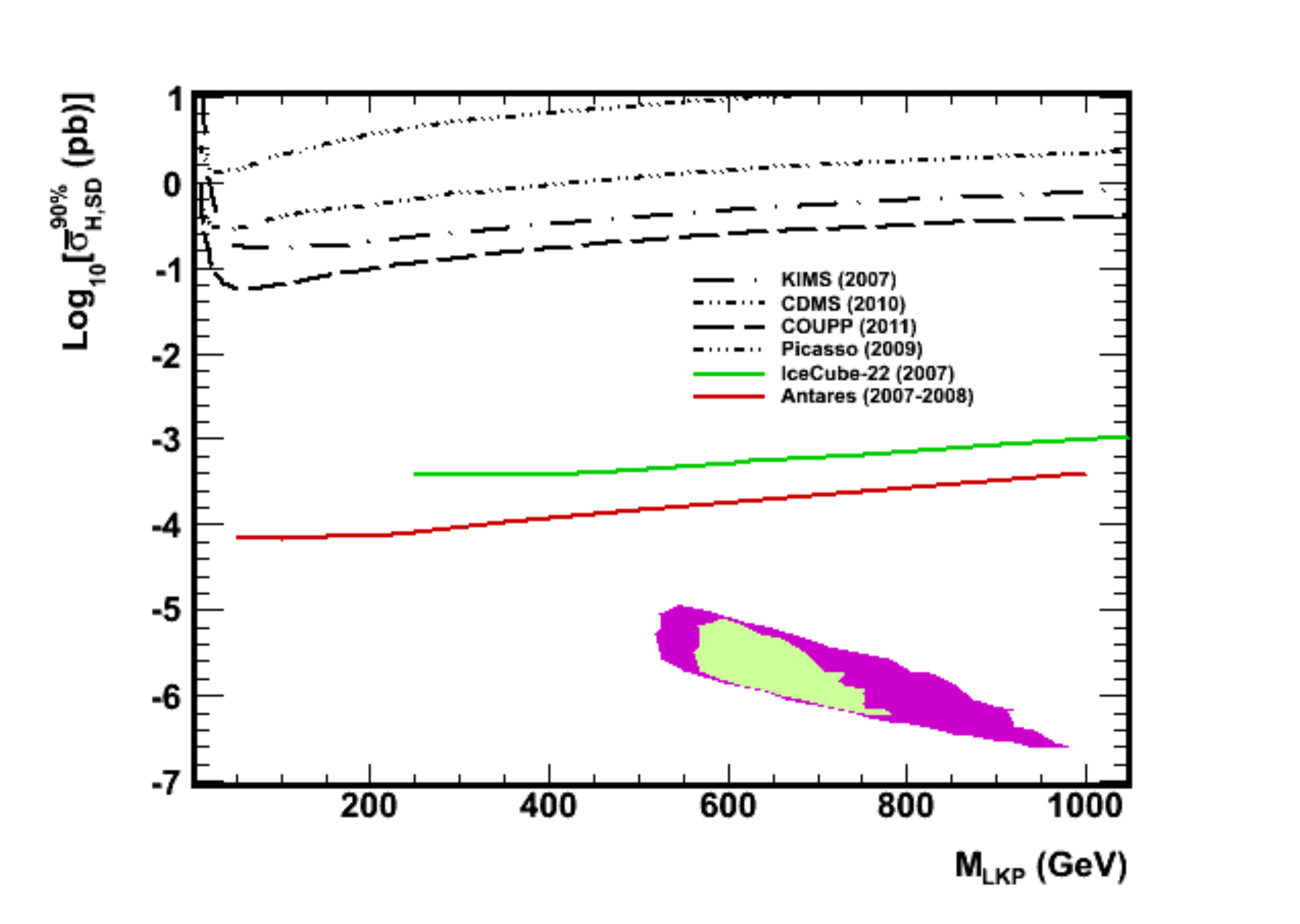}
\end{minipage}
\end{center}
\caption{Spin-dependent cross-section $\bar{\sigma}_{H,SD}^{90\%}$ as a function of the WIMP masses, in the range $M_{WIMP}/in$[$10$ GeV;$1$ TeV] and from the data 2007-2008, 
for the CMSSM framework, on the top, with a comparison to the present direct detection experiment as KIMS $2007$ (in dash-dot black line), CDMS $2010$ 
(in dash-two-dots black line), COUPP $2011$ (in dashed black line), or Picasso $2009$ (in dash-three-dots black line), and the usual indirect experiment like IceCube-22 
(in dashed green line for the $b\bar{b}$ channel, and in solid green line for the $W^{+}W^{-}$ channel), SuperKamiokande $1996-2001$ (in yellow solid line). The ANTARES 
$2007-2008$ contributions appear for the $b\bar{b}$ channel (in dashed red line), for the $W_{+}W_{-}$ channel (in solid red line), and for the $\tau\bar{\tau}$ channel 
(in red dot-dashed line). And on the bottom, the equivalent for the mUED framework, with IceCube-22 $2007$ (in green solid line), and ANTARES $2007-2008$ contribution 
(in red solid line).}
\label{aulimitsdcsFig}
\end{figure}

For the mUED case which appears in fig.~\ref{aulimitsdcsFig} (bottom), we took the same line style code than the previous one concerning the CMSSM (top) for the direct 
experiment. The contributions of the indirect search experiments is always a great complement to the direct detection experiments, with a best sensitivity with ANTARES on all the $M_{LKP}$ mass 
range.


\clearpage

\setcounter{figure}{0}
\setcounter{table}{0}
\setcounter{footnote}{0}
\setcounter{section}{0}
\newpage





\title{On neutrino oscillations searches with ANTARES}

\shorttitle{G.Guillard, J.Brunner --- neutrino oscillations in ANTARES}

\authors{Goulven Guillard$^{1\dag}$, J\"urgen Brunner$^{2}$, on behalf of the ANTARES Collaboration}
\afiliations{$^1$Clermont Universit\'e, Universit\'e Blaise Pascal,CNRS/IN2P3, Laboratoire de Physique Corpusculaire, BP 10448, F-63000 Clermont-Ferrand, France\\
$^2$Centre de Physique des Particules de Marseille (CNRS/IN2P3) \& Universit\'e de la M\'editerran\'ee, Marseille, France
}
\email{$^\dag$guillard@in2p3.fr}

\maketitle
\begin{abstract}Although the first evidence for neutrino oscillations came from measurements on atmospheric neutrinos in underground experiments, neutrino oscillations have yet to be demonstrated in high energy neutrino telescopes, whose energy threshold is significantly higher.
Recent studies have shown that a clean sample of atmospheric neutrinos with energies as low as 20\,GeV can be isolated in the ANTARES neutrino telescope.  Such a threshold is low enough to allow the observation of neutrino oscillation features.  A robust analysis method is presented which allows the extraction of atmospheric neutrino oscillation parameters.
\end{abstract}


\section{Introduction}

It is now well established that neutrinos can switch from a flavour to another, neutrino flavour eigenstates for the weak interaction being different from neutrino mass eigenstates.  This phenomenon, hypothesized in 1957 by Pontecorvo and described more precisely in the early sixties~\cite{osc_th}, is known as neutrino oscillations, and has been measured by a number of experiments~\cite{osc_exp}.

Although part of the scientific program of the ANTARES neutrino telescope, oscillations studies have been delayed because of the difficulty to reliably reconstruct muons at a sufficiently low energy.  Indeed, the granularity of ANTARES is rather coarse compared with the range of GeV muons in water~: its 885 photomultiplier tubes are distributed by triplets placed 14.5\,m apart along the detector lines, themselves spaced by about 60\,m~\cite{ant}.  As a consequence, the energy range where the effect of the atmospheric neutrino oscillations should be visible is situated at the very edge of the detector sensitivity, which makes the neutrino oscillations measurement very challenging.

The topic experienced a regain of interest since the completion of the detector in 2008.  As progress was steadily made in the detailed understanding of the detector and of its environment, it became clear that it should be possible to extract a clean sample of atmospheric neutrinos with energies around the first minimum in the survival probability of muon neutrinos propagating through the Earth.

Even though dedicated experiments are more sensitive for the measurements of neutrino oscillations, such an analysis is an important check for ANTARES.  The analysis being very sensitive not only to the telescope efficiency but also to the quality of the angular and energy reconstruction, it becomes a benchmark for the understanding of the detector.

\section{Purpose}

Neutrinos oscillations are commonly described in terms of $E/L$ dependence, where $E$ is the neutrino energy and $L$ its oscillation path length~\cite{osc_exp}.  For a neutrino telescope such as ANTARES, detecting neutrinos crossing the Earth, $L$ can be translated as $2R\sin\theta$, $R$ being the Earth radius and $\theta$ the anti-elevation (that is, the angle between the neutrino direction and the horizontal axis~: $\theta=\pi/2$ for a vertical upgoing neutrino).  Within the two-flavour approximation, the $\nu_\mu$ survival probability can then be written
$$P(\nu_\mu\rightarrow\nu_\mu)\simeq1-\sin^22\theta_{23}\sin^2\left(2.54R\Delta m_{32}^2\frac{\sin\theta}{E}\right),$$
$\theta_{23}$ and $\Delta m_{32}^2$ being respectively the mixing angle and the squared mass difference of the involved mass eigenstates (with $R$ in km, $E$ in GeV and $\Delta m_{32}^2$ in eV$^2$).

According to recent results from the \textsc{Minos} experiment~\cite{minos}, the first minimum in the muon neutrino survival probability $E/\sin\theta$ spectrum occurs at about 24\,GeV (figure~\ref{fig:esinth}).  This interesting region is in principle accessible for ANTARES.  Indeed, Monte-Carlo simulations have shown that it is possible to extract clean samples of atmospheric neutrinos with energies in this range, which should enable the observation of the first minimum.

\begin{figure}[!t]
\vspace{5mm}
\centering
\includegraphics[width=0.48\textwidth]{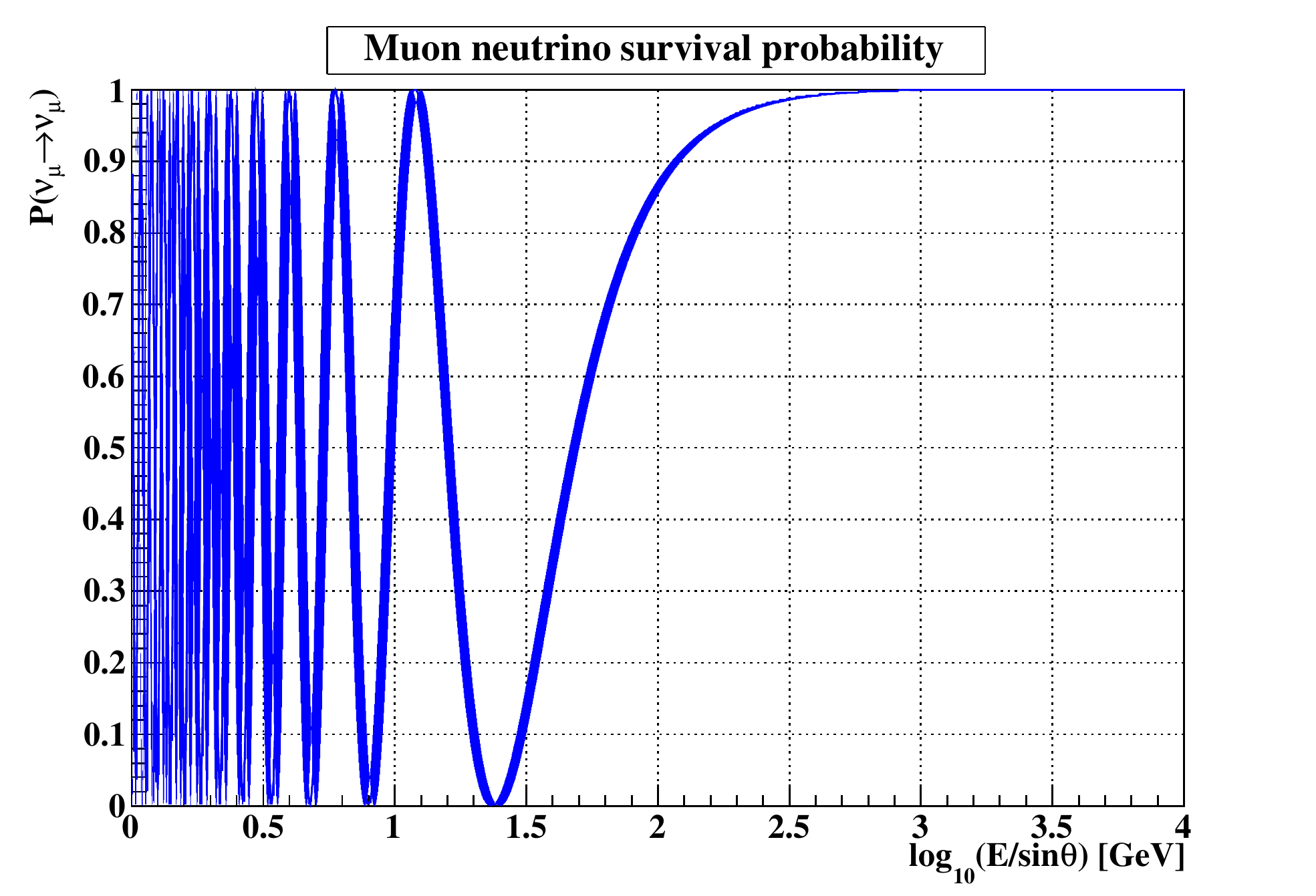}
\caption{Survival probability of muon neutrinos as a function of $E/\sin\theta$ in the two-flavour approximation, with $\sin^22\theta_{23}=1$ and $\Delta m_{32}^2=2.32\times10^{-3}\,\mathrm{eV}^2$ (the line width corresponds to \textsc{Minos} $1\,\sigma$ uncertainty on this parameter~\cite{minos}), neglecting matter effects.}
\label{fig:esinth}
\end{figure}

\section{Method}

A simple and robust method to measure neutrino oscillations in ANTARES consists in taking advantage of the different granularity of the detector along and across the detection lines.  Indeed the range of muons below about 20\,GeV makes them unlikely to give signal on different lines.  They can be well reconstructed however using only one line, especially if they are close to the vertical.  These low energy vertical tracks should be suppressed in the current oscillation scenarios compared to the more energetic and less vertical ones.  Consequently, the observed ratio between the number of tracks giving signal on a single detector line (1D tracks) and the number of tracks giving signal on several detector lines (3D tracks) should differ from the expected ratio in the case of no oscillations.  Many systematic effects such as the global flux normalisation or absolute detection efficiencies should cancel in this ratio, leading to a small residual systematic effect.  The observed ratio can be expressed in terms of Monte-Carlo events as
\begin{equation}\label{eq:ratio}
R=\frac{N_{\mathrm{1D}}^{no\ osc}-\sin^22\theta_{23}\cdot f_{\mathrm{1D}}}{N_{\mathrm{3D}}^{no\ osc}-\sin^22\theta_{23}\cdot f_{\mathrm{3D}}},
\end{equation}
where $N_{\mathrm{1D|3D}}^{no\ osc}$ is the number of expected 1D (3D) tracks without neutrino oscillations and
$$f_{\mathrm{1D|3D}}=\sum_i\sin^2\left(2.54R\Delta m_{23}^2\frac{\sin\theta_i}{E_i}\right)$$
is summed over each event $i$.

For each value of $\Delta m_{23}^2$ the mixing angle can be calculated analytically from the measurement of $R$ according to equation~\ref{eq:ratio}.  The statistical and systematic errors from $R$ propagate directly into an error on the mixing angle, which leads to a band in the $(\theta_{23},\Delta m_{32}^2)$ plane.

Tracks are reconstructed using a fast reconstruction algorithm~\cite{bbfit}, which has the advantage of being more efficient at lower energies than other existing reconstruction strategies.  A cut on the estimated quality of the fitted track and on the reconstructed angle is performed, in order to select well-defined events and to increase the upgoing neutrinos purity.  The 1D tracks are required to be reconstructed using hits from at least 8~detector storeys.  Lowering this cut would considerably enhance the number of 1D events and thus the sensitivity to oscillations, but would also increase dramatically the contamination from misreconstructed atmospheric muons.  Such a hard cut ensures a very low contamination and leads to a minimal track length of about 100\,m.  The energy threshold is thus roughly 20\,GeV, which is below the first oscillation minimum in the $E/\sin\theta$ spectrum.

This method is robust as it does not depend on many assumptions, nor does it rely on an energy estimator.  Furthermore the reconstruction algorithm is used only to assert the quality of the selected events.  Finally, assuming they affect similarly 1D and 3D events, the ratio should cancel most systematics, and in particular the large uncertainties on the atmospheric neutrino flux normalization and the uncertainties on the detector simulation.  The strict hit selection criteria and the strong cut on the number of hit storeys minimize the sensitivity of the ratio to potential biases in the optical background simulation.  Remaining systematic uncertainties are expected to be within a few percents, smaller than the current statistical errors.


\section{Expected sensitivity}

Table~\ref{tab:sens} shows, according to \textsc{Minos} results~\cite{minos}, the number of expected events for a 170 days Monte-Carlo sample (this corresponds to 2008 active time).  The atmospheric neutrino flux is weighted to match Bartol parametrization~\cite{bartol}.  The contamination of misreconstructed downgoing atmospheric muons is negligible.  The number of 1D events is suppressed by 16\,\% in the case of oscillations while the number of 3D events is suppressed only by 3.6\,\%.  The effect to be observed is however small~: the suppression of 1D tracks concerns only 26 events for this sample.  It is clear that the $E/\sin\theta$ spectrum cannot be reliably extracted under such conditions~: a larger statistics is needed.

\begin{table}[!ht]
\centering
\begin{tabular}{lccc}\hline
& no osc. & osc. & contamination\\\hline\hline
1D tracks & 186 & 160 & 4.7 \\\hline
3D tracks & 522 & 504 & 1.2 \\\hline
\end{tabular}
\caption{Number of expected $\nu_\mu/\bar\nu_\mu$ charged current events, after 170~days of lifetime, with and without oscillations, for 1D and 3D tracks, and number of misreconstructed atmospheric muons surviving the cuts.}
\label{tab:sens}
\end{table}

The sensitivity after 1000~days, which may be achieved after 4 or 5~years of real data taking depending on various external conditions, is presented in figure~\ref{fig:osc_sens}, extrapolating from the numbers given in table~\ref{tab:sens}.  The number of expected 1D events is roughly one per day, which leads for 1000~days to a statistical error of about 3\,\% ($\sim$30~events).  Studies where PMT and water parameters have been varied within their tolerance have shown that a 3\,\% systematic error on $R$ is realistic.  Consequently, a total standard deviation of about 5\,\% on $R$ has been used to draw the measurement contours of figure~\ref{fig:osc_sens}.

\begin{figure}[!t]
\vspace{5mm}
\centering
\includegraphics[width=0.48\textwidth]{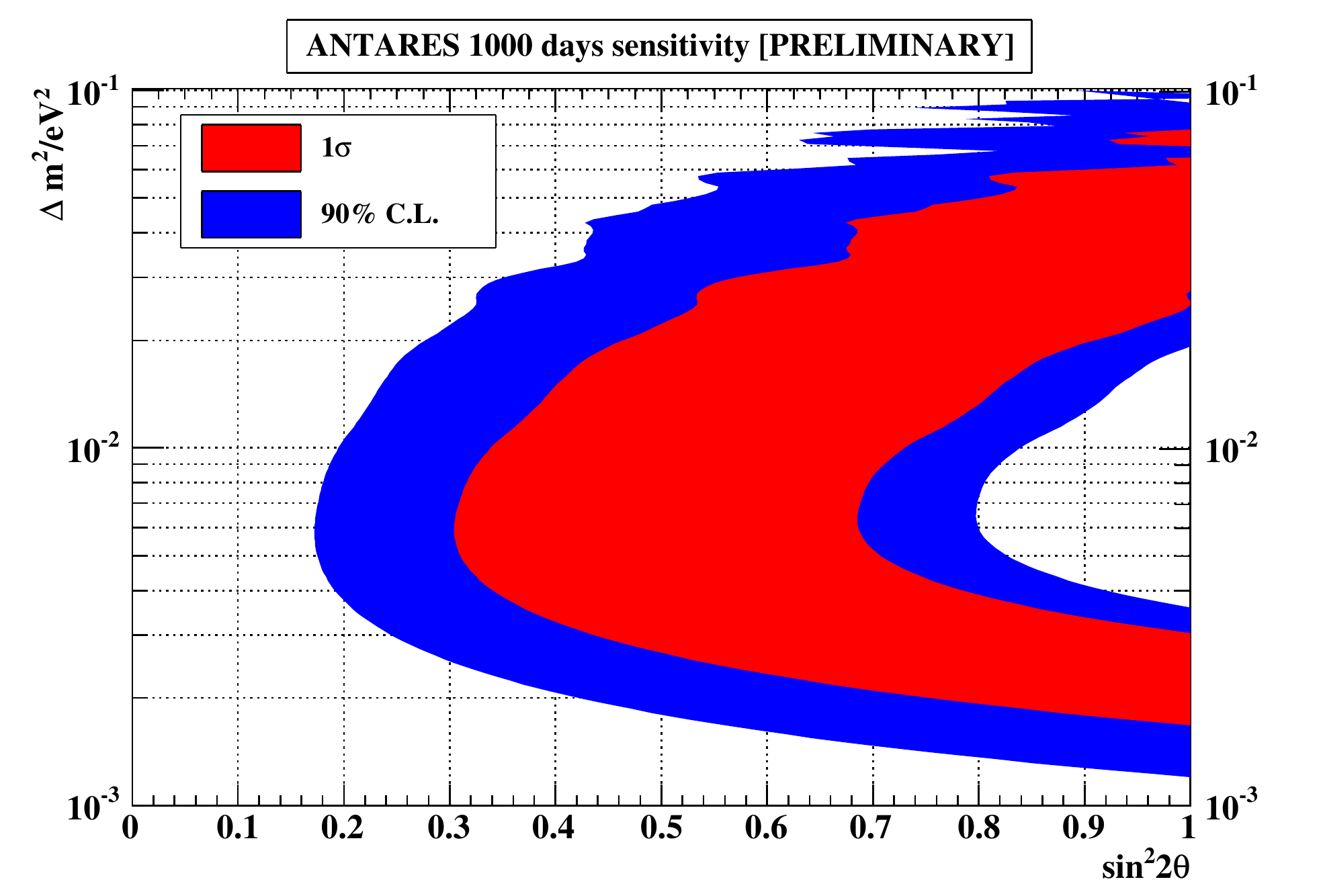}
\caption{ANTARES expected sensitivity to atmospheric neutrino oscillations after 1000~days.}
\label{fig:osc_sens}
\end{figure}

According to these results, although not competitive with dedicated experiments, ANTARES should be sensitive to neutrino oscillation parameters through disappearance of atmospheric muon neutrinos with the simple and robust analysis presented here.  Current preliminary results using a restricted data sample are compatible with world oscillation data.  Processing of the whole ANTARES data set is foreseen.

\section{Prospects}

Several directions are currently being investigated to improve ANTARES sensitivity to neutrino oscillations.  In a near future, it should be possible to increase significantly the number of low energy reconstructed muons thanks to a dedicated reconstruction algorithm coupled to an optimized hit selection, which would reduce statistical uncertainties using the same amount of data.  Furthermore the use of a different reconstruction algorithm would be an important cross-check of such an analysis.  Additionally, the proportion of low energy events can be improved by selecting events contained in the detector.

Such improvements should enhance ANTARES sensitivity to neutrino oscillation parameters.  They might also allow the extraction of the $E/\sin\theta$ spectrum, which would be a great opportunity to cross-check the understanding of the detector.

\section{Conclusions}

A robust method to extract atmospheric muon neutrino oscillation parameters from ANTARES data has been presented.  Although not competitive with dedicated experiments, ANTARES should be able to reach some sensitivity to these parameters, which would be a demonstration of the understanding of the ANTARES detector.  Preliminary analysis of a restricted data sample is compatible with existing constraints on neutrino oscillation parameters.  The complete analysis of ANTARES data is ongoing.


\clearpage

\setcounter{figure}{0}
\setcounter{table}{0}
\setcounter{footnote}{0}
\setcounter{section}{0}
\newpage
\mbox{}





\title{Search for magnetic monopoles with the ANTARES underwater neutrino telescope}

\shorttitle{Nicolas Picot-Clemente \etal Search for magnetic monopoles with the ANTARES underwater neutrino telescope}

\authors{Nicolas Picot-Clemente$^{1}$, on behalf of the ANTARES Collaboration}
\afiliations{$^1$Centre de Physique des Particules de Marseille, Marseille, France}
\email{picot@cppm.in2p3.fr}

\maketitle
\begin{abstract}The ANTARES underwater neutrino telescope, located in the Mediterranean Sea at a
depth of 2475 m, 40km off the Provencal coast, is composed by an array of 885 photomultipliers distributed on 12 vertical lines. The detector is fully operational
since May 2008. Besides the detection of neutrino-induced muons, the telescope is more generally sensitive to particles which emit Cherenkov light, and, thanks to its large volume,
offers new opportunities to improve the sensitivity to exotic particles, such as magnetic monopoles.
Magnetic monopoles are stable magnetically charged particles first introduced in quantum mechanics by Dirac in 1931, which showed that their existence would give
an explanation to the quantization of the electric charge. They would have been produced in the Early Universe, and would bring a first proof to the existence of
grand unified models. The paper introduces magnetic monopoles and their signal caracteristics in a
neutrino telescope, and then describes the analysis which uses a dedicated algorithm able to reconstruct the velocity of the incoming particles.
A new upper limit on the upward going magnetic monopole flux, extracted from ANTARES data taken in 2008, is presented. It is, at present, the best worldwide constraint in the velocity
range $\beta=[0.625,1]$, and provides, for the first time with a neutrino telescope, a constraint for monopole velocities below the Cherenkov threshold.
\end{abstract}


\section{Introduction}

Magnetic monopoles are hypothetical magnetically charged particles first introduced by Dirac in 1931. He demonstrated that the existence of magnetic monopoles naturally leads to the quantization of the electric charge~\cite{ref:Dirac}. In 1974, 't~Hooft and Polyakov discovered independently that, in certain spontaneously broken gauge theories, magnetic monopoles are not only a possibility, but a requirement~\cite{ref:thooft,ref:polyakov}.
However, despite intensive search efforts, no magnetically charged particles were detected up to now, and stringent experimental flux limits beyond the original Parker bound of flux $F_P \sim 10^{-15}$~cm$^{-2}$s$^{-1}$sr$^{-1}$~\cite{ref:Parker} were provided.
The development of huge detectors for performing neutrino astronomy gives rise to new expectations for the search for magnetic monopoles, and ANTARES offers new interesting opportunities in this field.

In this paper a search for upward going relativistic magnetic monopoles for one year of data taking in 2008 with the ANTARES detector~\cite{ref:AntaresDet} is presented.
Section~\ref{sec:1} will first introduce magnetic monopoles and their signature in the ANTARES neutrino telescope will be briefly presented. Then the simulation and reconstruction algorithm will be detailled in section~\ref{sec:2}, and finally the analysis method will be explained in section~\ref{sec:3}, and the final results will be given in section~\ref{sec:4}

\section{Magnetic monopoles and their signal in the ANTARES neutrino telescope}\label{sec:1}

\subsection{Magnetic monopoles}

Most of the Grand Unified Theories (GUTs) predicts the creation of magnetic monopoles in the early Universe. Indeed, in 1974, 't~Hooft and Polyakov showed independently that elements caracterising well a magnetic charge, as introduced by Dirac in 1931~\cite{ref:Dirac}, occur as solitons in Gran Unified Theories (GUT) in which a larger gauge group breaks down into a semi-simple subgroup containing the explicit U(1) group of electromagnetism~\cite{ref:thooft,ref:polyakov}.\\
These particles are topologically stable and carry a magnetic charge defined as a multiple integer of the Dirac charge $g_D=\frac{\hbar c}{2e}$, where $e$ is the elementary electric charge, $c$ the speed of light in vacuum and $\hbar$ Planck's constant. Depending on the specific model, the masses inferred for magnetic monopoles range over many orders of magnitude, from $10^4$ to $10^{20}$ GeV~\cite{ref:Preskill}.

Since fast monopoles have a large interaction with matter, they can loose large amounts of energy in the terrestrial environment. The total energy loss of a relativistic monopole with one Dirac charge is of the order of $10^{11}$ GeV after having crossed the full diameter of the Earth~\cite{ref:Derkaoui}. Because magnetic monopoles are expected to be accelerated in galactic coherent magnetic field domains to energies of about $10^{15}$ GeV~\cite{ref:Wick}, calculations suggest that monopoles with mass below $\sim 10^{14}$ GeV would be able to cross the Earth and reach the ANTARES detector as upward going signals.

\subsection{Signal in the ANTARES neutrino telescope}

The ANTARES detector is an underwater neutrino telescope immersed in the Western Mediterranean Sea at a depth of 2475~m~\cite{ref:Antares}. In its full configuration the detector consists
in 12 mooring lines, each comprising 25 storeys separated by 14.5 m. A storey consists of 3 glass spheres housing a photomultiplier.
Finally each line is connected to a Junction Box, which is connected to the shore station at La Seyne-sur-Mer by an electro-optical cable of 40 km length.

As for electric charged particles, magnetic monopoles crossing ANTARES with a velocity higher than the speed of light in the sea water of $\beta\sim0.74$, are expected to emit Cherenkov light. The expected light is however much more intense, with about 8550 times more Cherenkov photons than muons of the same velocity~\cite{ref:Tompkins}. 

In addition to the direct photon emission above the Cherenkov threshold, magnetic monopoles can produce indirect light below. Indeed, by ionizing the sea water along its path, a monopole with a velocity $\beta> 0.51$ will knock out electrons ($\delta-$rays), which will get enough energy for emitting Cherenkov light~\cite{ref:vanRens}. Fig.~\ref{deltaraychegraph} indicates the amount of light produced during the crossing of a monopole of charge $g_{D}$ as a function of its velocity, compared to the emission of a through going muon.
\begin{figure}[h!]
   \begin{center}
      \includegraphics[height=1.7in]{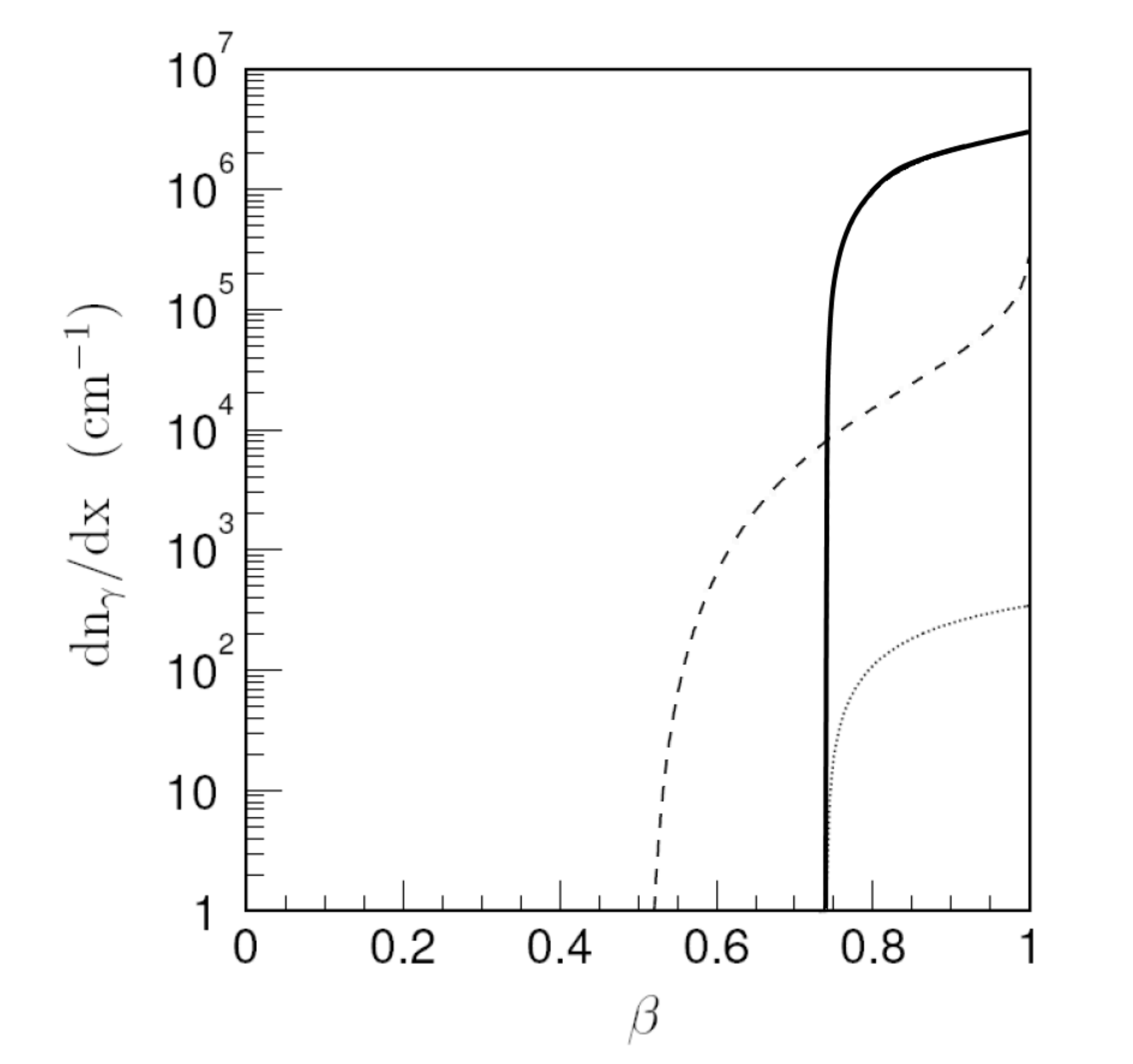}
   \end{center}
   \caption{\footnotesize Number of Cherenkov photons in the $300-600$ nm wavelength range per centimeter emitted by a monopole with $g=g_D$ (solid line), by the $\delta-$rays produced along its path (dashed line) and by a muon (dotted line), as a function of the particle velocity.}
   \label{deltaraychegraph}
\end{figure}

\section{Simulation and reconstruction}\label{sec:2}
\subsection{Simulation}
Upward going magnetic monopoles with one unit of Dirac charge ($g=g_D$) have been simulated using a Monte Carlo program based on GEANT3 \cite{ref:GEANT} for ten ranges of velocities in the region $\beta=[0.550,0.995]$ and with the incoming direction distributed isotropically over the lower hemisphere.
The simulation of emitted photons is processed inside a cylindrical volume surrounding the instrumented volume. A radius of $480$~m,  four times larger than that used for the standard ANTARES muon simulation, is chosen in order to take into account the large amount of light emitted by a magnetic monopole.

When searching for upward going magnetic monopoles, the main physical backgrounds are upward going atmospheric neutrino-induced muons and downward going atmospheric muons. For simulating atmospheric muons, the CORSIKA program~\cite{ref:Corsika} in combination with the QGSJET~\cite{ref:QGSJET} code was used, employing the Horandel model for the cosmic ray spectrum, whereas atmospheric neutrinos were simulated according to the Bartol flux~\cite{ref:Bartol, ref:Bartol2} and the RQPM model\cite{ref:RQPM}.

In order to match the real detector conditions, the simulations have been performed using the dead channel information and the optical background rates extracted from the data.

\subsection{Reconstruction algorithms}

Before the reconstruction step, the different trigger algorithms are applied to data before events are stored~\cite{ref:AntaresDAQ}. These conditions lead to less than 15 \% inefficiency for monopoles with a velocity greater than $\beta = 0.58$ that produced at least six hits in the detector.

The standard track reconstruction assumes that particles travel at the speed of light. In order to improve the sensitivity for magnetic monopoles travelling with lower velocities, the velocity was introduced as a free parameter in the track fit. The modified algorithm performs two independent fits, a track fit and a bright point fit. The former reconstructs the track of a particle crossing the detector at a velocity $\beta$, introduced as a free parameter, while the latter reconstructs the event as a point-like light source. Both fits minimize the same quantity $\chi^2$ defined as~:
\begin{equation}
 \chi^2=\sum_{i=1}^{N}\left[\frac{(t_\gamma-t_i)^2}{\sigma_i^2}+A_i\right]
\end{equation}
The first term on the right hand side is the sum for $N$ hits of the square of the time residual, where $t_\gamma$ is the expected arrival time of a hit, $t_i$ the measured time and $\sigma_i$ the time uncertainty for each hit. The second term $A_i$ corresponds to the contribution of hit charge weighted by its minimal approach distance. It brings a penalty to hits with large amplitude combined with a large minimal approach distance.

This algorithm yields a resolution $\delta \beta$ on the reconstructed velocity of magnetic monopoles of about $\delta \beta \simeq 0.025$ for velocities lower than the Cherenkov threshold, $\beta_C=0.74$, improving to $\delta \beta \simeq 0.003$ for higher velocities.

\section{Analysis}\label{sec:3}

The analysis has been performed for data taken from December 2007 to December 2008. Quality requirements are applied to the data to ensure low levels of bioluminescent activity and a well calibrated detector. After this selection the data is equivalent to a total of 136 days of live time~: 44 days with 12 lines, 46 days with 10 lines and 47 days with 9 lines.

For checking the agreement between simulated atmopheric events and real data, 15 \% of the whole dataset, corresponding to 20 days of data taking, was taken and used at each step of the analysis.
To avoid any bias, the sample of 15 \% of data has at the end been removed, and so has not been considered in the final result.
Despite the good agreement of shape between data and MonteCarlo atmospheric distributions observed, a normalisation factor of 1.8 is needed to be applied on simulated atmopheric muons, which is consistent with the expected uncertainties on the optical module angular acceptance for downgoing particles and on the input parameters of the model-dependent atmospheric muon flux.

\subsection{Preliminary cuts}


Only magnetic monopoles crossing the Earth are considered in the analysis, and a natural precut is then to only keep particules reconstructed as upward going, i.e. with a reconstructed zenith angle less than $90^{\circ}$. This cut is motivated by the large amount of background of downward going events coming from atmospheric muons.

Moreover a second preliminary cut consists in choosing only particles for which the track was reconstructed by using at least 2 lines, to improve the reconstruction quality.

Finally, the last preliminary cut consists in removing events for which the quality factor of the fit is better with the bright point hypothesis than with the track one.  Electromagnetic and hadronic showers dominate such events, and a large fraction of poorly reconstructed atmospheric muons are then removed.

\subsection{Discriminating variables}

The event selection for optimising the Model Discovery Factor was performed by using two discriminating variables.
As shown fig.~\ref{deltaraychegraph}, the first discriminant between atmospheric events and magnetic monopoles is the number of hits $nhit$ produced in the detector.
Fig.~\ref{nhitdist} represents distributions of the number of hits used by the reconstruction algorithm for events reconstructed in the velocity range $\beta_{rec}=[0.775,0.825]$.
A good agreement is observed between simulated atmospheric events and the 15 \%  data sample, whereas the distribution for magnetic monopoles simulated in the same velocity bin confirms the potential of discrimination for the $nhit$ variable.
\begin{figure}[h!]
   \begin{center}

      \includegraphics[height=1.55in]{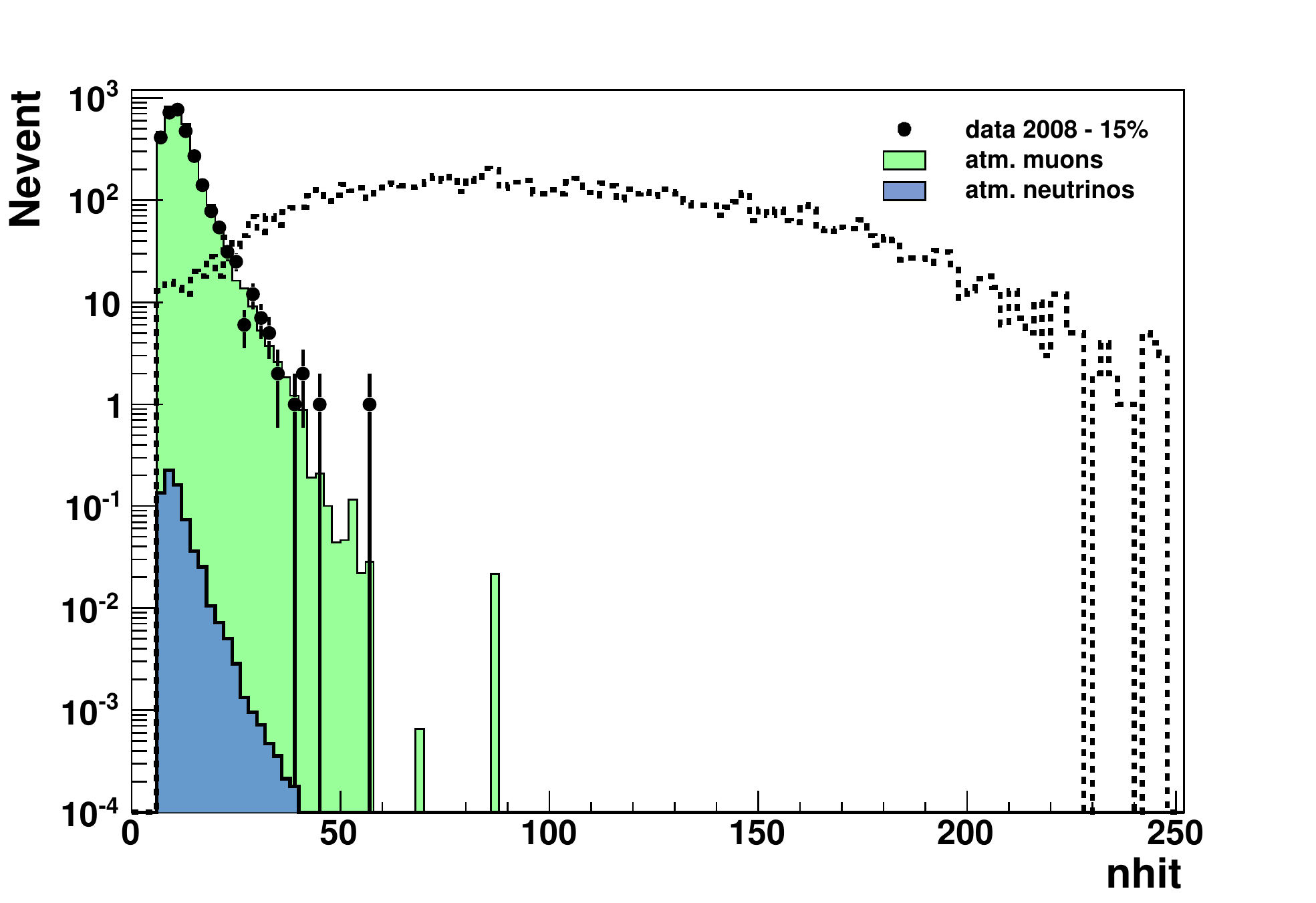}

   \end{center}
   \caption{\footnotesize Distributions of $nhit$ for events reconstructed with $\beta_{rec}=[0.775,0.825]$. Atmospheric down-going muons and up-going neutrinos (solid histograms) are compared to the 15 \% data sample (points). The dashed line stands for magnetic monopoles generated in the same velocity bin.}
   \label{nhitdist}
\end{figure}

However, for the lowest monopole velocities below the Cherenkov threshold, using only the number of hits is not sufficient for discrimating monopoles from atmospheric background events, and a new variable has thus been defined by applying two reconstruction algorithms.
The first one is the initial muon fit algorithm in which the velocity $\beta$ is fixed at 1, whereas the second modified algorithm allows $\beta$ as a free parameter in the procedure. The discriminating parameter $\lambda$ is then defined as~:
\begin{equation}
 \lambda=\log\left(\frac{\chi^2_t({\beta=1})}{\chi^2_t({\beta_{free}})}\right)
\end{equation}
where $\chi^2_t({\beta=1})$ and $\chi^2_t({\beta_{free}})$ are the track quality parameters for a fixed and free velocity $\beta$ respectively.
This new variable is then expected to be positive for monopoles and negative for atmospheric muon and neutrino events.
In fig.~\ref{lambdadist} are shown the $\lambda$ distributions of events reconstructed in the velocity range $\beta=[0.775,0.825]$, for atmospheric simulated muons and neutrinos, and for the 15\% data sample, as well as for magnetic monopoles simulated in the same range of velocity.
\begin{figure}[h!]
   \begin{center}

      \includegraphics[height=1.55in]{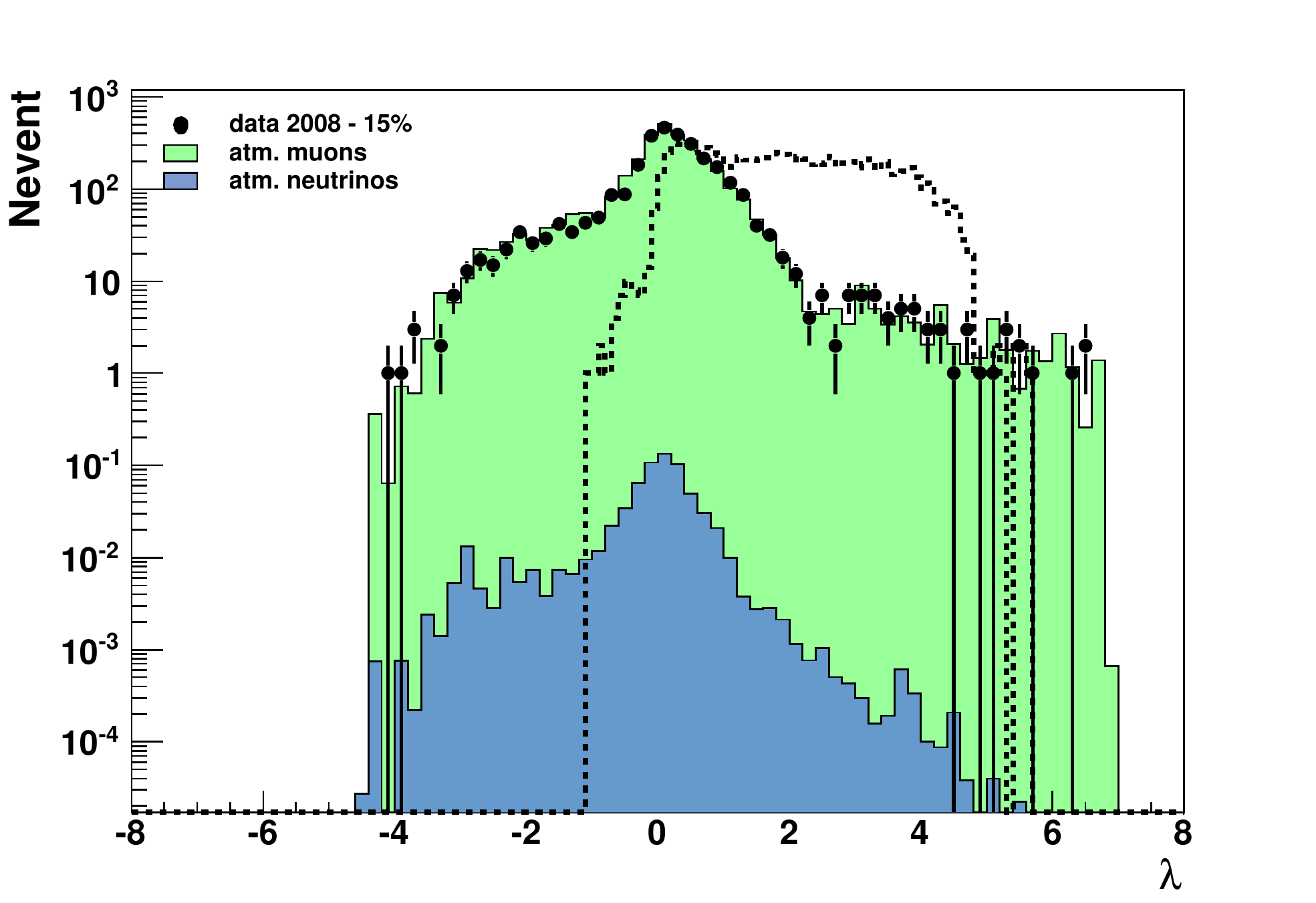}

   \end{center}
   \caption{\footnotesize Distributions of $\lambda$ for events reconstructed with $\beta_{rec}=[0.775,0.825]$. Atmospheric down-going muons and up-going neutrinos (solid histograms) are compared to the 15 \% data sample (points). The dashed line stands for magnetic monopoles generated in the same velocity bin.}
   \label{lambdadist}
\end{figure}

Finally, the optimisation of the model discovery factor for a $5\sigma$ discovery at $90\%$ probability was performed for 8 bins of monopole velocity between $\beta=[0.625, 0.995]$ by scanning the $nhit$-$\lambda$-plane.

\section{Results}\label{sec:4}

The [$nhit$, $\lambda$] cuts for every velocity bins, defined previously were then applied to the 85~\% data sample.
The number of atmospheric background events expected after 116 days of data taking is indicated in Table~1 for each bin of reconstructed velocity, as well as the number of observed events when the selection cuts are applied to the 85~\% data sample. No significant deviations from the background-only hypothesis are observed by comparing the number of expected events and the number of observed events. All the observed events are therefore considered as background, with no claim for a monopole discovery. The Feldman-Cousins 90~\% C.L. upper limit on the upward going magnetic monopole flux is reported in Table~1.
\begin{table}[!h]\label{tab:table1}
\begin{tiny}
   \begin{center}
     \begin{tabular}{ccccccc}
     \hline
     	\rowcolor[gray]{.8} $\beta_{rec}$    &  Number of exp.     & Number of  &  $90\%$ C.L. flux u. l. \\
	\cline{2-4}
     	\rowcolor[gray]{.8}      range         &  background events   &  observed events & (cm$^{-2}$ s$^{-1} $sr$^{-1}$)\\ \hline 
	$[0.625,0.675]$ &  $2.2\times10^{-2}$	&   0 & $7.5\times10^{-17}$ \\ 
	$[0.675,0.725]$ &  $1.3\times10^{-1}$ 	&   1 & $8.9\times10^{-17}$ \\
	$[0.725,0.775]$ &  $4.5\times10^{-2}$ 	&  0 & $4.0\times10^{-17}$ \\ 
	$[0.775,0.825]$ &  $1.1\times10^{-6}$	&  0 & $2.4\times10^{-17}$ \\ 
	$[0.825,0.875]$ &  $8.2\times10^{-7}$ 		&  0 & $1.8\times10^{-17}$ \\ 
	$[0.875,0.925]$ &  $6.9\times10^{-7}$ 	         &  0 & $1.7\times10^{-17}$ \\ 
	$[0.925,0.975]$ &  $2.3\times10^{-5}$ 	&  0 & $1.6\times10^{-17}$ \\ 
	$[0.975,1.025]$ &  $1.3\times10^{-2}$ 	&  0 & $1.3\times10^{-17}$ \\ \hline 
     \end{tabular}
    
   \end{center}
\caption{\footnotesize{Number of background events expected after 116 days, compared to the number of observed events from the 85 \% of data. The Feldman-Cousins 90~\% C.L. flux limit is also reported.}}
\end{tiny}
\end{table}

In the flux upper limit calculation, systematic uncertainties were considered. They mainly depend on uncertainties on the detector efficiency. The major contributions to the uncertainties are the OM angular acceptance which contributes to 15~\% and the uncertainty in the light absorption length in water which gives an uncertainty of 10~\%~\cite{ref:Antares5line}. To estimate the effect of the loss of detection efficiency, 18~\% of hits per event were removed randomly in Monte Carlo monopole simulation. This leads to a deterioration of the upper limit of 3~\% above the Cherenkov threshold and between 30~\% and 7~\% ffor velocities below. 

The flux upper limit for upward going magnetic monopoles is shown in Fig.~\ref{limite} as a function of the monopole velocity, and is compared to previous limits.

\begin{figure}[h!]
   \begin{center}
      \includegraphics[height=1.9in]{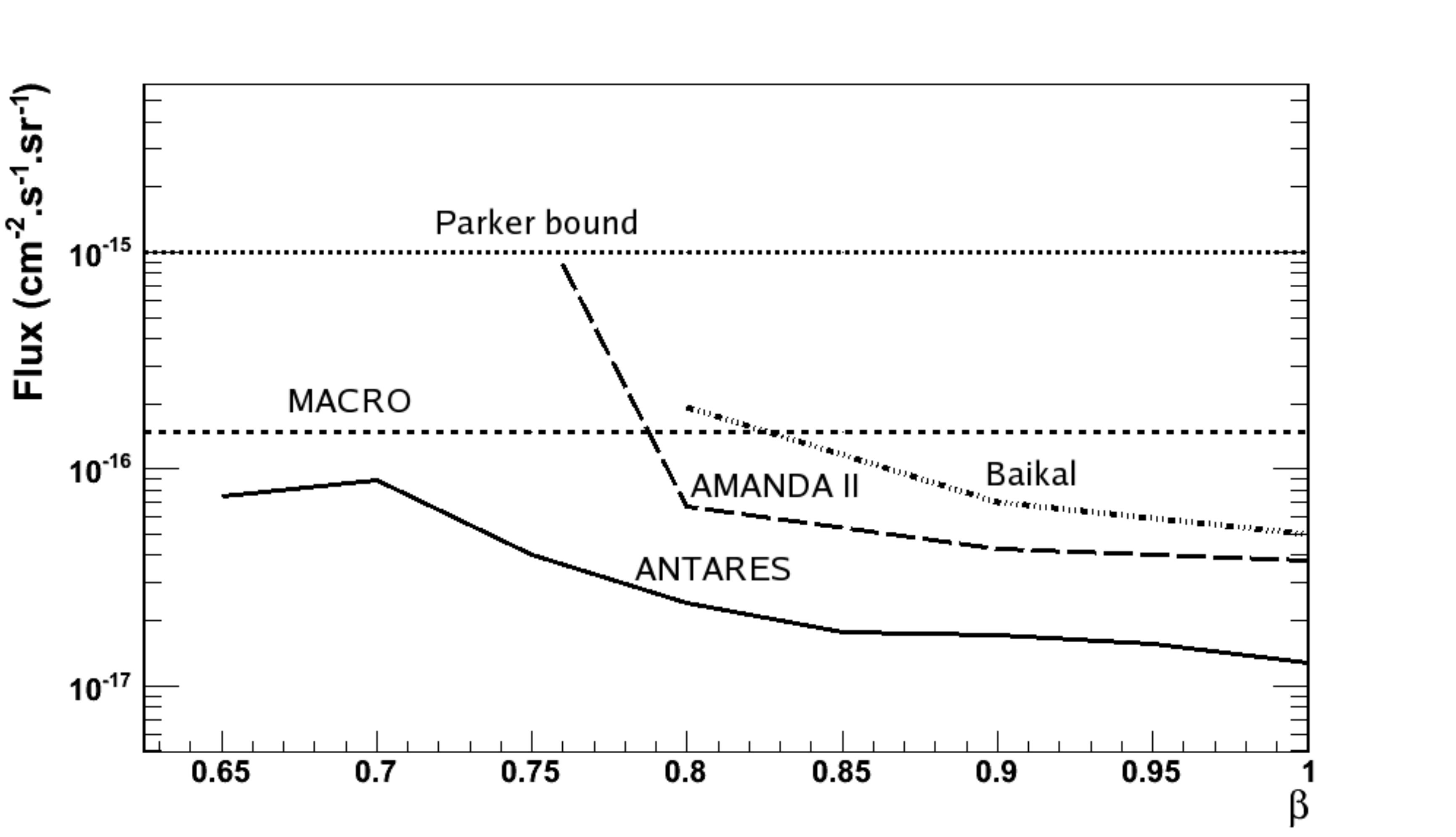}
   \end{center}
   \caption{\footnotesize The ANTARES flux upper limit (90~\% C.L.) on up-going magnetic monopole with $g=g_D$ for relativistic velocities $0.625\leq\beta\leq0.995$ is compared to the published upper limits set by MACRO~\cite{ref:MACRO} for an isotropic flux, and Baikal~\cite{ref:BAIKAL} and AMANDA~\cite{ref:AMANDA} for up-going monopoles, as well as the theoretical Parker bound~\cite{ref:Parker}.}
   \label{limite}
\end{figure}

\section{Conclusion}
The analysis performed on 116 days of data taking in 2008 by the ANTARES detector, for the search for magnetic monopoles yields to new constraints on the upward going magnetic monopole flux.
This new upper limit improves by a factor of about 3 the AMANDA's one over the range $0.75\le\beta\le0.995$ ($\gamma=10$), whereas for the first time for a neutrino telescope, an analysis is sensitive and improves the flux upper limit below the Cherenkov threshold.

\clearpage

\setcounter{figure}{0}
\setcounter{table}{0}
\setcounter{footnote}{0}
\setcounter{section}{0}
\newpage




\title{Nuclearite search with the ANTARES neutrino telescope}

\shorttitle{V. Popa \etal Nuclearites in ANTARES}

\authors{V. Popa$^{1}$, for the ANTARES Collaboration}
\afiliations{$^1$Institute for Space Sciences, Bucharest - M\u{a}gurele, Romania\\ }
\email{vpopa@spacescience.ro}

\maketitle
\begin{abstract}
We discuss the search for down-going nuclearites with the ANTARES neutrino telescope. After a very brief description of ANTARES, we explain the detection mechanism of slowly moving nuclearites. The search strategy, based on a blind analysis is then discussed. The search for nuclearites in data collected in 2007 and 2008 with different ANTARES configurations is ongoing.
\end{abstract}


\section{Introduction}
The ANTARES neutrino telescope was completely deployed in  2008 at a depth of 2475 meters, about 40 km South of Toulon, France. It is composed of 12 vertical lines, bearing a total of 885 optical modules (OMs) distributed in 25 storeys, at depths between 2050 and 2400 meters. A smaller instrumentation line is also present. The OMs, consisting of a glass sphere housing a 10" Hamamatsu photomultiplier (PMT) are arranged in triplets per storey, with the axes looking $45^{\circ}$ downwards. Each line is connected by an electro-optical cable to the main junction box, and then to the shore station \cite{a1}. A secondary junction box is also present, ensuring the connection of various sea science experiments. All data above a certain threshold are sent to shore, where they are filtered by various triggers before being stored \cite{a2}. This is the so called ``all data to shore" strategy.

ANTARES is optimized for the detection of upward going relativistic muons, but is also sensitive to a variety of exotic particles, as fast magnetic monopoles and slow nuclearites \cite{eu,gabi}. In this paper we report on the status of nuclearite search with the ANTARES detector.

\section{Nuclearite properties}

If Strange Quark Matter (SQM) is the ground state of Quantum Chromodynamics (QCD), nuggets of SQM (so called ``nuclearites") have to be present in the penetrating cosmic rays \cite{gl}. They could have been produced in the early Universe, or in violent astrophysical processes as binary strange star collapses \cite{mad}. The only phenomenological bound to the nuclearite flux in the galaxy is derived from the dark matter density \cite{gl}. The nuclearite detection in very large volume neutrino telescopes is possible through the black body emission from their over-heated path in water \cite{gl}. Nuclearites are known also as ``strangelets"; in most cases this denomination refers to nuclear SQM nuggets with mass similar to that of heavy nuclei, ``nuclearites", as used in this paper, refer to much heavier objects ($M \geq  10^{10}$ GeV). The density of SQM is estimated to be a little larger than that of ordinary nuclear matter, $\rho_N \simeq 3.6 \times 10^{14}$ g cm$^{-3}$. According to \cite{gl}, the chemical potential difference between $s$ and $u$ or $d$ quarks should induce a small residual positive charge for the finite size SQM.
Nuclearites moving with typical velocities of gravitationally trapped objects in the Galaxy ($\beta = v/c \simeq 10^{-3}$) should have the residual charge compensated by an electron cloud, or/and by electrons in weak equilibrium inside the SQM \cite{gl}.

\subsection{Nuclearite energy loss in matter}

For slowly moving nuclearites ($\beta \simeq 10^{-3}$) the dominant energy loss is through elastic collisions with the atoms in the traversed medium:
\begin{equation}
\frac{dE}{dx} = -\sigma \rho v^2
\end{equation}
where $\sigma$ is the nuclearite cross section, and $\rho$ the density of the medium. The cross section has the typical atomic value $ \sigma = \pi \times 10^{-16}$ cm$^2$ for nuclearite masses $M \leq 8.4 \times 10^{14}$ GeV, and $\pi(3M/4\pi \rho_N)^{2/3}$ for larger masses. Assuming that outside the atmosphere the velocity of a nuclearite is $\beta = 10^{-3}$, in order to cross the Earth its mass should be larger than about $10^{22}$ GeV \cite{gl}. As we expect that the nuclearite flux in the cosmic rays decreases with increasing mass (as for heavy nuclei), we concentrate for this analysis only on downgoing nuclearites.

\subsection{Principle of nuclearite detection in ANTARES}

The energy release through elastic collisions (Eq. 1) results in the over-heating of the nuclearite track in matter. In the case of  water, it is estimated that a fraction $\eta \simeq 3 \times 10^{-5}$ of the energy loss is dissipated as visible black body radiation emitted by the expanding cylindrical shock wave \cite{gl}. The number of visible photons emitted per unit path length may be estimated as:
\begin{equation}
\frac{dN_\gamma}{dx} = \eta \frac{dE/dx}{\langle E_\gamma \rangle}
\end{equation}
where $\langle E_\gamma \rangle \simeq \pi$ eV is the average energy of visible photons.

As a simple exemplification Fig. \ref{lum} shows the light yield at the ANTARES depth assuming a vertically down-going nuclearite, with the initial velocity (above the Earth atmosphere) $\beta = 10^{-3}$, as a function of its mass. The curves correspond (from up to down) to the light yield at the level of the upper, middle and respectively lower ANTARES storeys.
\begin{figure}[!t]
  \centering
    \includegraphics[width=3.in]{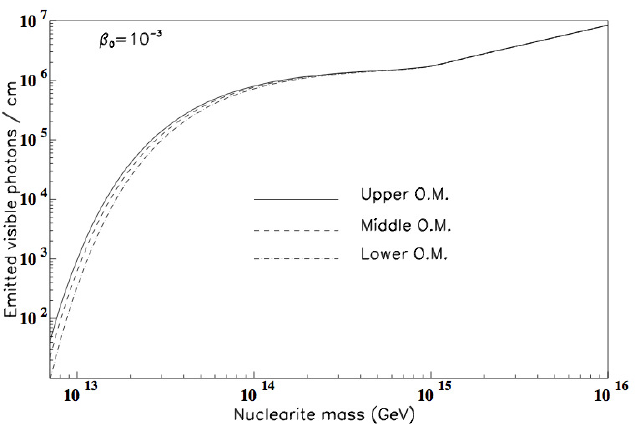}
  \caption{Light yield from a vertically down-going nuclearite at the level of the upper, middle and lower ANTARES storeys.}

  \label{lum}
 \end{figure}
Nuclearites with masses larger than few $10^{13}$ GeV produce enough light to be detectable with the ANTARES telescope. A complete Monte Carlo simulation describing the light production along nuclearite paths, considering initial isotropic arrival direction from the upper hemisphere and distances from the detector axis expanding to 5 absorbtion lengths in water was developed and used in the analysis described bellow.

\section{Nuclearite search strategy in ANTARES}

 The basic information in ANTARES is the ``hit", the time and charge (amplitude) information of a photon detected by a PMT. If the charge is over a pre-defined threshold, the hit is called ``L0" hit and buffered. A local coincidence (``L1" hit) is defined as two L0 hits in the same storey within 20 ns, or a single hit with large amplitude (3 photoelectrons (pe) or more, collected by a single PMT). The ``directional trigger" (DT) requires at least 5 L1  casually connected hits anywhere in the detector, within a 2.2 $\mu$s window.  A second trigger, implemented in a later phase, is the ``cluster trigger" (CT). It is based on the ``T3" cluster, a combination of two L1 hits in adjacent or next-to-adjacent storeys, and requires two of such clusters in a  2.2 $\mu$s window.
All PMT pulses in a 2.2 $\mu$s time window are conserved in a buffer.
If a trigger occurs, all hits with charges above a given threshold are saved for the actual time window, together with the previous buffered information. If, in the next time window another trigger fires, the data are merged together, so the minimum duration of an event ``snapshot" is 4.4 $\mu$s.

The data recorded in 2007 and 2008 were obtained during ANTARES completion, so they refer to various detector configurations. The variations in bioluminescence imposed also different L0 threshold values. In order to maximize the analysis efficiency, the ANTARES approach is to start with a ``blind analysis". This consists in defining the search strategy using Monte Carlo simulations, and validating them on  a small fraction (15\%) of the available data.

The search for nuclearites in ANTARES is based on the fact that the duration of a nuclearite event should be much longer (about 1 ms) than in the case of a muon. Furthermore, the luminosity produced by the passage of a nuclearite is expected to be orders of magnitude larger than from muons. The isotropic light emitted along the heated path should induce a long succession of fake muon signals, so we can use the same triggers as those implemented for relativistic ionizing particles. Monte Carlo simulations have been performed for different nuclearite masses, in all ANTARES configurations during 2007 and 2008. The background level was extracted from real runs. The simulated events were processed using the DT and CT (for the time period in which they were active) triggers. Triggers select only the hits compatible with the passage of a muon, so nuclearite events may expand on multiple adjacent snapshots. Downwards going muons are the main source of background for nuclearite searches, so we investigated the snapshot duration $dt$ for both simulated muons and nuclearites, defined as the time difference between the last and the first L1 hits that produce a trigger. Muons were simulated using the MUPAGE code \cite{mu}.
\begin{figure}[!t]
  \centering
  \includegraphics[width=3in]{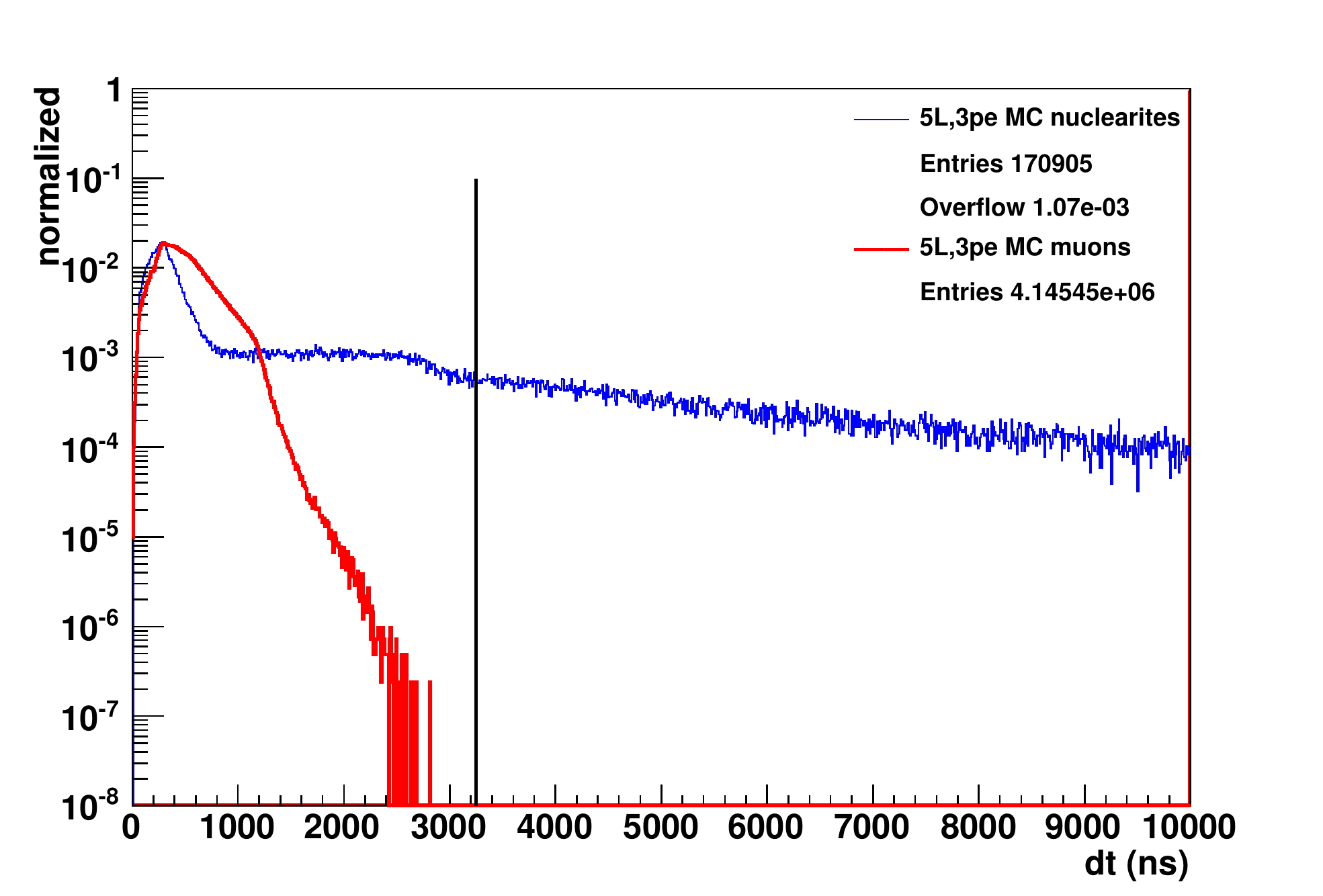}
  \caption{Snapshot duration distributions for simulated nuclearites and muons, broader and narrower curves respectively. The distributions are normalized to unity, and were obtained for a 5 line ANTARES configuration, DT only and a 3 pe threshold. The vertical line marks the optimal C1 cut for this configuration.}

  \label{cut}
 \end{figure}
Fig. \ref{cut} shows a comparison between the distributions of snapshot durations for simulated nuclearites and muons. This result corresponds to a 5 line ANTARES configuration, DT trigger and a 3 pe L0 threshold.  The vertical line (at a snapshot duration of 3250 ns) represents the optimized cut for this configuration. All configurations were analyzed in a similar way, and optimal cuts (ranging between 2500 and 4750 ns) were defined as to minimize the upper flux limits obtainable if no signal is found. In the following we refer to those first level selection, based only on the snapshot duration, as ``C1" cuts.
We tested the C1 cuts on 15\% of real data, selected to reproduce the proportion of different experimental configurations.

For the few events passing the C1 cut, the evolution in time of the event charge barycenter was studied.
 As the light emitted by nuclearites is isotropic, the charge barycenter offers an estimation of the position of the source at a certain moment. All events were consistent with static light sources, so they were interpreted as bioluminescence bursts. Characteristic to those events is that they were all single-snapshot events. In order to reduce the bioluminescence background, we introduced a second level cut C2, to be applied only to single-snapshot events, requiring an event duration twice the C1 cut value. With this second cut, no event in the 15\% data sample survived.

The efficiencies of the cuts for different nuclearite masses were computed and are presented in Table 1. They are defined as the fraction of the nuclearite events detectable after filtering the data.
\begin{table*}
 \label{table_effnucl}
 \centering
  \begin{tabular}{|c|c|c|c|c|c|}
  \hline
Detector  & 3 $\times 10^{13}$ GeV & $10^{14}$ GeV & $10^{15}$ GeV  & $10^{16}$ GeV & $10^{17}$ GeV\\
      configuration          &  eff. (\%)&  eff. (\%) & eff. (\%) & eff. (\%) & eff. (\%)\\
   \hline
  5L, DT,10 pe  & - & 5.4   & 65.6  & 80.1 & 88.6 \\
 5L, DT, 3 pe  & - &15.9   & 73.0 & 82.0 & 87.2 \\ \cline{1-6}
 10L, DT, 10 pe & - & 5.0   & 66.5 & 83.2 & 86.2 \\
 10L, DT, 3 pe & - & 18.8  &  75.5 & 80.2 & 91.4 \\
 10L, DT, CT, 3 pe & 3.0 &  70.3  &  83.8 & 84.0 & 92.6 \\  \cline{1-6}
  9L, DT, CT, 3 pe  & 4.6 & 70.0  &  83.2 & 83.7 & 93.5 \\
  9L, DT, CT, 10 pe &1.7 & 67.7   &  82.8 & 83.9 & 93.9 \\  \cline{1-6}
  12L, DT, CT, 10 pe & 5.4 & 68.5  &  82.4 & 82.6 & 94.0  \\
  12L, DT, CT, 3 pe & 2.1 & 70.4  &  83.0 &83.4 & 92.5  \\
    \hline
  \end{tabular}
  \label{eff}
  \caption{Efficiency of our analysis for different nuclearite mass values and various ANTARES configurations, used in 2007 and 2008.}
  \end{table*}

In terms of nuclearite mass, the threshold of our search (defined as the minimum mass of a nuclearite entering the Earth atmosphere with $\beta = 10^{-3}$) is about $3 \times 10^{13}$ GeV. This is in agreement with the results presented in Fig. \ref{lum}.

\section{Results}

After unblinding,
we performed the analysis of the experimental data collected during the 2007 and 2008 runs.
The analysis followed the pre-defined strategy: data are filtered applying the corresponding C1 cuts,
or the C2 cuts for the single-snapshot events.
No residual contamination was found from downward going atmospheric muons after this selection.

  Very few events survived our cuts, but they are not compatible with the hypothesis of slowly downgoing particles; they  might be safely interpreted as due to bioluminescence. Their topology was investigated  using the event charge barycenter versus time correlation. This approach is equivalent to a fast but less accurate geometrical reconstruction of the event.
 Consequently, we could compute a 90\% confidence level upper limit for a flux of downward going nuclearites, from the ANTARES 2007 and 2008 data. Our result is presented in Fig. \ref{lim}, compared with the MACRO limit \cite{macro}, as updated in \cite{laura}, and with the SLIM limit \cite{slim}. All limits in Fig. \ref{lim} are obtained in the same hypothesis: nuclearites are supposed to enter the Earth atmosphere with a velocity of 300 km/s, and refer only to downgoing nuclearites. In order to traverse the Earth, such slow moving objects should have masses larger than some $10^{22}$ GeV \cite{gl}.

\begin{figure}[!t]
  \centering

    \includegraphics[width=3.in]{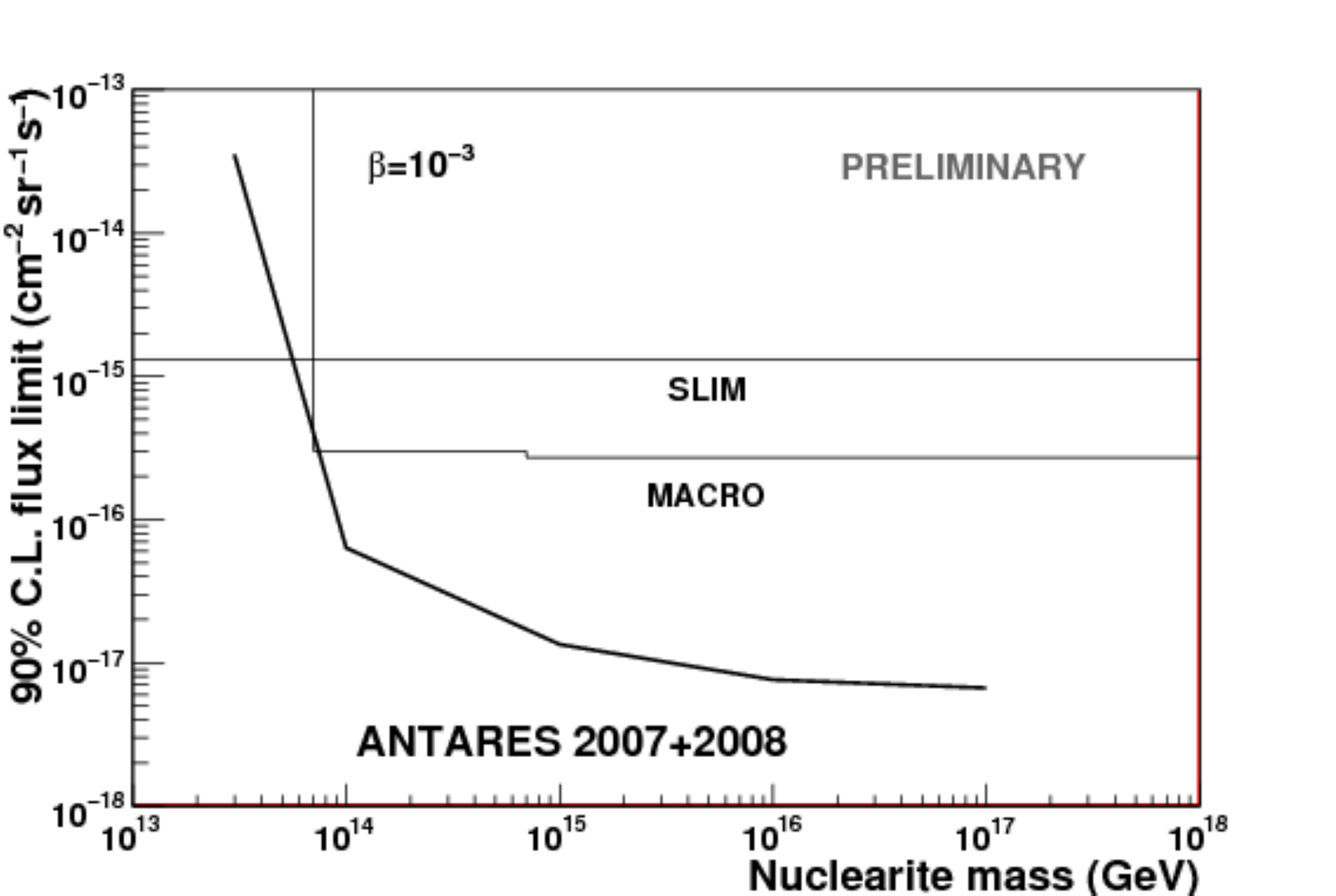}
  \caption{The ANTARES 90\% CL upper limit for a flux of downgoing nuclearites, obtained from the 2007 and 2008 data. The final limits reported by MACRO and SLIM are also shown.}

  \label{lim}
 \end{figure}

We recall that MACRO was an underground detector operated in the National Gran Sasso Laboratories of INFN, and SLIM was a passive detector operated at high altitude, at the Chacaltaya Cosmic Ray Laboratory.

The shape of our limit reflects the narrowing around the vertical direction of our solid angle acceptance as the nuclearite mass becomes smaller. The ANTARES preliminary limit improves significantly the MACRO result in all the common mass range; the SLIM limit remains better for nuclearite masses lower than about $5 \times 10^{-3}$ GeV.

\section{Conclusions}

We have presented the strategy to search for downward going nuclearites as well as the preliminary results obtained from the 2007 - 2008 data analysis with the ANTARES neutrino telescope. The results discussed were obtained after unblinding the data corresponding to the detector configurations during the 2007 and 2008 data tacking. No candidate event survived all criteria; the 90\% C.L. ANTARES upper flux limit improves significantly the MACRO result \cite{macro,laura}. In the lowest mass region, the SLIM limit \cite{slim} still remains the most stringent published result.

The same analysis procedure will be applied for the ANTARES data recorded after the completion of the detector (2009 and later on).

\vspace{\baselineskip}

{\it Acknowledgements.}  This work was partially supported (for the Romanian group) by CNCSIS Grant 539/2009.

\vspace{\baselineskip}

\clearpage

\setcounter{figure}{0}
\setcounter{table}{0}
\setcounter{footnote}{0}
\setcounter{section}{0}
\newpage




\title{Studying Cosmic Ray Composition around the knee region with the ANTARES Telescope}

\shorttitle{Ching-Cheng Hsu \etal Studying Cosmic Ray Composition around the knee region with the ANTARES Telescope}

\authors{Ching-Cheng Hsu $^{1}$, on behalf of the ANTARES Collaboration }
\afiliations{$^1$NIKHEF\\ }
\email{cchsu@nikhef.nl}

\maketitle
\begin{abstract}The composition of the cosmic rays in the \textit{knee} region ($\approx$ $10^{4}$ TeV/nucleus) of the all particle spectrum is considered to be the result of the particle acceleration and propagation from the astrophysical sources.  The steeply falling cosmic ray spectrum makes a direct measurement of the composition difficult, but it can be inferred from the measurements of the showers generated by the interaction of the primary cosmic ray with the Earth atmosphere. In particular the characteristics of the muon bundles produced in the showers depend on the primary cosmic ray nature.
The ANTARES telescope is situated 2.5 km under the Mediterranean Sea off the coast of Toulon, France. It is taking data in its complete configuration since 2008 with nearly 900 photomultipliers installed on 12 lines. The trigger rate is a few Hz dominated by atmospheric muons.  A method using a multiple layered neural network as a classifier was developed to estimate the relative contribution of proton and iron showers from the energy and multiplicity distribution of the muon tracks reaching the ANTARES detector. The performance of the method estimated from simulation will be discussed. 
\end{abstract}


\section{Introduction}
Although the main goal of ANTARES telescope is to look for high energy neutrinos coming from the deep space, it also provides us opportunities to study cosmic ray physics. One of the most important topics is to distinguish the different chemical compositions around knee region of its spectrum. The cosmic ray spectrum is known as a power law with power index about -2.7 up to few PeV. Then the slope changes to -3.1 until the energy around 4$\times10^{18}$ eV~\cite{CR}. The \textit{knee} was first observed by the MSU group in 1970s, then has been confirmed by many groups afterwards. The origin of the \textit{knee} could be generally summarized into either astrophysical origin or particle physics origin. Nevertheless it is still a puzzle and generally believed to be a key issue to the problem of the origin of galactic cosmic rays. \\ 
	Most people attributed the knee to the sudden reduction in Galactic trapping efficiency. A popular explanation is that the knee is associated with an upper limit of acceleration energy by galactic supernovae. Another popular scenario is the leakage of particles from the Galaxy, since the Larmor radius of a proton in the galactic magnetic field increases with its energy and finally exceeds the thickness of the galactic disk. Additionally, there is a minority of theorists who proposed that the knee is due to a single, recent and local supernova remnant (SNR) or a rapidly rotating pulsar interacting with radiation from its parent SNR\cite{source}.    \\
	If the knee is caused by the maximum energy attained during the acceleration process or it is due to leakage from the Galaxy, the energy spectra for individual elements with charge Z would exhibit a cut-off at an energy $E^{Z}_{c}$ = Z $\times$ $E^{p}_{c}$, where $E^{p}_{c}$ is the cut-off energy of protons. The sum of the flux of all elements with their individual cut-off makes up the all-particle spectrum. In this picture the knee is related to the proton cut-off  and the steeper spectrum above the knee is a consequence of the subsequent cut-off of heavier elements, resulting in a relatively smooth spectrum above the knee~\cite{Horandel}. 


\section{Composition Model} 

In the so-called polygonato model, the general form for the flux of primary nuclei of charge Z and energy $E_{0}$ is 
\begin{eqnarray*}
\frac{d\phi_{Z}}{dE_{0}} = \phi_{Z}^{0} [ 1 + (\frac{E_{0}}{E_{trans}})^{\epsilon_{c}}]^{\frac{-\Delta \gamma}{\epsilon_{c}}}
\end{eqnarray*}

where the transition energy $E_{trans}$  could be determined according to three different scenarios.
The parameter $\epsilon_{c}$ determines the smoothness of the transition, and $\gamma_{c}$ is the hypothetical slope beyond the knee. ${E_{trans}}$  is the cut-off energy. 
Under the polygonato model, three different scenarios are proposed : 
\begin{itemize} 
\item Rigidity dependent: from the astrophysical point of view, a rigidity dependent cut-off $E_{trans}$ = $\hat{Ep}$ $\cdot$ Z is the most likely description if we take into account the acceleration and propagation of the cosmic rays. 
\item Mass dependent: this model predicts that the change in the power law index depends on the mass $E_{trans}$ = $\hat{Ep}$ $\cdot$ A instead of the charge. This scenario leads to a steeper energy spectrum after the cut-off. However, the sharp cut-off would be hard to explain on astrophysical reasons. Maybe a nearby source or a new type of interaction in the atmosphere could yield such cut-off~\cite{Horandel}. 
\item Constant composition: $E_{trans}$ = $\hat{Ep}$. The knee is explained by a common steepening in the energy spectrum; it occurs for all the particles at the same energy.  
\end{itemize} 

So far, the best measurement of the compositions around the knee region was done by KASCADE~\cite{kascade}. The measured primary energy spectra show that the knee in the all particle spectrum is due to a steepening of the light elements spectra.    

\section{Analysis Strategies} 

	The ANTARES detector is located at 40 km off the coast of Toulon, France, at a depth of 2475 m in the Mediterranean Sea. It consists of 12 flexible strings, each with a total height of 450 m, separated by about 60 m. They are anchored to the sea bed and kept near vertical by buoys at the top of the strings. Each string carries a total of 75 10-inch Hamamatsu photo multipliers (PMTs) housed in glass spheres, the so-called optical modules (OMs)~\cite{ANTARES}. The OMs are arranged in 25 storeys (three optical modules per storey) separated by 14.5 m. The detector started taking data in 2007  and was fully completed in 2008. \\ 
	Since ANTARES is deeply buried under the sea, only the muon components from the air shower will survive at detector level. The muons will emit Cherenkov radiation only when passing through the sea water, which can be detected using photomultiplier tubes. To be registered by the ANTARES detector, muons have to travel at least 2.5 km of sea water and still be energetic enough to trigger the detector. At large zenith angles, the threshold increases because of the increasing depth of the sea water. The muon bundles properties (such as multiplicity) are strongly related to the primary energy and species of the nuclei. However, the ANTARES detector cannot resolve individual muons from a muon bundle. Hopefully, the topology of the hit distributions in space and time could give pertinent information about the properties of the muons in the bundles. Two useful analysis methods are combined and then applied on MC samples. The first one is a cluster finding algorithm, and the second one is an electromagnetic shower searching algorithm\cite{savi}.  \\
	In both analysis methods, we rotate our software coordinate system such that the z axis is along the reconstructed muon track. We define the plane which is passing through the detector center and perpendicular to the reconstructed track as \textit{detection plane}. Fig.~\ref{pattern} shows the snapshot of muon bundle patterns on the detector plane from the same primary energy of proton and iron nuclei.  The projections of all the hits positions on the detection  plane are also calculated, including the time correction information.    

\begin{figure}[!t]
\vspace{5mm}
\centering
\includegraphics[width=3.in]{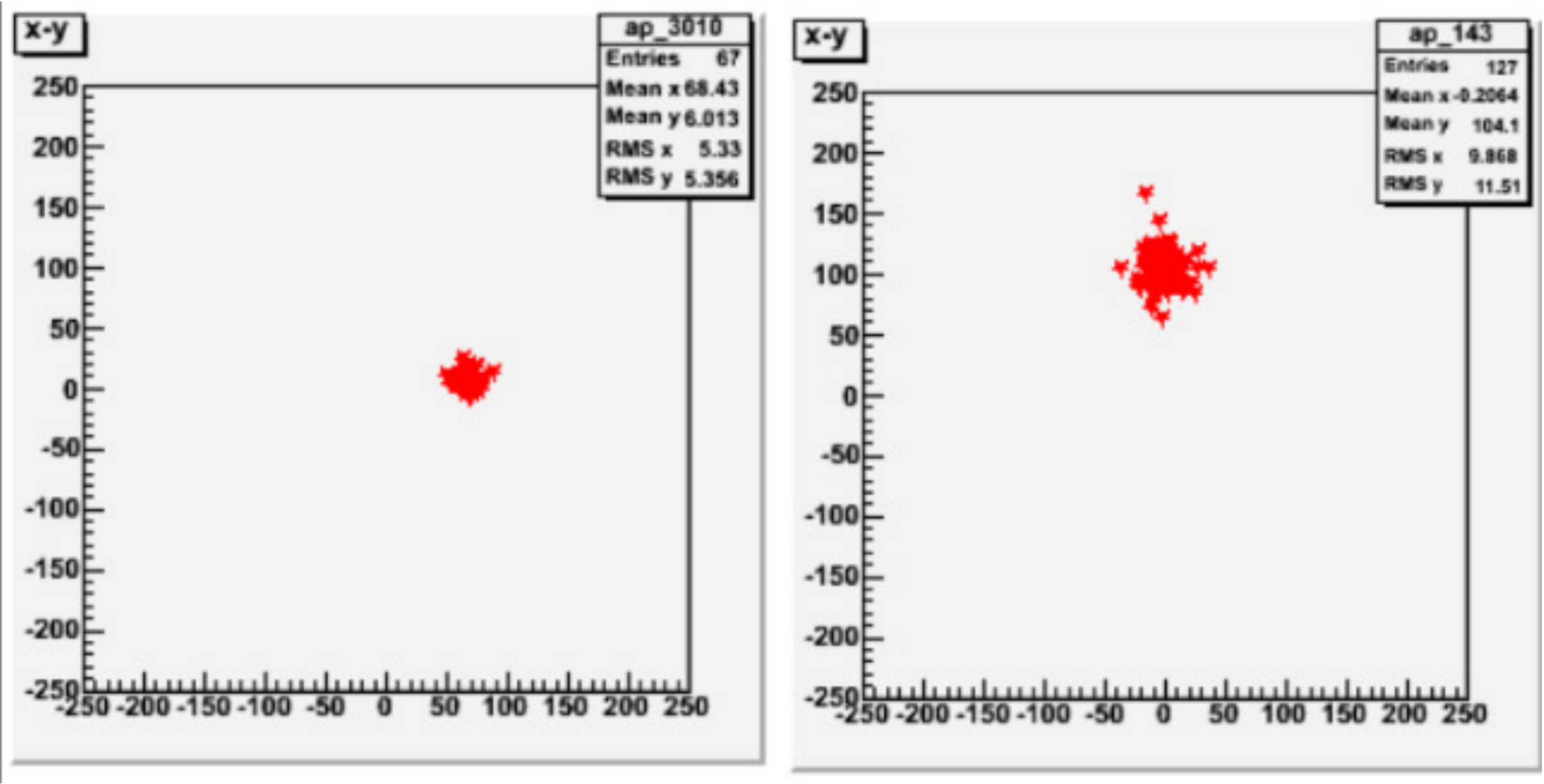}
\caption{Typical space distributions of muon bundles on the detector plane. The muon bunldes are from primary protons (left) and iron nuclei (right) with the same energy.}  
\label{pattern}
\end{figure}

In the cluster finding algorithm, the clusters are formed if the hits within the cluster fulfill the following three conditions:
\begin{itemize}
\item  $\mid$ $T_{i}$ - $T_{j}$  $\mid$ $\le$ $\mid$ $r_{i}$ - $r_{j}$$\mid$ /C/$n_{g}$  where $T_{i}$ and $T_{j}$ are the times for any two hit pairs in the hit cluster.  $r_{i}$ and $r_{j}$ are the associated positions of these two hits. $n_{g}$ is the index of refraction in the sea water and C is the speed of light. 
\item Any two hits within the cluster should be in the same or neighboring strings or floors. 
\item $\mid$  $T_{i}$ - $T_{j}$  $\mid$ $\le$ $\mid$($z_{i}$ - $z_{j}$)$\mid$/C + d $\times$ $tan\theta_{c}$ /c + $T_{ext}$, where $T_{ext}$ are maximum extra time, here we set 20 ns; $T_{i}$ and $T_{j}$ are again time informations of two hits.  $z_{i}$ and $z_{j}$ are the rotated Z positions of the two hits. $\theta_{c}$ is the Cherenkov angle in the sea water.     
\end{itemize} 

In order to quantify the cluster patterns from cosmic showers initiated by different groups of elements, we try to parametrize the hit patterns. There are two kinds of parameters in our analysis: 
\begin{itemize}
\item Cluster-wise parameters: the parameters which are related to each cluster. For instances, the "size" (total number of photoelectrons in one cluster).
\item Event-wise parameters: the parameters which are related to each event, such as the total number of clusters $N_{cluster}$.
\end{itemize} 
In total, we have about 25 event-wise and 12 cluster-wise parameters.  One example of pattern after parametrization superimposed with reconstruction position of the muon tracks (bundles) is presented in Fig~\ref{parameter}.  

\begin{figure}[!t]
\vspace{5mm}
\centering
\includegraphics[width=2.6in]{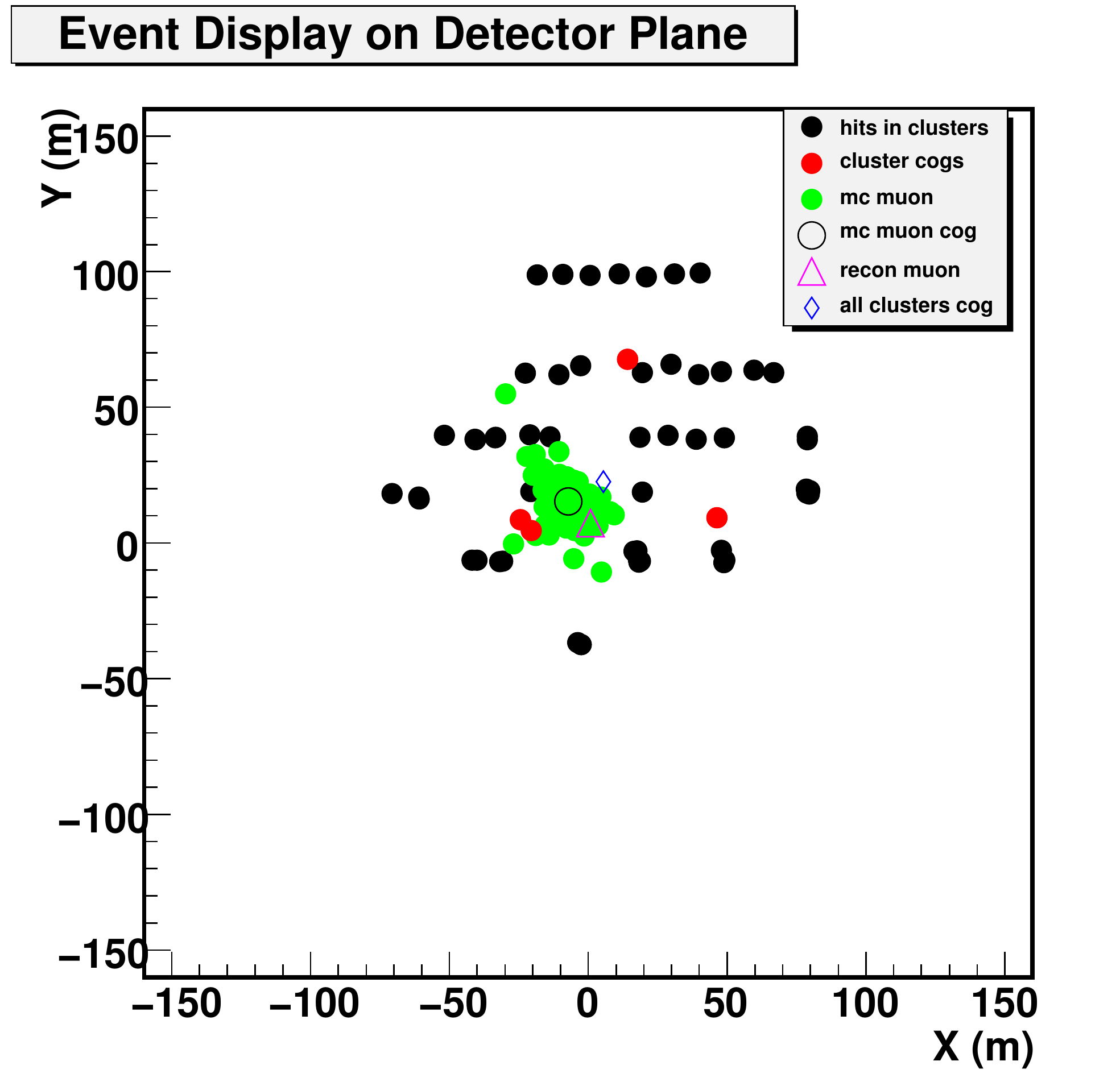}
\caption{The hit patterns projection on the detector plane and then parametrized by cluster algorithm. The black points are projection of hits. The red circle is the center of gravity (COG) of the four clusters. Green points are the positions of the muon in the bundles. Empty circle is the COG of the muons. The pink triangle is the reconstruced muon position. The blue diamond is COG of all the clusters. }  
\label{parameter}
\end{figure}

On the other hand, the electromagnetic shower searching method gives us additional information about the muon bundles. The high energy muons suffer from catastrophic energy losses. Once it happens, electromagnetic showers are initiated either by $\gamma$ or $e^{+}$ and $e^{-}$ pairs. 
Instead of projecting hits information on the detector plane, we project all the hits on the reconstructed axis of the event and search for the peaks using the TSpectrum function in the ROOT package. The details of the algorithm can be found in~\cite{savi}. Two more parameters are obtained by this method. They are the number of showers ($N_{shower}$) and the amplitude of the baseline ($N_{baseline}$) from the algorithm.   
   
\begin{figure*}[!t]
\centerline{\includegraphics[width=2.1in]{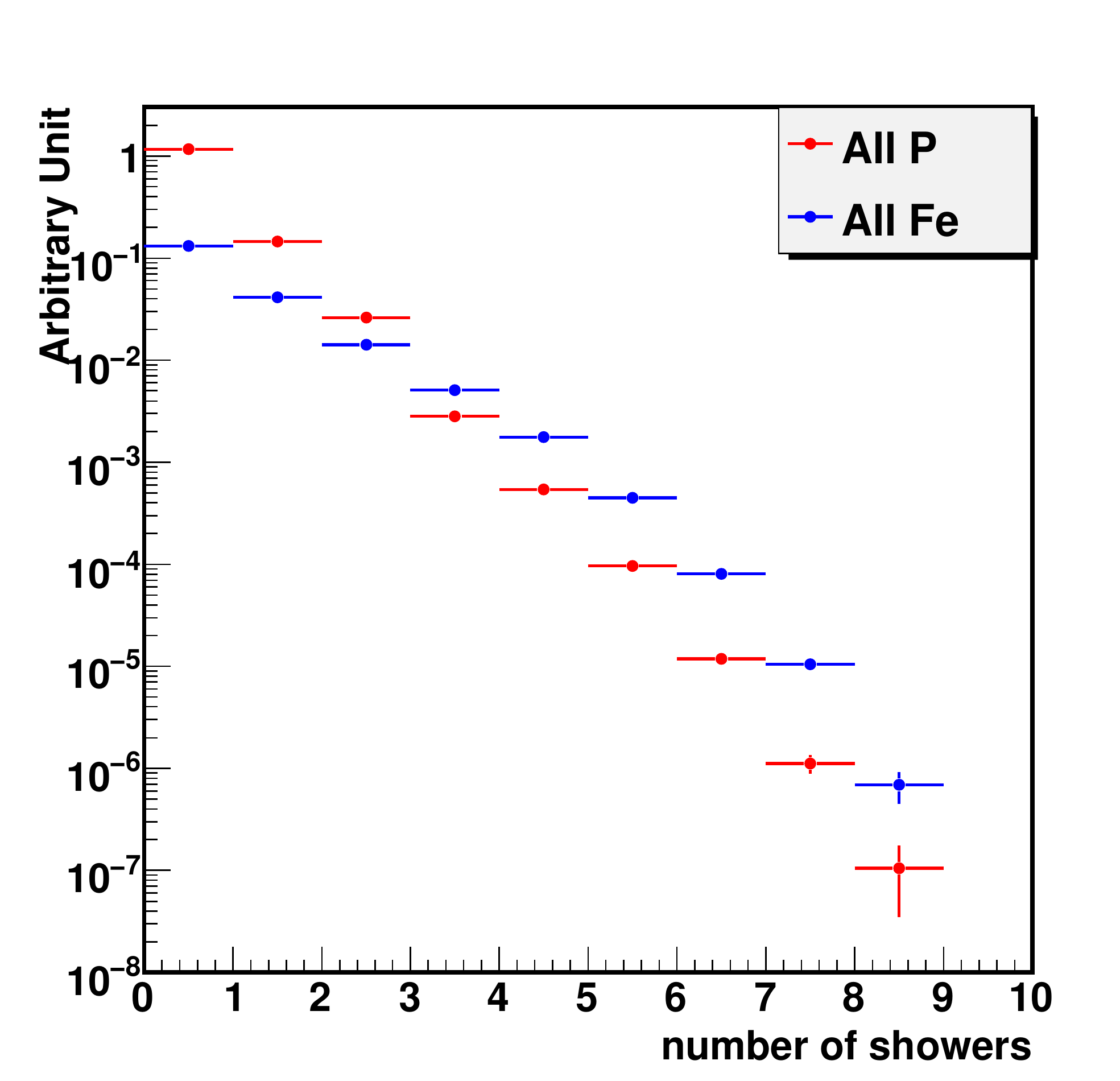}\label{p_dis}
 \hfil
 \includegraphics[width=2.1in]{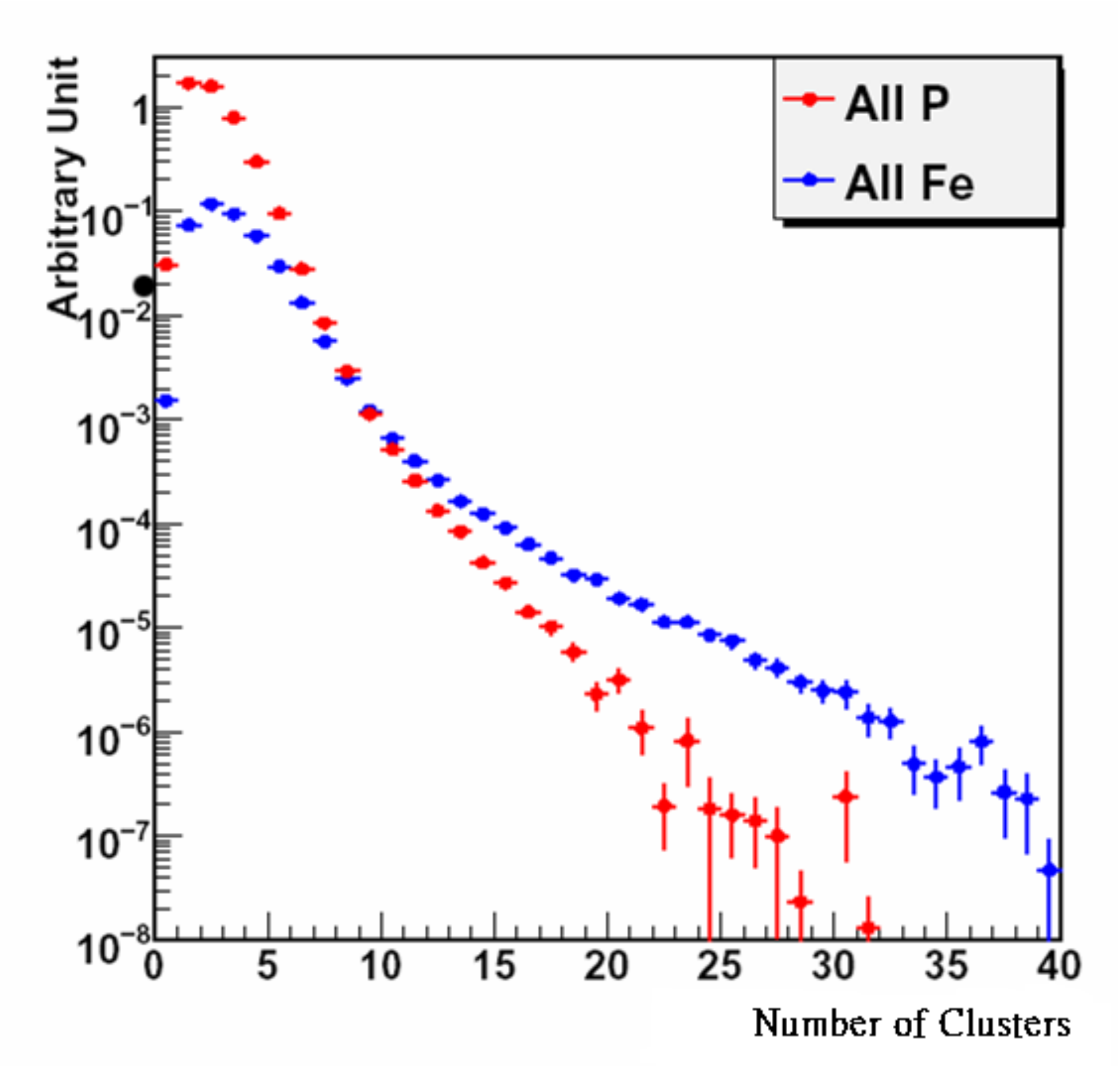} \label{p_dis}  }
\caption{ The relative distribution of $N_{shower}$ (left plot) and $N_{cluster}$ (right plot) parameters after the reconstruction quality cuts assuming that all the particles from cosmic rays are proton (red points) and iron (blue points), respectively. }
\label{p_dis}
\end{figure*}

	Combining the cluster-finding and shower-finding algorithms, we have in total 37 parameters. Each parameter has a different discrimination power for the chemical species. To achieve multi-dimensional comparisons, we use the existing package "TMVA" (Toolkit for Multivariate Data Analysis with ROOT ) for this analysis~\cite{TMVA}. TMVA is a toolkit which hosts a large variety of multivariate classification algorithms. Training, testing, performance evaluation and application of all available classifiers is carried out simultaneously. \\  
	Several methods are implemented inside TMVA packages. In order to cross check the results, four different methods were chosen for this analysis. They are Multilayer Perception (MLP), (MLPBNN), (TMlpANN)  and k-Nearest Neighbour (k-NN). The performances of multi-variate analysis are sensitive to the correlation of the training parameters. Thus, it is necessary to reduce the un-important parameters. We input each parameter into the analysis individually and then we kept only the five with the largest discriminating power. These five parameters are $N_{hit}$, $N_{cluster}$ and $N_{npe}$, representing number of hits, clusters, and photoelectrons from cluster-finding algorithm. $N_{shower}$ and $N_{base}$ are the numbers of showers and the number of photoelectrons of the baseline from EM shower-finding algorithm.

\section{Analysis on MC samples}

	A full MC simulation was adopted in this analysis. The air showers induced by the primary nuclei with energy ranging from 1 to $10^{5}$ TeV /nucleon and zenith angle between $0^{\circ}$ and $85^{\circ}$ are simulated using the CORSIKA (Version 6.2)~\cite{CORSIKA} and the hadronic interaction model QGSJET.01c. All muons reaching the sea level, with energies larger than the threshold energy, are propagated through sea water to the detector. At last, muons are transported through the ANTARES sensitive volume, Cherenkov light is produced and the detector response is simulated. Background noises were added afterwards. The standard ANTARES trigger is simulated.  The muon direction and position are reconstructed using a multi-stage fitting procedure, which basically maximizes the likelihood of the observed hit times as a function of the muon direction and position\cite{aafit}.   \\ 
	We further divide the MC samples into 3 independent subsets for training, testing and evaluating the TMVA analysis. The relative sizes of the three subsets are 40\%, 30\% and 30\%. Each event is weighted according to different models of CR composition. The iron component is tagged as signal whereas the protons are tagged as background. The "pure" signal distributions are obtained, if we assume that all the coming cosmic rays are iron. On the other hand, the background samples come from assuming all cosmic rays are protons. The distributions of $N_{shower}$ and $N_{cluster}$ assuming the all particle spectrum are, respectively, proton and iron, are shown in Fig~\ref{p_dis}. \\ 
	 We carefully checked the output of the neural network from test samples in order to avoid the over-training effects. 

\section{Results and Conclusions}
The training and test events fed into the neural network were subjected to a series of cuts. The main quality cut is the so-called $\Lambda$ cut~\cite{aafit}. Its purpose is to keep good quality on reconstructed events. For testing the method, we mixed proton and iron components from our pseudo-data set with $\approx$  98.2\%-1.8\%, individually. The output of the neural network training on the proton and iron components is shown on the Fig~\ref{nn_output}. The green curve corresponds to the true proton events, while the red curve corresponds to the true iron events. The blue curve in the plot shows the psedo-data events with the 98.2\%-1.8\% configuration. 
	We adjust proton and iron distibutions to fit the psedo-data distribution ("Template fitting" in roofit). We found that the $\chi^{2}$/ndf is 18.07/21. The fitted numbers from the two-component model are 98.5$\pm$0.7\% for proton. With this method, further calculating the efficiencies of triggers, different cuts and effective areas are necessary in order to find the ratio in the original spectrum.  
	Another approach is to put different weight in each event according to differnt physics models and check the output of different physics models from neural work. An example is shown on the Fig~\ref{models_nn}. From the plot, the differences between three extreme models are clearly seen. In the future, with this method, we are able to compute the compatibilities between different realistic physics models and data based on the neural work.   

	 In summary, we have developed a method to estimate the ratio of the heavy elements in the triggered cosmic ray events based on the information from the muon tracks and electromagnetic showers in ANTARES detector. To combine all the discrimination powers from established multi-parameters needs the help from multi-variate analysis (neural network). 
The analysis of ANTARES real data with the goal of deriving ratio between different groups of elements in the cosmic ray spectra is ongoing and will yield results in the near future.

\begin{figure}[!t]
\vspace{5mm}
\centering
\includegraphics[width=2.8in]{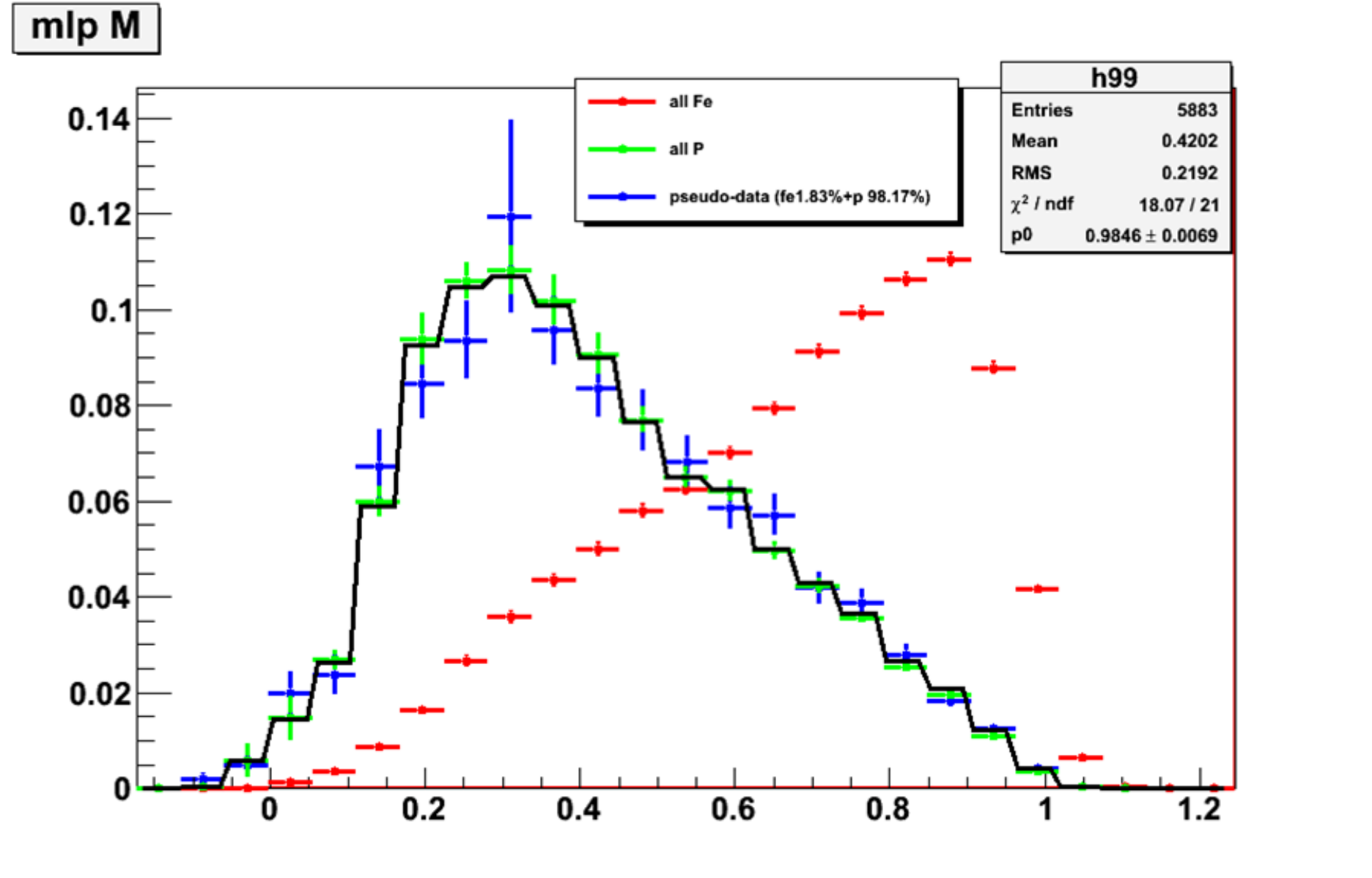}
\caption{Output computed by the neural network in the presence of a mixture of proton and iron components from the pseudo-data sets. The green curve region (left) corresponds to the true proton events, while the red curve region (right) corresponds to the true iron events. The artificial mixture of pseudo-data is in blue. }  
\label{nn_output}
\end{figure}

\begin{figure}[!t]
\vspace{5mm}
\centering
\includegraphics[width=2.5in]{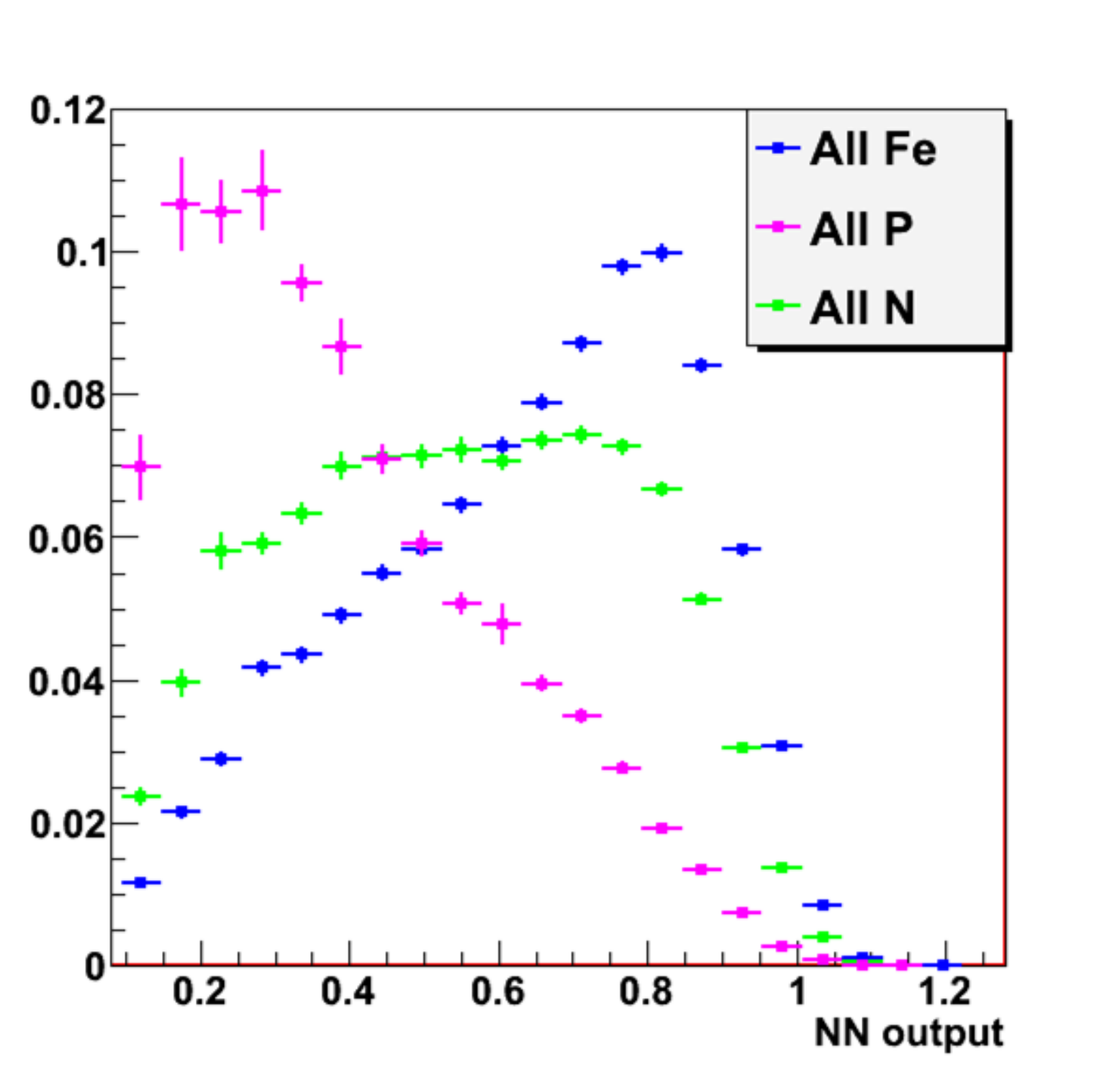}
\caption{Output computed by the neural network according to different cosmic ray knee models. Here we take three extreme composition hyphothesis: All particles are protons, nitrogen and irons. The built-up neural network gives different distributions from these three models. }  
\label{models_nn}
\end{figure}







\clearpage

\setcounter{figure}{0}
\setcounter{table}{0}
\setcounter{footnote}{0}
\setcounter{section}{0}
\newpage




\newcommand{\ant}{\textsc{Antares}}

\title{\ant\ sensitivity to steady cosmic gamma ray sources}

\shorttitle{G.Guillard --- \ant\ \& gamma rays}

\authors{Goulven Guillard$^{1}$, on behalf of the \ant\ Collaboration}
\afiliations{$^1$Clermont Universit\'e, Universit\'e Blaise Pascal,CNRS/IN2P3, Laboratoire de Physique Corpusculaire, BP 10448, F-63000 Clermont-Ferrand, France}
\email{guillard@in2p3.fr}


\maketitle
\begin{abstract}
Amongst the atmospheric muons recorded by neutrino telescopes are muons produced by the interaction of cosmic gamma rays with the Earth's atmosphere.  Although they are not numerous, it has been suggested that such muons could be distinguished by neutrino telescopes from the isotropic background by correlating their direction with known sources of gamma rays.

The \ant\ neutrino telescope is taking data at the bottom of the Mediterranean Sea in its full configuration since May 2008.  Its expected sensitivity to steady gamma ray sources is discussed, as well as the gamma ray induced neutrino contamination of cosmic neutrino signals.  It is shown that the expected signal from steady gamma ray sources is well below the \ant\ detection ability, and that gamma rays are a negligible source of atmospheric neutrinos background.

\end{abstract}



\section{Introduction}

A large number of astrophysical sources are known to emit photons above the TeV range.  When interacting with the Earth's atmosphere, such photons will produce electromagnetic air showers which may contain high energy muons.  Their detection by neutrino telescopes has long been proposed and investigated theoretically (see e.g. references in Ref.~\cite{icrc09}).  Although its location more than 2000\,m depth in the Mediterranean Sea is supposed to suppress most of atmospheric muons, which are considered as background for cosmic neutrino searches, some of these gamma ray induced muons may reach the \ant\ neutrino telescope.  Its configuration being optimized for the detection of upgoing muons~\cite{ant}, their detection is however challenging.

Even though it may not be competitive with atmospheric \v{C}erenkov telescopes and extended air shower arrays, the ability of \ant\ to detect gamma rays and correlate them to their source could be used as a calibration tool and would be an independant way to check the absolute pointing of the detector.

The following is an update of preliminary results presented earlier~\cite{icrc09}.  In particular, more accurate software is used.  Most notably, the treatment of muon pair production from photons has been corrected in the \textsc{Corsika} program~\cite{corsbug}, and the treatment of light dispersion in water as well as the trigger simulation have been improved in the \ant\ simulation tools.

\section{Monte-Carlo simulation}

Three Monte-Carlo productions of gamma rays have been generated in three contiguous energy ranges, following an $E^{-1}$ flux~: $2.5\times10^{9}$ photons between 1 and 100\,TeV, $2.5\times10^{8}$ photons between 100 and 1000\,TeV and $5\times10^{7}$ photons between 1 and 10\,PeV.  Interactions in the atmosphere are processed with the \textsc{Corsika} program (version 6.960)~\cite{corsika}, which makes use of the \textsc{Egs4} code system for the electromagnetic interactions~\cite{egs4}, and the selected hadronic model is \textsc{Qgsjet}~\cite{qgsjet}.  Sources are considered as fixed in the sky, with an azimut angle of 0$^{\circ}$ (North) and a zenith angle of 20 or 60$^{\circ}$ (downward vertical particles having a zenith angle of 0$^{\circ}$).

The propagation of muons in water is performed by a dedicated program which makes use of the \textsc{Geant3} program for the emission of secondary particles~\cite{geant3} and \textsc{Music} tables for the propagation of particles itself~\cite{music}, and which generate \v{C}erenkov photons using photon tables previously generated.

Optical background is added using standard medium quality real data, and hits on the detector are selected using a standard trigger strategy~\cite{ant}.

Muon tracks are reconstructed using a standard likelihood-based reconstruction algorithm relying on the minimization of hit time residuals~\cite{aart}.  A cut is performed on the reconstruction quality estimator so that most events with an angular resolution better than 2$^{\circ}$ are selected while most of badly reconstructed events are rejected.

\section{\ant\ effective area}

Dividing the number of surviving events by the incident flux, one can compute the effective area of the detector as a function of the incident particle energy, which expresses the detector efficiency.  Figure~\ref{fig:effective_area} shows the effective area of the \ant\ telescope to gamma rays at standard trigger level, which represents all remaining events after applying the standard trigger strategy, and at the reconstruction level, representing all reconstructed events surviving the reconstruction quality cut, for both simulated zenith angles.

 \begin{figure}[!t]
  \vspace{5mm}
  \centering
 \includegraphics[width=0.48\textwidth]{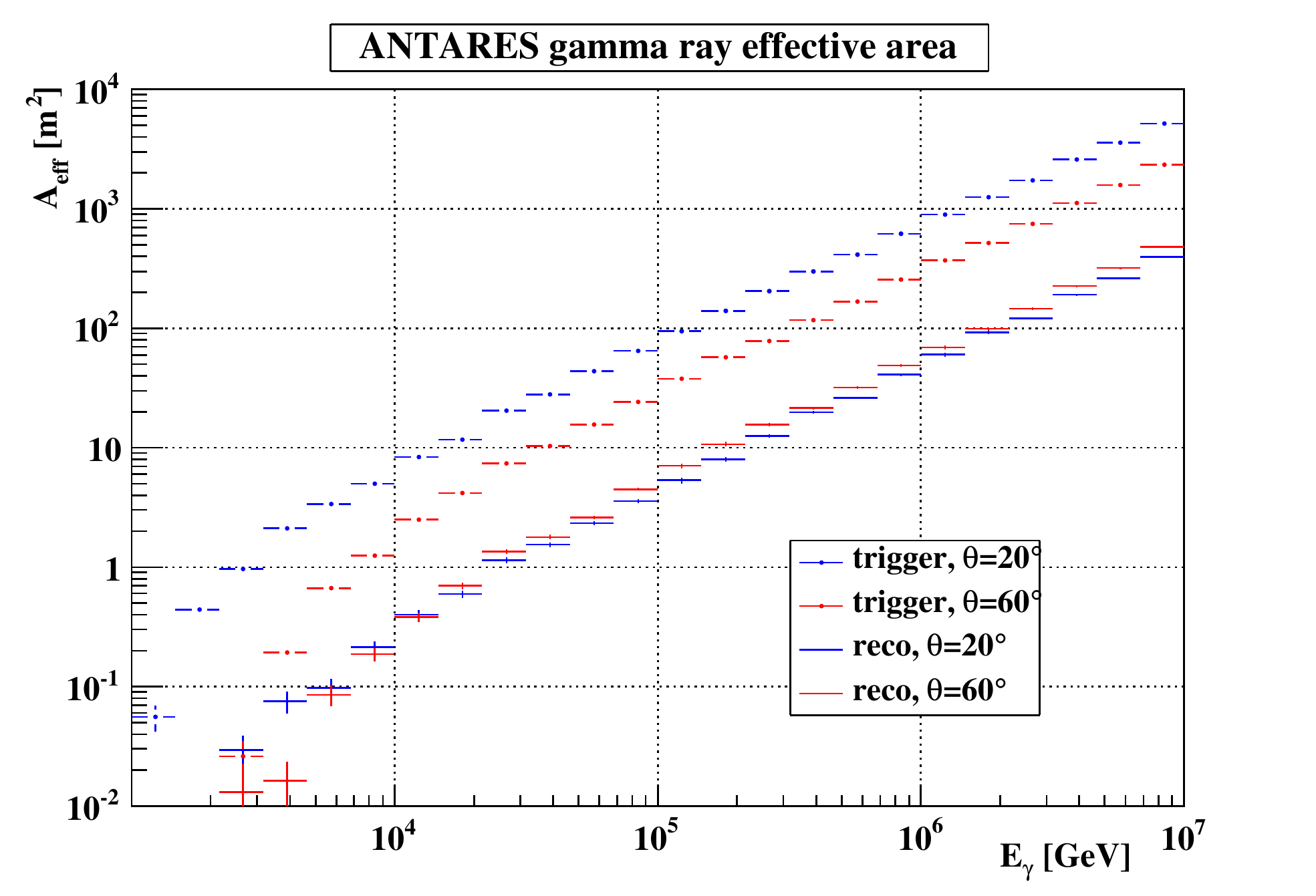}
  \caption{\ant\ gamma ray effective area at trigger and reconstruction level, for two zenith angles.}
  \label{fig:effective_area}
 \end{figure}

Although almost three times lower at trigger level, the effective area for a 60$^{\circ}$ zenith angle exceeds the effective area for a 20$^{\circ}$ zenith angle at reconstruction level.  The reason for this is comes from the granularity of the detector, which is more important in the horizontal plane than in the vertical dimension~: lines are separated by about 60\,m, while storeys on a line are separated from 14.5\,m.  Consequently, although muons are less numerous to reach the detector at 60$^{\circ}$ because of the increased distance to travel in water, their \v{C}erenkov light is detected by more lines than at 20$^{\circ}$.  Their direction can thus be reconstructed more efficiently.  This affects particularly the determination of the azimut angle.

\section{\ant\ sensitivity}

The expected number of events for a few sources of interest, assuming a 100\,\% visibility over one year and a fixed position in the sky, is presented in Table~\ref{tab:nevents}.  The sources are selected both for their visibility from \ant\ (Table~\ref{tab:sources}) and for their strong flux.  The case of a fictional source with a very powerful flux ($dN/dE=1000E^{-2}\,\mathrm{m^{-2}s^{-1}}$) is also considered.  The higher bound extrapolates gamma ray fluxes up to 10\,PeV and is thus quite unlikely since the validity of flux parametrizations is limited to a few tens of TeV at best and since interactions of ultra-high energy photons with photons from the extragalactic background light (EBL) limit their range to galactical distances.  The lower bound is more realistic and does not extrapolate above 100\,TeV.

\begin{table}
\centering
\footnotesize
\begin{tabular}{ccc}\hline
source		& $\mathrm{N_{\mathrm{trig}}}$	& 	$\mathrm{N_{\mathrm{reco}}}$	\\[0.1cm]\hline\hline
Crab		& 4-15			& 	0.1-0.3				\\	
		& \emph{0.6-2.6}	&	\emph{0.1-0.4}			\\\hline
1ES 1959+650	& 0.5-25		& 	0.01-0.5			\\	
		& \emph{0.1-6}		&	\emph{0.01-0.8}			\\\hline
Mkn 421		& 1-25			& 	0.01-1				\\	
		& \emph{0.1-2}		&	\emph{0.01-1}			\\\hline
Mkn 501		& 1-100			& 	0.03-10				\\	
		& \emph{1-50}		&	\emph{0.03-15}			\\\hline
fictional	& 820-1860		&	22-68			\\
		& \emph{226-660}	&	\emph{29-92}			\\\hline
\end{tabular}
\caption{Expected number of gammar ay events seen by \ant\ at trigger level ($\mathrm{N_{\mathrm{trig}}}$) and after reconstruction  ($\mathrm{N_{\mathrm{reco}}}$, with a cut on the reconstruction quality and reconstructed within 2$^{\circ}$ from source), for one year assuming a 100\,\% visibility, for a 20$^{\circ}$ zenith angle (straight font) and a 60$^{\circ}$ zenith angle (italic).  The lower bound assumes a cutoff at 100\,TeV while the upper bound assumes events as energetic as 10\,PeV.}
\label{tab:nevents}
\end{table}

\begin{table}[!ht]
\centering
\footnotesize
\begin{tabular}{lcc}\hline
source		& visibility	& $\langle\theta\rangle$\\\hline\hline
Crab		&	62\,\%	&	51.7$^{\circ}$\\\hline
1ES 1959+650	&	100\,\%	&	49.7$^{\circ}$\\\hline
Mkn 501		&	78\,\%	&	49.4$^{\circ}$\\\hline
Mkn 421		&	76\,\%	&	49.2$^{\circ}$\\\hline
\end{tabular}
\caption{Actual visibility and mean zenith angle of selected sources from the \ant\ down-going muons field of view.}
\label{tab:sources}
\end{table}

These results show that it is unlikely for \ant\ to detect steady gamma ray sources~: in the same optical background conditions, the number of reconstructed tracks surviving the selection cuts within 2$^{\circ}$ from source in a sample from actual data extrapolated to one year is about 53800 (17900) for a zenith angle of 20$^{\circ}$ (60$^{\circ}$), which is still far above the expected number of events detected from the fictional powerful source.  Reducing the observation cone does not improves the signal over background ratio enough to compensate the loss of signal it induces.

Results are reported in Figure~\ref{fig:sensitivity} as the power law flux normalization needed to obtain a 3 or 5 standard deviations sensitivity (after selection of events) as a function of the spectral index.  Also shown is the area covering Crab flux parametrizations and the position of the fictional powerful source.  It follows that \ant\ would need a source more than three orders of magnitude larger than the Crab or alternatively a powerful source with an extremely hard spectrum to obtain a reasonable sensitivity, again assuming a 100\,\% visibility and a fixed position in the sky.  Sources providing such characteristics are very unlikely to exist since similar features have yet never been measured even for short strong outbursts~\cite{tegasocat}.

 \begin{figure}[!t]
  \vspace{5mm}
  \centering
  \includegraphics[width=0.48\textwidth]{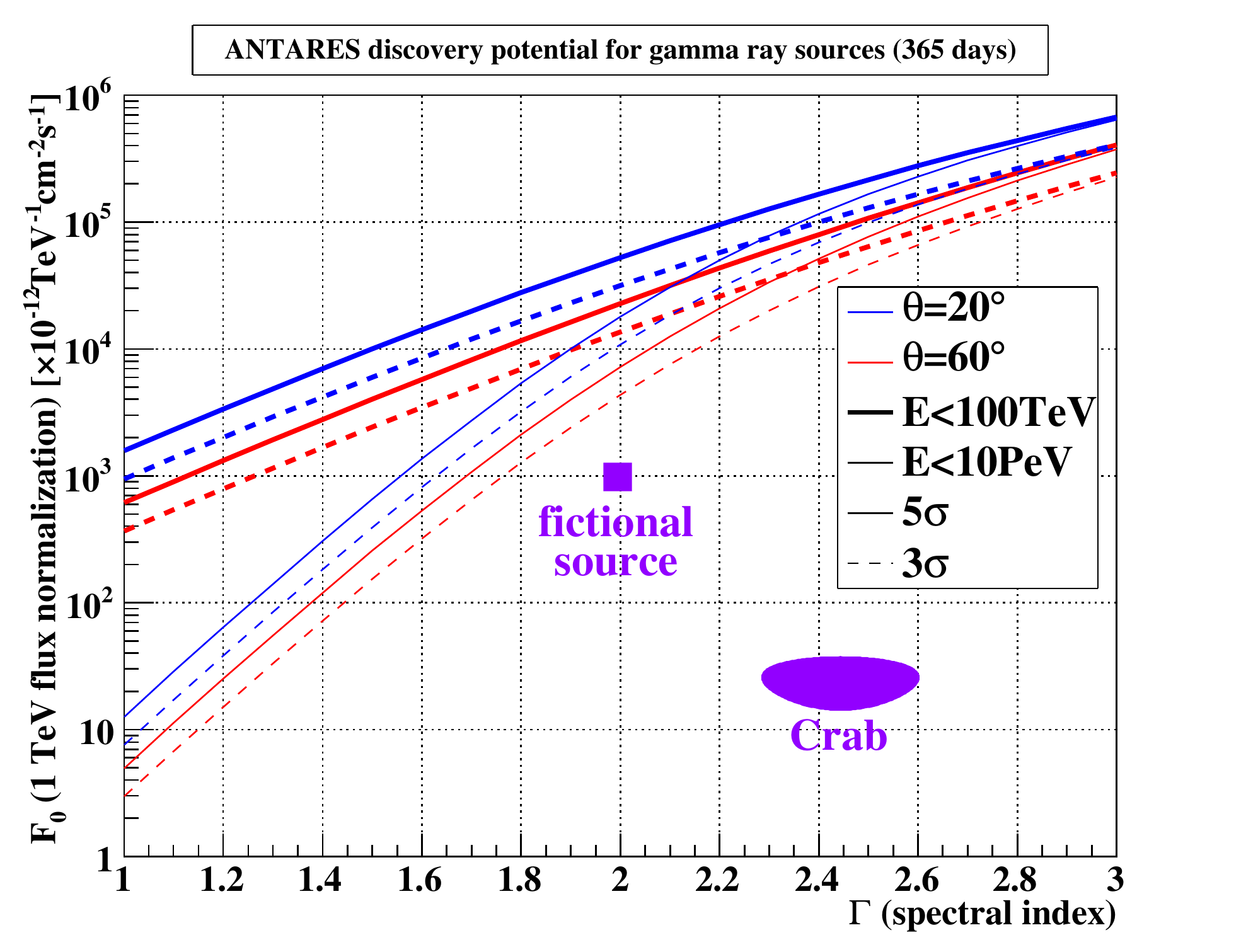}
  \caption{\ant\ one year sensitivity to steady gamma ray sources, as a function of flux normalization and spectral index assuming a power law spectrum.}
  \label{fig:sensitivity}
 \end{figure}

\section{Uncertainties}

This study suffers many sources of systematic uncertainties.  Given the low sensitivity of \ant\ to gamma ray induced muons, the various contributions are not computed in a quantitative way.  Many of the systematics have a negligible effect on the simulation.  This is for instance the case of the Earth magnetic field, the LPM effect~\cite{lpm} or the preshower effect~\cite{preshower}.

Important systematics are mainly due to the simulation of interactions in the atmosphere.  Charm photoproduction cross-section remaining unknown at very high energy, it is not included in \textsc{Corsika}.  It is small but increases with the energy and may not be negligible above a few tens of TeV~\cite{charm}.  Furthermore, the charm carries a large fraction of the incident photon energy and has thus a high probability to reach the detector, which makes this contribution difficult to estimate.  The lack of charm production is hence an important flaw of this simulation.  The hadronic model also introduces important uncertainties.  Switching the hadronic model to \textsc{Venus}~\cite{venus} can increase the number of muons with an energy higher than 500\,GeV at sea level by about 10\,\% and the number of events producing such muons by 7\,\%.  Finally, the photoproduction cross-section used in \textsc{Corsika} is quite conservative, and the muon yield might be increased by about 10\,\% by other realistic scenarios~\cite{cross-section}.  According to these effects, the present simulation may be thought as pessimistic.

On the other hand, the description of the sea water optical properties used for the light simulation in water is not the most accurate and overestimates the number of detected \v{C}erenkov photons and hence the number of reconstructed muons.  The number of detected muons could be increased by as much as 20\,\% with regard to reality.  The geometry of the detector also affects the detection of events~: the azimuth angles for which several detector lines are aligned are favoured, and the number of reconstructed events may be increased by a factor of two in the extreme cases.

Furthermore, a neutrino telescope is sensitive to variable sources of uncertainties.  The muon yield itself depends on atmospheric variations~: the higher the pressure and temperature, the lower the muon yield, by a few percents.  This affects both signal and muon background.  The sea water optical properties and most of all the optical background due to radioactive decays (mostly $^{40}$K) and bioluminescence are also strong sources of variability.  The present simulation is valid for standard conditions of optical background qualified for data analysis, but small degradation or improvement of these conditions may have an effect as large as 20\,\% on the number of reconstructed tracks.  Finally, the accuracy of the software used for the simulation of light propagation and detection has been greatly improved since versions used in this simulation.

Last but not least, in order to ensure statistical consistency of the simulation, the source zenith angle is fixed.  In reality \ant\ sensitivity to gamma ray sources is obviously affected by their motion in the sky.

\section{Discussion}

Although this simulation suffers large uncertainties, it is unlikely that the results would vary by an order of magnitude or more.  It is thus clear from this study that steady gamma ray sources are out of reach of the \ant\ neutrino telescope under the conditions discussed here.  

This analysis makes use of standard triggers and reconstruction algorithms, optimized for detection of upgoing neutrinos.  Dedicated strategies and background discrimination methods would have to be developped to improve significantly the sensitivity~: directional trigger, use of the muon-poor electromagnetic showers characteristic, muon pair discrimination\dots

In addition, the present study assumes that the gamma ray source flux at very high energy can be parametrized by a power law (possibly with an exponential cutoff).  A significant deviation of this assumption at higher energies would have a strong impact on the presented results.

\section{Gamma-induced neutrino background}

Gamma rays interacting with the atmosphere may also produce neutrinos.  These so-called atmospheric neutrinos form an irreducible background for neutrino telescopes, since there is no way to distinguish them from cosmic neutrinos.  Furthermore, such atmospheric neutrinos are localized from the direction of gamma ray sources, which are cosmic neutrino emitter candidates.

The present study is the opportunity to estimate the number of upgoing atmospheric neutrinos polluting the signal from the direction of a gamma ray source.  This is simply done by multiplying \ant\ neutrino effective area~\cite{nueffa} by the gamma ray induced neutrino flux at sea level.

It is found that the expected number of reconstructed events is smaller by several orders of magnitude than the expected cosmic neutrino signal for all simulated sources.  The contamination induced by steady gamma ray sources to a potential source of neutrino can thus be considered as negligible.

\section{Conclusions}

It has been demonstrated through a complete Monte-Carlo simulation that the observation and identification of very high energy gamma rays from steady sources is out of reach of the \ant\ neutrino telescope with its standard trigger and reconstruction strategies.  It is also clear from this study that the gamma ray flux of such sources will not interfere with the identification of a possible neutrino sources.


\clearpage

\setcounter{figure}{0}
\setcounter{table}{0}
\setcounter{footnote}{0}
\setcounter{section}{0}
\newpage




\title{Muon induced electromagnetic shower reconstruction in ANTARES neutrino telescope}

\shorttitle{S. Mangano \etal Electromagnetic shower reconstruction}

\authors{Salvatore Mangano$^{1}$, for the ANTARES collaboration}
\afiliations{$^1$IFIC - Instituto de F\'isica Corpuscular, Edificio Institutos de Investigati\'on, \\
                                 Apartado de Correos 22085, 46071 Valencia, Spain}
\email{manganos@ific.uv.es}

\maketitle
\begin{abstract}
The primary goal of the ANTARES telescope is the detection
of high energy cosmic muon neutrinos. The neutrinos are identified by the upward going muons
that are produced in charged current neutrino interactions in the vicinity of the detector.
The Cherenkov light produced by the muons in the detection volume is measured
accurately by an array of photosensors.
Muons that are going downward are background for neutrino searches.
These muons are the decay products of cosmic-ray collisions in the Earth's atmosphere.
The energy loss in water of a muon with an energy above a TeV is characterized
by discrete bursts of Cherenkov light originating mostly from pair production
and bremsstrahlung (electromagnetic showers). This paper presents
a method to identify and count electromagnetic showers produced by the muons.
The method can be used to select a sample of highest energy muons
with the ANTARES detector.
\end{abstract}


\section{Introduction}
\label{intro}

The ANTARES neutrino telescope is located on the bottom of the
Mediterranean Sea, roughly \mbox{40 km} off the French coast. 
The main objective is the observation of 
extraterrestrial neutrinos. 
Relativistic charged leptons produced by neutrino interactions in and around the detector produce 
Cherenkov light in the sea water. This light is detected by an array of photomultiplier
tubes (PMTs),  allowing the muon direction to be reconstructed. The muon energy loss
can be estimated from the sum of the measured number of photoelectrons.

The detector is installed at a depth of 2475 m and consists of twelve vertical
lines approximately \mbox{$450$ m} long
equipped with a total of 885 PMTs.
The lines
are separated from each other by about \mbox{$65$ m}, 
anchored to the sea bed by a dead weight and held taut by a buoy located 
at the top. The instrumented
part of the line starts \mbox{100 m} above the sea bed, with 25 storeys 
every \mbox{14.5 m} along the line. A storey consists of three 
PMTs pointing downward at an angle of $45^{\circ}$ with respect
to the vertical direction, in order to maximise the detection efficiency of upward going
tracks.

ANTARES is operated in the so called 
all-data-to-shore mode: all pulses above a threshold (typically 0.3 photoelectrons) 
are digitized off-shore
and sent to
shore to be processed in a computer farm. 
This computer farm applies a set of trigger criteria in order to separate muon-induced Cherenkov 
light from background light. The main sources of the background light are the decay of $^{40}$K 
nuclei and the bioluminescence of organisms in the sea water.
A detailed description of the ANTARES detector is given in \cite{Antares}.

Although ANTARES is optimised for upward going particle
detection, the most abundant signal is due to the atmospheric
downward going muons produced by the
interaction of primary cosmic-rays in the atmosphere. 
Being the most penetrating particles in such air showers,  
muons with enough energy can reach the detector and 
are reconstructed by the detection of the
Cherenkov light they emit when propagating through water.
The ANTARES detector 
has the capability to follow highly energetic muons  
over a few hundred metres.

The processes contributing to the energy loss of a muon in water
include ionisation, \mbox{$e^+e^-$ pair} production,
bremsstrahlung, and photonuclear interactions.
Below \mbox{1 TeV}, the muon 
energy loss is dominated by the continuous ionisation process.
Above \mbox{1 TeV}, the muon energy loss is dominated
by pair production and bremsstrahlung \cite{PDG}, which are
radiative processes classified as electromagnetic showers. They are 
characterized by large energy fluctuations and 
discrete bursts along the muon track. 
The average muon energy loss per unit track length due to these electromagnetic 
showers increases linearly with 
the energy of the muon allowing its energy to be determined.
Counting electromagnetic showers along muon tracks gives an estimate
of the muon energy \cite{Gandhi} and can help
in designing a better energy reconstruction algorithm.
A similar measurement technique as the one presented in this 
article has been
published recently by the Super-Kamiokande experiment \cite{Desai}.

\section{Algorithm for shower identification}
\label{algorithm}
\vspace{-2mm}
\begin{figure}[!t]
  \vspace{0mm}
  \centering
\includegraphics[width=2.8in]{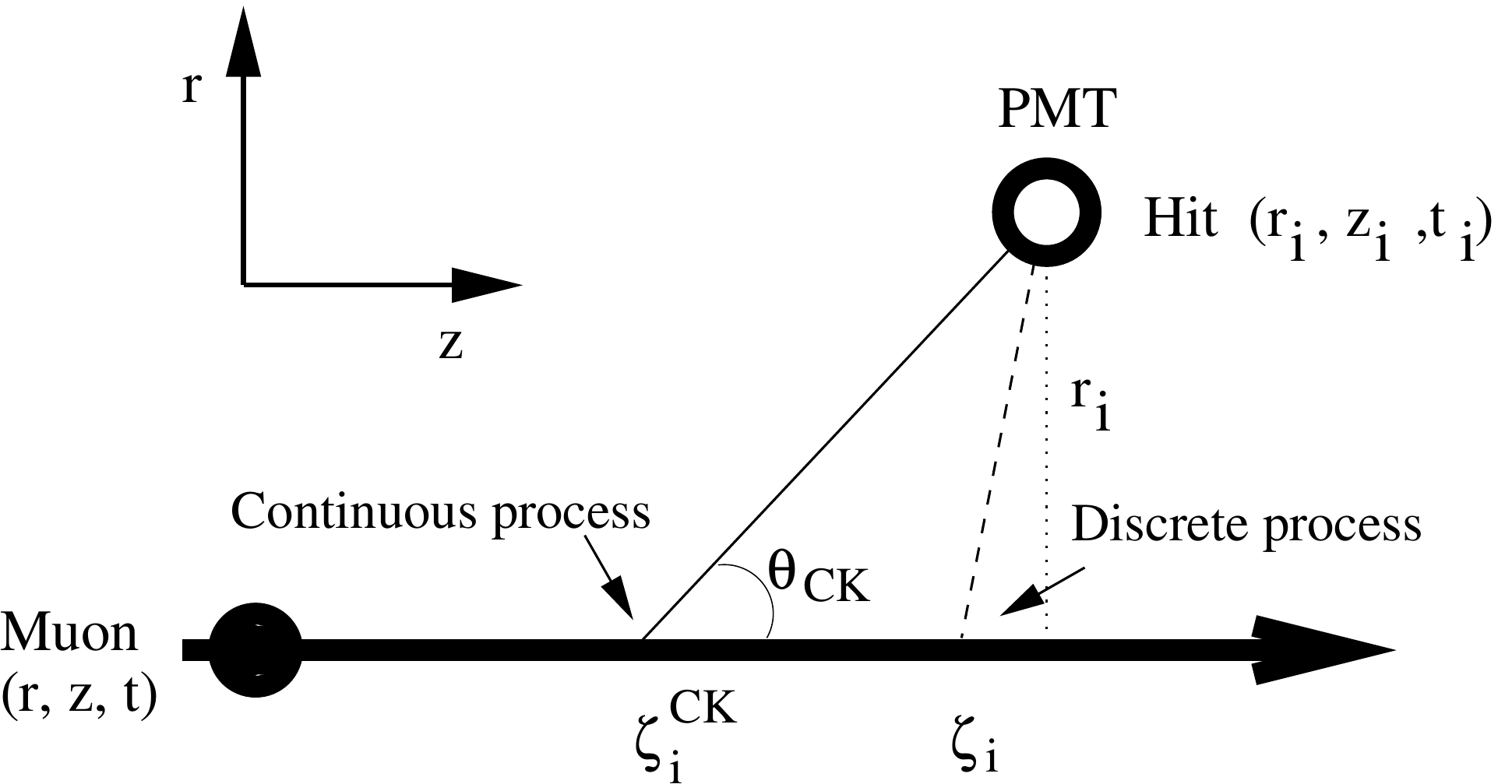}
\caption[]{\textit{{Schematic view of the Cherenkov light detection. The thick line represents the muon trajectory, the thin line the path of the Cherenkov light and the thin dashed line the path of the shower light. The muon goes through a reference point $(r, z, t)$. The Cherenkov light is emitted at an angle $\theta_{\rm CK}$ with respect to the muon track at point $\zeta_i^{\rm CK}$ and is detected by a PMT as a hit at point $(r_i, z_i, t_i)$. The shower light is emitted at point $\zeta_i$ and is detected by the same PMT at different time.}}}
\label{fig:calc}
\end{figure}
The purpose of the electromagnetic shower identification is to distinguish
Cherenkov photons emitted continuously along the muon track,
from the Cherenkov photons induced by electromagnetic showers.
Because of the radiation length in water ($X_0=35$ cm), these showers 
never extend more than a few metres and can be considered point-like
light sources in the ANTARES detector. The showers can be identified from 
a localised 
increase of the number of emitted photons above the continuous baseline of Cherenkov photons 
emitted by a minimum ionizing muon.

In what follows, a hit is a photomultiplier signal exceeding a threshold of 0.3 photoelectrons.
The shower algorithm consists of two steps. 
The first step is to identify and reconstruct downward going 
muon track candidates. In the second step, a distinct shower candidate is identified 
by looking for an 
accumulation of hits on a point along the muon path.
The criteria to isolate the accumulation are defined 
from a simulation code based 
on Corsika \cite{Corsika}. 
 
\subsection{Simulation}
\label{simulation}
\vspace{-2mm}
Cosmic-ray interactions in the atmosphere including shower development were
simulated for primary energies between 
\mbox{1 TeV} and \mbox{$10^5$ TeV}, and incident angles between zero (vertical) and  85 degrees with Corsika.
The primary cosmic-ray composition and flux model considered 
was a simplified version\footnote{The primary composition of the flux is subdivided into only five mass groups, namely proton, helium, nitrogen, magnesium and iron.} of the H\"orandel model \cite{Hoerandel}. 
The hadronic interaction 
model chosen was QGSJET \cite{QGSJET}. The result of the Corsika simulation
consists of muons with their positions and times and kinematic vectors on the surface of the sea. 
The muon propagation through water, the discrete energy losses 
at high energies, the 
Cherenkov light production and propagation, including scattering,  
and the response of the detector
was simulated using the ANTARES KM3 code \cite{Km3_0}.
KM3 uses tables generated from a simulation with GEANT 3 which 
parametrise the arrival time and the amount of light detected by individual PMTs. These tables 
take into account the angular dependence of the acceptance of the PMT
as well as the measured properties of the water at the ANTARES site. 
The muon propagation uses the MUSIC 
code \cite{MUSIC} and is done in short steps (1 m). If
the energy loss of the muon over the step exceeds a threshold (1 GeV),
an electromagnetic shower is initiated and shower photons are emitted, 
otherwise if the energy loss of
the muon over the step
is below the threshold muon Cherenkov photons are emitted. 
The optical background was assumed flat at a rate of 60 kHz on each 
photomultiplier.

\subsection{Algorithm}
\label{algorithm1}
\vspace{-2mm}
Muon events are reconstructed by using an existing algorithm \cite{line1, Ronald}
which provides an estimate of 
the direction and position of the muon at a given time.
Measured hit times are compared to the expected arrival time of direct Cherenkov photons.
The expected Cherenkov photon arrival time $t^{\rm CK}_i$ for each hit $i$ is calculated as (see \mbox{Figure \ref{fig:calc})}:
\begin{equation}
  t^{\rm CK}_i = t + \frac{1}{c} \Big(z_i-z - \frac {r_i}{\tan \theta_{\rm CK}}\Big) + \frac{n}{c} \frac{r_i}{\sin \theta_{\rm CK}},
\label{Eq:CKlight}
\end{equation}
where $t$ is the time where the muon passes point $(r,z)$, $c$ is the speed of light in vacuum, $n$ is the refraction index of water ($n$ is about 1.38), $\theta_{\rm CK}$ is the Cherenkov angle for a relativistic muon in water
($\theta_{\rm CK}=42^o$)
and $r_i$ is the perpendicular distance between the 
muon trajectory and the PMT. 
Equation (\ref{Eq:CKlight}) separates the direction along the track and
the direction perpendicular to the track. The direction along 
the track ($z$-coordinate) is given 
by the muon momentum vector. 
The direction perpendicular to the track ($r$-coordinate) 
is given by the photon momentum vector in water.
Hits too far in time from the expected muon hit time 
\mbox{$-t_{min} < t_i-t^{\rm CK}_i < t_{max}$}, are assumed 
to be background hits 
and are rejected, whereas direct hits have a roughly  
Gaussian distribution at zero with a width of 20 ns.
The value for $t_{min}= 20$ ns is given by mainly the dispersion of light in water
and the timing resolution of the PMT, 
whereas $t_{max}= 200$ ns is defined by the value 
where the number of signal hits 
approaches the level of background hits.  
Furthermore, the above defined time interval is subdivided into two intervals. 
The early interval 
contains mostly 
muon Cherenkov photons and is given by $ |t_i-t^{\rm CK}_i| < t_{min}$.
The Cherenkov photon emission position ($\zeta_i^{\rm CK}$) along
the muon trajectory is given by:
\begin{equation}
\zeta_i^{\rm CK}=z_i- z - \frac {r_i}{\tan \theta_{\rm CK}}.
\label{equ:cvpos}
\end{equation}
The late interval is defined by $t_{min} < t_i-t^{\rm CK}_i < t_{max}$ and contains 
mostly electromagnetic shower photons. These shower photons may not necessarily be emitted at the 
Cherenkov angle from the muon track. The emission angle is left as a free parameter and,
with the photon emission taking place at $\zeta_i$ (see Figure \ref{fig:calc}), the hit time is given by
\begin{eqnarray}
t_i = t + \frac{\zeta_i-z}{c} + \frac{n}{c} \sqrt{ r_i^2 + (z_i- \zeta_i )^2 }.
\label{equ:showertime}
\end{eqnarray}

Equation (\ref{equ:showertime}) can be solved for $\zeta_i^{\pm}$, yielding:
\begin{eqnarray*}
\zeta_i^{\pm}= \frac {-B_i \pm \sqrt { B_i^2-4AC_i} } {2A}, 
\label{equ:hittoz}
\end{eqnarray*}
where 
\begin{eqnarray*}
A     & = & 1-n^2,   \\
B_i   & = & 2(n^2z_i-z-c(t_i-t)),   \\ 
C_i   & = & c^2(t_i-t)^2 + 2cz(t_i-t) + z^2 - n^2(r_i^2 + z_i^2). 
\end{eqnarray*}

All calculated $\zeta_i^{\rm CK}$, $\zeta_i^{+}$ and $\zeta_i^{-}$  positions 
along the muon track are collected in a one-dimensional histogram. 
As an example of such a histogram, 
Figure \ref{fig:EvDisplayalone} 
shows all the calculated photon emission positions along the muon trajectory.
The electromagnetic showers are identified by an excess of photons above the continuous baseline of photons emitted by 
a minimum ionizing muon.  
Two excesses are visible that can be attributed to the two reconstructed showers.

 \begin{figure}[!t]
  \vspace{-3mm}
  \centering
 \includegraphics[width=3.2in]{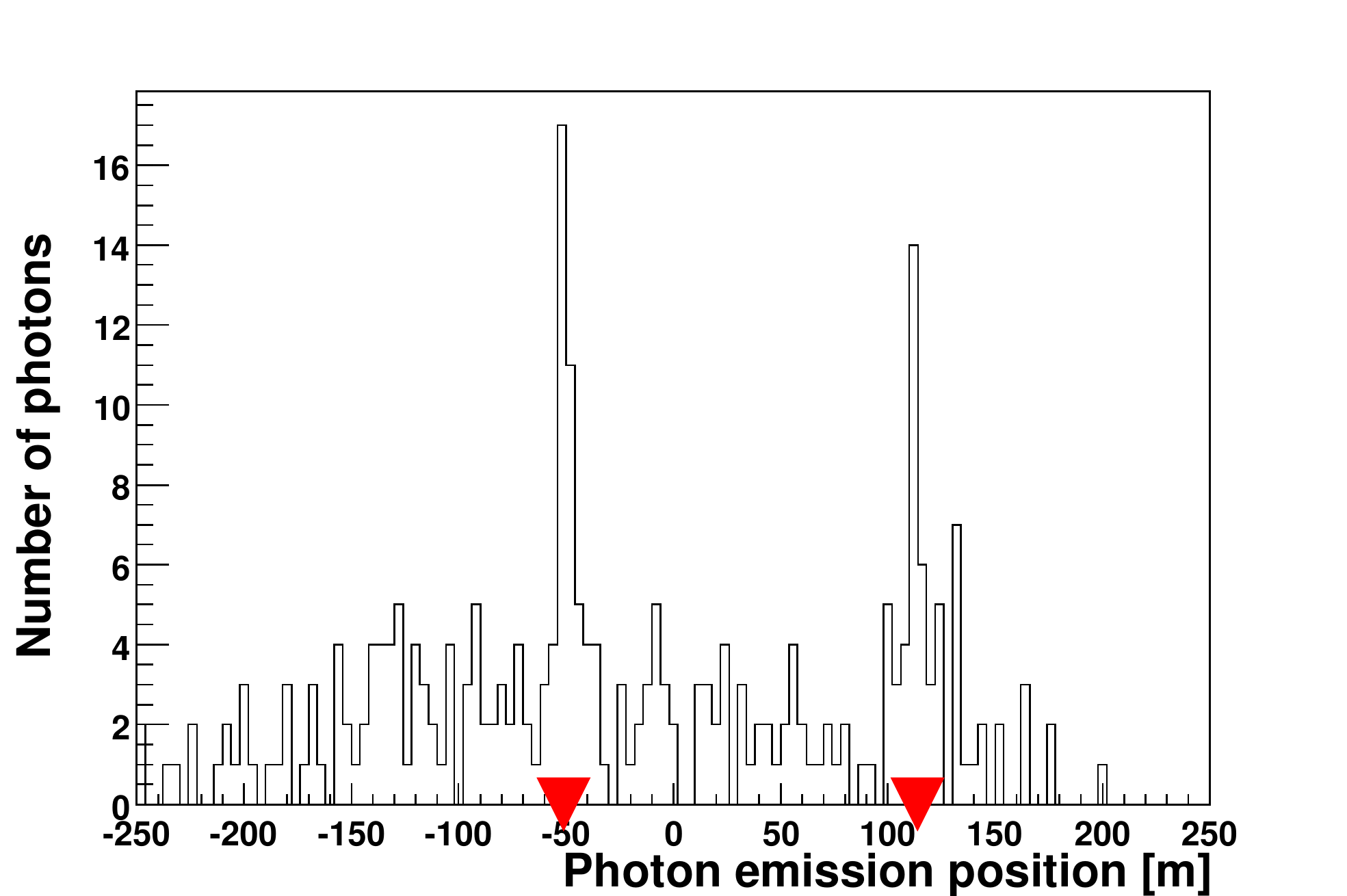}
\caption[Sc]{\textit{The number of detected photons projected along the muon trajectory for an atmospheric muon event with two electromagnetic showers. The peaks indicate the shower positions on the muon trajectory, with the reconstructed shower positions indicated by the triangles.}}
\label{fig:EvDisplayalone}
\end{figure}

 \begin{figure}[!t]
  \vspace{-3mm}
  \centering
  \includegraphics[width=2.6in]{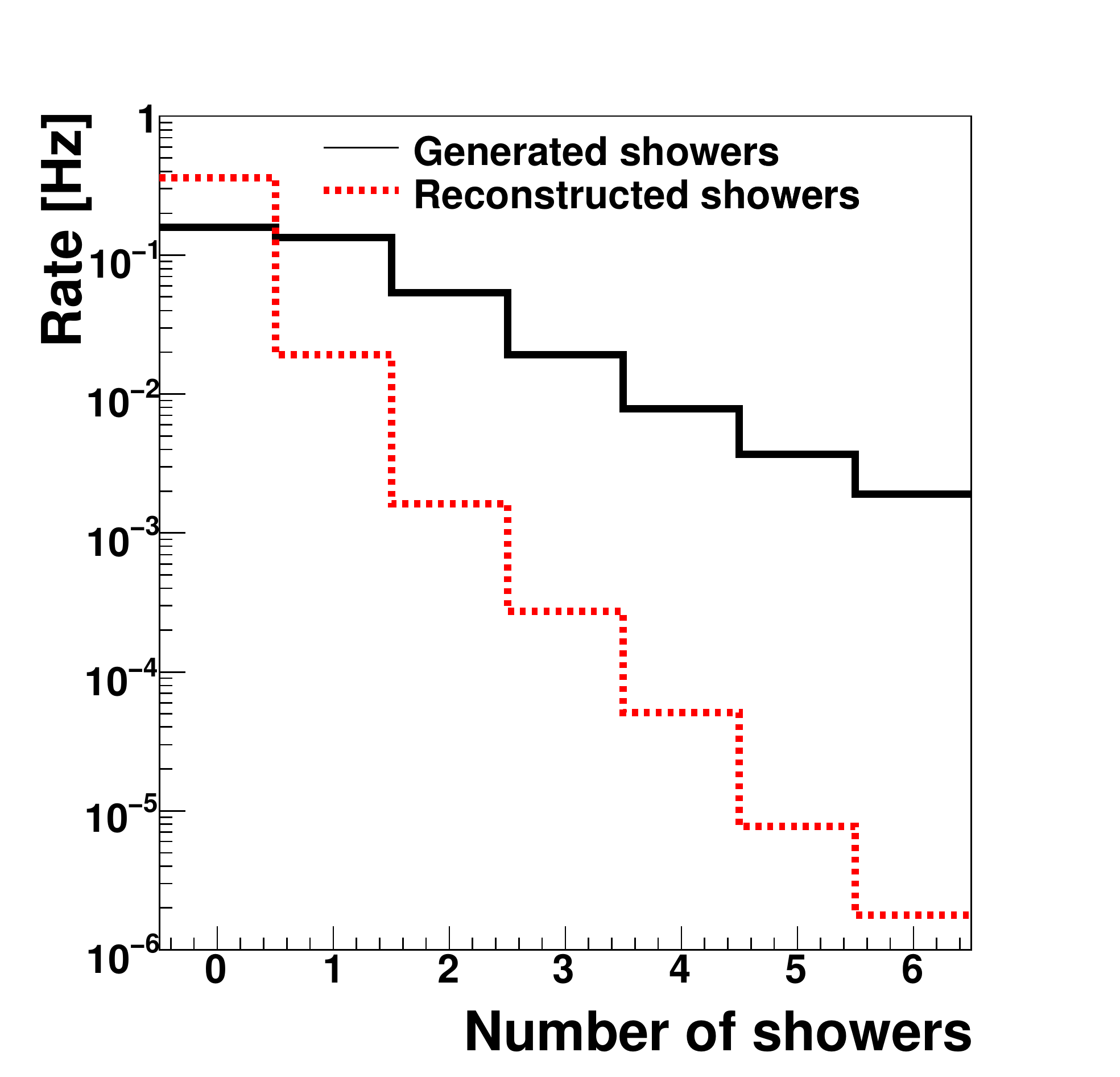}
\caption[Sc]{\textit{Muon event rate as a function of the shower multiplicity  for reconstructed showers  and for generated showers which have shower photons detected on at least five different storeys for the Corsika simulation.}}
\label{fig:recshower}
\end{figure}

\section{Selection and results}
\label{simulationandselection}
\vspace{-2mm}
The selection and performance of the shower identification algorithm has been studied 
and validated with a sample of simulated atmospheric multi-muons
with constant background light as described in section \ref{simulation}.

\subsection{Selection}
\label{sec:muonselect}
\vspace{-2mm}
The shower algorithm makes use of tracks fitted with the
muon reconstruction algorithm described in \cite{line1} with two 
additional criteria. These criteria require the tracks 
to be traced for at least 125 m and 
to have a 
minimum of twelve hits used in the track reconstruction.
These selected tracks account for around 65\% of all reconstructed tracks.
The advantage of the selected tracks 
is not only that the direction of the 
tracks is better reconstructed, but also that the tracks are long enough 
to have a high probability to emit showers.

The parameters in the shower reconstruction 
algorithm are the width and the height of the peak. 
The analysis has been tuned to select showers with a high level of purity,
possibly  at the expense of efficiency.
For each selected peak, the number of hits is integrated
in a $\pm 5$ m interval around the peak center.
Only peaks 
having at least 10 hits over the muon-track Cherenkov photon baseline 
in this interval of \mbox{10 m} are selected.
The number of baseline hits is defined as the 
average density of hits along the track times the interval width of 10 m.
In addition, in order to suppress fake identified showers, 
hits from at least 
five different storeys 
are required in each peak.

 \begin{figure}[!t]
  \vspace{-4mm}
  \centering
  \includegraphics[width=2.6in]{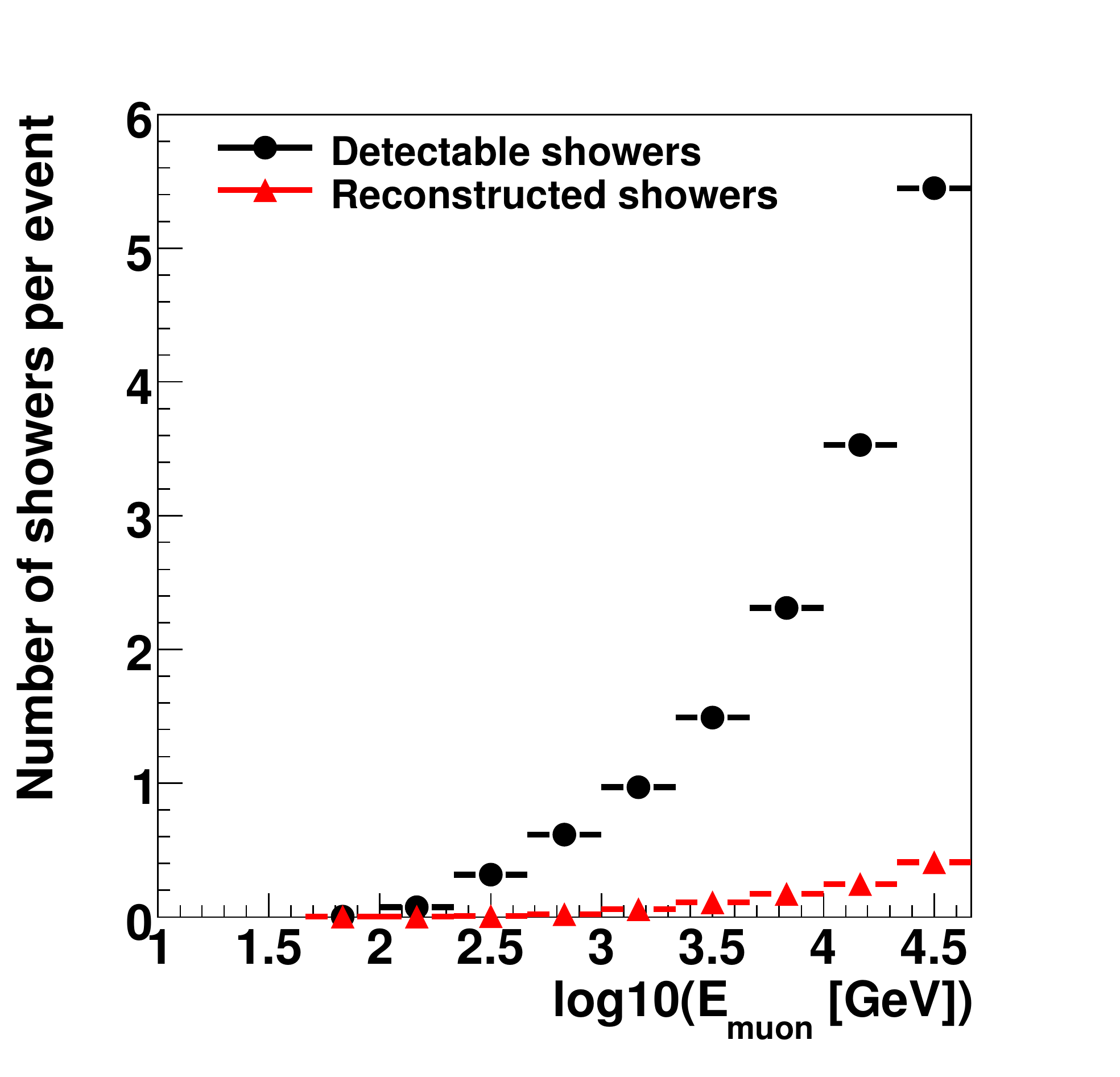}
\caption[Sc]{\textit{Average number of detectable showers which have shower photons detected on at least five different storeys per atmospheric muon event as a function of the muon energy. Also shown is the average number of reconstructed showers.}}
\label{fig:recshower1}
\end{figure}

\subsection{Results}
\vspace{-2mm}
The main result of the shower reconstruction algorithm is the
shower multiplicity per atmospheric muon event.
The atmospheric muon events are usually muons in a \mbox{bundle} with 
an average multiplicity of around 3.3.
\mbox{Figure \ref{fig:recshower}} shows the event rate
as a function of the number of
generated showers with shower photons detected on at least 
five different storeys. Also shown is the
number of reconstructed showers selected with the cuts mentioned in section \ref{sec:muonselect}. The average shower reconstruction 
efficiency over all shower energies is around 4\%. The reconstruction
algorithm starts to be efficient for shower energies above \mbox{300 GeV}. 

Figure \ref{fig:recshower1} shows the number of generated showers
with shower photons detected on at least 
five different storeys per atmospheric muon event as a function
of the muon energy. The muon energy refers to the muon
with the largest energy in the bundle. 
The number of generated shower and
reconstructed shower increases as a function of the muon energy.
Figure \ref{fig:energyspectrum} shows the energy distribution
of all muons as well as the one with at least one reconstructed shower. 
The muons have an average energy of 1.2 TeV, whereas muons with
at least one identified shower have on average 2.5 times higher energy.

\section{Conclusion}
\label{conclusion} 
\vspace{-2mm}
A method to identify showers emitted by
atmospheric muons has been applied to simulated data of the ANTARES
detector.
The main differences between the shower light and the muon 
Cherenkov light are that
the shower light is produced on a point along the muon path.
The essential element of the algorithm is that the selection 
of shower photons is
reduced to a one-dimensional problem. The performance of the identification
algorithm has been validated using a sample of simulated
atmospheric muon events.

The aim of this proceeding has been more to demonstrate the 
capability to detect electromagnetic showers than to
make precise measurements.
A more elaborated analysis is needed,
in order to use the
number of electromagnetic showers as a robust estimate of the muon energy.
Moreover, the method discussed here for selecting showers emitted by downward going
muons could be used also for upward going muons with the main purpose of selecting
the highest energy upward going muons.

 \begin{figure}[!t]
  \vspace{-4mm}
  \centering
  \includegraphics[width=2.6in]{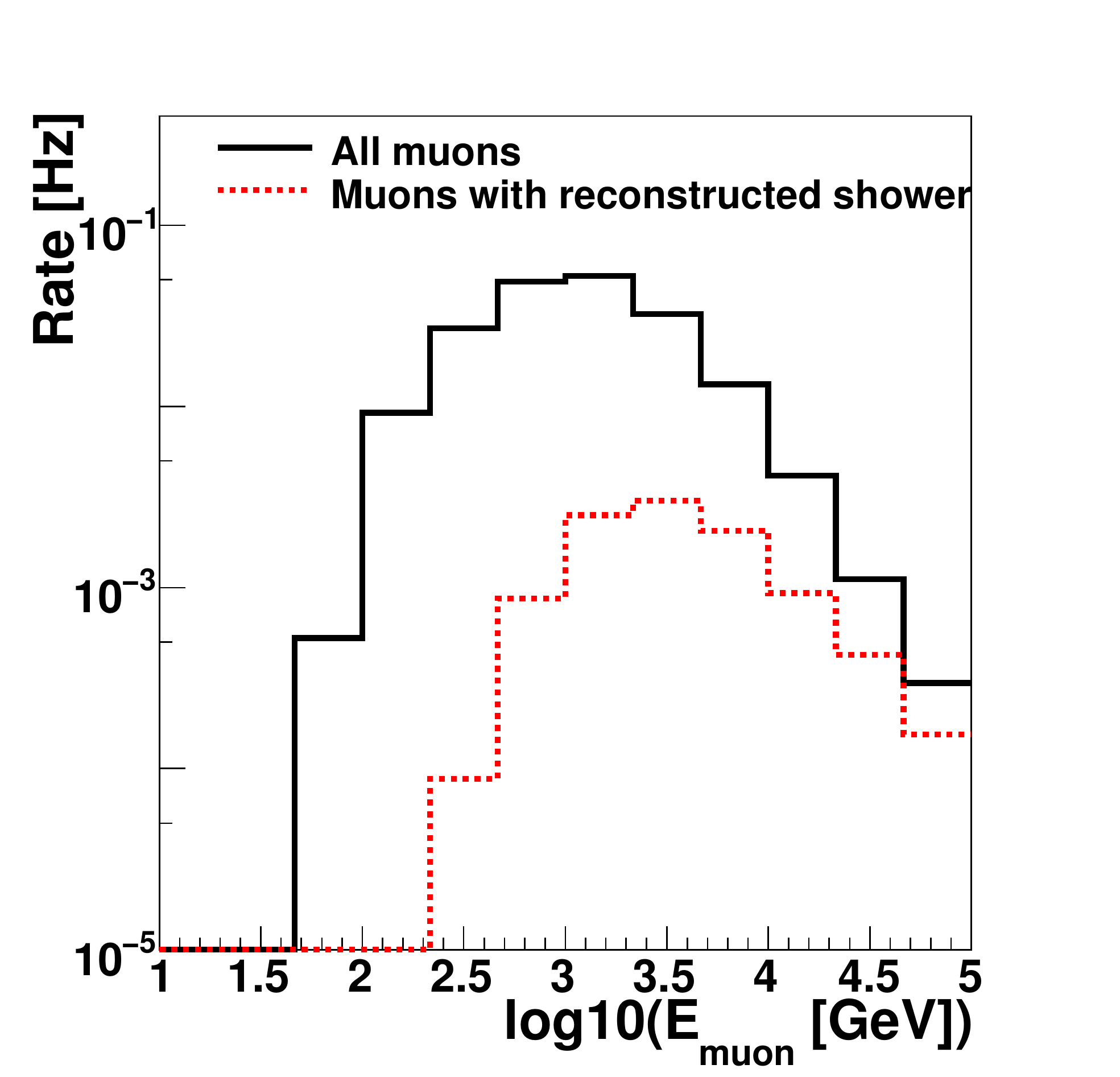}
\caption[Sc]{\textit{The solid line shows the energy distribution of all reconstructed atmospheric muons. The dotted lines  represent the muon energy distribution for the muons with at least one reconstructed shower.}}
\label{fig:energyspectrum}
\end{figure}

\section*{Acknowledgments}
 \vspace{-4mm}
I gratefully acknowledge the support of the JAE-Doc postdoctoral programme of CSIC.
This work has also been supported by the following
Spanish projects: FPA2009-13983-C02-01, MultiDark Consolider CSD2009-00064, ACI2009-1020 of MICINN and Prometeo/2009/026 of Generalitat Valenciana.

\clearpage

\setcounter{figure}{0}
\setcounter{table}{0}
\setcounter{footnote}{0}
\setcounter{section}{0}
\newpage




\title{Optical properties in deep sea water at the site of the ANTARES detector}

\shorttitle{S. Mangano \etal Optical properties}

\authors{Salvatore Mangano$^{1}$, for the ANTARES collaboration}
\afiliations{$^1$IFIC - Instituto de F\'isica Corpuscular, Edificio Institutos de Investigati\'on, \\
                                 Apartado de Correos 22085, 46071 Valencia, Spain} 
\email{manganos@ific.uv.es}

\maketitle

\begin{abstract}
The ANTARES neutrino telescope is located at a depth of 2475 m in the
Mediterranean Sea. Its main objective is the observation of
extraterrestrial neutrinos. Relativistic charged leptons
produced by neutrino interactions in and around the detector produce
Cherenkov light in the sea water detected by a three dimensional grid of
photomultiplier tubes. The propagation of Cherenkov
light depends on the optical properties of the sea water
and their understanding is crucial to reach the optimal performance
of the detector.
This paper presents the measurements made between 2008 and 2010 of
the light velocity and attenuation length
at the ANTARES site with a system of light
sources (LEDs and laser) at different
wavelengths (between 400 nm and 532 nm).
The time variability of the optical properties are presented
and the derived values are compared with theoretical predictions.
\end{abstract}


\section{Introduction}
The ANTARES neutrino telescope \cite{Ant1} is located at the bottom of the
Mediterranean Sea at a depth of 2475 m, roughly \mbox{40 km}
offshore from Toulon in France. Sea water is used as the detection medium for
the Cherenkov light emitted by relativistic charged particles resulting from interaction
of neutrinos around or inside the detector.
The particle direction is reconstructed from the arrival time of detected photons
through the array of photomultiplier tubes (PMTs).

The measurement of the refractive index
is performed with a pulsed light source shining through water
and the time of light distributions of photons
detected by PMTs at different distances from the source. 
The attenuation length is measured by the amount of light detected by these PMTs.
The optical properties
are measured at wavelengths between 400 nm and 532 nm
and are compared with
theoretical predictions.

A precise measurement of the optical properties minimizes the uncertainty on many physical results as seen in \cite{Coll2010}.
Moreover, the optical properties may change in time due to sea current.
Several measurements of those presented in this paper have been
performed in the past \cite{Agui,Danil,Lubs}.

\section{Experimental setup}
The ANTARES detector consists of a three dimensional array
of 885 PMTs
arranged in twelve vertical approximately 450 m long lines.
Along each line with a vertical separation of \mbox{14.5 m}, 
PMTs are grouped in triplets and oriented with their axis pointing downward at an angle of $45^o$ with
respect to the the vertical line direction.
The horizontal separation between lines is about \mbox{70 m}.

The PMTs are sensitive to single photons in the
wavelength range between 350 nm and 600 nm. They have a peak quantum efficiency
of about 25\% between 350 nm and \mbox{450 nm}. The PMT measures the
arrival time and charge amplitude of the detected photons. 

The Optical Beacon system consists of a series of pulsed 
light sources distributed through the detector.
The primary aim of the
Optical Beacon system is the time calibration between the PMTs
to reach the best angular resolution of the detector.
In addition the Optical Beacon system can be also used
to determine the optical properties of water.

There are four LED Beacons per line and one laser Beacon at the
bottom of the central line.
One LED Beacon contains 36 individual LEDs
distributed over six vertical faces
shaping an hexagonal cylinder.
On each face, five LEDs point radially and one upwards.
All the LEDs emit light at a nominal wavelength of 470 nm
except two LEDs located
on the lowest Beacon of line 12 which emit light
at nominal wavelength of 400 nm.

The LEDs emit light with a maximum intensity of $\sim$160 pJ and a
pulse width of $\sim$4 ns (FWHM).
The laser is a more powerful device and emits shorter pulses
than the LEDs. The laser
emits pulses of light with a maximum intensity of ($\sim$1 $\mu$J) and
pulse width of $\sim$0.8 ns (FWHM) at a nominal wavelength of 532 nm.
The LEDs and laser flash at a frequency of 330 Hz.
Further details about the Optical Beacon system can be found elsewhere
\cite{Ageron, Juanan, Timecalibration}.

The light spectrum of the three sources (two LEDs with nominal wavelengths of 470 nm 
and 400 nm and one laser with nominal wavelength of \mbox{532 nm}) 
were measured using a high resolution calibrated spectrometer from
Ocean Optics HR4000CG-UV-NIR.
The spectrometer was crosschecked with the Green Nd-YAG laser (532 nm).
The measured peak wavelengths of the LEDs in pulsed mode operation 
are \mbox{468.5 $\pm$ 14.4 nm} and 403.1 $\pm$ 6.9 nm respectively.

 \begin{figure}[!t]
  \vspace{3mm}
  \centering
  \includegraphics[width=2.8in]{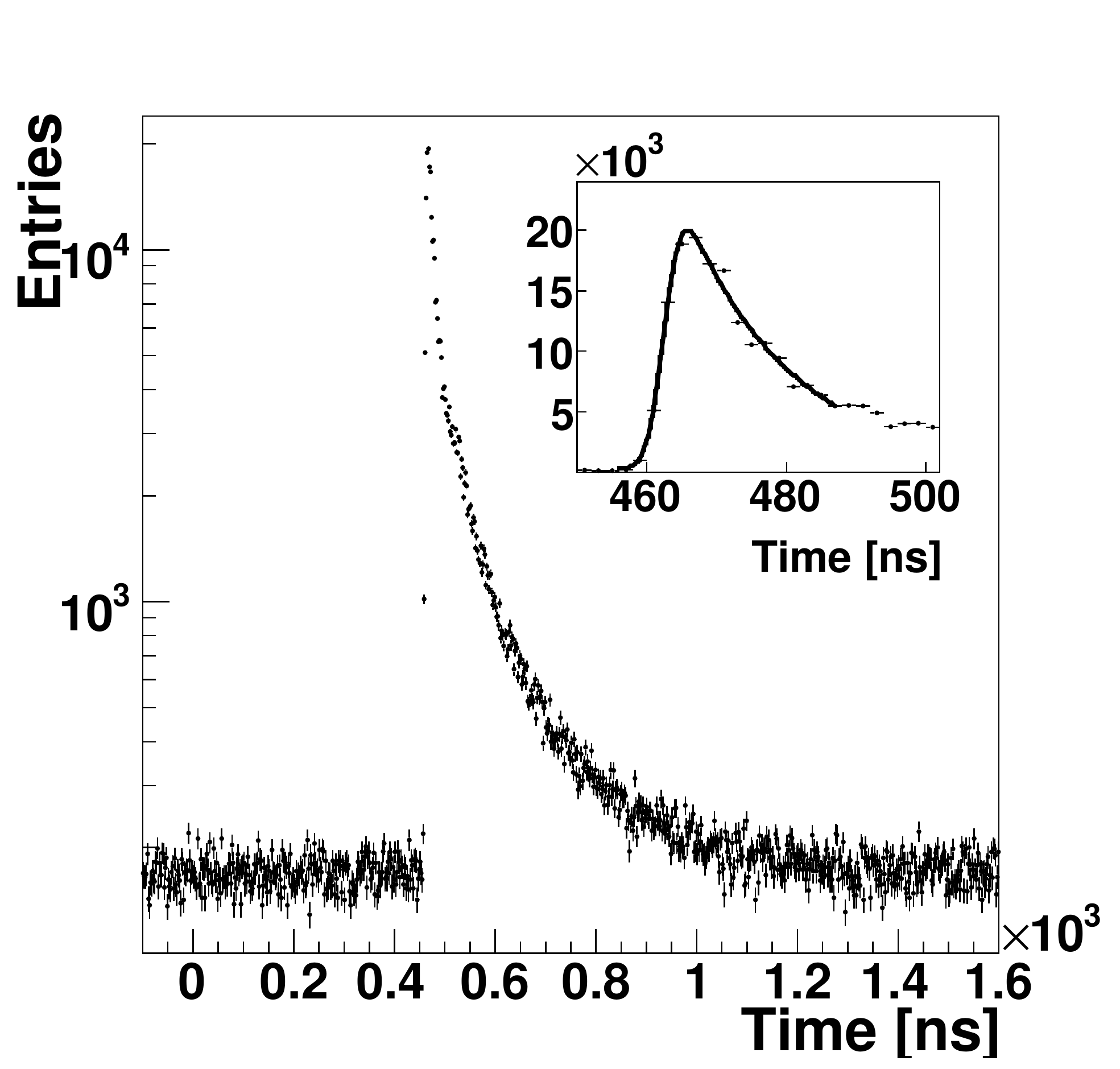}
  \caption{Arrival time distribution of LED  light detected by a PMT at around 100 m. The inset shows a zoom around the signal region. Superimposed is the result of the fit to the convolution of a gaussian distribution and an exponential distribution (see text).}
  \label{fig:time}
 \end{figure}

\section{Data acquisition and data analysis}
\label{sec:dataacqu}
In the special calibration data taken with Optical Beacon system, the emission
time and the position of the isotropic light flash,
as well as the arrival time and the position when the light reaches the PMTs are known.
From the time and position difference between the PMTs the refractive index is measured, whereas 
the amount of light collected by these PMTs gives an information about the attenuation length.

\subsection{Data acquisition}
The various measurements of the water optical properties
were taken from May 2008.
Only data with stable background rates and below 100 kHz have been analyzed.
The special calibration data duration takes only few minutes.
One single upward looking
LED of the lowest
Optical Beacon in the line emits more than $10^5$ light flashes
towards the PMTs above.
The photons are collected by the PMTs in the line.
Figure \ref{fig:time} shows the arrival time distribution of photons
for a LED Beacon illuminating a PMT located at 100 m. The small tail
of delayed photons is due to the light scattering between LED and PMT.
The zero time is defined by the illumination time of the LED.
The flat distribution before and after the peak shows the detected
optical background.

\subsection{Data analysis}
\label{sec:dataanalysis}
Since the light path used in the analysis (few 100 m) has the same order
as the scattering length the scattering has in some way
to be taken in account.
The arrival time distributions are fitted to a
function which is the convolution of a gaussian distribution and an exponential distribution \cite{Sigmund, Frodesen}.
The gaussian distribution models the transit time spread of the PMTs, the time width of the optical sources and the effect of the chromatic dispersion in water,
while the exponential distribution models the scattering
of photons in water.
The fit function is given by
\begin{equation}
  f(t)=b + h \cdot e^{-\frac{t-\mu}{\tau}} \cdot \textrm{Erfc} \Bigg(\frac{1}{\sqrt{2}} \Big(\frac{\sigma}{\tau} - \frac{t-\mu}{\sigma}\Big)\Bigg),
\label{equ:ciro}
\end{equation}
where $t$ is the arrival time of the photons and the fit parameters are
the \mbox{background ($b$)}, the height of the fit function ($h$), the mean and width of the gaussian 
distribution ($\mu, \sigma$) and the exponential constant ($\tau$).
The $\textrm{Erfc}(t)$ is the complementary error function distribution $\textrm{Erfc}(t) = \frac{2}{\sqrt{\pi}} \int_{t}^{\infty} e^{t'^2} dt'$.
An example of such a fit is shown in the zoom of Figure \ref{fig:time}.
The fit is made in the range from \mbox{-100 ns} to
the time of the most populated bin plus 20 ns.
The arrival time at each PMT
is given by the fitted mean value of the gaussian distribution.
The fit is stable with respect to changes
in the fit range and histogram binning.

 \begin{figure}[!t]
  \vspace{3mm}
  \centering
  \includegraphics[width=2.8in]{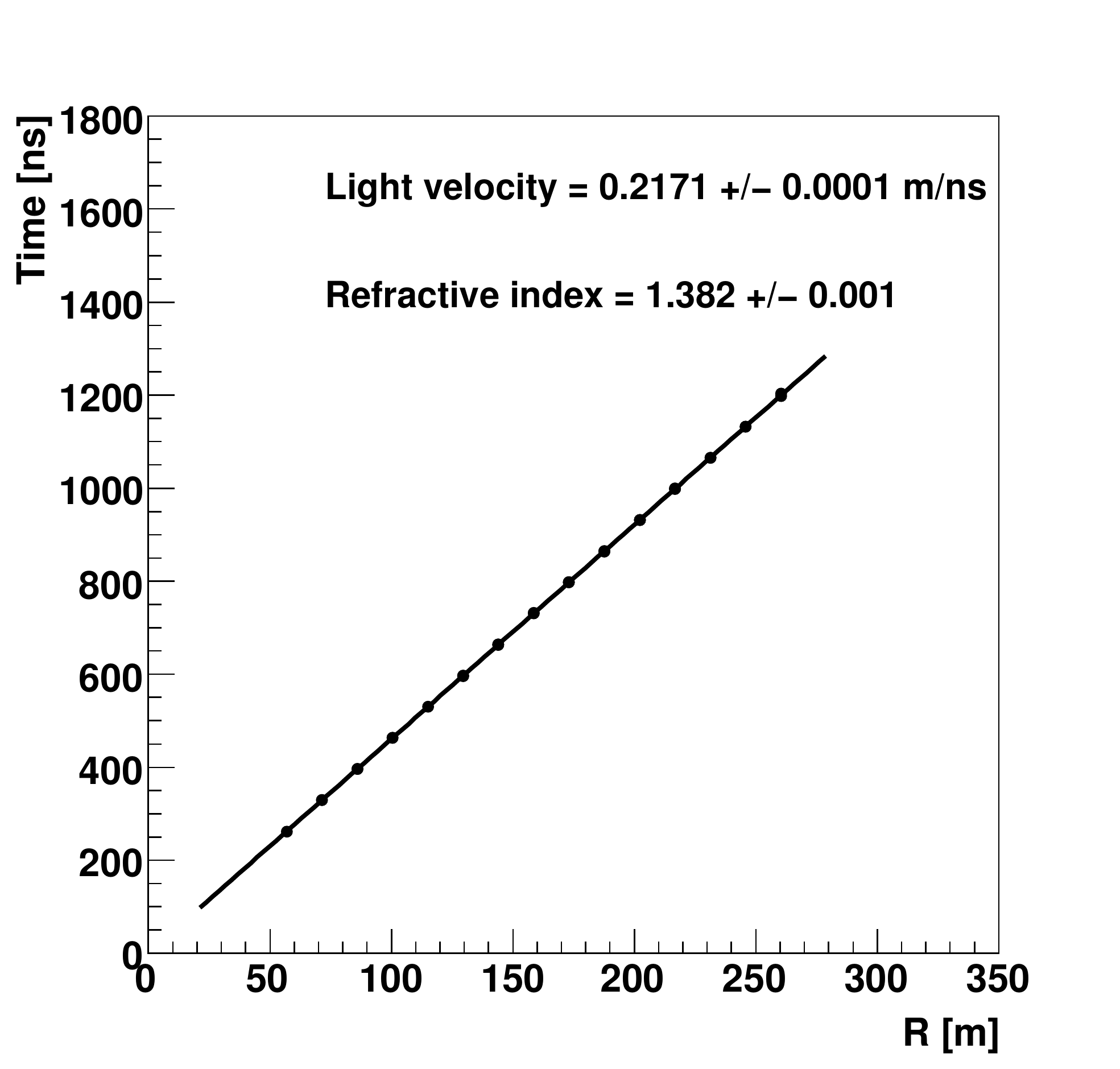}
  \caption {The figure shows the arrival time as a function of the distance between the LED and the different PMTs. The slope of a linear fit to the arrival time versus distance gives the inverse of the light velocity.}
  \label{fig:timevsr}
 \end{figure}

 \begin{figure}[!t]
  \vspace{3mm}
  \centering
  \includegraphics[width=3.0in]{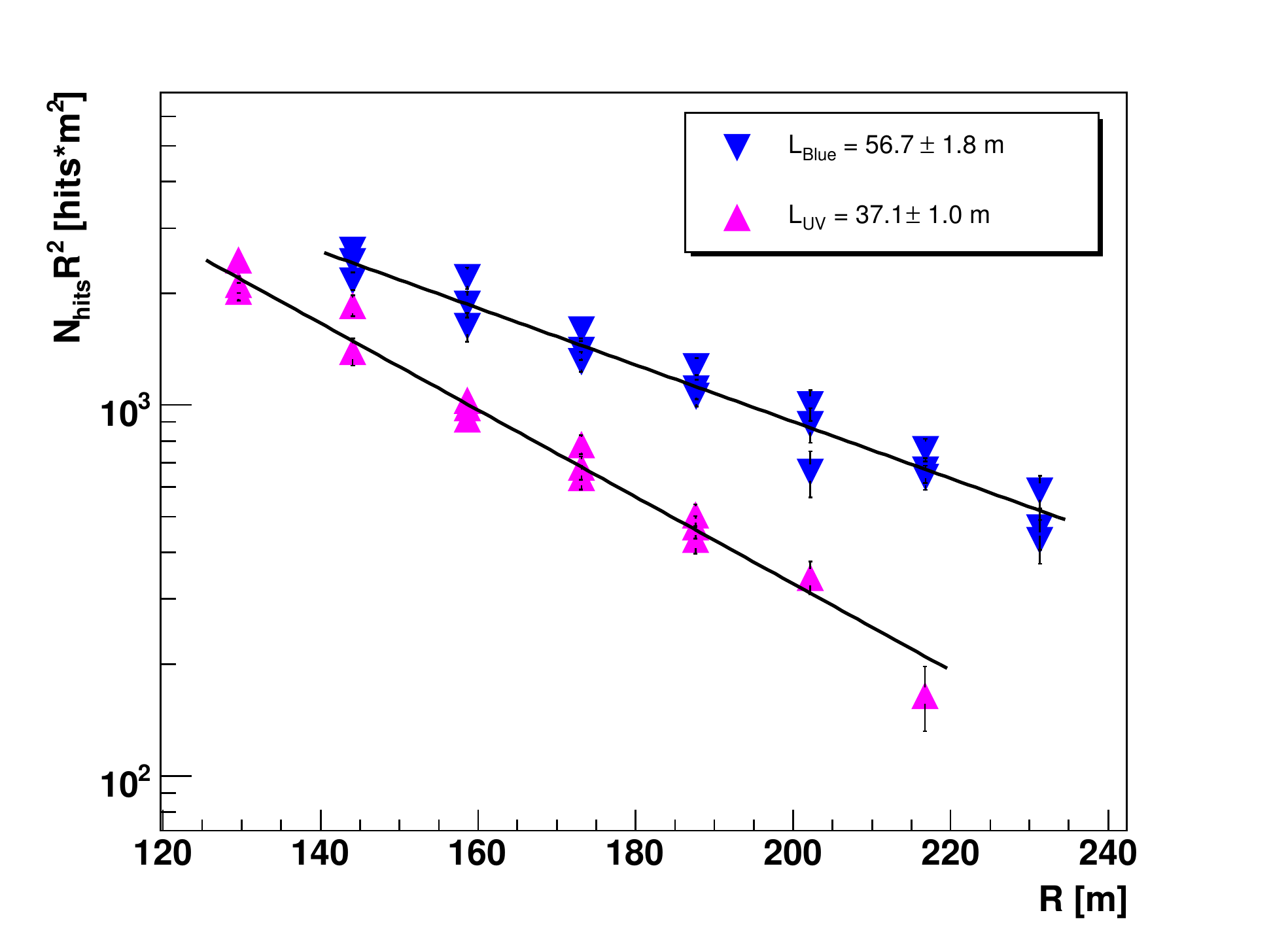}
  \caption {Amount of light collected by the PMTs as a function of the distance. The fit result of an exponential function is shown.}
  \label{fig:exp}
 \end{figure}

The distances between Optical Beacon and PMT versus the measured arrival
times are shown in Figure \ref{fig:timevsr} (note there can be up to
three PMTs at each distance).
Due to the small wavelength range of the individual LEDs the light dispersion is minimized.
By selecting only the PMTs
of the same line as the emission of light the uncertainty on the PMT angular acceptance and the PMT positions are reduced.
In this analysis, no correction of the PMT positions due to line movements have been applied.
The minimal distance is set to eliminate the PMT which
receives too much light and have an erroneous time estimation caused through
the early photon effect (for explanation see \cite{Timecalibration}) .
The minimal distance for the fit range is defined as the distance where the average collected
charge per hit (usually referred to as the amplitude) in the signal region
is below 1.5 photo-electrons.
The maximum distance is introduced due to avoid noise fluctuations.
The signal has to be significantly larger than the average background (above seven sigmas).
The slope of a
linear fit through the measured points gives the inverse of
the measured velocity of light in water ($v_{m}$).
The measured refractive index is defined as
\begin{equation}
n=c/v_{m}
\end{equation}
with $c=3 \cdot 10^8 m/s$.
The error of the refractive index given in Figure \ref{fig:timevsr} is the error estimated by the linear fit.

As seen in Figure \ref{fig:time} the PMTs perform a time resolved measurement of the collected light.
The amount of light detected depends on the attenuation length, whereas the shape of the 
arrival time distribution of the detected light is related to the photon path length distribution of the 
scattered photons.
For the attenuation length measurement a similar selection criteria as in the refraction index measurement is used.
An exponential fit to the collected charge as a function of the distance is shown in Figure \ref{fig:exp} 
for two runs with sources with wavelength of 470 nm (tagged as Blue) and wavelength of 400 nm (tagged as UV).
The attenuation length $L$ is obtained by
\begin{equation}
I\cdot R^2 \sim I_0 \cdot e^{\frac{-R}{L}},
\label{equ:intensity}
\end{equation}
where $I_0$ is the intensity at the source and $I$ the intensity detected by a
PMT at a distance $R$.

\section{Monte Carlo measurement}
\label{sec:mc}
The Monte Carlo simulation takes into account the geometry of the detector
and the optical properties of the water
(refraction index, scattering length and absorption length).
Monte Carlo generated time distributions
with a large variety of optical properties has been used to
analyze the time distributions and calculate
the refraction index and also to check the stability of the analysis method.
The analysis method was first validated with a Monte Carlo sample
without scattering at three
different refractive index.
The variation of the absorption length between 30 m and 120 m has
nearly no influence on the refraction index,
with variation of the refraction index of less than 0.1\%.
A variation in the scattering length between 20 m and 50 m produces an uncertainty of 0.3\%
in the measurement of the refractive index.

 \begin{figure}[!t]
  \vspace{3mm}
  \centering
  \includegraphics[width=3.2in]{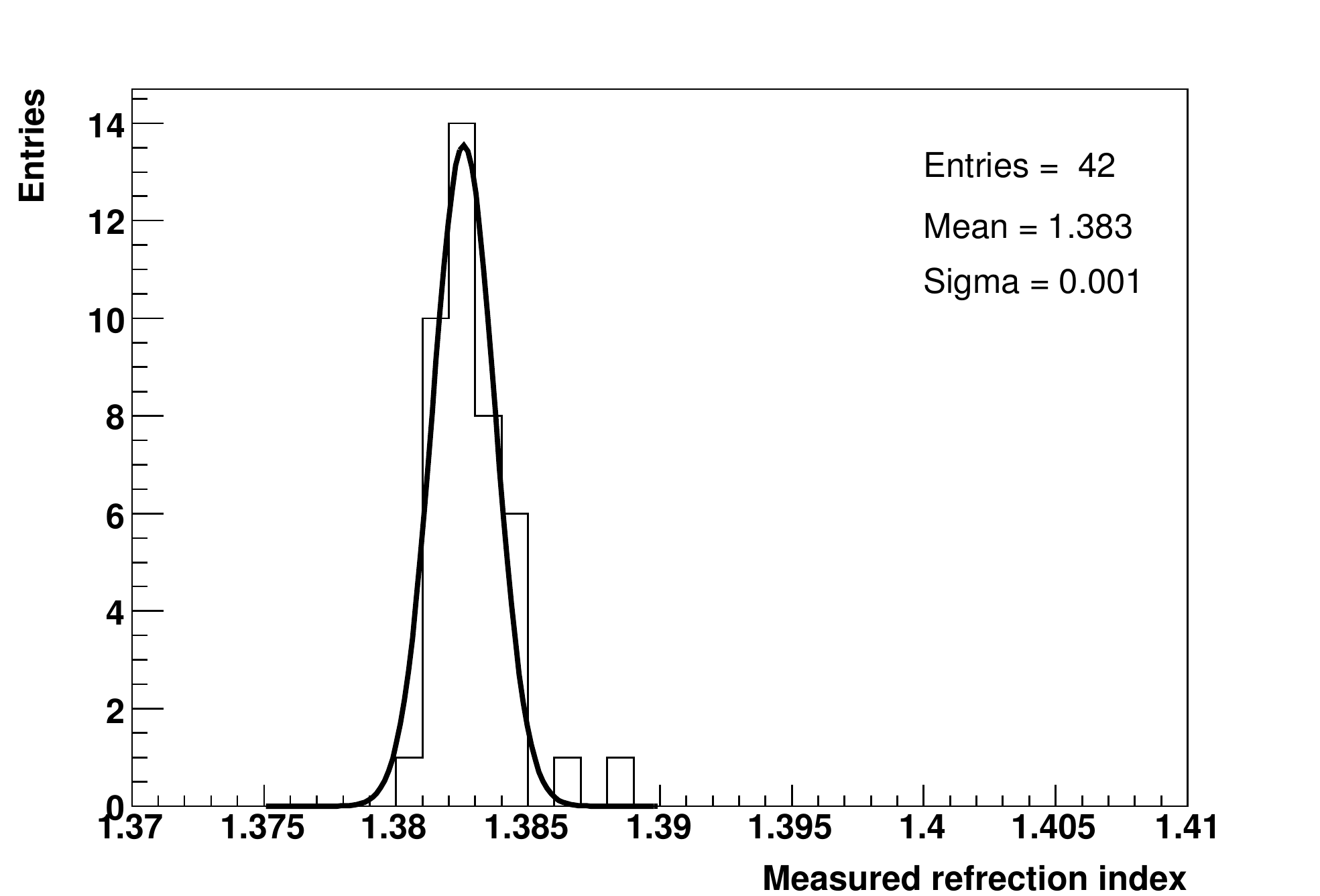}
  \caption {The measured refractive index for a total of 42 runs with an emission wavelength of \mbox{470 nm}. The result to a fit to a gaussian distribution is shown.}
  \label{fig:refdat}
 \end{figure}

\section{Data measurement}
\label{sec:data}
Between May 2008 and March 2010 a total of 42 runs were taken
with a nominal wavelength of 470 nm, 14 runs with 
a nominal wavelength of 400 nm
and 13 runs with a nominal wavelength of \mbox{532 nm} 
and have been analyzed according to the methods explained
in section \ref{sec:dataanalysis}.
The measured refraction of the 42 runs are shown
in the \mbox{Figure \ref{fig:refdat}} and the mean value is evaluated by fitting
the distribution with a gaussian.
The fitted attenuation lengths for some of these runs as function of the data period
are shown in Figure \ref{fig:stabil}. The fitted values are reasonably stable with time.

Since November 2010 runs with a modified
Optical Beacon have been collected
with light emission at six additional wavelengths with
nominal peak values of 385, 400, 440, 460, 505 and 518 nm.
First results are expected soon.

These measured refractive index with their
systematic errors estimated  in section \ref{sec:mc}
are shown in Figure \ref{fig:refr}.
Also shown is the parametric formula of the refractive index.
The refractive index at the ANTARES site depends on
the wavelength, on the temperature, on the salinity of the water 
and the pressure at the depth of the detector.
The parametric formula for the phase refractive index
of Quan and Fry \cite{Quanfry}, based on data from Austin and
Halikas \cite{Austin}, is modified with appropriate pressure corrections
as suggested in \cite{Agui}.
The phase refractive index for sea water as a
function of wavelength, temperature ($T$), salinity ($S$) and pressure ($p$)
is given by
\begin{eqnarray}
n_p (340<\lambda (nm)<560, S(\%)=3.844, \quad \quad \quad \quad  \nonumber \\ 
T(^oC)=13.2,200<p(atm)<240)= ~~~~~ \quad   \nonumber\\
1.32292+(1.32394-1.32292) \cdot \frac{p-200}{240-200}+ ~~ \quad   \nonumber \\
\frac{16.2561}{\lambda}-\frac{4382}{\lambda^2}+\frac{1.1455\cdot10^6}{\lambda^3}, \quad \quad 
\label{equ:nphase}
\end{eqnarray}
where $\lambda$ is the wavelength of light.

The group refractive index ($n$) \cite{Danil} is related
to its phase refractive index ($n_p$) through
\begin{equation}
n=\frac{n_p}{1+\frac{\lambda}{n_p}\frac{d n_p}{d \lambda}}.
\label{equ:n_g}
\end{equation}

The parametrization of the 
refractive indexes $n$ and $n_p$ is shown 
in Figure \ref{fig:refr} for the given values of
temperature, salinity and
for a pressure between 200 atm 
and 240 atm.
The measurements are in agreement with the parametrization of the group refractive index.

 \begin{figure}[!t]
  \vspace{7mm}
  \centering
  \includegraphics[width=3.5in]{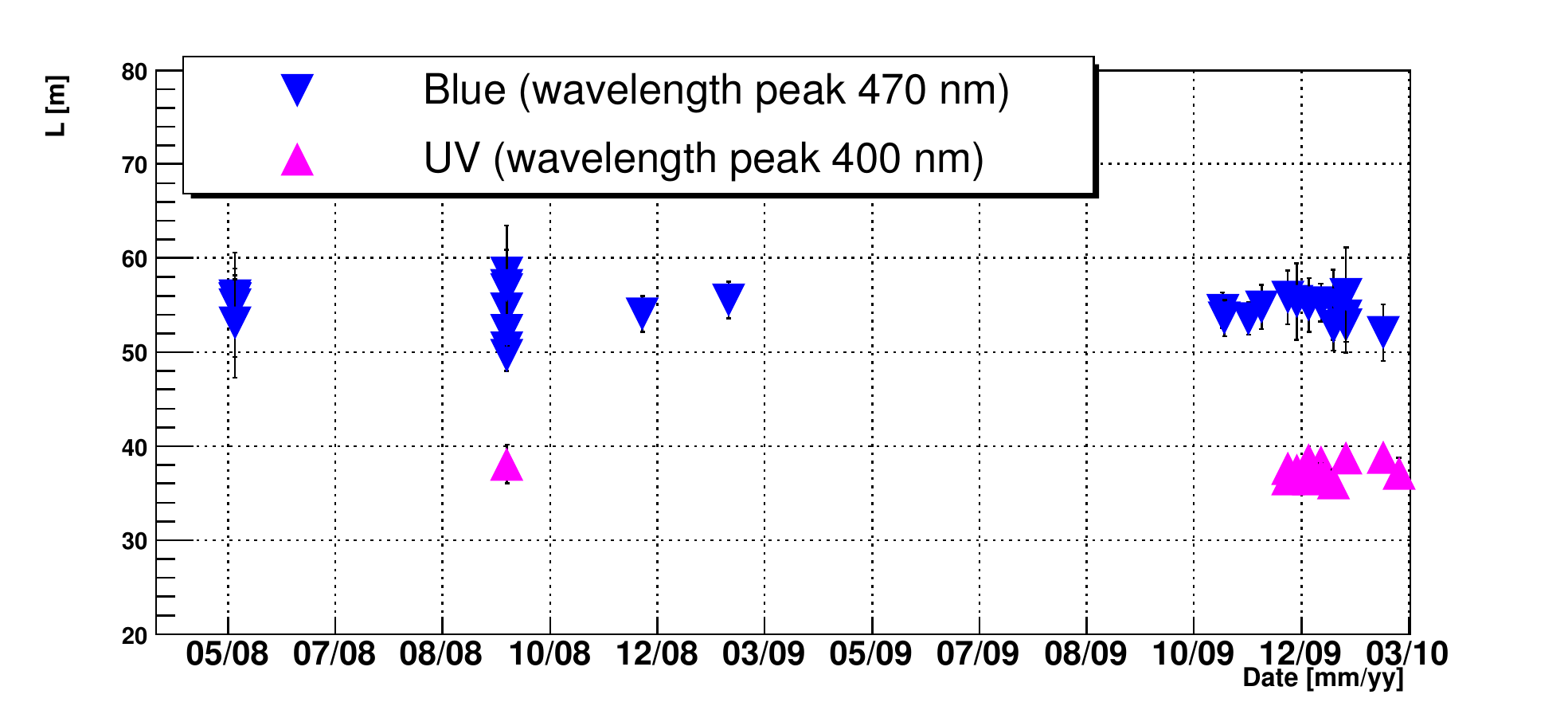}
  \caption {Time evolution of the attenuation length parameter for one and a half year of data taking for two different wavelengths.}
  \label{fig:stabil}
 \end{figure}

 \begin{figure}[!t]
  \vspace{-1mm}
  \centering
  \includegraphics[width=2.8in]{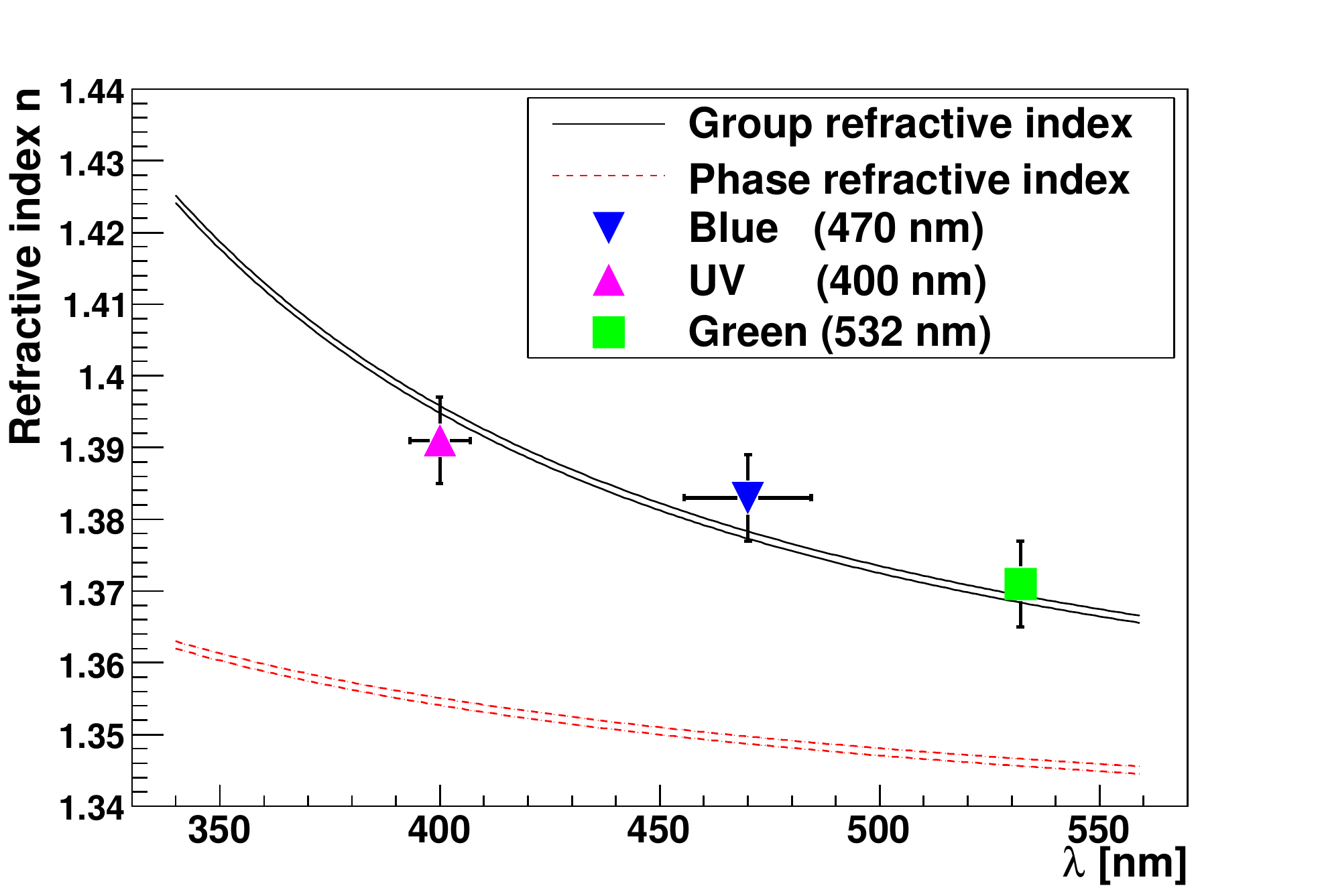}
  \caption {Group and phase refractive index as a function of the wavelength for a given temperature and salinity with measured data points and its systematical error bars. The upper curves (solid) show the group refractive index, the lower curves (dotted) give the phase refractive index.}
  \label{fig:refr}
 \end{figure}

\section {Conclusion}
\label{conclusion} 

Pulsed light sources with wavelengths between 400 nm and 532 nm
shining through sea water and the time of light distribution detected
by the PMTs at distances between few tens and few hundred meters from the
source have been used to measure the refraction index. Dedicated
Monte Carlo simulation has been used to validate the analysis method
and to evaluate the systematics.
The data results are compatible with the parametrization of the group refractive index.

\section*{Acknowledgments}
I gratefully acknowledge the support of the JAE-Doc postdoctoral programme of CSIC.
This work has also been supported by the following
Spanish projects: FPA2009-13983-C02-01, MultiDark Consolider CSD2009-00064, ACI2009-1020 of MICINN and Prometeo/2009/026 of Generalitat Valenciana.

\clearpage

\setcounter{figure}{0}
\setcounter{table}{0}
\setcounter{footnote}{0}
\setcounter{section}{0}
\newpage




\title{Moon shadow observation with the ANTARES neutrino telescope}


\shorttitle{ANTARES Collaboration, Moon shadow observation with ANTARES}

\authors{Colas Rivi\`ere$^{1}$, Carla Distefano$^2$, for the ANTARES Collaboration}
\afiliations{$^1$CPPM, Aix-Marseille Universit\'e, CNRS/IN2P3, Marseille, France\\
$^2$INFN - Laboratori Nazionali del Sud (LNS), Catania, Italy}
\email{riviere@in2p3.fr}

\maketitle
\begin{abstract}
The ANTARES neutrino telescope is operating in the Mediterranean sea in its full configuration since May 2008. While designed to observe up-going neutrinos, it also records many signals of down-going muons produced by the interaction of high energy cosmic rays in the atmosphere. The shadowing of cosmic rays by the Moon produces a deficit of muons coming from this direction at the ANTARES level. The observation of this deficit of events can be used to check the correct pointing of the detector with respect to a known object.

The strategies to observe this Moon shadow are discussed ant the current results are presented.
\end{abstract}



\section{Introduction}

The ANTARES neutrino telescope is operating in the Mediterranean Sea in full 12 lines configuration since May 2008~\cite{NIM2011}. Thanks to the good optical properties of the water of the deep sea, the median angular resolution obtained with the current reconstruction algorithm is estimated to be $0.4^\circ$ for cosmic neutrinos with an $E^{-2}$ flux~\cite{AART}.


The pointing performance of the detector relies on the knowledge of parameters, such as the relative delays of the optical sensors within each detector line as well as between the lines, the instantaneous shape of the detector lines which is influenced by the sea current or the absolute orientation of the detector.

 \begin{figure}
  \begin{center}
  \includegraphics[width=\columnwidth]{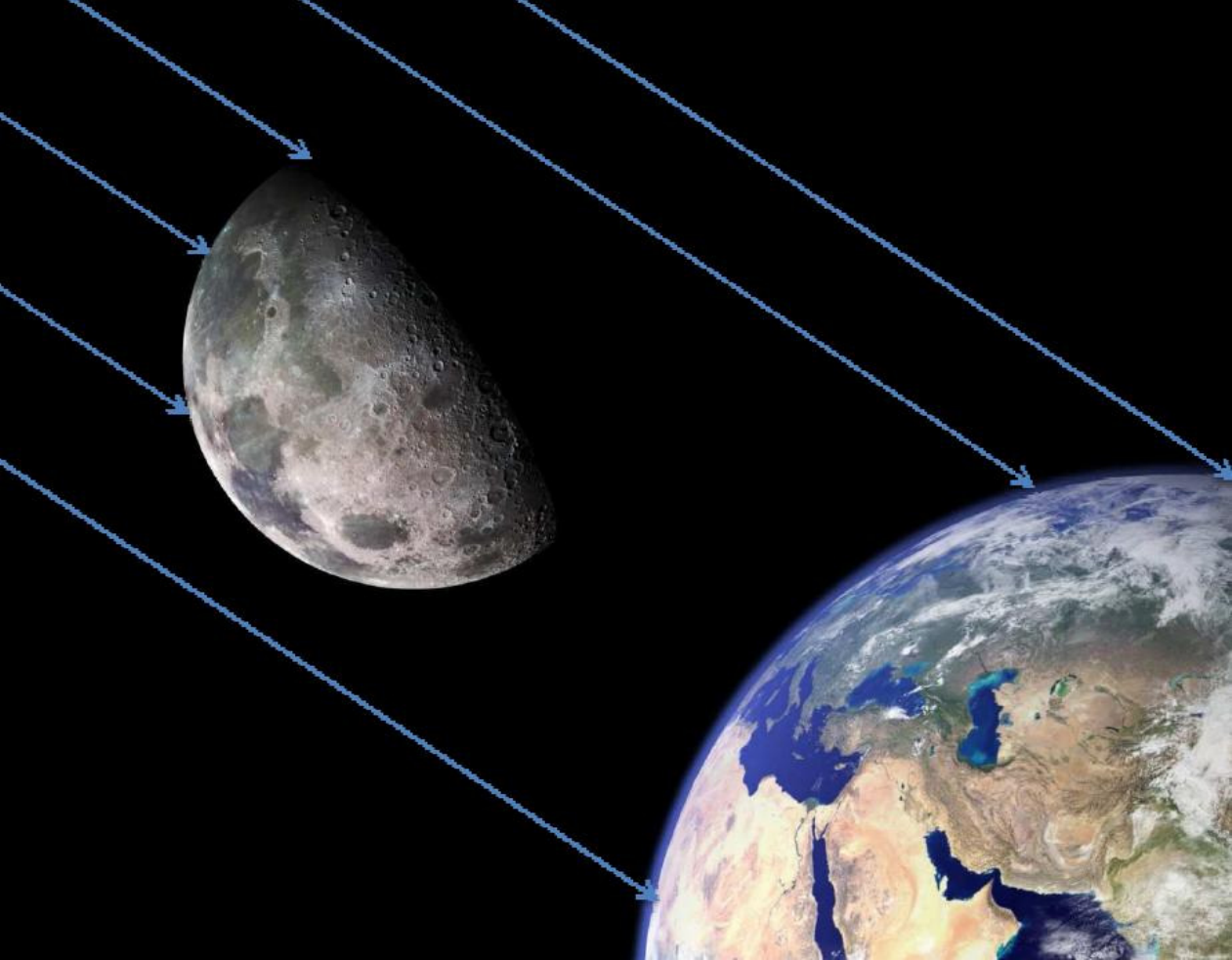}
  \caption{Illustration of the Moon blocking the cosmic rays, thus producing a deficit of muons originating from this direction.}
  \label{principle}
  \end{center}
 \end{figure}
The values of these parameters have been calibrated before immersion of the detector lines and are regularly measured in situ during operation. The relative time delays are measured using a laser beacon and LED beacons, and also with physical quantities such as the $^{40}$K decay or the atmospheric muon tracks~\cite{AP_Time}. The alignment measurements are performed continuously with acoustic emitters and transceivers, tiltmetres and compasses~\cite{NIM_Positionning}. The absolute orientation of the telescope is obtained by the triangulation of acoustic signals between anchors of the lines and the deployment vessel at the sea surface, positioned by GPS~\cite{garo}. The uncertainties on these values are small enough to ensure the optimal performances of the telescope, but it remains very important to verify the correct operation of the detector using a calibration source. While no such source exists, some possibilities remain:
\begin{itemize}
  \item The observation of a coincidence between an air shower at sea level with a surface array and the resulting muons at ANTARES level~\cite{surface}.
  \item The observation of the Moon shadow. This technique, first proposed in 1957~\cite{clark}, relies on the absorption of cosmic rays by the Moon, as illustrated on figure~\ref{principle}. As the cosmic rays are the source of the down-going muons at the telescope level, this absorption induces a deficit of the number of observed muons  in the Moon direction.
\end{itemize}


\section{Data and Monte Carlo}

\subsection{Data set and reconstruction}

The apparent Moon radius is on average $R_{Moon}=0.259^\circ$, which is smaller than the resolution of ANTARES, in particular for down-going atmospheric muons; thus the observation of the Moon shadow requires large statistics. For the work reported here, 884 days of data taken between 2007 and 2010 (including periods with 5, 10 and 12 lines configurations) have been analysed.
 \begin{figure}[t]
  \begin{center}
  \includegraphics[width=\columnwidth]{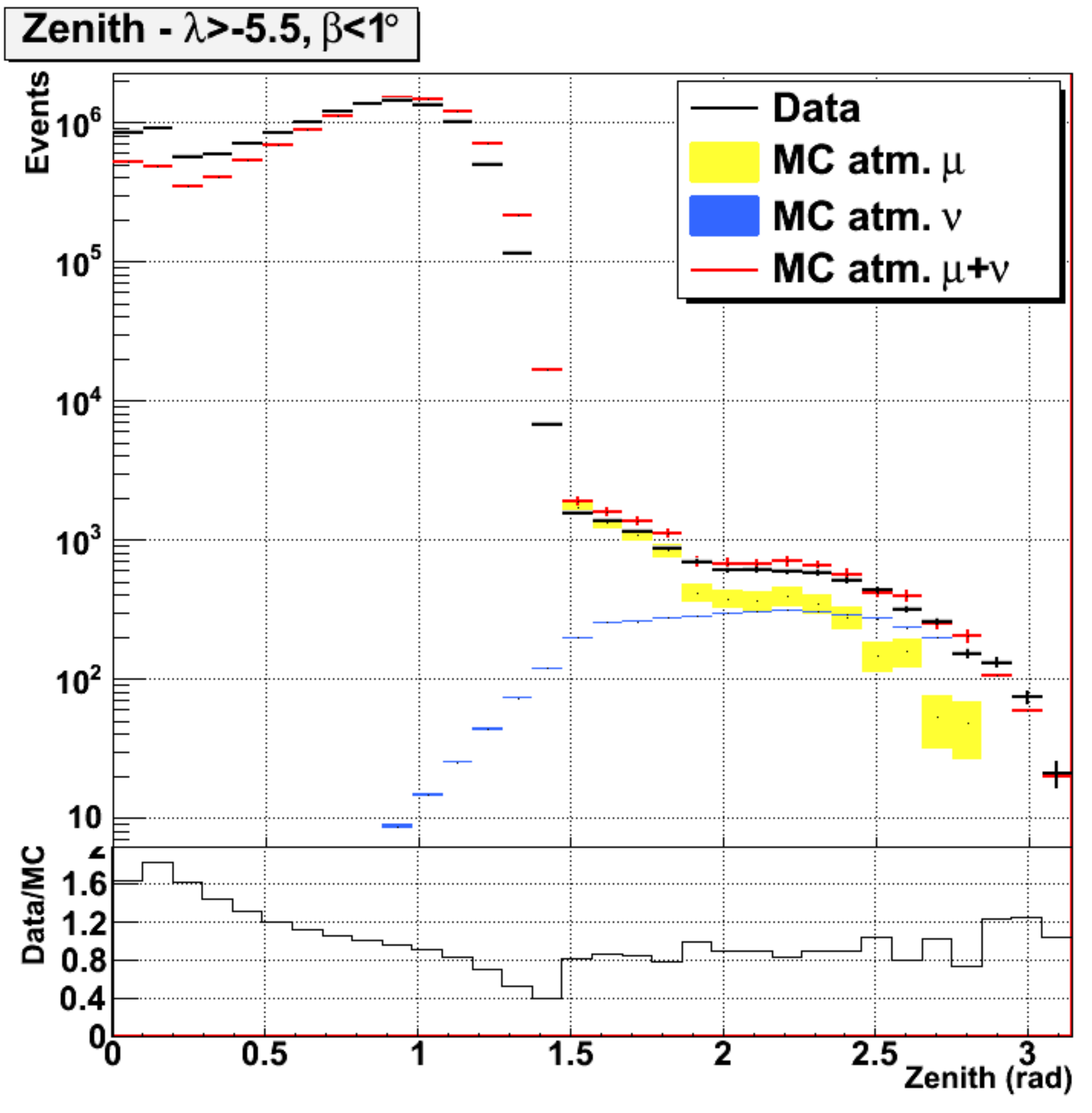}
  \caption{Top: Distribution of the reconstructed zenith of the events of the data set used in this analysis: data (black), Monte Carlo (red=yellow+blue). Bottom: Data/MC ratio.}
  \label{dataMC}
  \end{center}
 \end{figure}

These data have been processed using the standard track reconstruction algorithm of ANTARES. As this algorithm is optimised for up-going tracks of cosmic origin, an algorithm dedicated to the reconstruction of down-going atmospheric muons would improve the results presented in this paper.

The event selection criteria used here are the same as presented in~\cite{CarlaHEP}, i.e. an error estimate $\beta<1^\circ$ and a quality of reconstruction of the track $\lambda>-5.5$.

\subsection{Monde Carlo}

The point spread function (PSF) used to generate pseudo experiments and to compute our search strategy is obtained from Monte Carlo simulation (MC).
Each data run is simulated as close as possible to the experimental condition (same detector condition, bioluminescence rate, etc.), generating both atmospheric muons and atmospheric neutrinos. An example of data/MC comparison is represented in figure~\ref{dataMC}.

The PSF is extracted from this MC using all the events reconstructed in a region of $10^\circ$ around the Moon position (figure~\ref{PSFPDF}). With the chosen selection criteria, the median angular resolution is $0.75^\circ$.
 \begin{figure}[t]
  \begin{center}
  \includegraphics[width=\columnwidth]{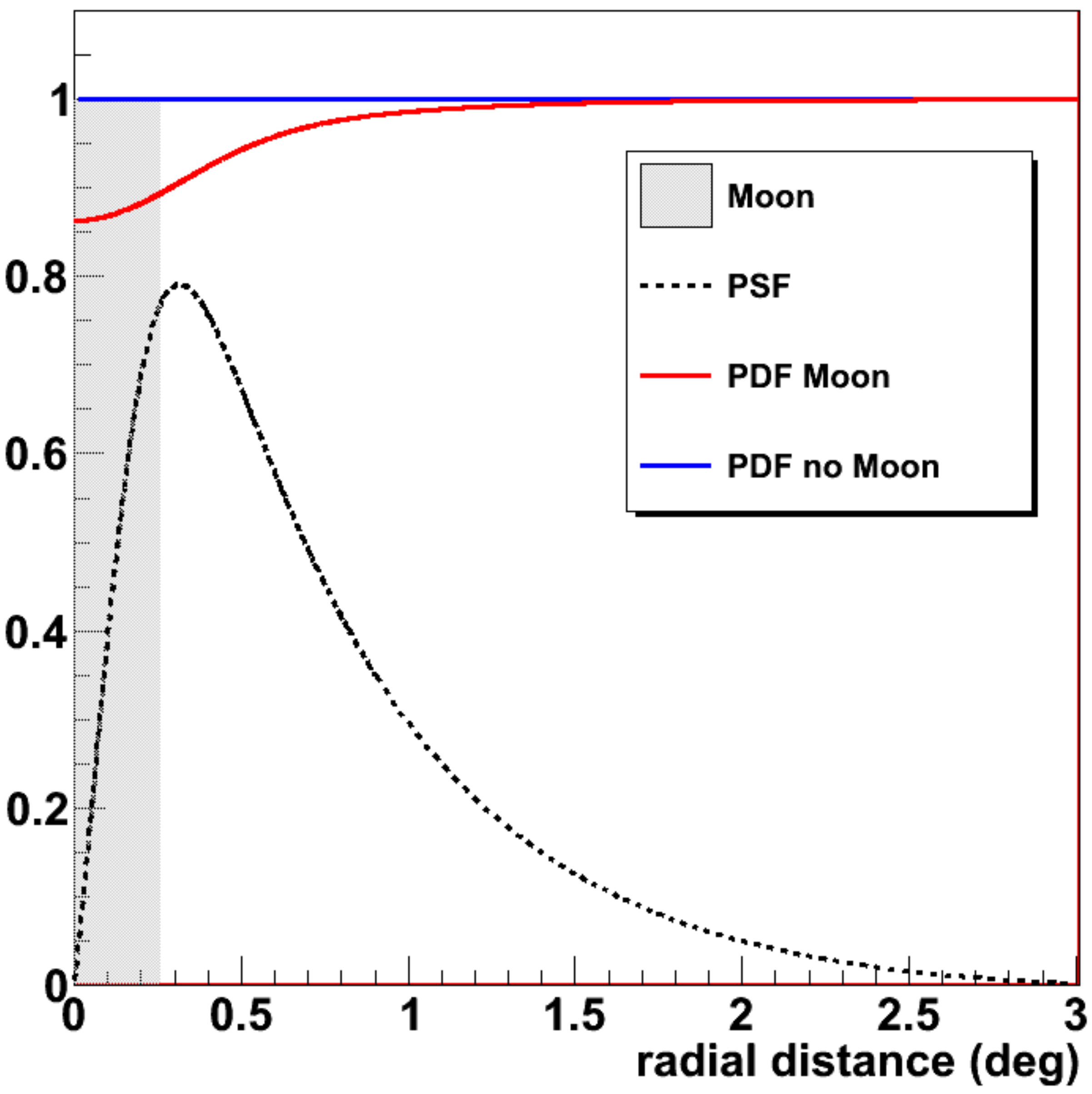}
  \caption{Black: Parametrization of the PSF of the events at less than $10^\circ$ from the Moon position. Blue: Constant PDF in the $H_0$ hypothesis. Red: PDF in the $H_1$ hypothesis.}
  \label{PSFPDF}
  \end{center}
 \end{figure}

Concerning the shape of the event density in the Moon region, we assume the event probability density function (PDF) is uniform in the abscence of Moon\footnote{If we consider a $10^\circ$ radius region around the Moon direction, there is actually a counting rate modulation of $\pm20\%$, but this is a dipolar modulation which cancels out during the likelihood computation, so it is neglected for simplicity. Only second or higher order modulations could change the likelihood.}, we call this hypothesis $H_0$. The PSF in the Moon hypothesis ($H_1$) is obtained by computing the 2D convolution product of the PSF with the Moon shape:
 \begin{figure*}[t]
  \begin{center}
  \includegraphics[width=0.8\textwidth]{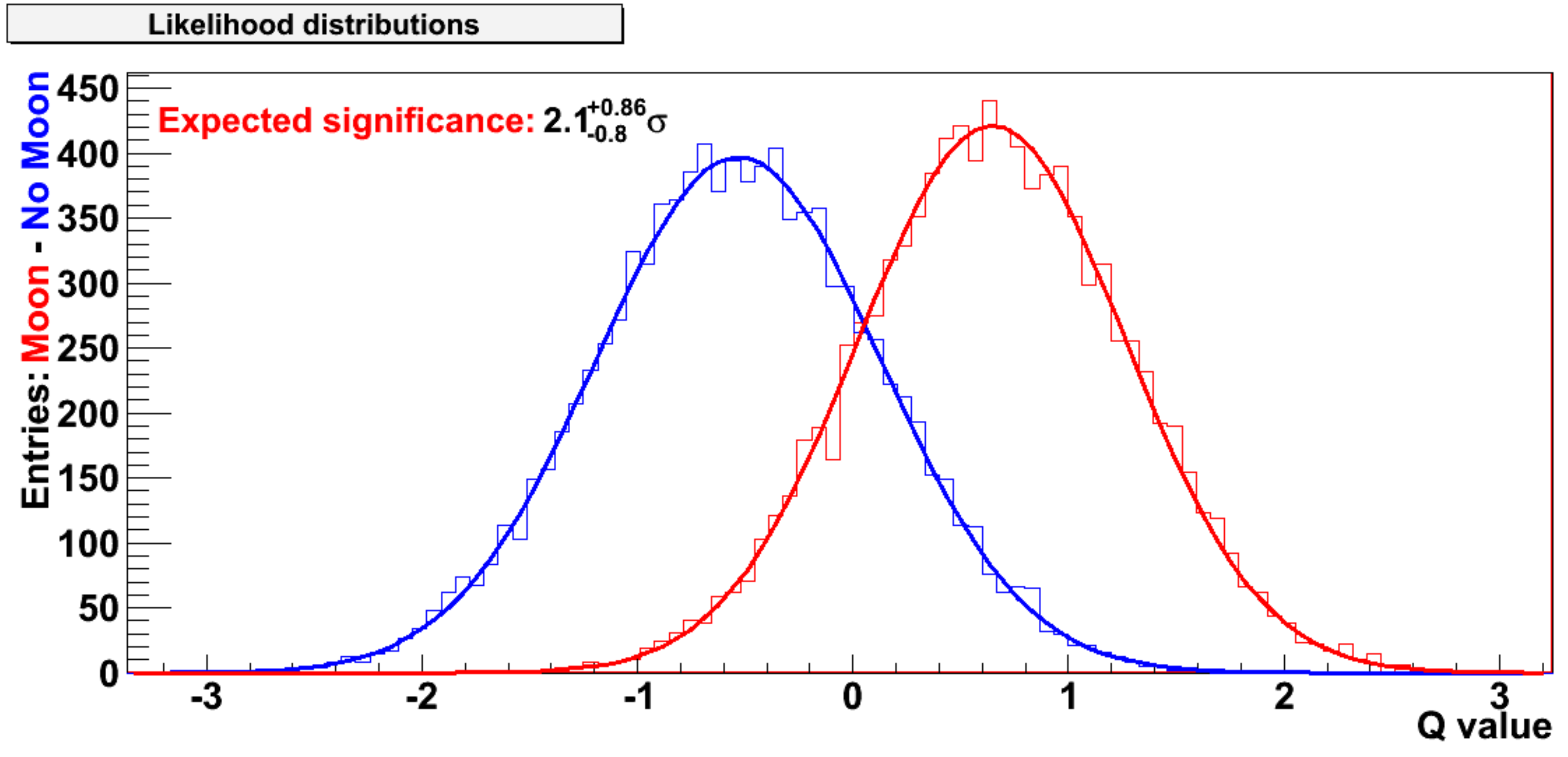}
  \caption{Distributions of the likelihood test function values, for the $H_0$ hypothesis in blue and for $H_1$ in red.}
  \label{like_pe}
  \end{center}
 \end{figure*}
\begin{gather*}
  PDF(x|H_0)\propto 1\\
  PDF(x|H_1)\propto (1-PSF\otimes \delta_{Moon})(x)
\end{gather*}

\section{Search strategy}
Different search strategies have been studied, a simple on/off search, a $\chi^2$ test of histograms~\cite{CarlaHEP} and the one presented here, a likelihood search. The results of these different approaches are compatible.

\subsection{Likelihood method}
For the likelihood search we build a test function based on the likelihood ratio using the expected distributions corresponding to the Moon hypothesis $H_1$ and no Moon hypothesis $H_0$.
The test function used is the logarithm of the likelihood ratio:
\begin{equation*}
t=\displaystyle\sum\limits_{events\ i} \log \frac{PDF(r_i|H_1)}{PDF(r_i|H_0)}
\end{equation*}

\subsection{Pseudo-experiments}
In order to test the search method, to estimate its power and to compute the significance, pseudo-experiments are performed.
After generating uniform event distributions in an area around the Moon position, a smearing is applied using the PSF, followed by a shadowing which is or is not applied for the events originating from the Moon direction. The pseudo experiments presented here are generated with the same event statistics as available in the 2007-2010 data sample.

 \begin{figure}[t]
  \begin{center}
  \includegraphics[width=\columnwidth]{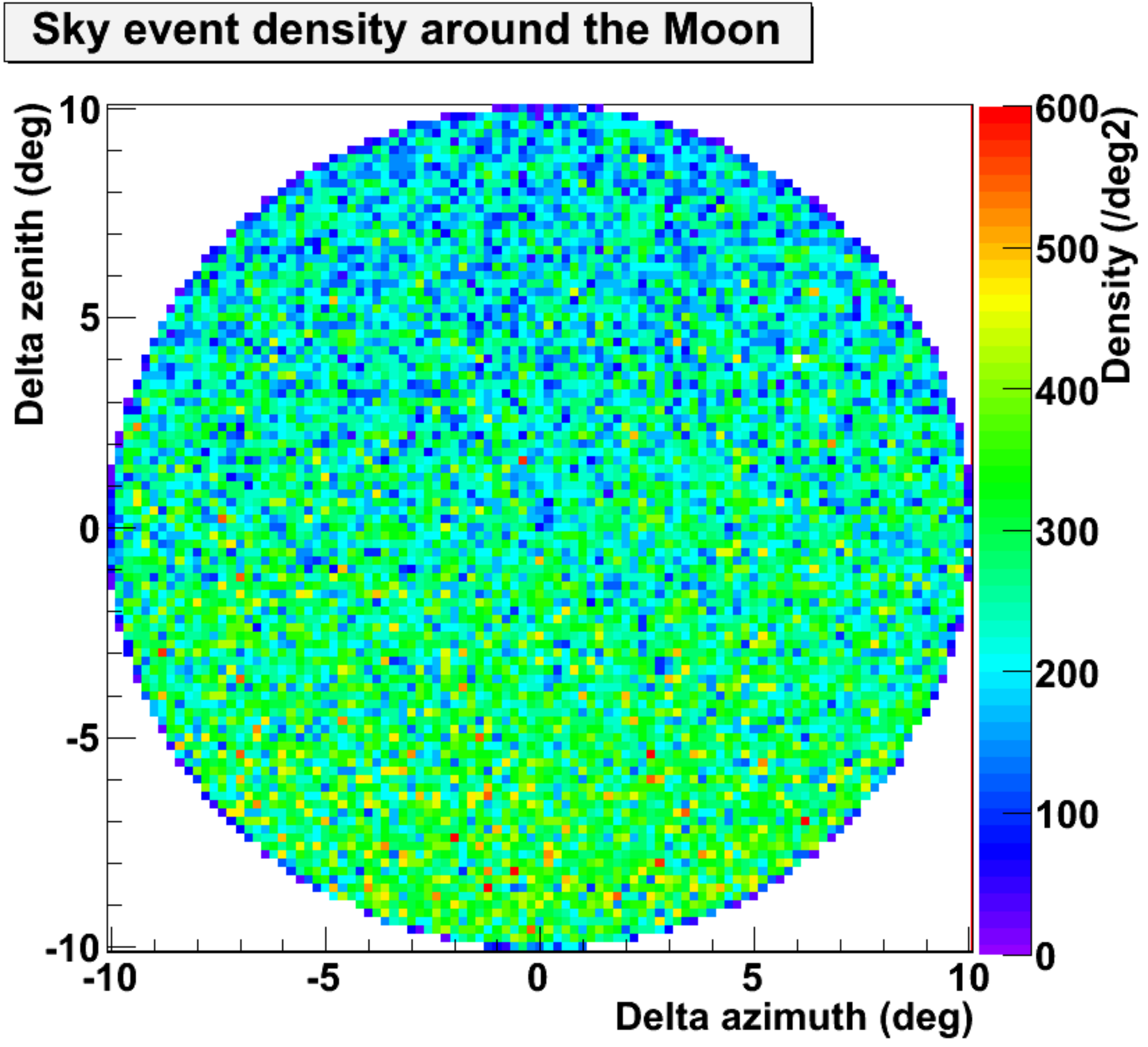}
  \caption{Event density around the Moon location $(0,0)$.}
  \label{density2D}
  \end{center}
 \end{figure}  
The distributions of the values of the test function obtained with the pseudo experiments are presented in figure~\ref{like_pe}. For a given value of the test function, its p-value (probability to obtain a value at least as important from a background fluctuation) is computed using the $H_0$ distribution (blue). The significance of this p-value can then equivalently be expressed in term of sigma.

The $H_1$ distribution (red) is used to estimate the power of the test. With the current event statistics, we could expect $2.1^{+0.9}_{-0.8}\sigma$. With 2, 5 and 10 times this statistics, we obtain respectively $2.8\pm0.9$, $4.2\pm0.9$ and $5.9\pm0.9\sigma$.

\section{Preliminary results}

The experimental event density in the Moon region is represented in figure~\ref{density2D} and the density as a function of the distance to the Moon position is shown on figure~\ref{density1D}.

 \begin{figure}[t]
  \begin{center}
  \includegraphics[width=\columnwidth]{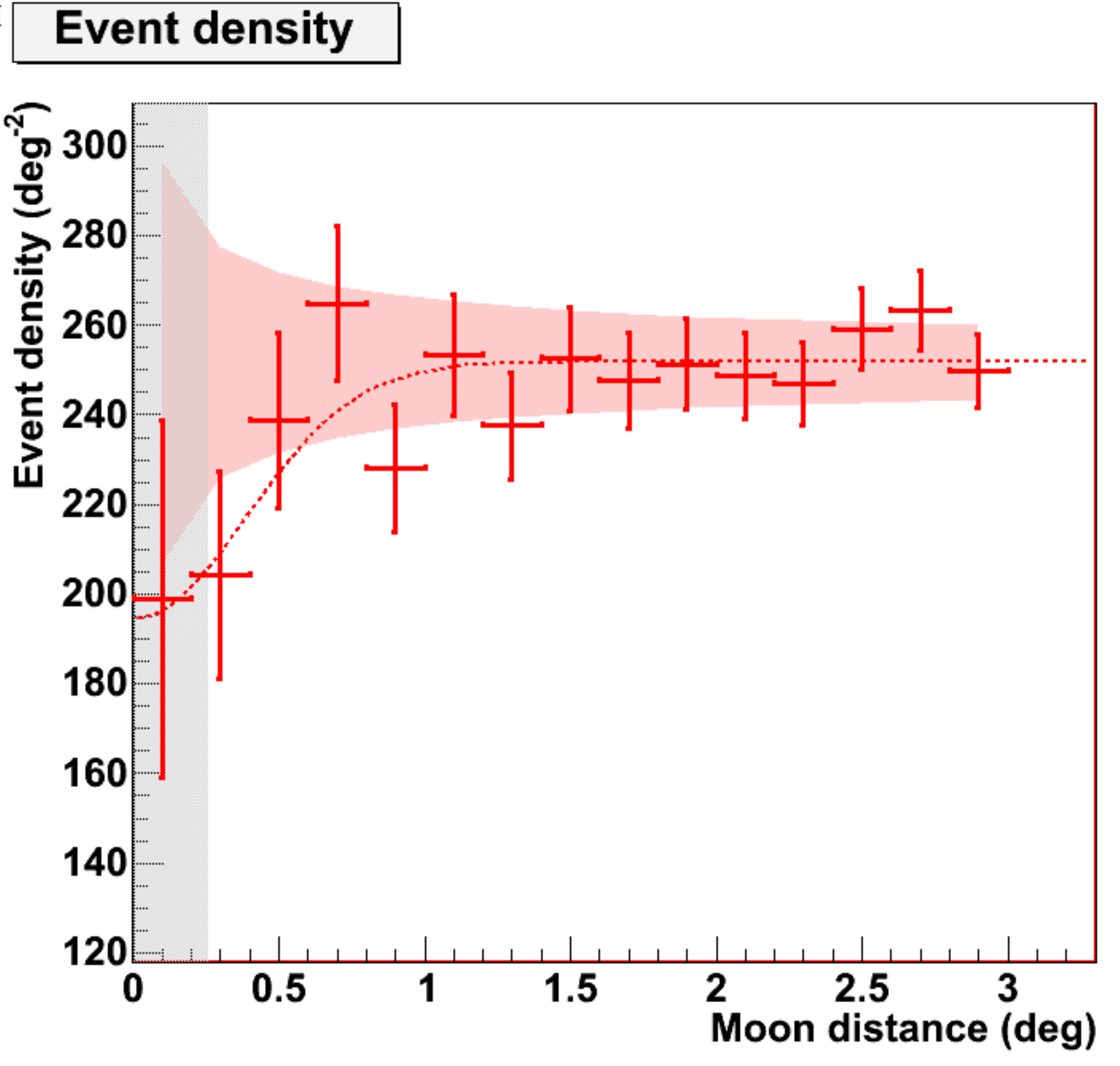}
  \caption{Event density as a function of the distance to the Moon.}
  \label{density1D}
  \end{center}
 \end{figure}

The value of the test function obtained experimentally for the Moon position is $Q=1.17$, which has a p-value of $0.007$. This corresponds to $2.7\sigma$, which is within the expected range. This is however not significant enough to unambiguously reject the $H_0$ assumption.

 \begin{figure}
  \begin{center}
  \includegraphics[width=\columnwidth]{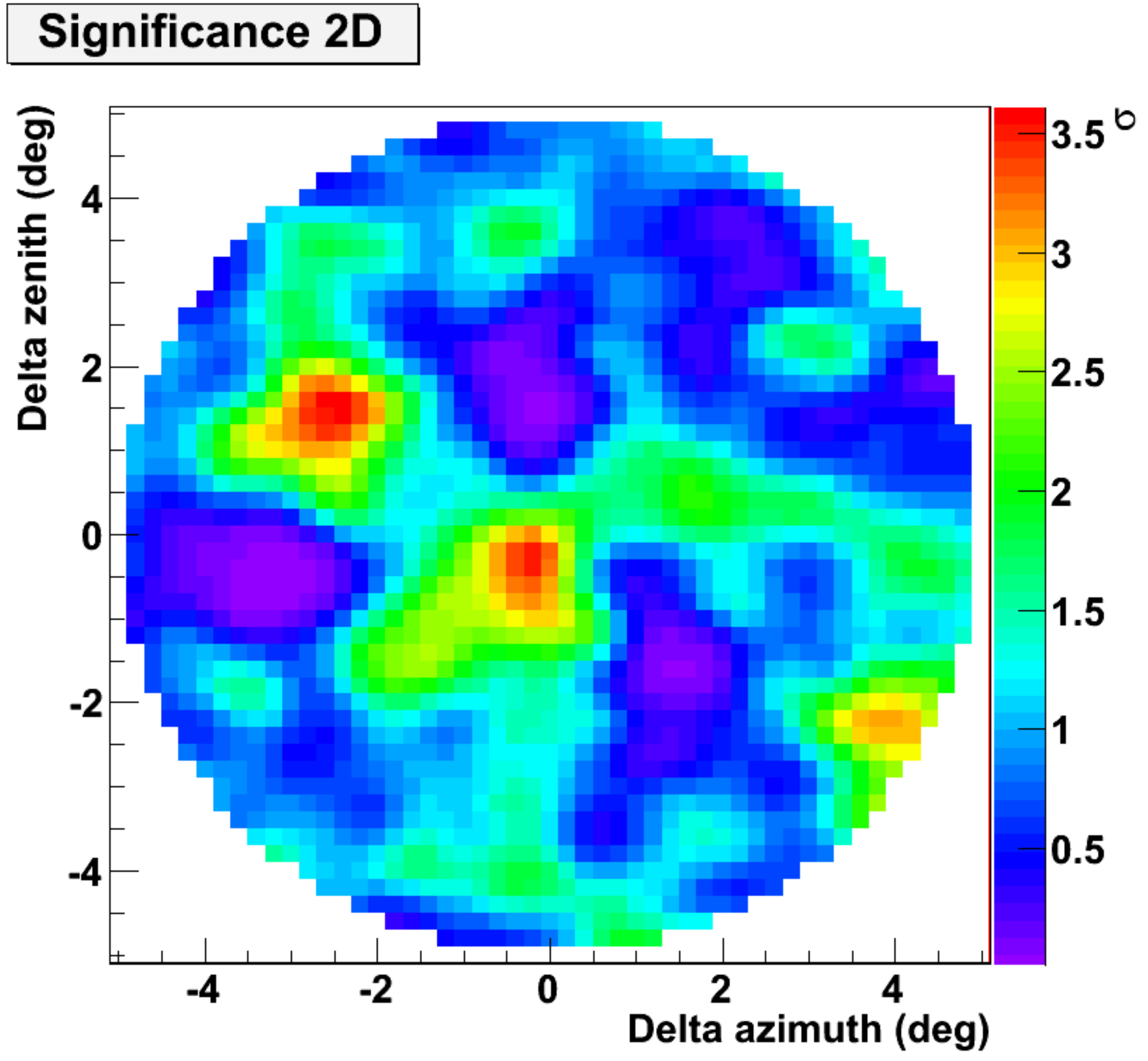}
  \caption{Significance of the search around the Moon location $(0,0)$.}
  \label{like2D}
  \end{center}
 \end{figure}

Scanning over directions around the Moon position (see figure~\ref{like2D}), we find similar hot spots; which are compatible with statistical fluctuations.

\section{Conclusion and outlook}
A search for the observation of the Moon shadow with the ANTARES  neutrino telescope has been performed on a dataset corresponding to 884 days of livetime. The resulting Moon shadow significance, $2.7\sigma$, is within the expected range and compatible with other analysis~\cite{CarlaHEP}. At this point, this result is not strong enough to put interesting constrains on ANTARES pointing capabilities.

The increase of statistics or the development of a track reconstruction algorithm optimized for down-going atmospheric muons should improve this results.

ANTARES will also perform campaigns of measurement of atmospheric showers by mean of a surface array installed on a boat above the detector, allowing to compare the direction of muons in coincidence with ANTARES. This will provide an additional check of the ANTARES pointing performance.

\clearpage

\setcounter{figure}{0}
\setcounter{table}{0}
\setcounter{footnote}{0}
\setcounter{section}{0}
\newpage




\title{Status and Recent Results of the Acoustic Neutrino Detection Test System AMADEUS} 


\shorttitle{Lahmann \etal Acoustic Detection with AMADEUS}

\authors{Robert Lahmann$^{1}$ for the ANTARES Collaboration}
\afiliations{$^1$Erlangen Centre for Astroparticle Physics (ECAP)}
\email{robert.lahmann@physik.uni-erlangen.de}

\maketitle
\begin{abstract} 
The AMADEUS system is an integral part of the ANTARES neutrino
telescope in the Mediterranean Sea. The project aims at the
investigation of techniques for acoustic neutrino detection in the
deep sea. Installed at a depth of more than 2000\,m, the acoustic
sensors of AMADEUS are based on piezo-ceramics elements for the
broad-band recording of signals with frequencies ranging up to 125kHz.
AMADEUS was completed in May 2008 and comprises six ``acoustic
clusters'', each one holding six acoustic sensors that are arranged at
distances of roughly 1m from each other. The clusters are installed
with inter-spacings ranging from 15\,m to 340\,m.
Acoustic data are continuously acquired and processed at a computer
cluster where online filter algorithms are applied to select a
high-purity sample of neutrino-like signals.  In order to assess the
background of neutrino-like signals in the deep sea, the
characteristics of ambient noise and transient signals have been
investigated.
In this article, the AMADEUS system will be described and recent
results will be presented.
\end{abstract}




\section{Introduction}
\label{sec:intro}

Measuring acoustic pressure pulses in huge underwater acoustic arrays
is a promising approach for the detection of cosmic neutrinos with
energies exceeding 100\,PeV.  The pressure signals are produced by the
particle showers that evolve when neutrinos interact with nuclei in
water.
The resulting energy deposition in a cylindrical volume of a few
centimetres in radius and several metres in length leads to a local
heating of the medium which is instantaneous with respect to the
hydrodynamic time scales.  This temperature change induces an
expansion or contraction of the medium depending on its volume
expansion coefficient.  According to the thermo-acoustic
model~\cite{bib:Askariyan2,bib:Learned}, the accelerated expansion of the
heated volume---a micro-explosion---forms a pressure pulse of bipolar
shape which propagates in the surrounding medium.
Coherent superposition of the elementary sound waves, produced over the
volume of the energy deposition, leads to a propagation within a flat
disk-like volume (often referred to as {\em pancake})
in the direction perpendicular to the axis of the particle shower.
After propagating several hundreds of metres in sea water, the pulse
has a characteristic frequency spectrum that is expected to peak
around 10\,kHz~\cite{bib:Sim_Acorne,bib:Sim_Acorne2,bib:Bertin_Niess}.
As the attenuation length in sea water in the relevant frequency range
is about
 one to two orders of magnitude larger than that for visible light,
a potential acoustic neutrino detector would require
a less dense instrumentation of a given volume
than an optical neutrino telescope.

The AMADEUS project~\cite{bib:amadeus-2010} 
was conceived to perform a feasibility study for a
potential future large-scale acoustic neutrino detector. For this purpose, 
a dedicated array of acoustic sensors was integrated into the
ANTARES neutrino telescope~\cite{bib:ANTARES-paper}. 
In the following, the AMADEUS device will be described and recent results
will be presented.

\section{The ANTARES Detector}
\label{sec:antares_detector}

The ANTARES neutrino telescope was designed to detect neutrinos by
measuring the Cherenkov light emitted along the tracks of relativistic
secondary muons generated in neutrino interactions.
A sketch of the detector, with the AMADEUS modules highlighted, is
shown in Figure~\ref{fig:ANTARES_schematic_all_storeys}.  The detector
is located in the Mediterranean Sea at a water depth of about 2500\,m,
roughly 40\,km south of the town of Toulon at the French coast at the
geographic position of 42$^\circ$48$'$\,N, 6$^\circ$10$'$\,E.  ANTARES was
completed in May 2008 and comprises 12 vertical structures, the {\em
  detection lines}.  Each detection line holds up to 25 {\em storeys}
that are arranged at equal distances of 14.5\,m along the line,
starting at about 100\,m above the sea bed and interlinked by
electro-optical cables.  A standard storey consists of a titanium
support structure, holding three {\em Optical Modules}
(each one consisting of a photomultiplier tube inside a
water-tight pressure-resistant glass sphere) and one 
cylindrical electronics container

\begin{figure}[tb]
\centering
\includegraphics[width=7.0cm]{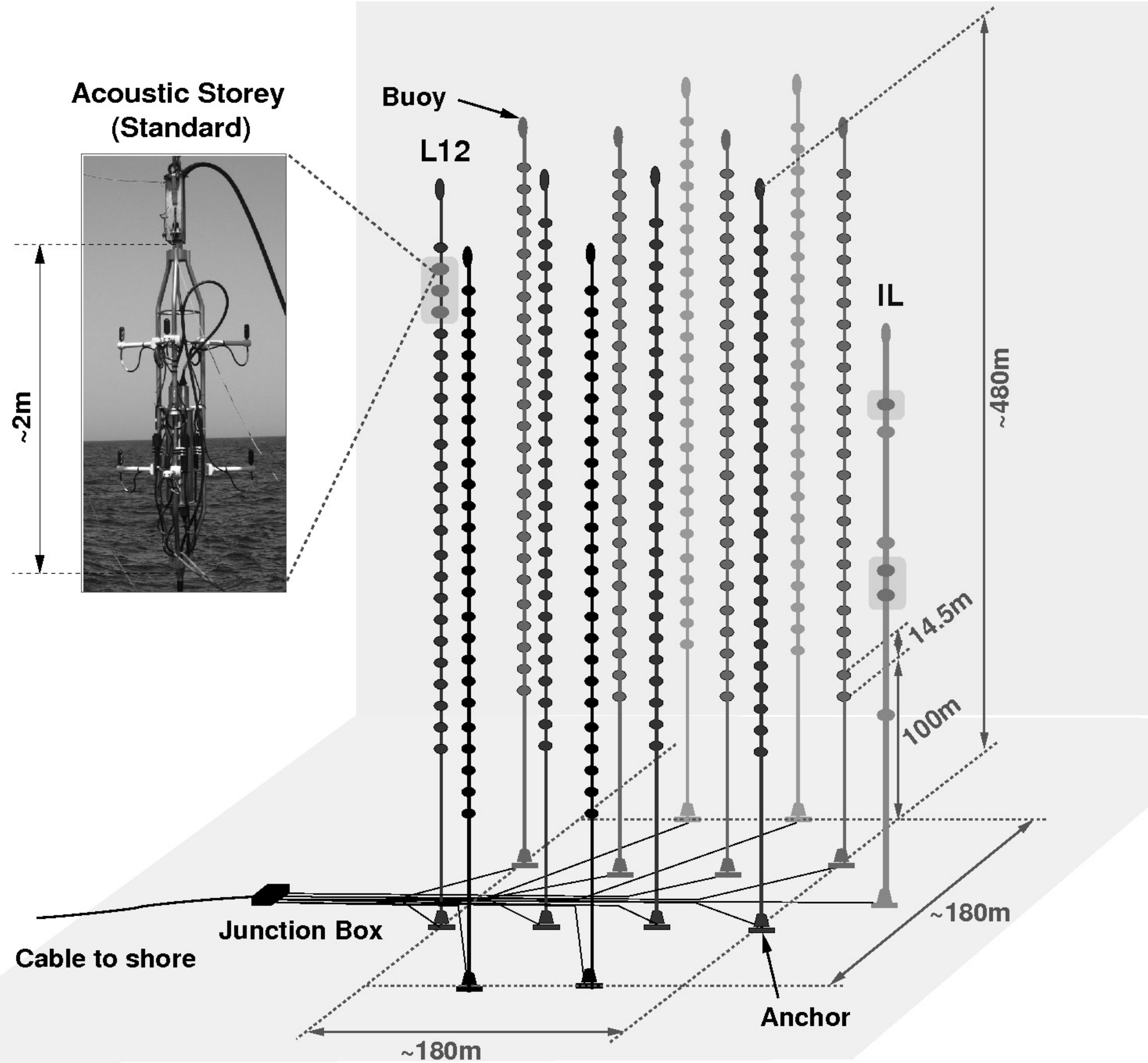}
\caption{A sketch of the ANTARES detector.  The six acoustic storeys
  are highlighted and a photograph of a storey in standard
  configuration is shown.  L12 and IL denote the 12th detection line
  and the Instrumentation Line, respectively.}
\label{fig:ANTARES_schematic_all_storeys}
\end{figure}

A 13th line, called {\em Instrumentation Line (IL)}, is equipped with
instruments for monitoring the environment. It holds six storeys.
For two pairs of consecutive storeys in the IL, the vertical distance
is increased to 80\,m.
Each line is fixed on the sea floor by an anchor equipped with
electronics and held taut by an immersed buoy.  An interlink cable
connects each line to the {\em Junction Box} from where the main
electro-optical cable provides the connection to the shore station.

\section{The AMADEUS System}
\label{sec:amadeus}
\label{sec:data_processing}
Within the AMADEUS system~\cite{bib:amadeus-2010}, acoustic sensing is
integrated in the form of {\em acoustic storeys} that are modified
versions of standard ANTARES storeys, in which the Optical Modules are
replaced by custom-designed acoustic sensors.  Dedicated electronics
is used for the amplification, digitisation and pre-processing of the
analogue signals.  Figure~\ref{fig:acou_storey_drawing} shows the
design of a standard acoustic storey with hydrophones.  Six acoustic
sensors per storey were implemented, arranged at distances of roughly
1\,m from each other.  The data are digitised with 16 bit resolution
and 250\,k samples per second.

\begin{figure}[ht]
\centering
\hspace*{-10mm}\includegraphics[width=6.0cm]{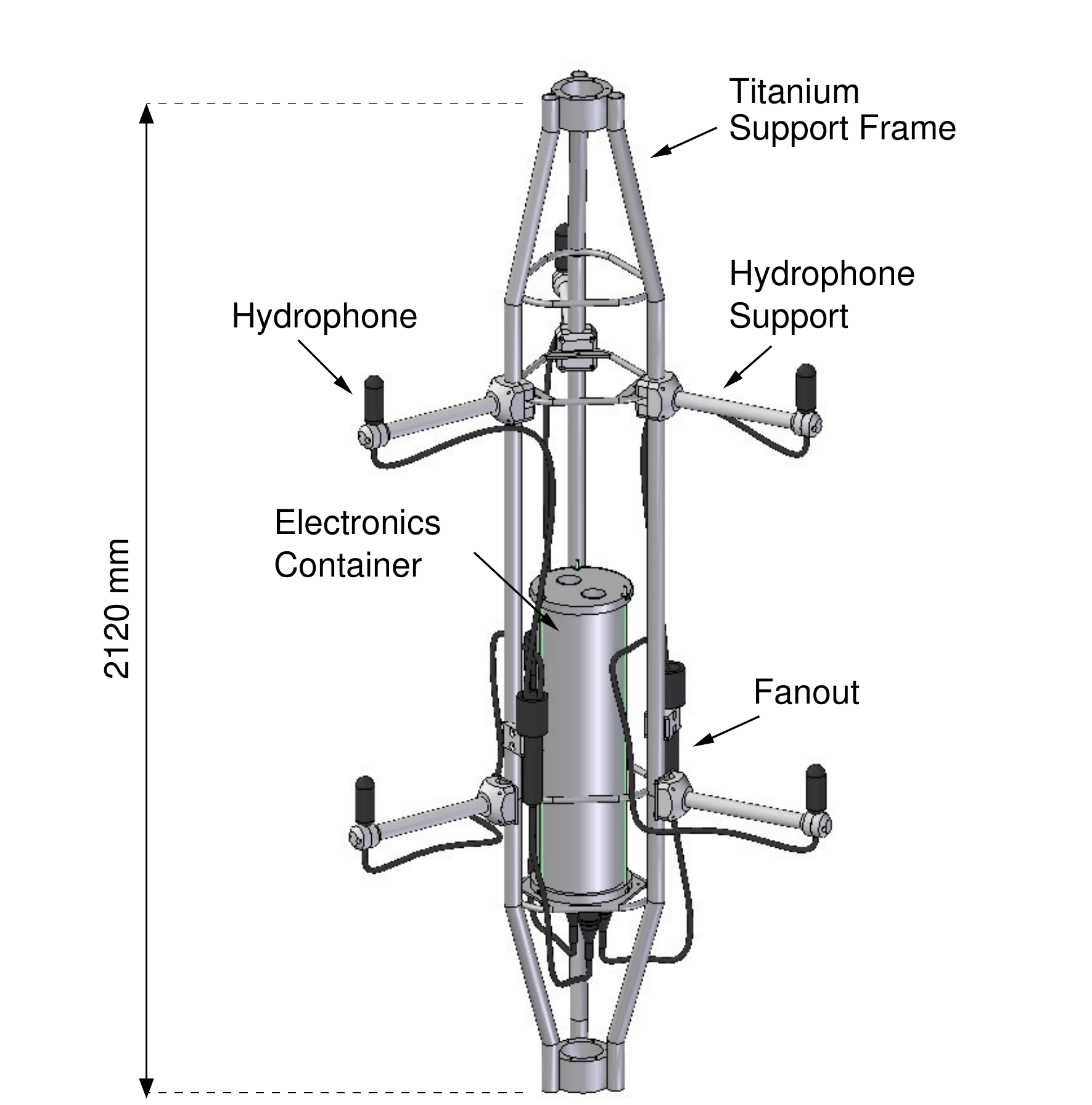}
\caption{
Drawing of a standard acoustic storey,  or acoustic cluster, 
with hydrophones.
\label{fig:acou_storey_drawing}
}
\end{figure}

The AMADEUS system comprises a total of six acoustic storeys: three on
the IL, which started data taking in December 2007, and three on the
12th detection line (Line 12), which was connected to shore in May
2008.  AMADEUS is now fully functional and routinely taking data. 

Two types of sensing devices are used in AMADEUS: hydrophones and {\em
  Acoustic Modules}~\cite{bib:amadeus-2010}. 
The acoustic sensors employ in both cases piezo-electric elements for
the broad-band recording of signals with frequencies ranging up to
125\,kHz. 
For the hydrophones, the piezo elements are coated in polyurethane,
whereas for the Acoustic Modules they are glued to the inside of
standard glass spheres which are normally used for Optical Modules.

The measurements presented in this article were done with the hydrophones.
Their calibration will be discussed in Sec.~\ref{sec:ambient-noise}.

The AMADEUS on-shore trigger\footnote{While this functionality might be more
  commonly denoted as filtering, it is ANTARES convention to refer
  to the ``on-shore trigger''.} searches the data by an adjustable
software filter; the events thus selected are stored to disk. This way
the raw data rate of about 1.5\,TB/day is reduced to about 10\,GB/day
for storage.
Currently, three trigger schemes are in 
operation~\cite{bib:amadeus-2010}:
A minimum bias trigger which records data continuously for about 10\,s
every 60\,min, a threshold trigger which is activated when the signal
exceeds a predefined amplitude, and a pulse shape recognition
trigger. For the latter, a cross-correlation of the signal with a
predefined bipolar signal, as expected for a neutrino-induced shower,
is performed. The trigger condition is met if the output of the
cross-correlation operation exceeds a predefined threshold.  
For the latter two triggers,
the thresholds are automatically 
adjusted to the prevailing
ambient noise and the condition must be met in at least four sensors of a 
storey.

\section{Ambient Noise}
\label{sec:ambient-noise}

Ambient noise, which can be described by
its characteristic power spectral density (PSD), is caused
by environmental processes and 
determines the minimum
pulse height that can be measured, if a given signal-to-noise ratio (SNR)
can be achieved with a search algorithm. 
To measure the ambient background at the ANTARES site, data from one
sensor on the IL07 taken from the beginning of 2008 until the end of
2010 were evaluated.
After quality cuts, 27905 minimum bias samples (79.9\% of the total
number recorded in that period) were remaining for evaluation, each
sample containing data continuously recorded over a time-span of
$\sim$10\,s.
For each of these samples, the noise PSD
(units of $\mathrm{V^2/Hz}$)  was integrated in the
frequency range $f = 10 - 50$\,kHz, 
yielding the square of the ambient noise
for that sample, as quantified by the output voltage of the hydrophone.  
Preliminary studies using the
shower parametrisation and algorithms from~\cite{bib:Sim_Acorne2}
indicate that this range optimises the SNR for
the expected neutrino signals.

The frequency of occurrence distribution of the resulting noise
values, relative to the mean noise over all samples, is shown in
Fig.~\ref{fig:noise_distr}.  Also shown is the corresponding
cumulative distribution.  For 95\% of the samples, the noise level is
below $2\ave{\sigma_\mathrm{noise}}$, demonstrating that the ambient
noise conditions are stable.

\begin{figure}[ht]
\centering
\includegraphics[width=8.0cm]{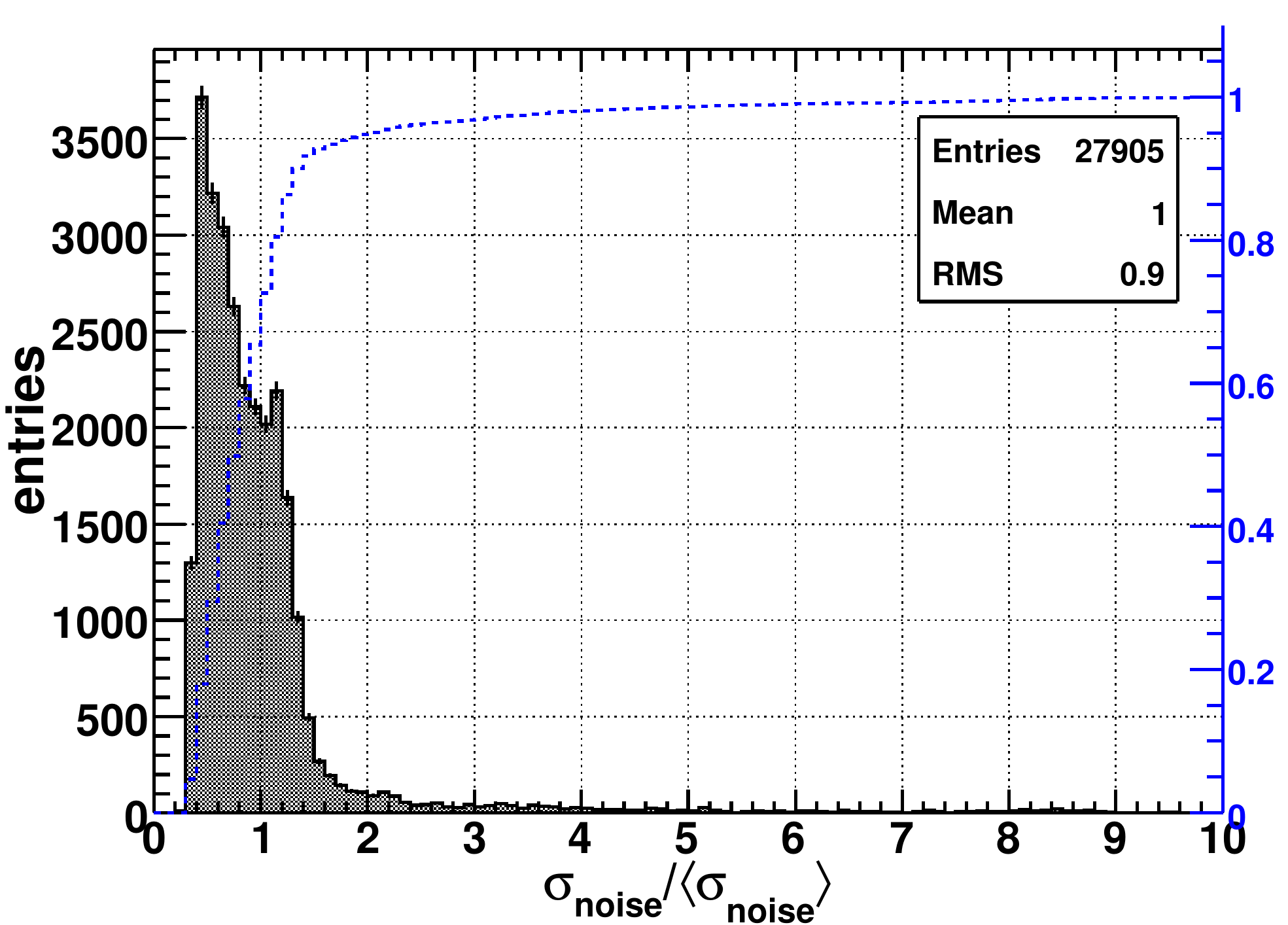}
\caption{
Frequency of occurrence distribution
for the ambient noise in the range $10 - 50$\,kHz, relative to the mean ambient 
noise   recorded over the 
complete period of three years that was used for the analysis
(left scale, filled histogram).
Also shown is the cumulative distribution, normalised to
the total number of entries of the distribution (right scale, dotted
line). 
}
\label{fig:noise_distr}
\end{figure}

All sensors have been calibrated in the laboratory prior to
deployment.  The absolute noise level can be estimated by assuming a
constant sensor sensitivity\footnote{ The ambient noise is originating
  mainly from the sea surface and hence displays a directivity which
  has to be folded with the variations of the sensitivity over the
  polar angle to obtain an effective average sensitivity.  For the
  results presented here, the noise has been assumed to be isotropic.
} of $-145\pm2$\,dB\,re\,1V/$\upmu$Pa.  With this value, the mean
noise level is $\ave{\sigma_\mathrm{noise}} = 10.1^{+3}_{-2}$\,mPa
with the median of the distribution at $8.1$\,mPa.

Currently, the detection threshold for bipolar signals corresponds to
a SNR of about 2 for an individual hydrophone.  For this SNR, the
median of the noise distribution corresponds to a signal amplitude of
$\sim$15\,mPa, equivalent to a neutrino energy of $\sim$1.5\,EeV at a
distance of 200\,m~\cite{bib:Sim_Acorne}.  By applying pattern
recognition methods that are more closely tuned to the expected
neutrino signal, this threshold is expected to be further reduced.

\section{Transient Sources}
Transient sources, e.g. from sea mammals, may create signals containing 
the characteristic bipolar pulse shape that is expected from 
neutrino-induced showers. 
Furthermore, 
the pulse shape recognition  trigger (see Sec.~\ref{sec:data_processing})
selects events with a wide range of shapes. Therefore,
a classification scheme is being developed that selects
neutrino-like events and suppresses background events with high efficiency.
After selecting 
neutrino candidates on the level of a storey, measurements from
multiple storeys can be combined to search for patterns that are
compatible with the characteristic ``pancake'' pressure field resulting
from a neutrino interaction.

\subsection{Source Position Reconstruction}
\label{sec:source_dir_reco}
The sensors within a cluster allow for efficient triggering of
transient signals and for direction reconstruction.  The combination
of the direction information from different acoustic storeys yields
the position of an acoustic source.
\begin{figure}[ht]
\centering
\includegraphics[width=9.0cm]{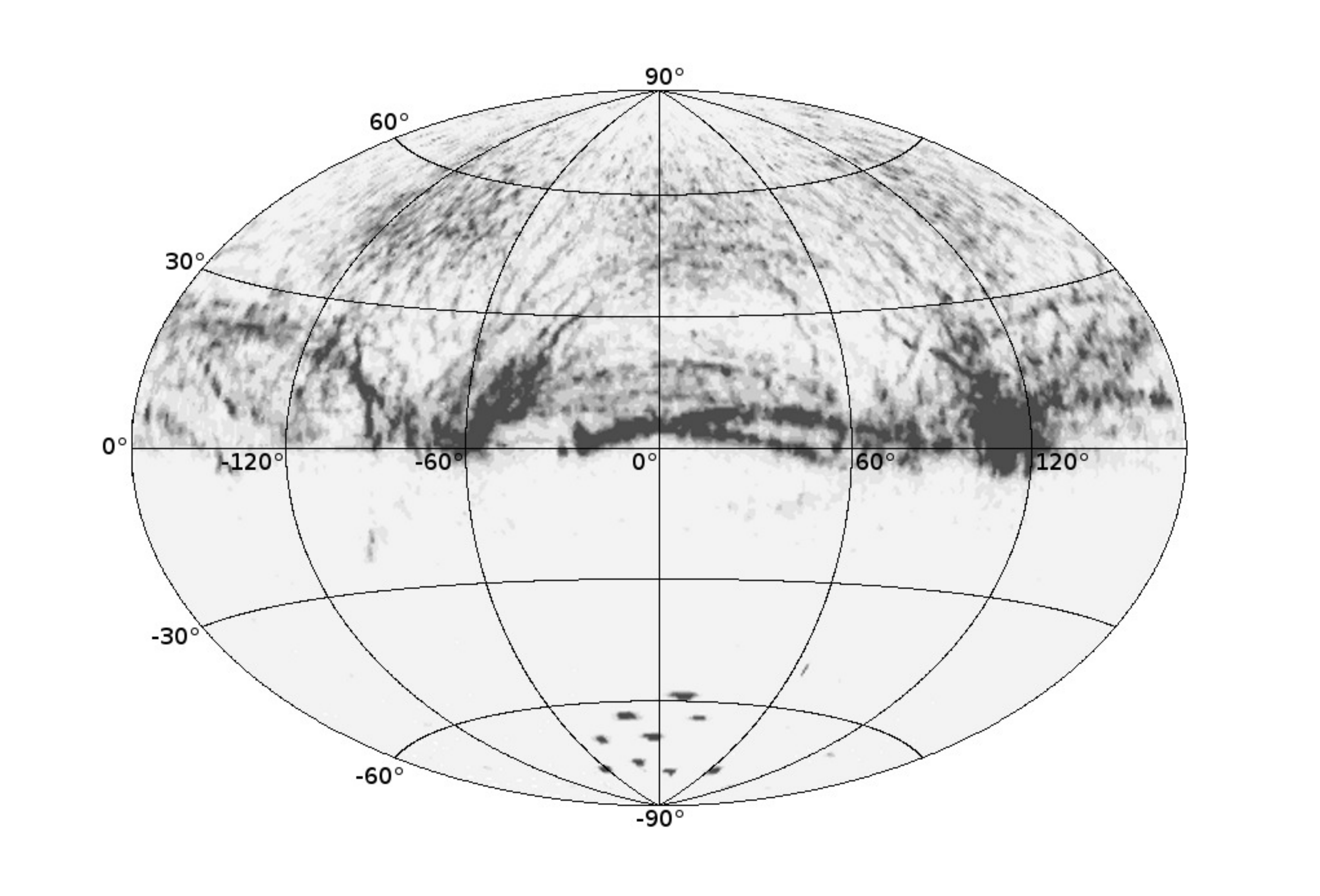}
\vspace{-7mm}
\caption{
Map of directions of sources as reconstructed with an acoustic storey on
Line 12. Zero degrees in azimuth correspond to the north direction, the 
polar angle of zero corresponds to the horizon of an observer on the
acoustic storey. At the bottom, the signals of the emitters
of the ANTARES positioning system are visible. 
}
\label{fig:skymap}
\end{figure}
Figure~\ref{fig:skymap} shows the reconstructed directions of all
sources that were triggered during a period of one month. 
The dark bands of increased acoustic activity 
can be associated with shipping routes and  points of
high activity with the directions of local sea ports. 
It is obvious from Fig.~\ref{fig:skymap} that a fiducial volume for the 
determination of the background rate of bipolar events must exclude the 
sea surface.

\subsection{Signal Classification}
The pulse shape recognition trigger described in 
Sec.~\ref{sec:data_processing} selects a wide range of events each of which 
can be allocated to one of four classes: 
Genuine bipolar events that are compatible
with signals expected from neutrinos (``neutrino-like events''), 
multipolar events, reflections of
signals from the acoustic emitters of the ANTARES positioning system and random 
events, where the latter class
contains all events that do not fit into any of the other classes.
For the classification, simulated signals representing the four
classes in equal proportions were produced and a set of features
extracted which 
are highly discriminant between the classes.
This feature vector is then
fed into a machine learning algorithm~\cite{neff:arena2010}.
Classification is performed for the signals from individual hydrophones.
Subsequently,
the results from individual hydrophones are combined to derive a 
classification for a given acoustic storey.
Several algorithms were investigated 
, the best of which yielded a failure rate (i.e. wrong decision
w.r.t. simulation truth) at the 1\%-level when applied to the two
signal classes ``neutrino-like'' and ``not neutrino-like''.

\section{Monte Carlo Simulations}
Monte Carlo simulations based on
~\cite{bib:Sim_Acorne,bib:Sim_Acorne2} are currently 
being implemented
for the AMADEUS detector setup.
Figure~\ref{fig:shower} shows 
the simulated density of the energy deposition of a
$10^{10}$\,GeV 
hadronic shower, projected into the $xz$-plane.
The $z$- and $x$-coordinates denote the
directions along the shower axis and a direction orthogonal to the shower axis,
respectively.
It is mostly the radial energy distribution within the shower which is
responsible for the shape and amplitude of the acoustic pulse that is
observed in the far field.
\begin{figure}[thb]
\centering
\includegraphics[width=8.5cm]{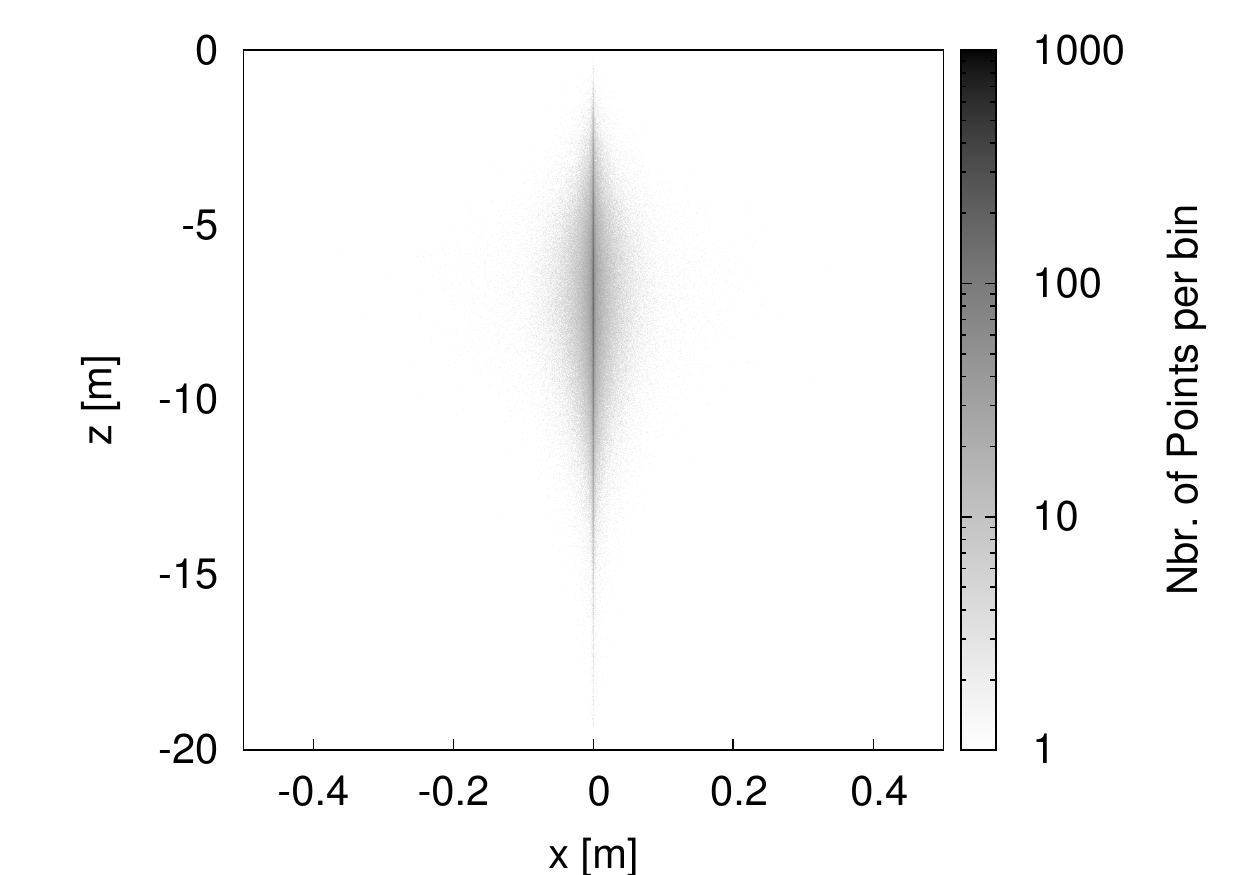}
\caption{ Density of the energy deposition of a $10^{10}$\,GeV
  hadronic shower resulting from a neutrino interaction, projected
  from a three-dimensional distribution upon the $xz$-plane.  Bin
  sizes are 0.01\,m in $x$ and 0.1\,m in $z$.  }
\label{fig:shower}
\end{figure}
The resulting pulse is shown in Fig.~\ref{fig:bip}. The corresponding
neutrino interaction was generated such that the centre of the
hadronic shower for a vertically downgoing neutrino lies within the
same horizontal plane as a storey denoted ``Storey 2'', at a distance
of 200\,m.
This way, the storey lies within the ``pancake'' of the pressure
field.  On a storey 14.5\,m below that storey, denoted ``Storey~1'',
no signal is observed.  This configuration corresponds to two adjacent
acoustic storeys on L12 or the two lowermost storeys on the IL07, see
Fig.~\ref{fig:ANTARES_schematic_all_storeys}.  This simulation
illustrates the characteristic three-dimensional pattern expected from
neutrino-generated pressure waves.

\begin{figure}[tbh]
\centering
\includegraphics[width=8.4cm]{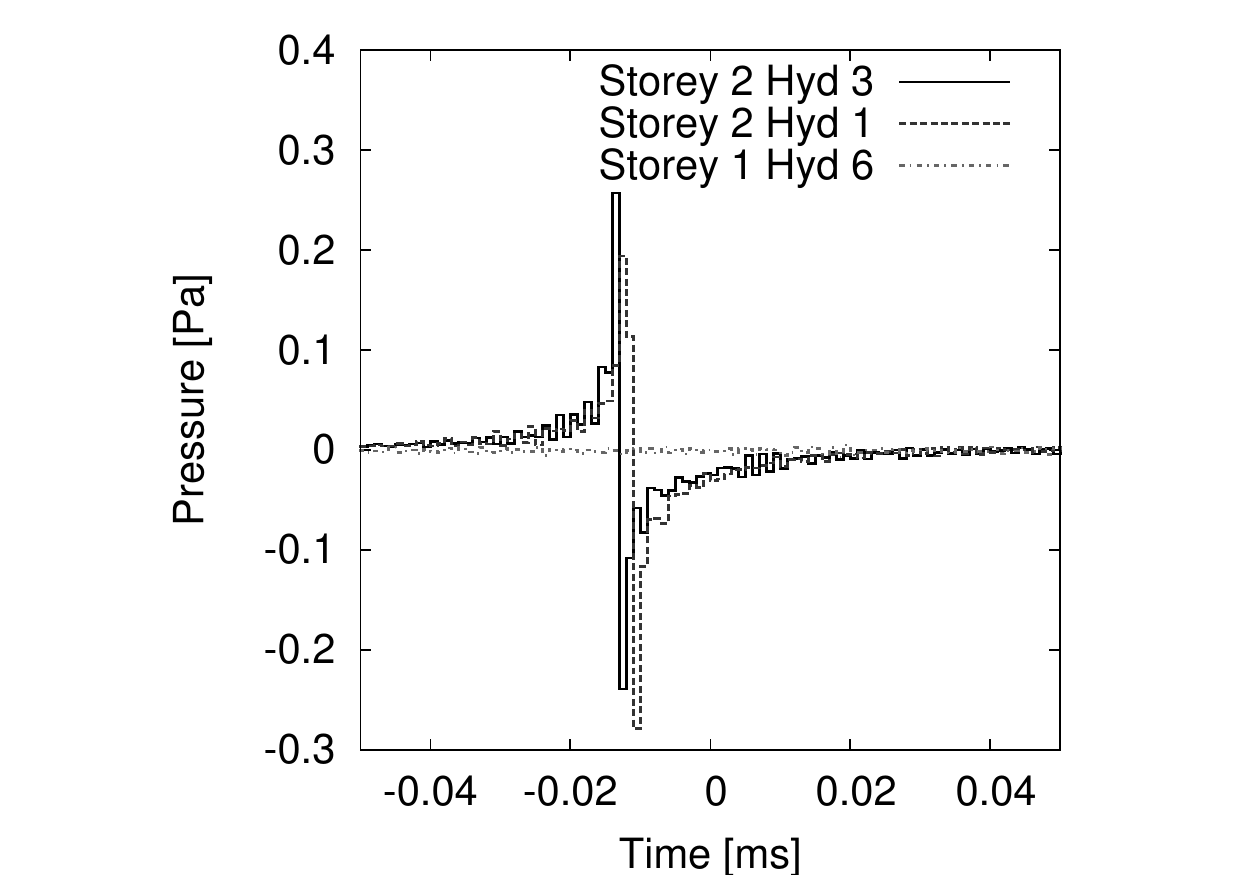}
\caption{Simulated acoustic signals as recorded with hydrophones in
  two acoustic storeys with a vertical spacing of 14.5\,m. See text
  for details. For Storey 2, signals from two different hydrophones are
  shown.  }
\label{fig:bip}
\end{figure}

\section{Summary and Conclusions}
Recent results from the acoustic neutrino detection test system AMADEUS, 
an integral part of the ANTARES detector in the Mediterranean Sea, 
have been presented.
Measurements of the ambient noise 
at the ANTARES site
show that the noise level is very stable and at the 
expected level, allowing for measurements of neutrino energies down to 
$\sim$1\,EeV.
The current focus of the analysis work is on the classification
of transient bipolar events to minimise the irreducible background for
neutrino searches.
In addition, 
Monte Carlo Simulations are under development.
AMADEUS is excellently 
suited 
to assess the background conditions for the measurement of 
bipolar pulses expected to originate from neutrino interactions. 

\section{Acknowledgements}
This study was supported by the German government through BMBF grants
5CN5WE1/7 and 05A08WE1.

\clearpage

\setcounter{figure}{0}
\setcounter{table}{0}
\setcounter{footnote}{0}
\setcounter{section}{0}
\newpage
}

\end{twocolumn}
\begin{onecolumn}
\section*{Acknowledgements}

The authors acknowledge the financial support of the funding agencies:
Centre National de la Recherche Scientifique (CNRS), Commissariat
\`{a} l'\'{e}nergie atomique et aux \'{e}nergies alternatives (CEA), Agence
National de la Recherche (ANR), Commission Europ\'{e}enne (FEDER fund
and Marie Curie Program), R\'{e}gion Alsace (contrat CPER), R\'{e}gion
Provence-Alpes-C\^{o}te d'Azur, D\'{e}parte-ment du Var and Ville de
La Seyne-sur-Mer, France; Bundesministerium f\"{u}r Bildung und
Forschung (BMBF), Germany; Istituto Nazionale di Fisica Nucleare
(INFN), Italy; Stichting voor Fundamenteel Onderzoek der Materie
(FOM), Nederlandse organisatie voor Wetenschappelijk Onderzoek (NWO),
the Netherlands; Council of the President of the Russian Federation
for young scientists and leading scientific schools supporting grants,
Russia; National Authority for Scientific Research (ANCS), Romania;
Ministerio de Ciencia e Innovaci\'{o}n (MICINN), Prometeo of Generalitat
Valenciana (GVA) and MultiDark, Spain. We also acknowledge the
technical support of Ifremer, AIM and Foselev Marine for the sea
operation and the CC-IN2P3 for the computing facilities.
\end{onecolumn}


\begin{thebibliography}{}
\bibitem{ICRC11_PointSources} C. Bogazzi for the ANTARES Collaboration, Searching for point sources of high energy cosmic neutrinos
with the ANTARES telescope, these proceedings, 2011
\end{thebibliography}
\end{document}